\documentclass[a4paper, 12pt, twoside]{report}

\pdfoutput=1
\usepackage{graphicx}
\usepackage{amssymb}
\usepackage{amsmath}

\usepackage{tikz}
\usetikzlibrary{shapes,arrows}

\usepackage[toc,page]{appendix}

\usepackage{multirow}

\usepackage{fancyhdr}
\usepackage{braket}

\usepackage{bm}

\newcommand{\rf}[1]{(\ref{#1})}
\newcommand{\beq}{\begin{equation}}
\newcommand{\beql}[1]{\beq\label{#1}}
\newcommand{\eeq}{\end{equation}}
\newcommand{\bea}{\begin{eqnarray}}
\newcommand{\eea}{\end{eqnarray}}

\newcommand{\e}{\mbox{e}}

\newcommand{\lam}{\lambda}

\renewcommand{\a}{\alpha}
\newcommand{\n}{\nu}
\newcommand{\m}{\mu}

\newcommand{\eps}{\epsilon}

\newcommand{\Del}{\Delta}

\renewcommand{\k}{\kappa}

\newcommand{\oh}{\frac{1}{2}}

\newcommand{\tr}{\mathrm{tr}\,}
\newcommand{\ra}{\rangle}
\newcommand{\la}{\langle}

\newcommand{\mi}{\!-\!}

\newcommand{\plu}{\!+\!}

\newcommand{\cM}{{\cal M}}

\newcommand{\cT}{{\cal T}}
\newcommand{\cTT}{{\cal T}^{(3)}}

\newcommand{\cN}{{\cal N}}

\newcommand{\cR}{{\cal R}}

\def\a{\kern+.6ex\lower.42ex\hbox{$\scriptstyle \iota$}\kern-1.20ex a} 
\def\e{\kern+.5ex\lower.42ex\hbox{$\scriptstyle \iota$}\kern-1.10ex e}

\begin{document}

\begin{titlepage}

\font\bigfat=ecbx12 scaled\magstep5 \pagestyle{empty}
\enlargethispage{2cm} \centering \vspace*{0.01cm} {\large Jagiellonian 
University\\ Faculty of Physics, Astronomy and Applied Computer Science\\ } \vspace*{0.1cm}

\begin{figure}[h]
    \begin{center}
  \scalebox{0.25}{\includegraphics{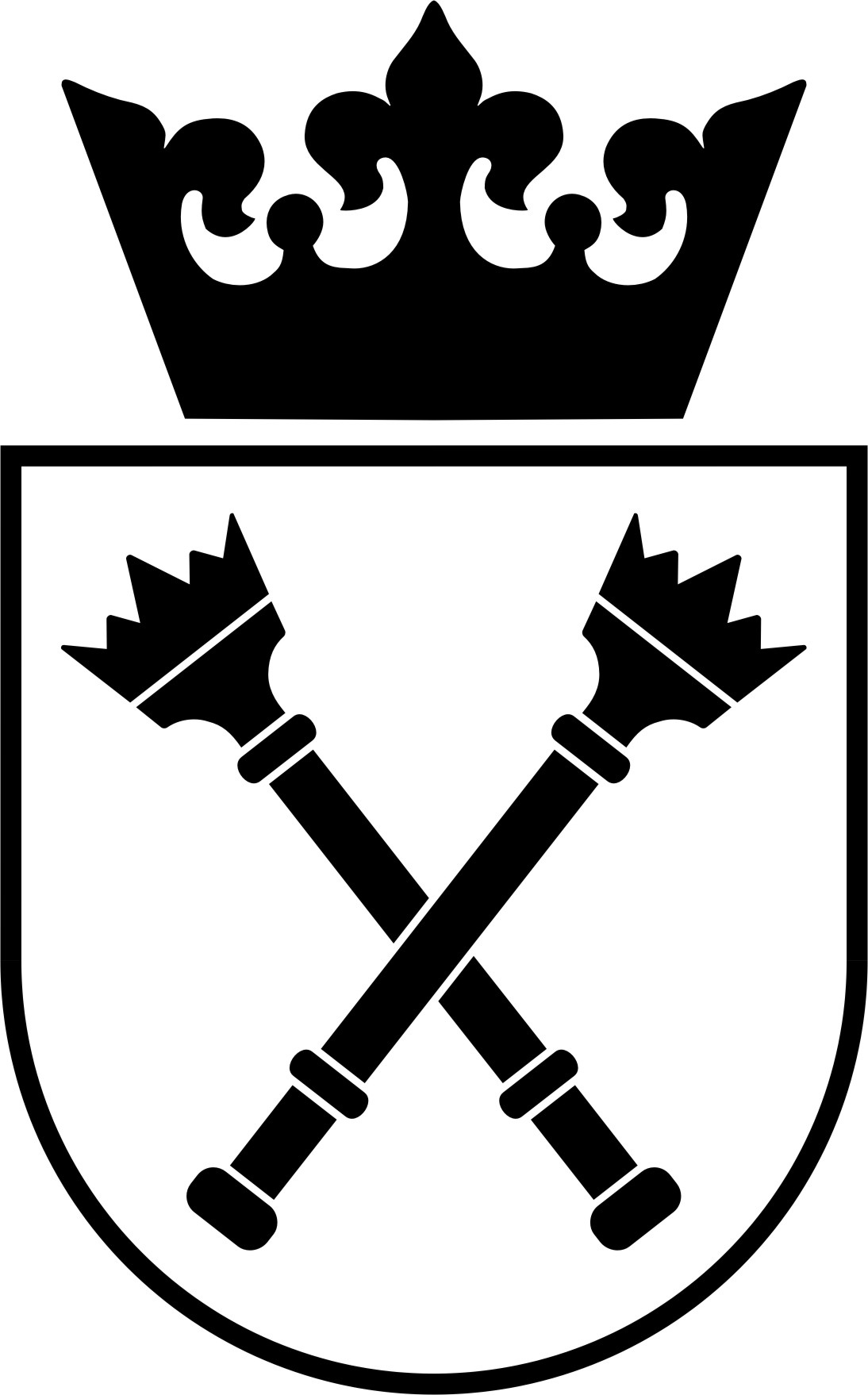}}
    \end{center}
\end{figure}
\vspace{2.0cm}
{\Large Jakub Gizbert-Studnicki }\\
\vspace*{1cm}
\vspace*{0.01cm} \Large {\bf  The effective action in four-dimensional CDT}\\ \vspace{3cm}
{\textit{\LARGE{}}}\\
\vspace*{2.5cm}
\vspace*{-4cm}

{\large A thesis submitted for the degree of \\
Doctor of Philosophy in Physics at \\
Jagiellonian University\\
\vspace*{0.3cm}
written under the supervision of\\ 
\vspace*{0.3cm}
prof. dr hab. Jerzy Jurkiewicz \\
and dr Andrzej G\"orlich
}\\
\vspace*{4cm}
\vspace{5ex} {\large Krak\'{o}w, 2014 }

\end{titlepage}

\newpage
\thispagestyle{empty}
\mbox{}


\newpage
\thispagestyle{empty}
\vspace{2cm}

\noindent Wydzia\l{} Fizyki, Astronomii i Informatyki Stosowanej

\noindent Uniwersytet Jagiello\'nski
\newline

\begin{center}
\bf \large O\'swiadczenie
\end{center}
\vspace{1cm}

Ja \ ni\.zej \ podpisany \ {\it Jakub \ Gizbert-Studnicki} \ (nr indeksu: 1001910)
 doktorant  Wydzia\l{}u Fizyki,  Astronomii  i  Informatyki Stosowanej  Uniwersytetu  Jagiello\'nskiego
o\'swiadczam, \.ze przed\l{}o\.zona 
przeze mnie rozprawa doktorska 
pt. {\it ``The effective action in four-dimensional CDT"}
jest oryginalna i przedstawia wyniki bada\'n wykonanych przeze mnie osobi\'scie,
pod kierunkiem {\it prof. dr. hab. Jerzego Jurkiewicza} \  i \  {\it dr. Andrzeja  G\"orlicha}. Prac\e  \ napisa\l{}em samodzielnie.
\newline

O\'swiadczam, \.ze moja rozprawa doktorska zosta\l{}a opracowana zgodnie z Ustaw\a \
o prawie autorskim i prawach pokrewnych z dnia 4 lutego 1994 r. (Dziennik Ustaw
1994 nr 24 poz. 83 wraz z p\'o\'zniejszymi zmianami).
\newline

Jestem \'swiadom, \.ze niezgodno\'s\'c niniejszego o\'swiadczenia z prawd\a \ ujaw-niona
w dowolnym czasie, niezale\.znie od skutk\'ow prawnych wynikaj\a cych z ww. ustawy,
mo\.ze spowodowa\'c uniewa\.znienie stopnia nabytego na podstawie tej rozprawy.
\newline
\vspace{1cm}

\noindent Krak\'ow, dnia 9.07.2014
\newline

\begin{flushright}
..............................

podpis doktoranta
\end{flushright}

\newpage
\thispagestyle{empty}
\mbox{}

\newpage
\vspace*{5.0cm}
\setcounter{page}{1}
\begin{center}
\section*{Streszczenie}
\end{center}

\noindent W niniejszej pracy przedstawiono najnowsze wyniki pomiaru  i analizy dzia\-\l{}ania efektywnego w czterowymiarowym modelu Kauzalnych Dyna\-micznych Triangulacji (CDT). Dzia\l{}anie efektywne opisuje kwantowe fluktuacje obj\e\-to\'sci przestrzennej wszech\'swiata (\'sci\'sle zwi\a zane z fluktuacjami czynnika skali) obserwowane
 po ``wyca\l{}kowaniu" innych stopni swobody. Do po\-miaru i parametryzacji dzia\l{}ania efektywnego w fazie ``de Sittera" (zwanej r\'ownie\.z faz\a \,``C") u\.zyto metody opartej na analizie macierzy kowariancji obj\e to\'sci przestrzennej w poszczeg\'olnych li\'sciach foliacji zdefiniowanej przez globalny czas w\l{}asny. Pokazano, \.ze mierzone dzia\l{}anie efektywne jest zgodne z~prost\a \ dyskretyzacj\a \ dzia\l{}ania minisuperspace (z odwr\'oconym znakiem). Przeana\-lizowano r\'ownie\.z mo\.zliwe   poprawki do  dzia\l{}ania efektywnego oraz przedsta\-wiono spos\'ob pomiaru i parametryzacji dzia\l{}ania \l{}\a cz\a cego  warstwy prze\-strzenne ca\l{}kowitego i po\l{}\'owkowego dyskretnego czasu w\l{}asnego. 
 W pracy wprowadzono tak\.ze   now\a \ metod\e \ pomiaru dzia\l{}ania efektywnego  opart\a \ na macierzy transferu. Wykazano, \.ze w fazie ``de Sittera" wyniki nowej metody s\a \ w pe\l{}ni zgodne z poprzednio u\.zywan\a \ metod\a \ opart\a \ na macierzy kowa\-riancji. Now\a \ metod\e \ pomiaru wykorzystano do zbadania dzia\l{}ania efektywnego w obszarze ma\l{}ych obj\e to\'sci przestrzennych oraz do opisu kwantowych fluktuacji obj\e to\'sci mierzonych w odpowiadaj\a cym im obszarze trian\-gulacji. Dokonano  r\'ownie\.z pomiaru i parametryzacji dzia\l{}ania efektywnego w pozosta\l{}ych fazach (``A" i ``B") czterowymiarowego modelu CDT oraz przeprowadzono analiz\e \ przej\'s\'c fazowych. Uzyskane wyniki wskazuj\a  \ na obecno\'s\'c nowej, wcze\'sniej nieodkrytej fazy ``bifurkacji", oddzielaj\a cej ``tradycyjne" fazy ``B" i ``C". Dokonano analizy w\l{}a\'sciwo\'sci geometrii w nowej   fazie oraz wyznaczono nowy diagram fazowy.

\newpage
\thispagestyle{empty}
\mbox{}
\newpage
\vspace*{5.0cm}

\begin{center}
\section*{Abstract}
\end{center}

\noindent In this dissertation we present recent results concerning the measurement and analysis of the effective action in four-dimensional Causal Dynamical Triangulations. The action describes quantum fluctuations of the spatial volume of the CDT universe (or alternatively the scale factor) after integrating out other degrees of freedom. We use the covariance of volume fluctuations to measure and parametrize the effective action inside the ``de Sitter" phase, also called the   ``C" phase. We show that the  action is consistent with a simple discretization  of the minisuperspace action (with a reversed overall sign). We discuss possible  subleading corrections and show how to construct a more complicated effective action comprising both integer and half-integer discrete proper time  layers. We introduce a new method of the effective action measurement based on the  transfer matrix. We show that the results of the new method are fully consistent with the covariance matrix method inside the ``de Sitter" phase. We use the new method to measure the effective action in the small volume range and to explain the  behaviour of the ``stalk" part of the CDT triangulations. Finally we use the transfer matrix method to measure and parametrize the effective action inside the ``A" and ``B" phases, and to analyze the phase transitions. The results lead to an unexpected discovery of a new ``bifurcation" phase separating the ``old" ``C" and ``B" phases. We analyze  geometric properties of triangulations inside the new phase and draw a new phase diagram.

\newpage
\thispagestyle{empty}
\mbox{}
\newpage

\newpage
\vspace*{5.0cm}
\begin{center}
\section*{Acknowledgements}
\end{center}
I am very grateful to  Prof. Jerzy Jurkiewicz and Dr Andrzej G\"orlich for 	introducing me into a fascinating world of Quantum Gravity described by the model of Causal Dynamical Triangulations. Their scientific inspiration and support led me to  results presented in this thesis. I would also like to thank all my collaborators, especially  Prof. Jan Ambj\o rn and  Prof. Renate Loll. I also acknowledge the collaboration and many interesting discussions with   MSc Tomasz Trze\'sniewki and Dr Jakub Mielczarek. Last but not least I would like to thank Dr Daniel Coumbe for many fruitful comments and a careful   proofreading.
Finally I wish to acknowledge  financial support by the  Polish National
Science Centre (NCN)  via the grant
DEC-2012/05/N/ST2/02698.

\newpage
\thispagestyle{empty}
\mbox{}
\newpage

\newpage

\tableofcontents
\fancyhf{}
\fancyhead[RE]{\bf Contents}
\fancyhead[LE,RO]{\thepage}

\chapter*{Introduction}
\addcontentsline{toc}{chapter}{Introduction}

\fancyhf{}
\fancyhead[RE]{\bf Introduction}
\fancyhead[LE,RO]{\thepage}

Almost one hundred years have past since the foundation of two major theories of  twentieth century physics. In 1915, Albert Einstein and David Hilbert published gravitational field equations \cite{Einstein1915,Hilbert1915} describing a geometric theory of gravity, known as  {\it General Relativity}.  General Relativity  successfully explained many observed phenomena (e.g. gravitational time dilatation,  gravitational lensing, gravitational redshift and the expansion of the Universe) and paved the way for the development of  modern astrophysics and cosmo\-logy.
At the same time the work of  Max Planck, Albert Einstein, Arnold Sommerfeld, Niels Bohr, William Wilson, Otto Stern, Walther Gerlach, Max Born, Werner Heisenberg, Wolfgang Pauli, Louis de Broglie, Erwin Schr\"o\-dinger,  Paul Dirac and others led to  the formulation of  {\it Quantum Mechanics} in the mid 1920's \cite{QMbeg}. Quantum Mechanics describes the nature of  (sub)atomic  scale physics and became the basis of  atomic, nuclear and condensed matter physics. Further attempts to merge quantum mechanics with special relativity and to explain the  creation and annihilation of particles finally led to the development of   {\it Quantum Field Theory}. This theory had  enormous success in explaining  fundamental particle physics and in particular  led to a formulation of the Standard Model, whose final confirmation came in 2012 after the  discovery of the Higgs boson at the Large Hadron Collider \cite{LHC,LHCATL,LHCCMS}.

Despite many successes of both theories, there are still many open questions. Just to mention few of them: What are the quantum origins of space and time?  What is the microstructure of space-time, and  can we use it to explain the macroscopic gravitational interactions and the large-scale structure of the Universe? Are space, time and causality fundamental or emergent concepts? Is it possible to unify gravity with the other three fundamental forces? To answer these (and other) questions   theoretical physicists struggled  for over half a century to merge Quantum Mechanics and General Relativity into a single theory of {\it Quantum Gravity}.

The problem of combining Quantum Mechanics and gravity becomes an issue  at very high energies or equivalently very short distances. In this {\it ultraviolet} limit  new degrees of freedom or symmetries may potentially occur. The issue lies in extremely small length scales at which  quantum gravitational effects can play an important role  and there is presently no confirmed experimental evidence of them.\footnote{The Planck length $\sqrt{\hbar G / c^3}\approx1.6\times 10^{-35}$ m is roughly 20 orders of magnitude shorter than length scales available in current high energy physics experiments. The most promising  tests for Quantum Gravity concern imprints of quantum gravitational effects in the Cosmic Microwave Background (in particular its polarization) \cite{BICEP2,BICEP2a}.}  As a result all candidate theories are at the moment mostly theoretical considerations.

The  challenge to formulate a {\it predictive} theory of Quantum Gravity using  traditional field theory techniques is not an easy task. The main issue lies in the fact that  Einstein's theory of General Relativity 
is  {\it perturbatively nonrenormalizable}. When describing graviton interactions in pure gravity (without matter fields)  Feynman diagrams with two (or more) loops lead to ultraviolet divergences  that cannot be removed by a finite number of counter terms \cite{Nonren1,Nonren2}.\footnote{This is due to the fact that the Newton's constant $G$, which plays a role of the coupling constant of gravity, has in space-time dimension four a dimension $[G] = -2$ in  mass  units (where  $\hbar=c= 1$). Inclusion of matter fields makes the situation even worse as Feynman diagrams diverge already on the one-loop level.} 
As a result the infinite number of parameters need to be fixed to describe (perspective) experimental  results at high energy scales and such a theory  looses all predictive power. Therefore, any perturbative expansion can  be  treated only as an effective theory applicable for energies $E^2 \ll 1/G$ (in units where $\hbar = c =1$) \cite{EinstEff}.

Perturbative nonrenormalizability is an important drawback, however  there is still a chance that gravity can be renormalizable in a  {\it non-perturbative} way. The idea was  put forward by S. Weinberg in his {\it asymptotic safety} scenario \cite{Weinberg}. The proposal is formulated in a framework of the Wilsonian renormalization group and stipulates that the coupling constants of  Quantum Gravity flow (in  abstract coupling space) toward a nontrivial ultraviolet fixed point. In the vicinity of this point the co-dimension of the critical surface  is finite and only  few {\it relevant couplings} are needed to  describe physics in that region. As a result the theory regains its predictive power although the perturbative expansion is not reliable as the values of the relevant dimensionless couplings are not necessarily  small. Recent renormalization group studies for different Quantum Gravity models  support the asym\-ptotic safety conjecture \cite{RenGr2,RenGr4}. The solutions of the renormalization group equations flow toward the ultraviolet fixed point characterized by a three-dimensional critical surface, even in models with more than three independent coupling constants \cite{RenGr3,RenGr5,RenGr6}. This suggests that the dimension of  the critical surface is finite.

There are many non-perturbative approaches to Quantum Gravity. An important example is  Loop Quantum Gravity \cite{LQG1,LQG2,LQG3} which implements  Dirac's
procedure of canonical quantization to General Relativity. As a result the holonomies of connections  become quantum objects. In the simplest case of spatially homogenous and isotropic geometries, the theory can be reduced to the model of Loop Quantum Cosmology \cite{LQC} which describes quantum evolution of the scale factor using the effective Hamiltonian.

In this dissertation we will focus on another group of non-perturbative approaches, namely    {\it Lattice Quantum Gravity} models. Lattice field theory gained a rising interest  in the mid 1980's when the increasing power of computers made numerical calculations feasible. An example  is  lattice QCD, which provides solutions for low-energy problems of strong interactions not tractable by means of  analytic or perturbative methods \cite{QCD3, QCD4}.
This approach uses a discrete grid of space-time points  to approximate path integrals of the continuum theory. The finite lattice spacing $a$ provides a high momentum cut-off of the order $1/a$ making the formulation  well defined. At the same time it is possible to approach the continuum limit by taking the lattice spacing $a\to0$.

{Lattice Quantum Gravity} is formulated in the same spirit, however there is one important difference: conventional Quantum Field Theories assume the existence of a fixed Minkowski\footnote{It is also possible to formulate quantum field theories on curved (non-Minkowskian) space-times \cite{QFTCurved1,QFTCurved2}.} background  while in General Relativity   space-time geometry itself is a degree of freedom. This feature should be incorporated into lattice models. Consequently, a fixed regular grid is replaced by a dynamical lattice which can evolve in a quantum sense. In this approach the  gravitational path integral is defined by a sum over different microscopic {\it space-time histories} weighted by a quantum amplitude defined by (the exponent of) the gravitational action. As a result the underlying theory becomes {\it background independent} and the requirement of the correspondence principle to recover classical physics in a regime of large quantum numbers becomes a crucial task. For Quantum Gravity it translates into the need to reproduce the solutions of  Einstein's General Relativity in the low energy (or equivalently large distance) limit. Among many different approaches of this kind\footnote{For example: Causal Sets \cite{CausalSets}, Spin Foam models \cite{SpinFoams}, Regge calculus in lattices with variable edge lengths  \cite{Hamber1,Hamber2,Williams}.} 
the one that has been most successful at addressing the low energy problem is the model of {\it Causal Dynamical Triangulations} (CDT).

In Causal Dynamical Triangulations  $D$-dimensional (pseudo-)Riemannian  manifolds are approximated by  lattices (called {\it triangulations}) constructed from  $D$-dimensional  simplices with fixed edge lengths. The interior of each simplex is  isomorphic to (a subset of)  Minkowski space-time and the geometry is  encoded in the way the building blocks are "glued" together. The continuous path integral of the theory is approximated by a sum over such simplicial manifolds weighted by (the exponent of) the  Einstein-Hilbert action.

This idea originates from earlier methods of  (Euclidean) Dynamical Triangulations (EDT), which assume Euclidean (local SO(4) symmetry) instead of Lorentzian (local SO(3;1) symmetry) space-time geometry. In EDT the time direction is not distinguished and all simplices are equilateral and identical. According to  Regge calculus \cite{regge} the curvature of  a simplicial manifold is defined through a deficit angle located at  $(D-2)$ subsimplices and depends on  the number of $D$-simplices sharing a given subsimplex. As a result the gravitational action  takes a very simple form of a linear combination of the total number of $D$-simplices and $(D-2)$ sub-simplices with two coupling constants  related to the Newton's constant and the cosmological constant. The model turned out to be analytically solvable in $D = 2$  dimensions both for pure gravity \cite{ET2d1,ET2d2,ET2d3,ET2d4} and gravity coupled to simple matter systems \cite{ET2Dmatter1,ET2Dmatter2,ET2Dmatter3,ET2Dmatter4,ET2Dmatter5,ET2Dmatter6}, and tractable by numerical methods in three \cite{ET3D1,ET3D2,ET3D3,ET3D4,ET3D5} and four dimensions \cite{ET4D1,ET4D2,ET4D3,ET4D5,ET4D6}.  Computer simulations showed that in $D=4$ there are two phases, none of which can be associated with the four-dimensional semiclassical General Relativity. Detailed studies revealed that the phases are separated by a first-order phase transition \cite{ET1stOrder1,ET1stOrder2} making it impossible to define a suitable  continuum limit.\footnote{In a lattice formulation of an asymptotically
safe field theory, the fixed point would appear as a second-order critical point, the
approach to which would define a continuum limit. The divergent correlation length
characteristic of a second-order phase transition would allow one to take the lattice
spacing to zero while keeping observable quantities fixed in physical units.} The attempts to revive  the theory by introducing a third coupling constant which could potentially enrich the phase structure were also not successful  \cite{EDTRec1,EDTRec2,EDTRec3}. The  issue most likely lies in the definition of  EDT which assumes Wick-rotated (Euclidean) space-time  geometry from the outset. This leads to problems with the  conformal mode of the theory which may encounter very large (potentially unbounded) fluctuations \cite{ConformalModeProblem}.  

To overcome this problem the  model of Causal Dynamical Triangulations was proposed by J. Ambj\o rn, J. Jurkiewicz and R. Loll in the late 1990's \cite{CDT1998}. CDT starts with the Lorentzian setup and distinguishes between space-like and time-like links. As a result it is possible to impose the causality constraint on the set of triangulations over which the path integral is calculated. In CDT each space-time history has a well defined causal structure which restricts triangulations to only such configurations of  simplices whose $D-1$ dimensional space-like faces form a sequence of spatial  hyper-surfaces ({slices}) of fixed topology. This is a simplicial equivalent of  {\it globally hyperbolic} manifolds with spatial slices playing a role of Cauchy surfaces of equal (discrete) proper time. The distinction between ``space"  and ``time" together with causality constraints  require that in $D=4$ dimensions four different types of  building blocks (simplices) are necessary. It also introduces a new degree of freedom to the theory in a form of the third coupling constant which depends on a possible asymmetry between the edge lengths in the spatial and time directions. As a result the structure of the theory is different and much richer than that of  EDT. 

The model of Causal Dynamical Triangulations for pure gravity can be solved analytically in $D=2$ dimensions \cite{CDT1998,CDT2DAnal,CDT2DAnal2,CDT2DAnal3} and some analytical results can be obtained in $D=3$ dimensions \cite{CDT3DAnal1,CDT3DAnal2,CDT3DAnal3}. Inclusion of matter fields or higher dimensions require numerical methods \cite{CDT2DAnal2,CDT3DNum,CDTRegge,CDT4D2004, CDT2Dscalar}.  Numerical simulations in $D=4$ dimensions showed that the pure  gravity model  can be in one of three different phases \cite{CDTphases1,CDTphases2}.  There is evidence that two of these phases are separated by a second (or higher) order phase transition \cite{phaseTrans1,phaseTrans2} which in principle could allow one to define a continuous ultraviolet limit of the theory where the lattice spacing $a \to 0$. The model also provides a well behaved  infrared limit (inside the so called {\it ``de Sitter"} phase) in which fluctuations of geometry occur around a dynamically generated semiclassical background \cite{deSitter1,deSitter2,deSitter3,deSitter5}. The emergent average geometry  is consistent with the (Wick-rotated) de Sitter solution of the Einstein's gravitational equations.  Quantum fluctuations of the the scale factor around this semiclassical solution are well described by the effective action being a simple discretization of the minisuperspace action proposed by S.W. Hawking and J.B. Hartle \cite{HawkingHartle}.

In this dissertation we focus on the effective action of four-dimensional Causal Dynamical Triangulations. We present recent results of studies concerning the form of the action inside the ``de Sitter" phase and discuss possible corrections to a simple discretization of the minisuperspace action. We introduce a new method of analysis based on the effective transfer matrix  parametrized by the spatial volume of Cauchy surfaces of constant (discrete) time which can be measured in numerical simulations. The method provides a very useful tool for precise measurement of the effective action both in a large and a small volume limit. We use this concept to measure the effective action in all phases of CDT and to analyze phase transitions. The results lead to an unexpected discovery of a new phase characterized by a bifurcation of the  kinetic term. We analyze geometric properties of generic triangulations inside the new phase and  study new phase diagram.
\newline

This thesis is organized as follows. In Chapter 1 we provide a detailed introduction to concepts and methods of four-dimensional Causal Dynamical Triangulations. In Chapter 2 we  briefly  summarize  the state of the art of the theory and  the author's contribution to the field. In Chapter 3 we analyze the effective action in  the ``de Sitter" phase (also called the ``C" phase) using the semiclassical effective propagator  approach. We start with the action parametrized by  spatial volumes of   Cauchy surfaces 
of integer (discrete) time and show that it is consistent with a discretized minisuperspace action. We discuss possible corrections to the minisuperspace action and show how the volume contribution from  half-integer time spatial layers can be  implemented to the effective  action. 
In Chapter 4 we introduce a new method of the effective action measurement based on a transfer matrix approach. We show that the results of the new method not only agree with the previous ones in the large volume limit, but also enable a detailed analysis of a small-medium volume range, where discretization effects are large. We study self-consistency of the transfer matrix method and argue that one can reconstruct the CDT results by using a simplified model based on the measured transfer matrix alone. In Chapter 5 we use the  transfer matrix approach   to measure and parametrize the effective action in phases ``A" and ``B",  and to analyze phase transitions. We show that the transfer matrix data suggest the existence of a new, previously undiscovered ``bifurcation" phase separating the ``B" and ``C" phases. In Chapter 6 we study the properties of the new ``bifurcation" phase and show preliminary results concerning  the new phase diagram structure. Finally, in Conclusions we briefly summarize the  main results and discuss  prospects for future developments.

It is important to note that in the whole thesis we use the natural Planck units, in which the reduced Planck constant $\hbar=1$ and the vacuum speed of light $c=1$, but we keep an explicit dependence of expressions on the Newton's constant $G$. In $D=4$ dimensions we have: $[G] = [M]^{-2}=[L]^2$, where $[M]$ and $[L]$ are mass and length units, respectively.

\newpage
\thispagestyle{empty}
\mbox{}
\newpage

\chapter{Causal Dynamical Triangulations in four dimensions}

Causal Dynamical Triangulations is a background independent, non-pertur\-bative approach to Quantum Gravity formulated in the framework of a ``traditional" Quantum Field Theory. It is based on the path integral  approach. The method of path integrals was introduced by R. Feynman in the context of the Lagrangian (action) formulation of Quantum Mechanics \cite{Feynman1,Feynman2}. In this framework the {\it propagator} of a particle, defined as a quantum amplitude of the transition between  initial $\ket{i}$ and final  $\ket{f}$ states, can be expressed as 
\beq\label{Kprop}
{\cal K}(f,i) \equiv \bra{f} e^{-{i}T \hat H}\ket{i} =
\eeq
$$ =\int D[x(t)] \exp\left[{i}\int_{0}^{T} dt \ L(x,\dot x) \right] = \int D[x(t)] \exp{\Big[ i S[x(t)] \Big] } \ , $$
where $\hat H$ is the (quantum) Hamiltonian, and $L$ and $S$ are the (classical) Lagrangian and action, respectively, and the integral is taken over all (including non-classical) trajectories $x(t)$ connecting the initial and final position in time $T$. To define the integration measure $D[x(t)]$ one usually discretizes the time evolution into $N$ equal periods of length $\tau = T/N$, and expresses (\ref{Kprop}) as a product of $N$ integrals  and ultimately takes $N\to \infty$ limit.

The idea was further generalized in  Quantum Field Theory where the amplitude, also called the {\it partition  function} (or generating function),  can be defined for $T \to \infty$ and $\ket{i}=\ket{f}=\ket{0}$ (vacuum state) by taking the  path integral over all possible field configurations $\phi(x)$ playing the role of a trajectory:
\beq
{\cal Z}= \int D[\phi(x)] e^{ i S[\phi(x)]  } \ .
\eeq
This amplitude  provides  complete information about the theory, and in principle enables one to calculate (vacuum) expectation values and correlation functions of all observables:

\beq
\braket{{\cal O}_1 ... {\cal O}_n}= \frac{1}{\cal Z}\int D[\phi(x)] {\cal O}_1(\phi(x)) ...  {\cal O}_n(\phi(x)) e^{ i S[\phi(x)] } \ .
\eeq

The path integral method can be also used in the quest for Quantum Gravity. In this case the space-time geometry plays a role of the  field and the gravitational amplitude (partition function) can be written as:
 \beq\label{Zgr}
{\cal Z}= \int D[g_{\mu\nu}] e^{ i S_G[g_{\mu\nu}] } \ ,
\eeq
where $g_{\mu\nu}$ is a metric tensor and $S_G$ is a gravitational action. To get a well defined theory one should give mathematical meaning to the (formal) expression (\ref{Zgr}) by defining:
\begin{enumerate}
\item{a space of metrics (geometries) contributing to the path integral,}
\item{the measure $D[g_{\mu\nu}]$, and}
\item{the form of the gravitational action $S_G$. }
\end{enumerate}
 One should also provide an explicit prescription how to calculate the path integral. It usually requires the introduction of some regularization (ultraviolet cut-off) and also possibly some topological ``cut-off", e.g. in a form of the topological restrictions on the ensemble of admissible geometries. Consequently, one should also define a way of approaching the continuum limit where the ultraviolet cut-off is finally removed.  Causal Dynamical Triangulations is a research program which fulfills all these requirements.

\fancyhf{}
\fancyhead[RE]{\leftmark}
\fancyhead[LO]{\rightmark}
\fancyhead[LE,RO]{\thepage}
\section{CDT geometries}

Causality  seems to be one of the most basic features of the observed  Universe and provides strong  limits  on  properties of Quantum Gravity models. The open question remains if its nature is emergent or fundamental. The model of  Causal Dynamical Triangulations follows the second approach by restricting the space of admissible (quantum) geometries to {\it globally hyperbolic pseudo-Riemannian manifolds} ${\cal M}$. In such manifolds one can use the  Arnowitt-Deser-Misner (ADM) decomposition of the metric field:
\beq\label{CDTgauge}
g_{\mu\nu}= \left( \begin{array}{cc}
-N^2+ g_{ij}N^iN^j& N_i\\ 
N_j& g_{ij}\\ 
\end{array} 
\right) \ ,
\eeq
where $g_{ij}$ is the induced spatial metric with $N$ and $N_i$ called the lapse function and  shift vector, respectively.
In such a parametrization  four-dimensional  space-time is  diffeomorphic to  ${\cal M}=R \times \Sigma$, where $\Sigma$ is a three-dimensional Cauchy surface of  constant proper time. Since  changes of topology between different quantum states (geometries)  presumably violate causality, in CDT one  assumes that {\it the topology is preserved}  between different Cauchy surfaces $\Sigma$ in a given manifold,  as well as  between different manifolds  ${\cal M}$ (quantum states). 

In numerical simulations of four-dimensional CDT one usually uses time-periodic boundary conditions $S^1$ and assumes  that each Cauchy surface is topologically isomorphic to a three-sphere $S^3$. The resulting manifolds have  topology $S^1 \times S^3$. Such manifolds can be approximated with arbitrary precision by {\it  simplicial manifolds}, called  {\it triangulations} ${\cal T}$.  CDT triangulations are constructed from {\it 4-simplices} with fixed edge lengths.  A  4-simplex is a generalization of a triangle to four dimensions and consists of 5 vertices (0-simplices), 10 links (1-simplices), 10 triangles (2-simplices) and 5 tetrahedra (3-simplices). All these elements are shared between neighbouring 4-simplices which are  ``glued"  along their common three-dimensional faces (tetrahedra), i.e. each 4-simplex is directly connected to five other 4-simplices. Due to the imposed proper time foliation, in order to construct CDT triangulations  one must use four kinds of  ``building blocks" which will be explained later. 

In CDT it is assumed that the simplicial manifolds are {\it piecewise linear}, which means that a metric tensor inside each simplex is {flat}, i.e. the interior of a 4-simplex is isomorphic to a subset of  3+1 dimensional Minkowski space-time.  The geometry of each triangulation is entirely defined by the way in which 4-simplices are connected. In particular, curvature can be defined by  a deficit angle ``around" 2-simplices (triangles) and depends on a number of 4-simplices sharing a given triangle (see Chapter 1.3 for details).
 
To construct the simplicial manifold described above one starts by introducing a discrete global proper time foliation, parametrized by  $\tau = t \cdot \delta \tau$ with constant $\delta \tau$ and 
``discrete time" labels $t$.    Each three-dimensional Cauchy surface for  integer $t$ is built from identical equilateral tetrahedra (3-simplices) with fixed edge length  $a_s > 0$. The tetrahedra are ``glued" together along their two-dimensional faces (triangles) to form a {\it spatial layer}  with topology $S^3$. Each such spatial layer ${\cal T}^{(3)}_t$   is itself a three-dimensional (Euclidean) piecewise linear simplicial manifold.

The spatial layer  ${\cal T}^{(3)}_t$, at (integer) discrete time $t$, must by causally  connected to a spatial layer ${\cal T}^{(3)}_{t+1}$ at time ${t+1}$ and ${\cal T}^{(3)}_{t-1}$, at time ${t-1}$. This can be done by introducing four types of 4-simplices with time-like links. The length of these links $a_t$ is assumed to be constant but in general can be different than  the length $a_s$ of the spatial links. Consequently, one defines an asymmetry  parameter $\alpha$, such that:
\beq\label{asat}
a_t^2=-\alpha \cdot a_s^2\quad , \quad \alpha > 0
\eeq
in Lorentzian signature. The asymmetry parameter $\alpha$ is an important feature of Causal Dynamical Triangulations and can be promoted  to a new coupling constant in the theory, which enriches its phase structure. 
 
 \begin{figure}[h!]
\centering
\scalebox{0.7}{\includegraphics{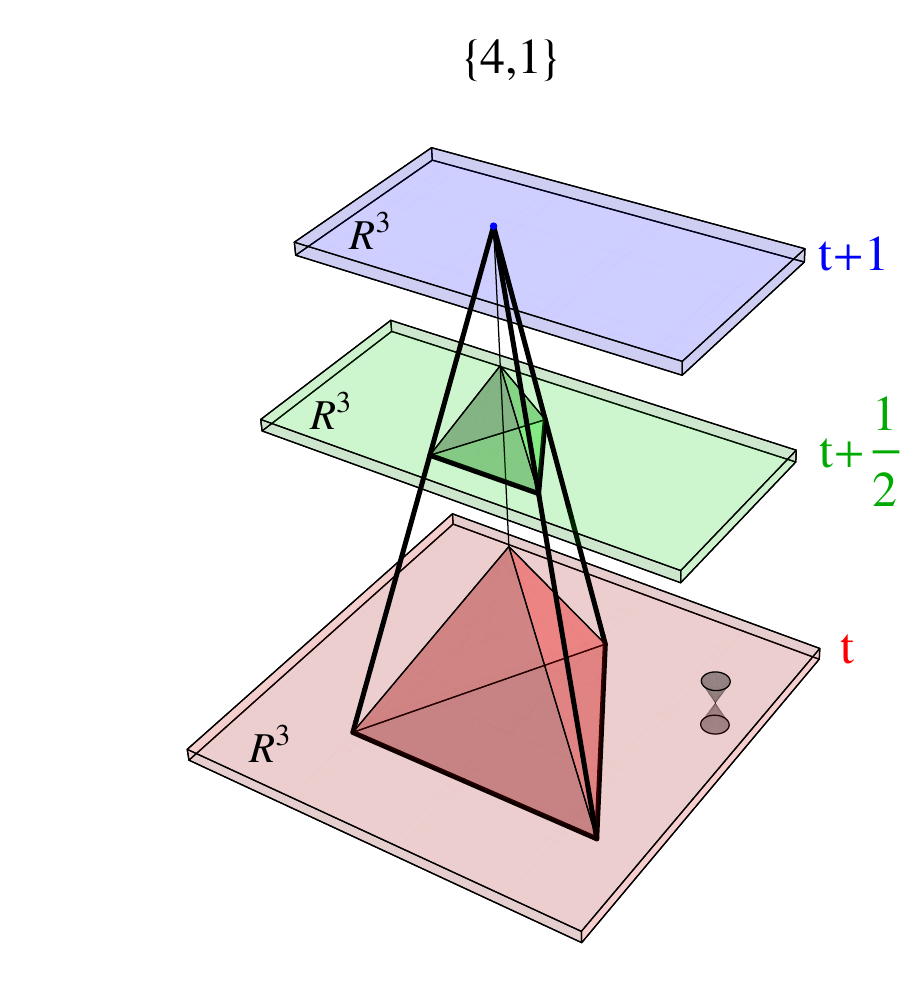}}
\scalebox{0.73}{\includegraphics{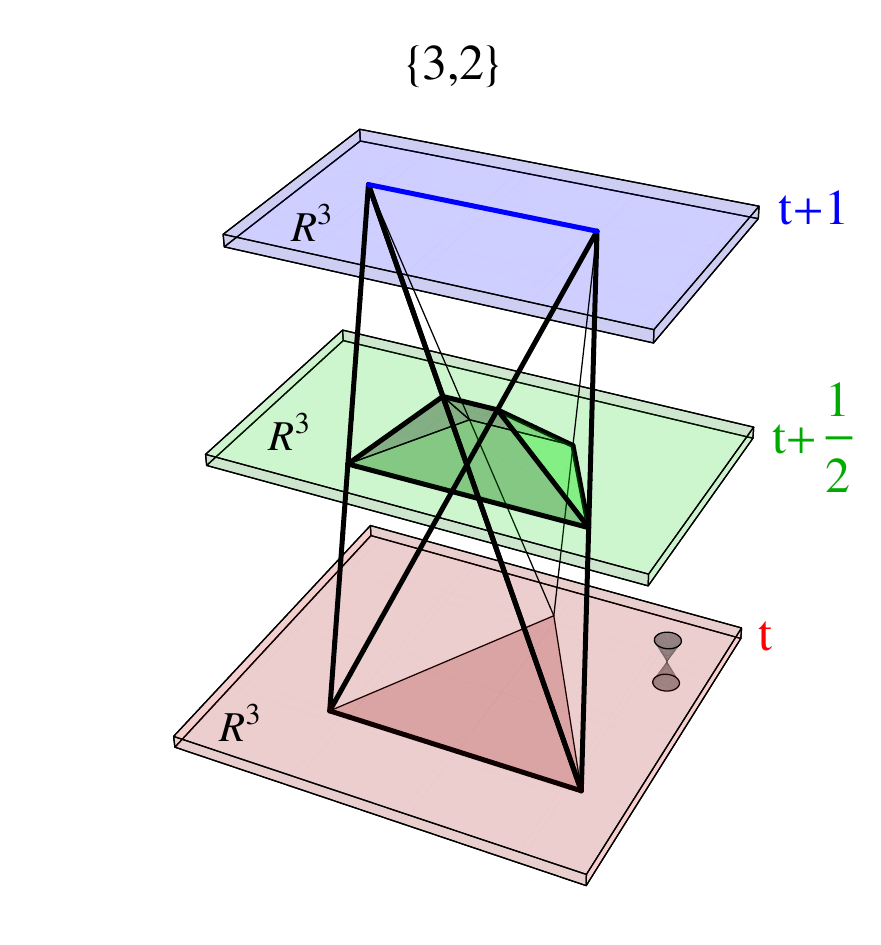}}
\caption{Visualization of  fundamental building blocks in four-dimensional CDT. A (4,1) simplex (left) has four vertices at (discrete) proper time $t$ (forming a tetrahedron) and one vertex at time $t+1$. A (3,2) simplex (right) has three vertices at time $t$ and two at $t+1$. Cauchy surfaces for $t+\oh$ are built from a combination of tetrahedra (obtained by slicing \{4,1\} simplices with hyperplanes of constant $t+\oh$) and triangular prisms (from \{3,2\} simplices). }
\label{Fig4simplices}
\end{figure}

Each three-dimensional tetrahedron in a spatial slice ${\cal T}^{(3)}_t$ at time $t$ has  four vertices which can be connected by  time-like links to a single vertex at time ${t+1}$. As a result one obtains a so called {\it (4,1) simplex} (see Fig. \ref{Fig4simplices}).  Analogously  one can define a {\it(1,4) simplex} with one vertex at time $t$ and four vertices at time ${t+1}$. As both 4-simplices are time-reversed version of each other we will treat them together as simplices of type \{4,1\}, if not  stated otherwise. 

All spatial tetrahedra building a Cauchy surface ${\cal T}^{(3)}_t$ have the same volume. Consequently the total three-volume $V_3(t)$  of a given spatial layer  is proportional to the number $N_3(t)$ of tetrahedra building this slice. As each spatial tetrahedron belongs to exactly one $(4,1)$ and one $(1,4)$ simplex, $N_3(t)$ is by construction equal to half the number of $\{4,1\}$ simplices with four vertices belonging to this slice. Hence:
\beq\label{3vol}
V_3(t) \propto  N^{\{4,1\}}(t) \ .
\eeq

To construct CDT's simplicial manifold one needs two additional building blocks. These are: a {\it(3,2) simplex}, with three vertices at time $t$ and two vertices at time ${t+1}$, and its time reversed counterpart called a {\it(2,3) simplex}. Again we will treat them together as \{3,2\} simplices if not  stated otherwise.

Let us consider a (1,4) simplex with one vertex at time ${t-1}$ and four vertices at $t$. It  is by construction ``glued" to exactly one  (4,1) simplex with the same vertices at  $t$ and  one vertex at time ${t+1}$. The (4,1) simplex   also has four other direct neighbours. These can be other (4,1) simplices or  a (3,2) simplex with three vertices at $t$ and two at ${t+1}$.  The $(3,2)$ simpex may be in turn connected  to a (2,3) simplex with two vertices at $t$ and three at $t+1$. Finally the (2,3) simplex can share a common tetrahedron with a (1,4) simplex in the next spatial layer. Consequently, connecting  any (4,1) simplex at time $t$ to a (4,1) simplex at time ${t+1}$ requires at least the following steps: (4,1)$\to$(3,2)$\to$(2,3)$\to$(1,4)$\to$(4,1). 

According to CDT assumptions, the 4-simplices interpolate between consecutive spatial layers in such a way that the  topological constraints (usually $S^1 \times S^3$) are also preserved  for all  global proper times $\tau$: $\tau(t-1)<\tau<\tau(t)$ and $\tau(t)<\tau<\tau(t+1)$. As a result one can construct Cauchy surfaces for any $\tau$ by slicing 4-simplices with three-dimensional hyperplanes of constant $\tau$, e.g. for $t+\oh$. Such Cauchy surfaces are built from a combination of tetrahedra (obtained by slicing $\{4,1\}$ simplices)  and  triangular prisms  (from $\{3,2\}$ simplices) - c.f. Fig. \ref{Fig4simplices}. These building blocks are again  ``glued" together, and by construction  form a slice  topologically isomorphic to $S^3$.

The structure described above is repeated for consecutive $t$. In time-periodic boundary conditions one can fix the (discrete) period of the time axis $T$ by setting: $t+T \equiv t$. As a result the spatial layer  ${\cal T}^{(3)}_{t+T-1}$ is directly connected to the spatial layer ${\cal T}^{(3)}_t$.
 \newline
 
One should stress that the Causal Dynamical Triangulations model is formulated in a coordinate-free way since all geometrical properties of simplicial manifolds, including curvature,  are encoded in the connectivity of the elementary ``building blocks" (pieces of flat space-time) and there is no need to introduce any coordinate system.  Consequently, despite the simplicial manifolds being   constructed from lattices with fixed edge lengths they do not break diffeomorphism-invariance since the diffeomorphism group does not act on the triangulation data.  
In this sense one only considers ``physical" geometries which is  in the spirit of Einstein's original idea of ``general covariance". As a result the lattice spacing is not a coordinate length but a physical spacing. It is also important to stress that  one does not assume the discreteness of space-time in principle. The edge lengths of simplices play only a role of a cut-off which tames ultraviolet divergencies of the path integral and thus regularizes the theory. This cut-off should be finally removed by taking the edge lengths to zero while increasing the number of simplices to infinity in a continuum  limit, if such a limit exists for CDT.


\section{CDT measure}

As was explained in the previous section, one can approximate any smooth globally hyperbolic pseudo-Riemannian manifold  $\cal M$ with a piecewise linear simplicial manifold, called a triangulation $\cal T$. The path integral  (\ref{Zgr}) can now be defined as  a sum over a (countable) set of such triangulations: 
\beq\label{ZCDT1}
\int D[g_{\mu\nu}] e^{i S_G[g_{\mu\nu}] }\ \ \ \to \ \ \ {\cal Z}= \sum_{\cal T} \frac{1}{C_{\cal T}}e^{i S_G[{\cal T}] } \ .
\eeq
The amplitude (partition function) $\cal Z$ can be understood as a regularization of the
formal gauge-fixed continuum expression. To define the integration measure $D[g_{\mu\nu}]$ 
one conventionally assumes that the path integral is taken over geometries $\cal M$, i.e. equivalence classes of metrics $g_{\mu\nu}$ with respect to the diffeomorphism group $Diff({\cal M})$. The space of metrics is much larger than that of the geometries. As a result if one integrates over metrics the measure should be divided  by the volume of   $Diff({\cal M})$:
\beq\label{Dgmn}
D[g_{\mu\nu}] \propto \frac{1}{Vol[Diff({\cal M})]} \ .
\eeq
In other words, any geometry (physical space-time) should contribute to the path integral only once, independently  of the number of  its different parameterizations which are linked by the coordinate system transformations. The factor ${1}/{C_{\cal T}}$ which defines the CDT integration (summation) measure is a remnant of this approach. $C_{\cal T}$  is equal to the order
of the automorphism group of a triangulation  ${\cal T}$, i.e. it counts symmetries of the triangulation. A triangulation may be indirectly compared to an {\it unlabelled graph}. In that case $C_{\cal T}$ would be the (inverse) of the symmetry factor of the graph. There is no straightforward way to compute $C_{\cal T}$ for a general triangulation. However the problem may be solved by considering {\it labelled triangulations} ${\cal T}_l$. The easiest way to do this is to assign labels to vertices. Each link, triangle, tetrahedron and 4-simplex in a triangulation can be subsequently  defined by an (unordered) list of its vertices. If a triangulation $\cal T$ consists of $N_0[{\cal T}]$ vertices there are in general $N_0[{\cal T}]!$ different ways to perform  such labelling.

Two labelled  triangulations  ${\cal T}_l$ represent the same unlabelled triangulation  ${\cal T}$ if there is a one-to-one map between
the labels, such that  links are also mapped to links, triangles to triangles, etc. If we denote the number of such maps by ${\cal N(T})$ we can compute $C_{\cal T}$ as:
\beq\label{CT}
C_{\cal T}=\frac{N_0[{\cal T}]!}{{\cal N(T})} \ .
\eeq
Therefore the partition function (\ref{ZCDT1}) can be represented as a sum over labelled triangulations:
\beq\label{ZCDTl}
{\cal Z}=\sum_{\cal T} \frac{1}{C_{\cal T}}e^{i S_G[{\cal T}]} = \sum_{{\cal T}_l} \frac{1}{N_0[{\cal T}_l] !}e^{i S_G[{\cal T}_l] } \ .
\eeq

The $N_0[{\cal T}_l]!$ factor appears because one only wants to count physically distinct triangulations, independent of the  the number of  different labelling methods. This is a discrete
analogue of dividing by the volume of the diffeomorphism group, as also
in the continuum formulation one only counts geometries, not the
number of their parameterizations.

In a numerical algorithm used in computer simulations it is not necessary to consider all possible $N_0[{\cal T}_l] !$ alternative ways of  labelling. Instead, one usually fixes some labelling method which leads to further simplification. In that case the $N_0[{\cal T}_l]!$  factor is taken into account automatically and the measure term becomes trivial:
\beq\label{ZCDT}
{\cal Z} = \sum_{{\cal T}}e^{i S_G[{\cal T}] } \ .
\eeq
In the above expression we omitted the index ``$l$", but from now on we assume that one works with labelled triangulations and with some arbitrarily chosen  labelling method.

\section{CDT action and Wick rotation}

As was explained in the previous  sections, in Causal Dynamical Triangulations the gravitational path integral is defined as a sum over triangulations, where the weight assigned to each geometry depends on the gravitational action. When considering Quantum Gravity models one usually starts with the Einstein-Hilbert action\footnote{In general one can also take into account the Gibbons-Hawking-York boundary term. In numerical simulations of four-dimensional CDT there is no such need as one usually assumes $S^1\times S^3$ topology and the resulting manifold is compact and without a boundary.}:
\beq\label{SHE}
S_{HE}[g]=\frac{1}{16 \pi G}\int{d^4x \sqrt{-g}(R-2 \Lambda)} \ ,
\eeq
where $G$ is the Newton's constant, $g$ is the determinant of the metric tensor, $R$ is the Ricci scalar and $\Lambda$ is the cosmological constant.

The idea of defining the action (\ref{SHE}) in an entirely geometric way originates from Regge \cite{regge} and it may be implemented for any  $D$-dimensional ($D\geq2$) \ piecewise \ linear\  simplicial \ manifold. The \ cosmological \  term \ ($\Lambda  \int{d^4x \sqrt{-g}}$) is trivial as it is just proportional to the total volume of the manifold,  given by the sum of volumes of the individual building blocks.  In four-dimensional CDT,  the simplicial manifolds are constructed from two kinds of  building block, namely the \{4,1\} and \{3,2\} 4-simplices. Let us denote the total number of  simplices in a triangulation by $N^{ \{4,1\}}$ and $N^{ \{3,2\}}$, respectively. Consequently:
\beq\label{SRcosmol}
 \Lambda \int{d^4x \sqrt{-g}} \quad = \quad \Lambda \ \left( N^{ \{4,1\}} \  V_4^{\{4,1\}} +   N^{ \{3,2\}} \ V_4^{\{3,2\}} \right) \ ,
\eeq
where the 4-volumes of the simplices:  $V_4^{\{4,1\}}$ and  $V_4^{\{3,2\}}$ are analytic functions of the asymmetry parameter $\alpha$ between the length of space-like and time-like links (the expressions are explicitly given in Appendix A). Of course $V_4^{\{4,1\}}$ and  $V_4^{\{3,2\}}$ are also proportional to $a_s^4$, where $a_s$ is the lattice spacing in  spatial direction. Since we would  like to study our model in computer simulations we should work with dimensionless variables. Consequently we set $a_s\equiv 1$ and express the dimensionful  constants ($G$ and $\Lambda$) in lattice units. 

The part with scalar curvature ($\int{d^4x \sqrt{-g} R}$) is more complicated but can be expressed in terms of the deficit angle. To illustrate this let us start with the simplest case of $D=2$ with the Euclidean metric. In that case the curvature is singular and the Ricci scalar $R$ is a distribution, whose  support is localized  on vertices. The integral of the curvature over a small circle around a vertex will be proportional to a deficit  angle around the vertex. If a triangulation is built from equilateral triangles ``glued" together along their sides, the vanishing (local) curvature will be associated with exactly six triangles meeting at the vertex (the sum of internal angles of the triangles adds to $2 \pi$). If the number of triangles meeting at a vertex is smaller, the resulting integrated curvature is positive, while if the number of triangles is bigger the curvature is negative. The integral of the Ricci scalar over the manifold can be computed as the sum of the curvature contributions for individual vertices.\footnote{For $D=2$ dimensions the sum is just a topological constant.}

The concept can be generalized to higher dimensions. In general, the curvature is localized at  $(D-2)$ dimensional {\it hinges} (in $D=4$   these are  two-dimensional triangles). For $D>2$ the integrated curvature is also proportional to  the volume of the
hinges  (area of triangles in $D=4$). Generalizations of  internal angles  are called {\it dihedral angles}. For $D=4$ these are angles between three-dimensional tetrahedra forming the faces of  4-simplices.  If one considers pseudo-Riemannian  manifolds one must additionally  distinguish between Lorentzian angles (``boosts") that appear in rotations around spatial triangles and Euclidean angles in rotations around triangles containing time-like links. As a result \cite{CDTRegge}:

\beq\label{SRcurv}
\int{d^4x \sqrt{-g} R} \ \ \ =
\eeq
$$
2  \sum_{TL\triangle}V_2^{TL}\left(2 \pi - \sum_{S \ at \ TL\triangle} \Theta_{TL} \right) + \frac{2}{ i} \sum_{SL\triangle}V_2^{SL}\left(2 \pi - \sum_{S \ at \ SL\triangle} \Theta_{SL} \right) \ ,
$$
where SL stands for space-like, TL - time-like and the first sum is over triangles ($\triangle$) while the second one is over simplices sharing a given triangle ($S \ at \ \triangle$).  $V_2$ is the volume (area) of a triangle and  $\Theta$ is the dihedral angle ($\Theta_{SL}$ is a Lorentzian angle which in general is a complex number). Both $V_2$ and   $\Theta$  are analytic  functions of the asymmetry parameter $\alpha$. Since there are two types of  4-simplices and each of them consists of 10 triangles one has to  consider different types of dihedral angles,  nevertheless the Einstein-Hilbert action can be simplified to the following Regge action (see Appendix A):
\beq\label{SR} 
S_R [{\cal T}]=  -(\kappa_0+6\Delta) N_0 + K_4 \left( N^{\{4,1\}}+ N^{\{3,2\}} \right)+ \Delta  N^{\{4,1\}} \ ,
\eeq
where  $N_0$, $N^{ \{4,1\}}$ and $N^{ \{3,2\}}$ are  the total numbers of vertices, \{4,1\} simplices and \{3,2\} simplices  in a triangulation $\cal T$, respectively. These numbers are weighted by three dimensionless   bare coupling constants: $\kappa_0$, $K_4$ and $\Delta$ which are analytic functions of the asymmetry parameter $\alpha$,  Newton's constant  $G$ and the cosmological constant $\Lambda$. The  actual functional relation between the bare coupling constants $\kappa_0$, $K_4$, $\Delta$ and $G$, $\Lambda$, $\alpha$ is quite complicated (see  Appendix A), but for future reference we will call $\kappa_0$ - the bare (inverse) ``gravitational constant'', $K_4$ - the bare ``cosmological constant" and $\Delta$ - the bare ``asymmetry parameter". To justify these names one may check that for $\alpha=1$ which corresponds to equal length of time-like and space-like links one obtains: $\Delta = 0$ and $\kappa_0 = const \cdot G^{-1}$. At the same time in Eq. (\ref{SR}) $K_4$ multiplies the total number of 4-simplices which is related to the total volume of the simplicial manifold just as $\Lambda$ multiplies the total volume in the original Einstein-Hilbert action (\ref{SHE}).\footnote{Strictly speaking the total 4-volume of the simplicial manifold is a linear combination of  $N^{\{4,1\}}$ and $N^{\{3,2\}}$ with coefficients which depend on the 4-volumes of the simplices of both types. For $\alpha=1$ the volumes are identical and $K_4$ simply multiplies the  total volume.}

It is important to note that the Regge action (\ref{SR}) is purely geometrical and does not require introduction of any coordinate system. As triangulations are constructed from only two types of the building blocks the action is also very simple.   One should  stress that the Regge action is  exactly equal to the Einstein-Hilbert action computed for the triangulation. Therefore, the CDT approximation   of the smooth manifold lies in a triangulation itself, not in the value of the action computed for the triangulation.

The  bare coupling constants appearing in ($\ref{SR}$)  are obviously  real for $\alpha > 0$, but they can be analytically continued to $\alpha < 0$ by considering a rotation $\alpha \to -\alpha$ in the  lower half of the complex $\alpha$ plane,
such that $\sqrt{-\alpha} = -i \sqrt{\alpha}$. Treating the square roots in these expressions for the bare couplings, one can show that the action (\ref{SR}) becomes purely imaginary for  $\alpha < -\frac{7}{12}$ (see Appendix A). Such a value of  $\alpha$ simply corresponds to the  {\it Wick rotation} from Lorentzian to Euclidean signature used in standard Quantum Field Theory. To show this  recall Eq.\,(\ref{asat}), where the asymmetry parameter $\alpha$ is defined via:
$$a_t^2 = - \alpha \cdot a_s^2\ .$$
The rotation  $\alpha \to -\alpha$ from positive to negative values changes time-like links into space-like links which is  consistent with 
\beq\label{Wick}
d t_L \to dt_E = i \ dt_L \ ,
\eeq
where $t_L$ is the real (Lorentzian) time and $t_E$ is the imaginary (Euclidean) time. The condition $\alpha<-\frac{7}{12}$ additionally ensures that all triangle inequalities are fulfilled in the Euclidean regime, which means that all 4-simplices and their building blocks become real parts of the Euclidean space with well defined positive volumes. 

As for $\alpha<-\frac{7}{12}$ the Regge action is purely imaginary, let us denote:
\beq\label{SRR}
S_R^{(L)} = i \cdot  S_R^{(E)} \ ,
\eeq
where $ S_R^{(E)}$ has exactly the same form as $ S_R^{(L)}$ (see Eq. (\ref{SR})) and in which all the bare coupling constants ($\kappa_0$, $K_4$ and $\Delta$) are purely real for $\alpha<-\frac{7}{12}$ (see Appendix A). Consequently, after Wick rotation   $\alpha \to -\alpha$ ($|\alpha|>\frac{7}{12}$) the CDT quantum amplitude  (\ref{ZCDT}): 
\beq\label{ZCDTW}
{\cal Z} = \sum_{{\cal T}}e^{{i}S_R^{(L)}[{\cal T}] } \quad \to \quad {\cal Z} = \sum_{{\cal T}}e^{- S_R^{(E)}[{\cal T}]}
\eeq
becomes a  partition function of  the statistical theory of  triangulated (Euclidean) 4-dimensional surfaces. Such a theory can be studied by numerical methods. As the functional form of $S_R^{(L)}$ and $  S_R^{(E)}$ is the same, we will skip the indices in further considerations. 


\section{Numerical simulations}

By performing  a Wick rotation (\ref{Wick}) the quantum amplitude of CDT becomes a partition function of the statistical field theory (\ref{ZCDTW}) in which
\beq\label{pMC}
P({\cal T})= \frac{1}{\cal Z}e^{- S_R[{\cal T}]}
 \eeq
is the probability to obtain a given triangulation $\cal T$. A similar theory formulated in two dimensions  can be solved analytically  (a comprehensive review of 2D analytical methods can be found in \cite{CDT2DAnal3}), but higher dimensions require numerical methods. In particular,   Causal Dynamical Triangulations in four dimensions can be studied  using Monte Carlo simulations. 

The idea of Monte Carlo simulations is to probe the space of all possible triangulations with a probability given by Eq. (\ref{pMC}). As a result one obtains a sample of triangulations $\{ {\cal T}_1, {\cal T}_2, ..., {\cal T}_{N_{MC}} \}$ which can be used to estimate expectation values or correlation functions of observables:
\begin{eqnarray}\label{correlsamp}
\braket{{\cal O}_1 ... {\cal O}_n} &= & \frac{1}{\cal Z} \sum_{{\cal T}}   {\cal O}_1({\cal T}) ...  {\cal O}_n({\cal T})    e^{- S_R[{\cal T}]}   \nonumber \\
&\approx& \frac{1}{N_{MC}}\sum_{i=1}^{N_{MC}}  {\cal O}_1({\cal T}_i) ...  {\cal O}_n({\cal T}_i)  \ . \end{eqnarray}

The  description of the Monte Carlo algorithm can be found in Appendix  B. In general
we use a set of  seven {\it  Monte Carlo moves} which transform triangulations from one to another. 
The transformations define a Markov chain in the space of triangulations as a new configuration  depends only on a previous configuration  and the type of the  move performed. Each of the moves is applied locally, which means that it considers only a small number of adjacent (sub)simplices at a given position in the triangulation. We use the  {\it detailed balance} condition to ensure that the probability  distribution  of  the Monte Carlo  simulation approaches the stationary distribution given by Eq. (\ref{pMC}). After a large number of Monte Carlo steps (so called {\it thermalization}) the approximation of  (\ref{pMC}) is very good, and one can use generated triangulations to compute expectation values or  correlation functions of observables according to Eq. (\ref{correlsamp}).

The Monte Carlo moves obey the {\it causal} structure of CDT, i.e. they preserve both the  topology $S^3$ of the spatial slices and the global  topology $S^1\times S^3$ of the whole simplicial manifold, and therefore preserve the global proper time foliation. They are also believed to be {\it ergodic}, which means that any final topologically equivalent triangulation can be possibly reached from the initial one by applying a series of the moves.\footnote{There is currently no rigorous mathematical proof of the ergodicity condition. The ergodicity conjecture is based on the fact that the moves used in Monte Carlo simulations are a combination of the three-dimensional  Pachner moves acting in the spatial slices alone (which are proven to be ergodic and
preserving the $S^3$ topology \cite{Pachner1,Pachner2})  and   additional ``Lorentzian" moves acting within the adjacent spatial slices (these moves 
do
not affect connections between tetrahedra building the spatial slices and are a combination of the four-dimensional Pachner moves compatible with the discrete time slicing of CDT
\cite{CDTRegge}). Nevertheless, it is in general  not known  what class
of geometries  should be considered in Quantum Gravity.
The set of the moves used in CDT can be understood as an additional condition which defines the theory.}
As a result one can  start a  simulation  with some simple {\it minimal  triangulation} \footnote{In such minimal triangulation with topology $S^1\times S^3$ each spatial slice of integer  time $t$ is built from five tetrahedra ``glued" one to each other. Consecutive spatial slices are connected by a minimal possible number of:  five (4,1) simplices,  ten (3,2) simplices, ten (2,3) simplices and  five (1,4) simplices. This structure is continued in the (discrete) time direction until $T$ period is reached and the last spatial layer is connected to the first one.} and ``enlarge" it by  applying  the Monte Carlo algorithm. After a sufficiently long thermalization time (a large number of Monte Carlo steps) one  eventually gets  complicated triangulations sampled from the requested probability distribution (\ref{pMC}). 

To make an accurate, unbiased approximation of expectation values or correlation functions (\ref{correlsamp}), one should generate a suitably large sample of {\it statistically independent} triangulations. Statistical independence can be achieved by defining a so called {\it sweep}, i.e.\,\,the number of Monte Carlo steps separating  triangulations taken into account in the measurement of observables. The minimal length of the sweep can be evaluated by monitoring the autocorrelation (in Monte Carlo steps) of some slowly changing parameters characterizing generated triangulations. If the length of the sweep is longer than the autocorrelation time the consecutive triangulations are considered to be statistically independent. The length of the sweep in our simulations varies from a few thousand (for small systems) to a  few million (for large systems) attempted Monte Carlo moves. The number of statistically independent triangulations (number of sweeps) used to calculate correlation functions vary from a few million (for small systems) to a few thousand (for large  systems).
\newline

To investigate the properties of four dimensional Causal Dynamical Triangulations one usually  performs numerical simulations for different points in  the bare coupling constant space $(\kappa_0, \Delta, K_4)$. The results of such simulations show that for fixed values of $\kappa_0$ and $\Delta$, to leading order the partition function behaves as:
\beq\label{K4crit}
{\cal Z}(\kappa_0, \Delta, K_4) = \sum_{{\cal T}}e^{- S_R(\kappa_0, \Delta, K_4)[{\cal T}]} \ \propto \ e^{(K_4^{crit}-K_4) N_4} \ ,
\eeq 
where $N_4 = N^{\{4,1\}}+ N^{\{3,2\}} $. The factor $e^{-K_4 N_4}$ comes directly from the bare Regge action (\ref{SR}) as $ S_R =... + K_4 N_4 $.  The factor  $e^{K_4^{crit} N_4}$ comes from the entropy of states with $N_4$ simplices and a given value of the bare action, as typically the number of possible configurations grows exponentially with the size of the system (at least to leading order). The value of $K_4^{crit}$ is some, {\it a~priori} unknown, function of $\kappa_0$ and $\Delta$ which can be estimated in Monte Carlo simulations.
If $K_4<K_4^{crit}$ the partition function is divergent and the theory becomes ill-defined. For $K_4 > K_4^{crit}$ the size of the system remains finite. Therefore taking the infinite volume limit ($N_4 \to \infty$) requires at the same time taking  $K_4\to K_4^{crit}$. For practical reasons, in numerical simulations we fix the total size of the system and perform the measurements for a range of values of $N_4$. For each $N_4$ the  $K_4$ coupling constant is fine-tuned  to the critical value up to finite size effects: $K_4\to K_4^{crit}(N_4)$. In the limit $N_4\to \infty$ the finite-size effects vanish and one effectively obtains  $K_4 \to  K_4^{crit}\equiv K_4^{crit}(\infty)$. By fixing $N_4$, one in fact studies the properties of ${\cal Z}(\kappa_0,\Delta, N_4)$, which is linked with ${\cal Z}(\kappa_0,\Delta, K_4)$ by the Laplace transform:
\beq
{\cal Z}(\kappa_0,\Delta, K_4)=\int_0^\infty dN_4 e^{-K_4 N_4}   {\cal Z}(\kappa_0,\Delta, N_4) \ .
\eeq
As a result, for given $N_4$ the partition function effectively depends only on two bare coupling constants $\kappa_0$ and  $\Delta$, while the third one is set to $K_4\approx K_4^{crit}(N_4)$.

To perform Monte Carlo simulations efficiently it is convenient  to introduce some volume fixing  method. {In order to make the system oscillate around the desired number of 4-simplices one could  in principle  dynamically adjust the value of $K_4$ to compensate for changes in $N_4$. However such a procedure is unstable and causes additional measurement errors.} Therefore one usually adds to the original Regge action (\ref{SR}) a volume fixing term:
\beq\label{SRSVF}
 S_R \to  S_R + S_{VF} \ .
\eeq 

In our simulations we typically use a {\it global} volume fixing method related to the total number of \{4,1\} simplices.\footnote{In the measurements of the effective transfer matrix in Chapter 3 we use a {\it local} volume fixing method: $S_{VF} = \eps \sum_{t=1}^{T} \left( N^{\{4,1\}}(t) - \bar N_{3}  \right)^2$, where the number of \{4,1\} simplices in each spatial slice at (integer) time $t$ oscillates around the same average $\bar N_3$.}
This can be done by defining either  a quadratic or a linear potential:
 \beq\label{SVFGQ}
S_{VF} = \eps \left( N^{\{4,1\}} - \bar V_{4}  \right)^2
\eeq
$$\text{or}$$
\beq\label{SVFGL}
S_{VF} = \eps \left| N^{\{4,1\}} - \bar V_{4}  \right | \ ,
\eeq
where $\epsilon$ is a small parameter controlling the amplitude of  $N^{\{4,1\}}$ fluctuations around the fixed average $\bar V_4$. As will be explained later the (average) total number of 4-simplices $\braket{N_4} \propto \braket{N^{\{4,1\}}}$ (the ratio  $1+1/\rho\equiv \braket{N_4} / \braket{N^{\{4,1\}}} $ is a function of the bare coupling constants $\kappa_0$ and $\Delta$ - see Chapter 3.4 for details) and for a given number of  \{4,1\} simplices   the total number of 4-simplices  has an approximately Gaussian distribution centered at $\braket{N_4} = (1+ 1/\rho)\bar  V_4$. Therefore, fixing  $N^{\{4,1\}}$ is equivalent to fixing $N_4$. The impact of the additional volume fixing term on measured observables can be easily removed (the details are explained in Chapters 3-5). We checked that the corrected results do not depend on the value of $\epsilon$ if one takes $\epsilon$ sufficiently small. Consequently, the volume fixing method is just technical and does not influence the final results. 

\chapter{State of the art and the author's contribution to CDT}

Causal Dynamical Triangulations is a relatively new approach to Quantum Gravity. The theory was formulated in the late 1990's \cite{CDT1998}, and the first results of numerical simulations in four dimensions were published in 2004 \cite{CDT4D2004}. In this Chapter we briefly describe  the most important results. 

\section{State of the art}

In this section we summarize  the  main results of four-dimensional Causal Dynamical Triangulations excluding the results obtained by the author, himself.
A comprehensive description can be found in \cite{CDTreviews1,CDTreviews2,CDTreviews3,CDTreviews4}.  A short description of the the author's contribution will be presented in the next section and the details in the following Chapters.

\subsection*{Phase structure}
\begin{figure}[h!]
\centering
\scalebox{0.8}{\includegraphics{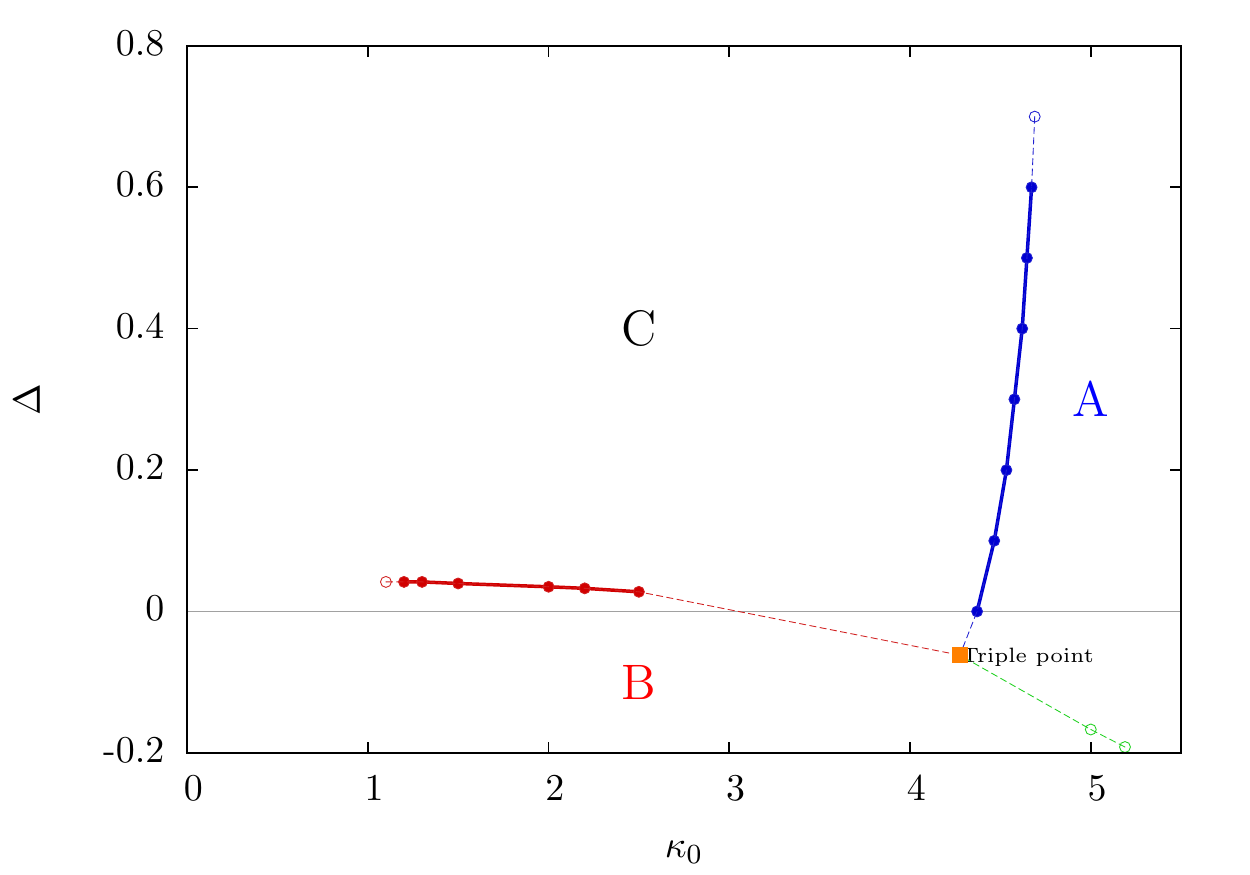}}
\caption{Phase diagram of the four dimensional CDT in ($\kappa_0, \Delta$) plane ($K_4$ is set to the critical value as explained in the previous section). Typical configurations in different phases are presented in Fig.\,\ref{FigPhaseConf}. }
\label{FigPhases}
\end{figure}

Depending on the values of the bare coupling constants $\kappa_0$ and $\Delta$ ($K_4\to K_4^{crit}(\kappa_0,\Delta)$) one  observes three  different phases. The phase 
structure  of CDT was first qualitatively described in \cite{deSitter2} where the three phases were labelled ``A", ``B" and ``C", and the first detailed phase diagram was published in \cite{CDTphases2}. It is presented in Fig. \ref{FigPhases}
 \begin{figure}[h!]
\centering
\scalebox{0.45}{\includegraphics{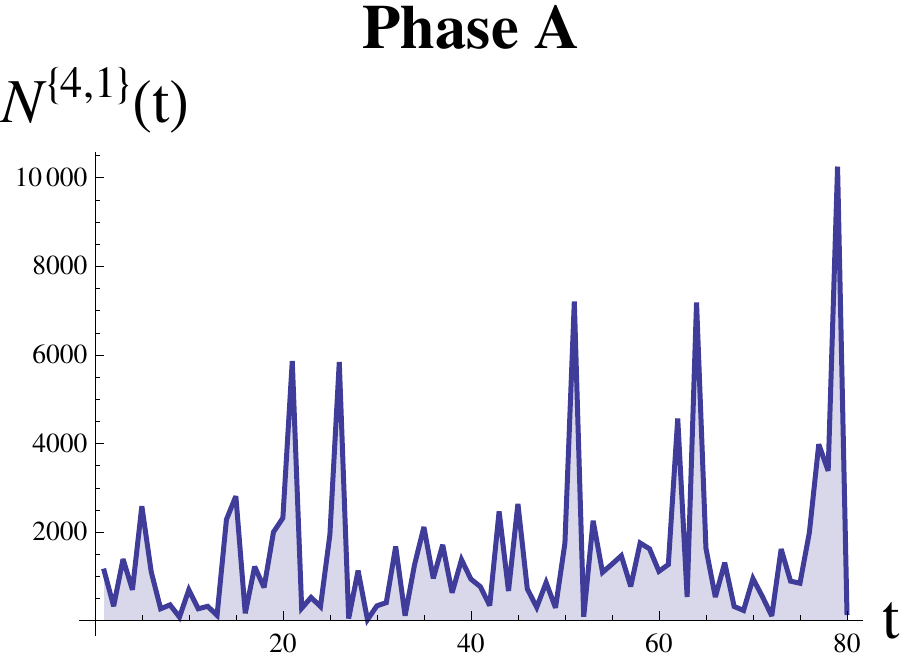}}
\scalebox{0.45}{\includegraphics{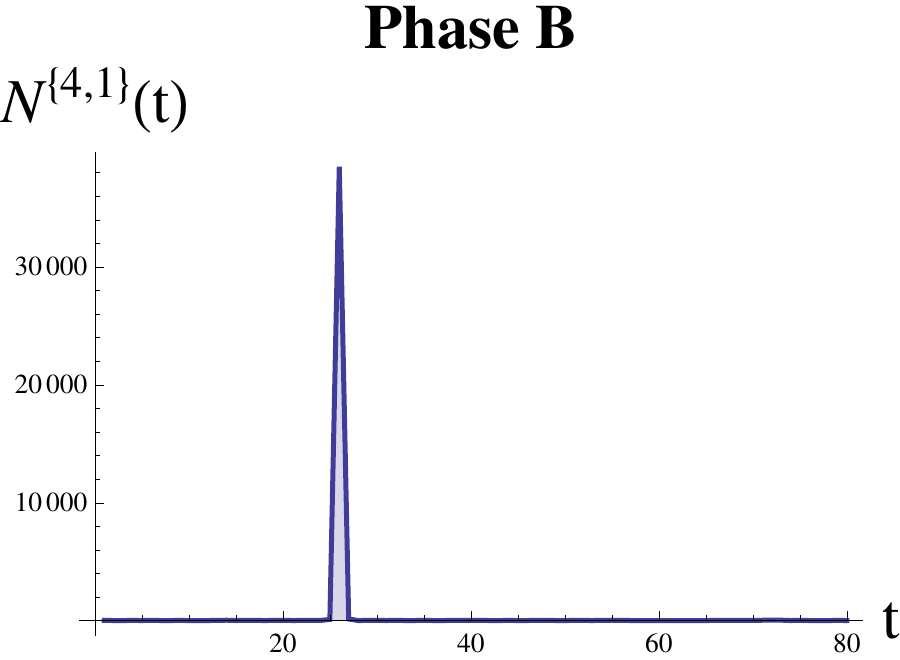}}
\scalebox{0.5}{\includegraphics{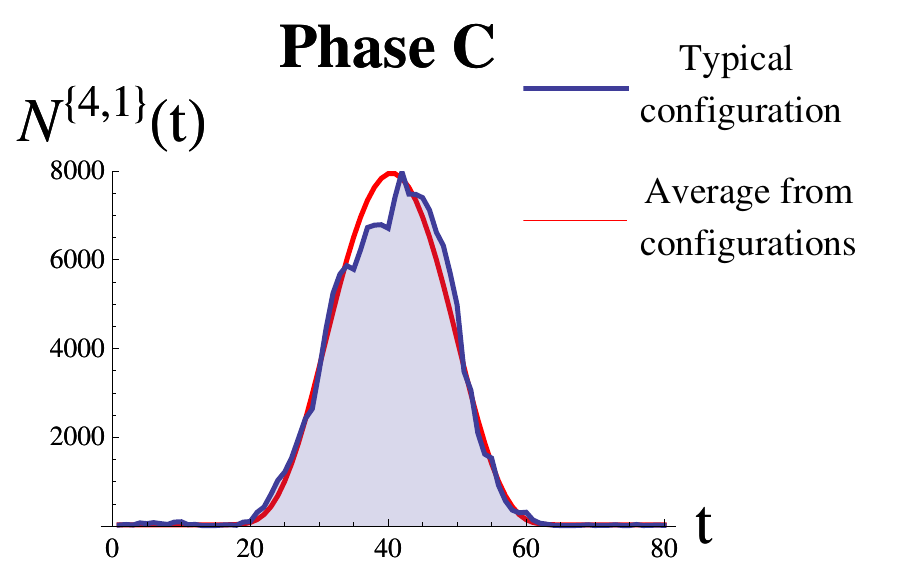}}
\caption{Typical (temporal) distribution of the spatial volume (the number of \{4,1\} simplices) in phases ``A" (left), ``B" (middle) and ``C" (right).}
\label{FigPhaseConf}
\end{figure}

Phase ``A" is observed for sufficiently large values of the  bare (inverse) cosmological constant $\kappa_0$.  A typical configuration consists of  many disjoint ``baby universes"  with time extension of approximately three time slices and uncorrelated spatial volumes. A typical spatial volume profile is presented in Fig.\,\ref{FigPhaseConf} (left) where $N^{\{4,1\}}(t)$ forms an irregular sequence of  maxima and minima. The maxima vary in an unpredictable way and the minima are of the cut-off size (due to $S^1\times S^3$ manifold restrictions  $N^{\{4,1\}}(t)\geq10$).  This phase is a CDT analogue of the ``branched polymer" phase  observed earlier in the Euclidean Dynamical Triangulations (EDT).

Phase ``B" is realized  for small values of the bare asymmetry  parameter $\Delta$. 
In contrast to  phase ``A", inside  phase ``B" the whole manifold ``collapses" into a single spatial slice containing almost all \{4,1\} simplices. The slice ends in the ``past" and the ``future"  in a vertex of very high order (belonging to almost all 4-simplices). The spatial volume $N^{\{4,1\}}(t)$ outside the collapsed slice is close to the cut-off size (see Fig.\,\ref{FigPhaseConf} (middle)). This phase is a CDT analogue of the ``crumpled" phase of EDT.

For sufficiently small values of $\kappa_0$ and large values of $\Delta$ one can observe   phase ``C". A typical triangulation in this phase consists of the extended {\it``blob"} where most 4-simplices are placed and the number of 4-simplices corresponding to discrete time slices $N^{\{4,1\}}(t)$ changes quite smoothly from slice to slice (see Fig.\,\ref{FigPhaseConf} (right)). The ``past" and the ``future" of the "blob" are connected by a thin {\it ``stalk"} formed from the (almost) minimal  number of 4-simplices in each time slice. 

Phases ``A" and ``B" do not appear to have an appropriate physical interpretation, while  phase ``C" has   non-trivial physical properties which will be explained below.

\subsection*{Phase transitions}

There is strong evidence that  the ``A"-``C" transition is a first order transition while the ``B"-``C" transition is a second (or higher) order transition. These results are based on extensive numerical studies described in detail in \cite{phaseTrans1,phaseTrans2}. The transitions were analyzed  by studying properties of triangulations in  particular paths in the bare coupling constant space $(\kappa_0,\Delta)$. The ``A"-``C" transition was considered for fixed $\Delta = 0.6$ by changing $\kappa_0$ while the  ``B"-``C" transition for fixed $\kappa_0=2.2$ by changing $\Delta$. The order parameters were defined as variables conjugate to the changed bare coupling constants in the Regge action (\ref{SR}). These are: $\text{conj}(\kappa_0) = N_0$ for the ``A"-``C" transition and $\text{conj}(\Delta) =N^{\{4,1\}}-6 N_0$ for the ``B"-``C" transition.
By looking at the susceptibility of the order parameters one could identify the position of the phase transition point with very high precision. Analysis of histograms of the order parameters measured at the phase transitions pointed to the  order of the phase transition in question. This was reaffirmed by measuring critical exponents which quantified  the phase transition point shift as a function of   the size of the system.

\subsection*{Geometric properties of   phase ``C"}

A study of  geometric properties of  phase ``C" was  first published in \cite{CDT4D2004} and described in detail in \cite{CDTphases1}, where it was shown that generic triangulations in  this phase can be attributed to the physical four-dimensional  universe.    More precisely, the emerging average geometry is  consistent with a  four-dimensional elongated  spheroid \cite{CDTgeom}. 
This result is  non-trivial since even the effective dimension  of four is not obvious, despite one  using four-dimensional building blocks, and for example this is not the case in the other two phases. 

To get this result one has to define how to measure the effective dimension and  check if the results obtained from different definitions coincide.  The analysis  used the Hausdorff dimension, related to the scaling properties of volume distribution within the manifold, and spectral dimension, measured by  running a diffusion process  of a point particle inside the triangulation (see Chapter 6.2 for details). It was shown that the  Hausdorff dimension $d_H=4$ \cite{CDT4D2004}.  The 
 spectral dimension  $d_S\to 4$ for large distances (diffusion  times) while  $d_S \approx 2$ for small distances \cite{CDTgeomSD}.

Other results point to a fractal structure of individual quantum geometries (triangulations). This  was measured by studying geometric properties of  individual spatial slices  of a given time $t$, where the  spectral dimension for large diffusion times $d_S\approx 1.5$ is significantly
smaller than the Hausdorff  dimension $d_H=3$ {\cite{CDTphases1}. The difference between $d_H$ and $d_S$  is a clear
indication of the fractal nature of  spatial layers, which was also measured directly.

This behaviour is a quantum gravitational analogue of conventional Quantum Mechanics where individual trajectories in the path integral are highly non-trivial, e.g. they are nowhere  differentiable, but the ``average" semiclassical trajectory is smooth.

\subsection*{Semiclassical limit}
As explained above, the geometry of the simplicial manifold in the  phase ``C" averaged over many triangulations is consistent with a four-dimensional regular spheroid. This statement can be made even stronger if one focuses on the behaviour of the spatial volume (or alternatively the scale factor)  of the CDT universe by integrating over other degrees of freedom.  The three-volume of a spatial slice at a (discrete)  integer time $t$ is proportional to the number of \{4,1\} simplices with four vertices at $t$, denoted $n_t \equiv N^{\{4,1\}}(t)$, and in the following  we disregard all local information about the geometry of a spatial slice at time $t$ except its volume. 
\begin{figure}[h!]
\centering
\scalebox{0.85}{\includegraphics{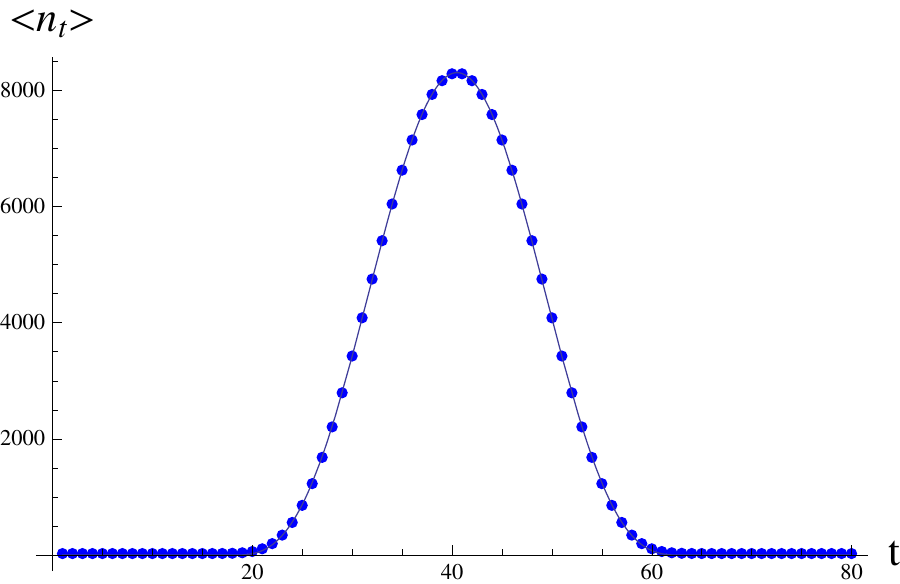}}
\caption{Average volume profile $\braket{n_t}$ (points) and the fit of Eq. (\ref{deSitter}) (line). The data were measured for $\kappa_0 =2.2$, $\Delta=0.6$ inside the ``de Sitter" phase, also called the ``C" phase. The fit was done in the extended ``blob" region (for $20 \leq t \leq 60$).}
\label{FigavPhC}
\end{figure}

\noindent In the extended ``blob" region the average $\braket{n_t}$ can be fitted very well  to \cite{deSitter3}:
\beq\label{deSitter}
 \braket{n_t} = \frac{3}{4}\cdot \frac{ V_4}{A \,V_4^{1/4}}\cdot \cos^3\left(\frac{t-t_0}{A\, V_4^{1/4}}\right) \ ,
 \eeq
 where $V_4=\sum_t n_t =  \braket{N^{\{4,1\}}}$ - see Fig.\,\ref{FigavPhC}.\footnote{To obtain this relation one has to redefine the discrete time coordinate of each individual triangulation in order to get rid of the translational zero mode which moves the centre of volume along the periodic proper time axis (see Chapter 3.1 for details).}
 This semiclassical trajectory  is fully consistent with the  (Wick rotated) {\it de Sitter} solution of 
 Einstein's  equations, describing a maximally symmetric four-dimensional universe with positive cosmological constant. This is why the phase ``C" is also called the ``de Sitter" phase. This solution  corresponds to a low energy (infrared) limit of Quantum Gravity defined by CDT. 

\subsection*{Quantum fluctuations}

The semiclassical de Sitter solution can be obtained from equations of motion derived from the effective {\it minisuperspace action}. The action originates  from the usual  (Euclidean) Einstein-Hilbert action for the  spatially homogeneous and isotropic  space-time, with the following infinitesimal line element:
\beq\label{ds2}
ds^2=d\tau^2+a^2(\tau)d\Omega_3^2 \ ,
\eeq
where $a(\tau)$ is the scale factor depending on the proper time $\tau$ and $d\Omega_3^2$ denotes the line
element on $S^3$. In   phase ``C",   quantum fluctuations of the spatial volume around the semiclassical average  (\ref{deSitter})  are (in the ``blob" range) described very well  by the following effective action \cite{deSitter2,deSitter5}\footnote{The original coupling constants have been changed in order to be consistent with notation used in Chapters 3-6.}:
\begin{equation}\label{Seff1}
S_{eff}=\frac{1}{\Gamma}\sum_t \left( \frac{\left( n_t-n_{t+1}\right)^2}{2 \ n_t } - \lambda \, n_t + \mu \, n_t^{1/3} \right) \ , 
\end{equation}
which is (up to the overall sign) a simple
discretization of the  minisuperspace action (see Appendix C). In this expression $\Gamma$ controls  the amplitude of  quantum fluctuations, $\mu$ is related to  the (temporal) width of the semiclassical solution  $A$,  while $\lambda$ is a Lagrange multiplier fixing the total volume $V_4$. The value of $\Gamma$ is proportional to $a_s^2/G$, where $G$ is the physical Newton's constant and $a_s$ is the lattice spacing. Therefore the measurement of $\Gamma$ together with assumption (\ref{ds2}) enable one to restore physical dimensions to the system and thus estimate the physical lattice spacing \cite{deSitter5}.The simulated CDT universes have radii of the order of 10 Planck's lengths.

The behaviour of the spatial volume is highly  non-trivial. In Causal Dynamical Triangulations one does not {\it  a priori} freeze any spatial degrees of freedom (as is done in the effective minisuperspace  model). Both models only use the same observable, i.e. the temporal distribution of the spatial volume (or alternatively the scale factor). What is more, in the original minisuperspace formulation one obtains the effective  action with a reversed overall sign which makes the action unbounded from below. This issue  is known as the conformal mode problem, and is a major obstacle in   Euclidean Quantum Gravity models \cite{ConformalModeProblem}. In CDT the problem is  fixed dynamically as a result of a very subtle interplay between the entropy of states and the bare Regge action which both contribute to the path integral in such a way that the effective action  sign is corrected.

\section{The author's contribution to the field}
The authors contribution to the development of four-dimensional Causal Dynamical Triangulations can be summarized in the following points. A detailed description will be provided in Chapters 3 - 6. 

\subsection*{The zero mode problem}
Analysis of  quantum fluctuations described in the previous section is a starting point for further studies of the effective action. The form of the effective action was  first determined by considering its semiclassical  approximation and measuring the covariance matrix $C_{tt'}$ of  spatial volume fluctuations $ C_{t t'} =  \braket{\delta n_{t} \delta n_{t'}}$ where: $\delta n_t \equiv n_t - \braket{n_t}$. The previous measurement method (with a constraint $V_4 = \braket{N^{\{4,1\}}}$ constant)   introduced an artificial {\it zero mode}, which had to be projected out before one could invert the covariance matrix and use it to reconstruct the action \cite{deSitter5,CDTreviews4}. This procedure was quite complicated and introduced additional measurement errors. In his Master's thesis \cite{JGSMSc}  the author dealt with the problem of the zero mode by changing the measurement method, namely  by allowing for Gaussian fluctuations of $V_4$ around $\bar V_4$.
 The relaxation of the volume constraint  made the covariance matrix invertible and enabled high precision measurements of the effective action. A short summary   of the effective action analysis using the new volume fixing  method  is presented in the ``Toy model" section in Chapter 3.2. High precision measurements of the effective action paved the way for posing additional questions.

\subsection*{Curvature corrections}
First of all, one should ask  if there are any  {\it corrections} to the discretized minisuperspace action  (\ref{Seff1})? If such corrections exist one should check if they are finite-size  or  physical effects? This problem is analyzed in detail in Chapter 3.3, where possible  curvature corrections of $R^2$ type are discussed. We found that   subleading terms in  the measured effective action can be associated with  some $R^2$ terms but other such  terms are not present. Consequently, the observed  corrections of the effective action seem to be  discretization effects. 

\subsection*{The role of \{3,2\} simplices}
The second group of questions considers the role of the \{3,2\} simplices. So far the analysis of the semiclassical solution and the  quantum fluctuations was limited  to the spatial slices of integer (discrete)  proper time $t$ where the spatial volume depends only on $N^{\{4,1\}}(t)$. One should ask if it is possible to  refine this time spacing and to analyze the role of the intermediate spatial  layers (spatial slices of non-integer $t$)? It turned out that, after proper rescaling, the (temporal) distribution of   \{3,2\} simplices can be associated with spatial volume of half-integer time layers: $n_{t+\oh}\equiv N^{\{3,2\}}(t)$  and the effective action comprising both \{4,1\} and \{3,2\} simplices can be constructed. To construct such an effective action one has to measure and analyze the covariance matrix of volume fluctuations in both integer and half-integer time layers. The results of such analysis are presented in Chapter 3.4.  We show that there is a direct interaction between the  $t \leftrightarrow (t\pm1)$ and $t \leftrightarrow (t\pm\oh)$ spatial layers, but not between  the $(t-\oh) \leftrightarrow (t+\oh)$ layers ($t \in \mathbb Z$). Both interactions are well described by the (discretized) minisuperspace action but have opposite signs. 
The pure \{4,1\}$\leftrightarrow$\{4,1\} part of the action has a negative sign (as in the original minisuperspace model) while  the part  including \{3,2\}  simplices has a corrected sign (which effectively  stabilizes the whole system). Nevertheless, one can  show that after integrating over the \{3,2\} simplices one recovers the effective action for integer $t$ layers alone, with corrected (positive) sign.

\subsection*{Transfer matrix}
The results summarized  above were achieved by measuring the inverse of the covariance matrix of volume fluctuations, which allowed one to analyze the effective action indirectly (using a semiclassical approximation). A natural question arises if it is possible to measure the effective action directly  and therefore to analyze some subtle effects which may vanish in the previous measurement method? In Chapter 4 we introduce a new method of the effective action measurement based on the  transfer matrix parametrized   by the spatial volumes. The transfer matrix can be measured in numerical simulations of CDT. We show that the measured transfer matrix and the resulting effective action are consistent with  previous results. The  advantage of the transfer matrix method is threefold. Firstly, the effective action is measured directly, without resorting to the semiclassical  approximation. Secondly, both the kinetic and the potential parts of the effective action can be measured with high precision.  Last but not least,   the method can be used not only in the extended ``blob" region  but also in the ``stalk",  where finite-size effects are very large. We provide evidence that despite strong discretization effects one can still see a remnant of the effective minisuperspace action  in the  ``stalk". We also used the directly  measured  transfer matrix to define a simplified effective model  which  reconstructed the spatial volume profile and quantum fluctuations observed in  CDT.

\subsection*{Effective action in phases ``A" and ``B" }
All the results summarized so far concerned  the effective action measured ``deep inside" the de Sitter phase ``C". In Chapter 5 we extend the analysis of the effective action to the other two phases. We argue that the effective action in these phases can be measured by the transfer matrix method even though it is not possible using the ``traditional" covariance matrix approach.  In particular, we wanted to verify a conjecture \cite{burda} that the discretized minisuperspace action  could be used to explain a generic  spatial volume  behaviour in all three phases of four-dimensional CDT (if one allowed vanishing or negative kinetic term). We used the transfer matrix measurements to parametrize the effective action in phases ``A" and ``B",  and found that the conjectured scenario was not realized in the CDT data. In fact, in  phase ``A" the measured kinetic term vanishes, and the potential term is also corrected. In  phase ``B" we observe a bifurcation of the kinetic term. Our results also point to a possibility of a richer  phase structure of the model (see below).

\subsection*{Effective action and phase transitions}

Phase transition studies described in the previous section were based on the order parameters which were some global characteristics of the CDT triangulations, e.g. the total number of vertices  $N_0$. A change in such order parameters does not necessarily give much insight into the ``microscopic" nature of the phase transitions, which is an
obvious drawback of this approach. In Chapter 5 we use the effective action  to obtain additional information about
the phase transitions.  We argue that the phase transition point may be identified with a change of the kinetic term in the effective action. Indeed, for the ``A"-``C"  transition, the  vanishing  kinetic term coincides with the phase transition. The situation is dramatically different for the ``B"-``C"  transition, where  bifurcation of the kinetic term persists for much higher values of the bare coupling constant $\Delta$ than the critical value measured in the previous studies. 

\subsection*{Discovery of a new ``Bifurcation" phase}
We argue that the above results point to the  existence of a new, previously undiscovered phase in four-dimensional CDT. Due to the bifurcation of the kinetic term observed in this region of the phase diagram we call it a ``bifurcation" phase.
In Chapter 6 we study the properties of the new phase and compare it to the generic ``C" phase. We constructed a simple model based on the effective  action which explained a  (temporal) narrowing of the spatial volume profiles observed in the bifurcation region (compared to the ``C" phase). We also measured the Hausdorff and spectral dimensions and argued that they are much higher than in the ``C" phase and tend to infinity as one approaches  the generic ``B" phase. We also present  preliminary results of extensive numerical studies of a new phase diagram structure.
\newline

All Monte Carlo simulations and analysis of numerical data presented in the thesis were performed by the author, himself.  The measurements required many CPU years  of simulations in total, and were carried out on a  computer cluster ``Shiva" at the
Institute of Physics of Jagiellonian University. The author used the GNU Compiler Collection (gcc) and the Intel C++ Compiler (icc) to compile the computer code written in C. In data analysis and visualization the author used Wolfram Mathematica. Some of the plots were prepared in Gnuplot and the thesis was written in \LaTeX. 
The computer simulations required various adjustments of the original computer code performed by the author, some  in cooperation with  Dr Andrzej G\"orlich. Discussion of the results was done in cooperation with Prof. Jerzy Jurkiewicz, Prof. Jan Ambj\o rn and Dr Andrzej G\"orlich. The analysis of the curvature corrections and the role of \{3,2\} simplices was also done in cooperation with Prof. Renate Loll and  MSc. Tomasz Trze\'sniewski.

\chapter{Effective action in the de Sitter phase}
This Chapter is partly based on the article: {\it J. Ambj\o rn, A. G\"orlich, J. Jurkiewicz, R. Loll, J. Gizbert-Studnicki, T. Trze\'sniewski, ``The Semiclassical Limit of Causal Dynamical Triangulations", 
Nucl. Phys. B 849: 144-165, 2011} 
\newline
\newline

In this Chapter we present the results of studies of the effective action   in the  ``de Sitter" phase  (also called  the ``C" phase)  of four-dimensional Causal Dynamical Triangulations. The analysis is based on the measured covariance matrix of spatial volume fluctuations. We start with a short description of the method and  present the results of the transfer matrix measurements in a generic point inside the ``de Sitter" phase, for $\kappa_0 =2.2$, $\Delta=0.6$ ($K_4=0.9222$).  We describe a  ``toy model" with the simplest possible form of the effective action consistent with the CDT numerical data to explain the analysis method. We proceed with a more complicated  form of the   effective action parametrized by the spatial volumes of  integer $t$  layers (which depend only on \{4,1\} simplices) and discuss possible corrections to the minisuperspace action. Finally, we analyze the role of half-integer $t$  layers (which also depend  on \{3,2\} simplices). We construct the effective action comprising  both \{4,1\} and \{3,2\} simplices and show that such an action is consistent with our previous results.
\newline

Analysis of  quantum fluctuations of spatial volumes is a useful tool to determine the form of the  effective action. Let us consider a continuous model of quantum fluctuations  around a well defined semiclassical trajectory $\bar n$. In a semiclassical  approximation 
the quantum fluctuations $\delta n$ around this trajectory are described by a Hermitian operator $P(t,t')$ obtained by a  quadratic expansion of the effective action:
\beq\label{SExpCont}
S_{eff}\big[\bar n(t) + \delta n(t)\big] = S_{eff}\big[\bar n(t)\big] + \frac{1}{2}\int dt\,dt'\ \delta n(t) P(t,t')  \delta n(t') + {O}(\delta n ^3) \ .
\eeq
In a discretized model:
\beq\label{Sexpansion}
S_{eff}\big[\bar n_t + \delta n_t\big]  =  S_{eff}\big[\bar n_t\big]+ \frac{1}{2} \sum_{t t'} \delta n_{t} P_{t t'} \delta n_{t'}+ {O}(\delta n^3)
\eeq
$P$ becomes a matrix parametrized by a (discrete) time variable:
\beq\label{Ptt}
P_{t t'}= \left.\frac{\partial^2 S_{eff}}{\partial n_t\partial n_{t'}}\right|_{{n_{ t}}=\bar{n}_{t}}.
\eeq
 For  such an expansion quantum fluctuations around the semiclassical average are Gaussian and the covariance matrix of quantum fluctuations is given by:
 \beq
 C_{tt'}\equiv \braket {\delta n_t \delta n_{t'}} =  \left(P^{-1}\right)_{t t'} \ .
\eeq 
 
 The problem can simply be  inverted. In numerical simulations  one can measure the covariance matrix $ C$. The inverse of the covariance matrix defines  fluctuation matrix $ P$ and thus the effective action (or at least its second derivatives at the semiclassical solution). 
 
 To adopt such a method one first  has to check if the semiclassical expansion (\ref{Sexpansion}) is valid, i.e. if  quantum fluctuations are (approximately) Gaussian. In CDT one can do this by  measuring the probability  distributions (histograms) of the spatial volumes $n_t$ obtained during Monte Carlo simulations. In the ``de Sitter" phase, the  generic triangulations consist of an extended ``blob" region,  which can be associated with  the physical universe, and a ``stalk" which is present due to the CDT topological restrictions.  In the ``blob",   quantum fluctuations are indeed Gaussian (Fig.\,\ref{FigHistFluctC}, left). The situation is dramatically different in the ``stalk"  where strong discretization effects are clearly visible (Fig.\,\ref{FigHistFluctC}, right). In this Chapter we restrict the analysis to the ``blob" range and we will come back to the ``stalk" problem in the next Chapter.

\begin{figure}[h!]
\centering
\scalebox{0.55}{\includegraphics{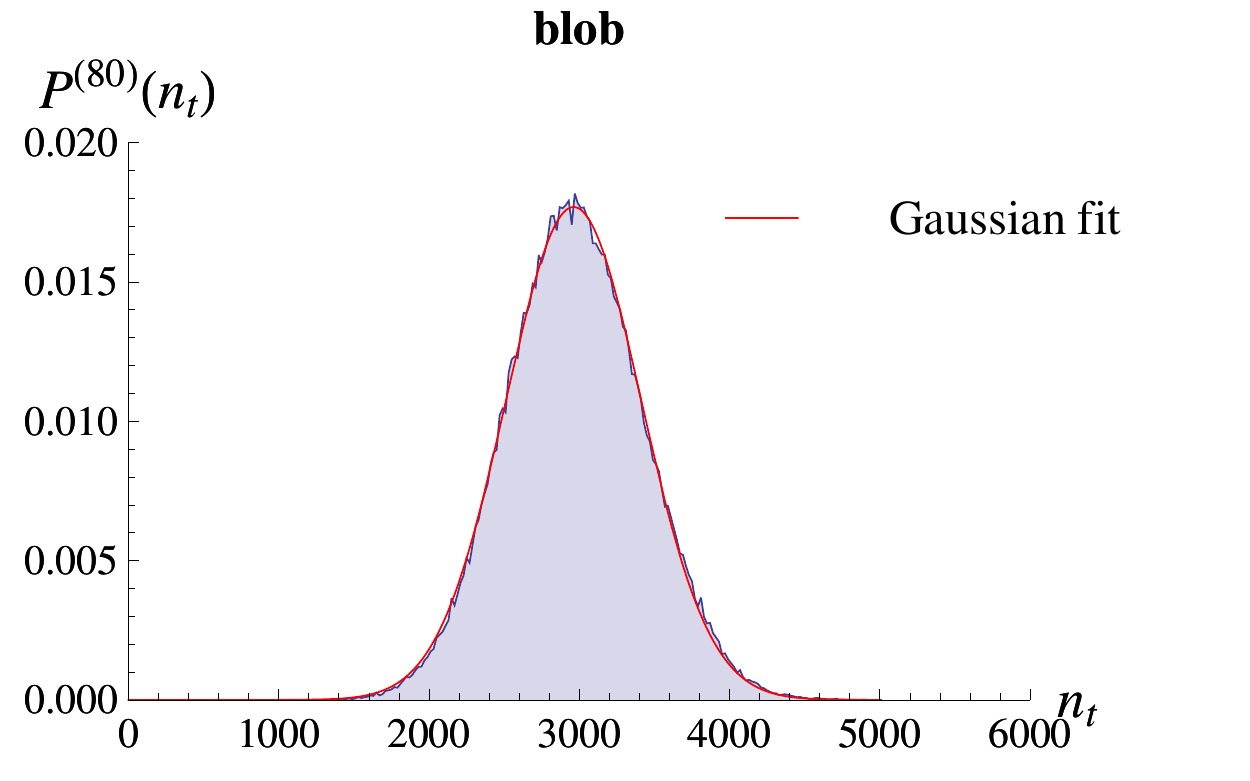}}
\scalebox{0.5}{\includegraphics{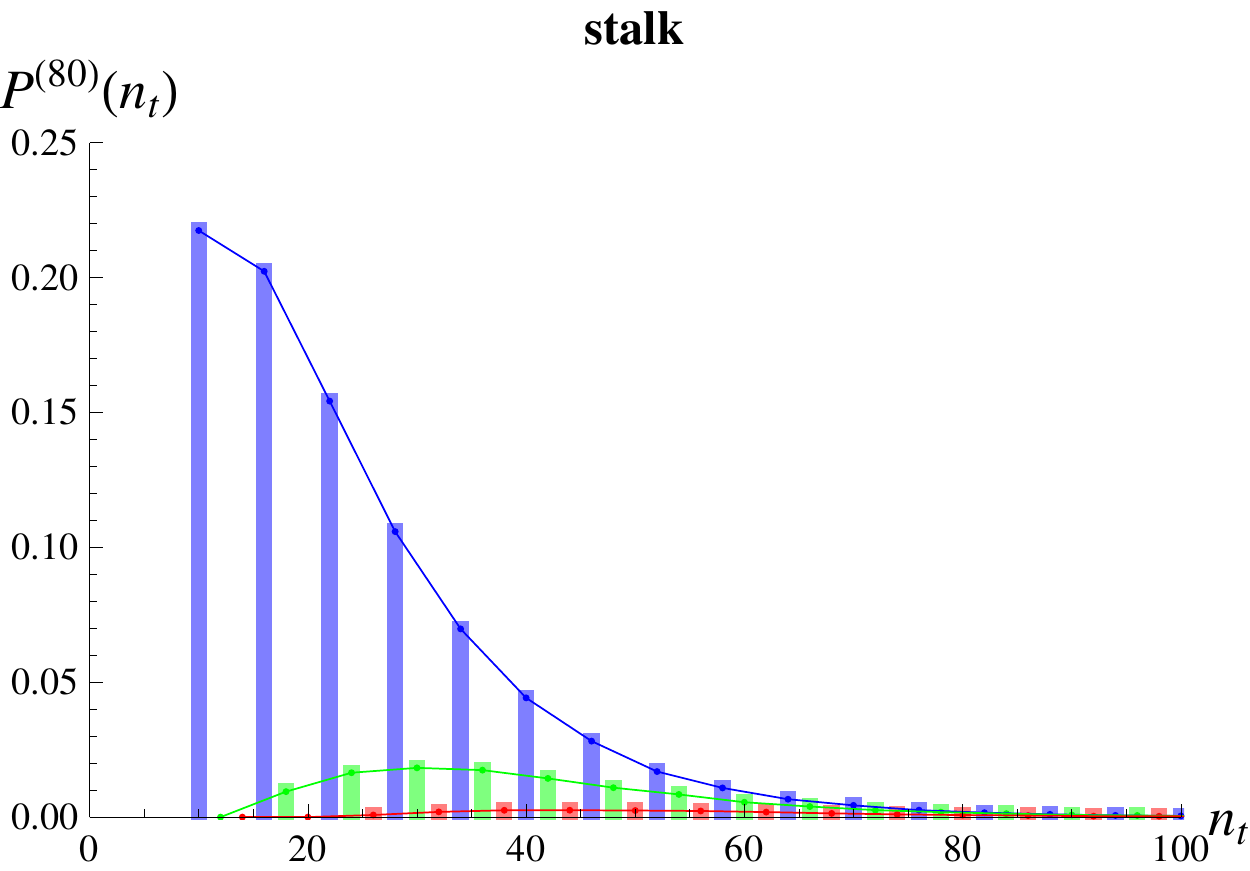}}
\caption{Probability distribution (histogram) of three-volumes $n_t$ measured in the ``de Sitter" phase (for $\kappa_0=2.2$, $\Delta=0.6$, $T=80$) inside  the ``blob" (left) (for fixed $t=29$) and in the ``stalk" (right) (for every $1 \leq t\leq 17 $ or $64 \leq t \leq 80$). In the ``blob" the probability distribution is approximately Gaussian (the mean and dispersion depend on $t$). In the ``stalk" the probability distribution is the same for each $t$ and the data fall into three families, marked by different colours  in the graph, each one with a
different behaviour (see Chapter 4.4 for details).}
\label{FigHistFluctC}
\end{figure}

\section{Measured covariance matrix}

As described in detail in  Chapter 1, in four-dimensional CDT the Cauchy surfaces of integer discrete proper time $t$ (spatial slices) are built from identical equilateral tetrahedra. The volume of each spatial slice is proportional to the number $N^{\{4,1\}}(t)$ of \{4,1\} simplices whose tetrahedral faces form this slice. For simplicity in further considerations we  will drop the proportionality factor  ($\oh$ the volume of a spatial tetrahedron) and will call  $n_t \equiv N^{\{4,1\}}(t)$ ($t \in \mathbb{Z}$) a ``spatial volume in time $t$". Spatial volumes are  well defined observables which can be measured in numerical  simulations.

In the ``de Sitter" phase the physical universe is represented by the ``blob" range on generated triangulations. We observe that during Monte Carlo simulations the position of the ``blob" changes from triangulation to triangulation as the center of the ``blob" performs a slow random walk around the periodic proper time axis. In order to obtain a meaningful average over geometries one should get rid of this translational mode. To do this, for each measured triangulation one can calculate the position of the centre of volume\footnote{There are many possible definitions of the centre of volume. Here we use a definition introduced in \cite{deSitter5}, which takes into account the time-periodic boundary conditions. We find such $t_{CV}\in \{1, 2, .., T\}$ which minimizes:
$
CV(t_{CV})=\left| \sum_{t=-T/2}^{T/2-1} (t+0.5)n_{(t_{CV}\oplus t)}\right|
$, where $t_{CV}\oplus t\equiv{1+\text{mod}(t_{CV}+t-1,T)}$ is  the addition modulo the time period $T$. Additionally, we request $t_{CV}$ to be inside the "blob" range and if more than one minimum exists we take a $t_{CV}$ where the spatial volume is bigger.  
Alternative definitions may shift the centre of volume by one time-step in either direction.} and redefine the time axis by shifting it by an integer number of time steps in such a way that the centre of volume is fixed closest to  some chosen position $t_0$ (we usually choose $t_0 = T/2 + 0.5$). The (time) shifted triangulation data ${\cal T}_i$ are subsequently used to estimate the (average) semiclassical solution $\hat n_t$ and the covariance matrix $\hat C$  using (\ref{correlsamp}):
\begin{eqnarray}\label{ntctt}
\hat n_t \ \equiv \ \braket{n_t}  &= & \frac{1}{N_{MC}}\sum_{i=1}^{N_{MC}}  {n_t}({\cal T}_i) \ , \\
 \hat C_{tt'}\equiv \braket {\delta n_t \delta n_{t'}}  &= & \frac{1}{N_{MC}}\sum_{i=1}^{N_{MC}} \big( {n_t}({\cal T}_i) - \bar n_t \big)
 \big( {n_{t'}}({\cal T}_i) - \bar n_{t'} \big) \ , \nonumber
\end{eqnarray}
which will  later be referred   to as the ``measured" (or ``empirical") quantities.

Originally, the empirical covariance matrix had an {\it artificial zero mode} which was an artifact of the measurement method in which only triangulations with a constant total volume  constraint   ($\sum_t n_t \equiv N^{\{4,1\}} = \bar V_4 $) were taken into account while calculating expressions (\ref{ntctt}). As a result:
\beq\label{zero mode}
\hat C \ket{e_0} = 0 \quad \text{for a constant vector:} \quad \ket{e_0}_t = \frac{1}{\sqrt{T}}
\eeq
and the zero mode  had to be projected out before one could invert  the covariance matrix and reconstruct the effective action \cite{deSitter5,CDTreviews4}. An unpleasant feature of this projection was a mixing of the discretization effects from the ``stalk" with physical effects from the ``blob". In \cite{JGSMSc} we proposed to change the measurement method to get rid of the zero mode by allowing for  Gaussian fluctuations of $ N^{\{4,1\}}$ around $\bar V_4$ with a controlled amplitude. This was done by introducing an additional quadratic volume fixing potential (\ref{SVFGQ}) to the bare Regge action:
$$
S_R \to S_R+ S_{VF}\quad , \quad S_{VF} = \eps \left(\sum_t n_t - \bar V_{4}  \right)^2 \ .
$$ 
Such additional volume fixing term clearly affects the measured effective action but can be 
easily corrected from the empirical data  by subtracting a constant shift
$$
\frac{\partial^2 S_{VF}}{ \partial n_t \partial n_{t'}} = 2 \eps
$$
from the measured $\hat P$ matrix:
\beq\label{corPtt}
\hat P_{t t'} \equiv  \left(\hat C^{-1}\right)_{t t'} \quad \to \quad \hat P_{t t'} =\left(\hat C^{-1}\right)_{t t'} - 2 \eps \ .
\eeq

\begin{figure}[h!]
\centering
\scalebox{0.5}{\includegraphics{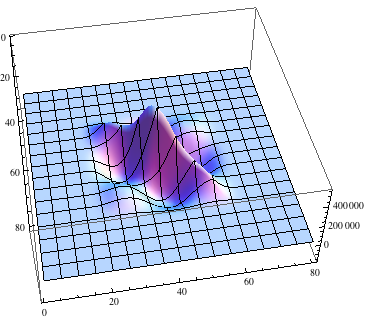}}
\scalebox{0.57}{\includegraphics{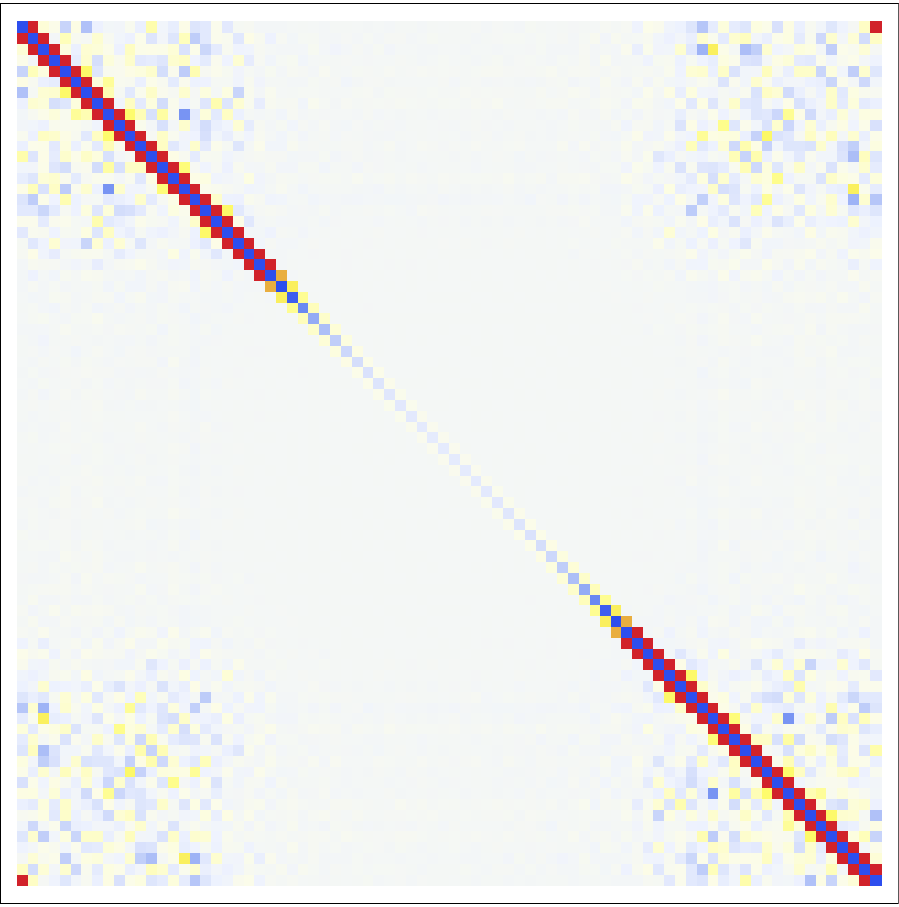}}
\caption{Left: the covariance matrix $\hat C$  of three-volume fluctuations measured in the ``de Sitter" phase  (for $\kappa_0=2.2$, $\Delta=0.6$). Right: the fluctuation matrix $\hat P = \hat C ^{-1} - 2 \eps$. Colours indicate a shift from zero towards positive (blue) or negative (red) values. The matrix has positive diagonal and negative sub- and super-diagonals (with periodic boundary conditions). In the region corresponding to the "stalk" (corners of the measured matrix) some numerical noise is visible. }
\label{FigCovAndP}
\end{figure}
We checked that in a range of the parameter $\eps$ used in our Monte Carlo simulations, the corrected matrix (\ref{corPtt}) does not depend on $\eps$ and we will use it in further analysis. The results of the measurements for the generic point inside the de Sitter phase: $\kappa_0 =2.2$, $\Delta=0.6$ ($K_4=0.9222$) are presented in Fig. \ref{FigCovAndP}. The (corrected) $\hat P$  matrix has a simple tridiagonal structure with positive diagonal and negative neighbouring sub- and super-diagonals, and zero elements elsewhere up to numerical noise (non zero elements in the ``corners" of $\hat P$  are due to the  time periodic boundary conditions). This structure suggests that the effective action is quasi-local in time, which means that there is only a direct interaction between the adjacent  spatial slices. As a result, the effective action describing the behaviour of the spatial volumes can be expressed as:
\beq\label{SeffGeneral}
S_{eff} = S_{kin}+S_{pot} =\sum_t  K[n_t,n_{t+1}] + \sum_t  V[n_t] \ ,
\eeq
where the  functional form of the kinetic term $K[.]$ and the potential term $V[.]$ need to be determined from the empirical  data.

\section{Toy model}
The effective action defining  quantum fluctuations of the spatial volumes in four-dimensional CDT is {\it a priori} unknown. Nevertheless it is natural to assume that  quantum fluctuations are governed by some discretization of the minisuperspace action which also generates the average semiclassical de Sitter solution (Fig. \ref{FigToyAv}):
\beq\label{nttt}
 \bar{n}_t \equiv \braket{n_t}= \frac{3}{4}\cdot \frac{ \bar V_4^{3/4}}{A}\cdot \cos^3\left(\frac{t-t_0}{A \bar V_4^{1/4}}\right) \ .
\eeq
\begin{figure}[h!]
\centering
\scalebox{0.95}{\includegraphics{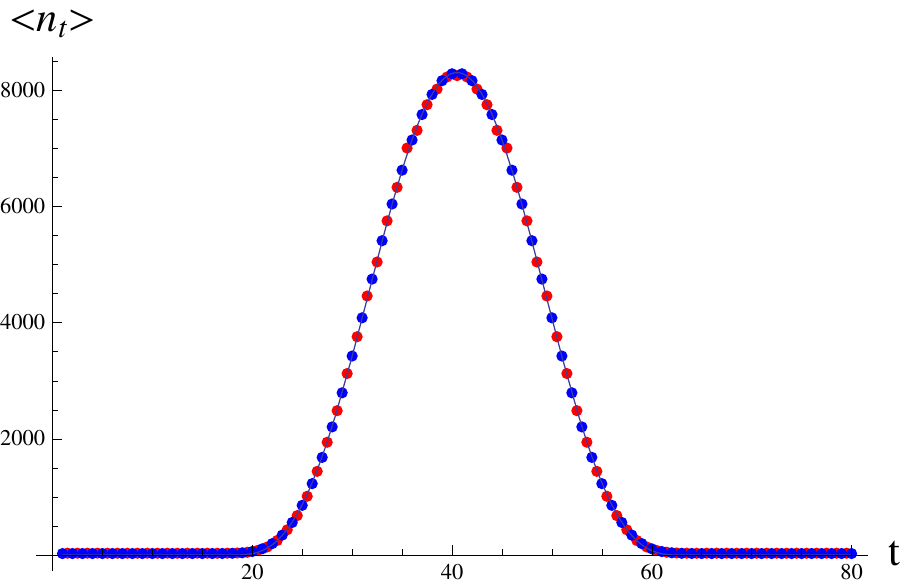}}
\caption{The semiclassical solution measured in the  "de Sitter" phase (for $\kappa_0=2.2$, $\Delta=0.6$). The directly measured average $\hat n_t\equiv \braket{n_t}$ for integer $t$ (blue points)  can be fitted very well  using Eq. (\ref{nttt})  (blue line). Red points correspond to $\hat { n}_{t+\oh} \equiv( {\Gamma} \cdot \hat k_t)^{-1}$  derived from the measured fluctuation matrix - Eq. (\ref{kToy}) - for appropriately   fixed $\Gamma$.}
\label{FigToyAv}
\end{figure}

\noindent The original minisuperspace action is derived  for the maximally symmetric (Euclideanized) metric: 
\beq\label{dss}
ds^2 = d\tau^2 + a^2(\tau) d\Omega_3^2  
\eeq
$$d\Omega_3^2 = d \theta ^2+ \sin^2\theta\left(d \phi^2_1 + \sin^2\phi_1d \phi^2_2\right)$$
and written in terms of the spatial volume $V_3(t)$ yields (for derivation see Appendix C):
\beq\label{SMS}
S_{MS}=- \frac{1}{24 \pi G} \int d\tau \left( \frac{ \dot V_3^2}{V_3} + \beta  V_3^{1/3} - 3  \Lambda V_3 \right) \quad, \quad \beta = 9 (2 \pi^2)^{2/3} \ .
\eeq 
The simplest discretization has the following form:
\begin{equation}\label{SToy}
S_{eff}= \frac{1}{\Gamma}\sum_t \left( \frac{\left( n_t-n_{t+1}\right)^2}{2 \   n_{t+\oh}}+ \mu \, n_t^{1/3}  - \lambda \, n_t  \right) 
=
\end{equation}
$$
=\sum_t  \underbrace{\frac{1}{2 \Gamma}\ \frac{\left( n_t-n_{t+1}\right)^2}{ n_{t+\oh}}}_{K[n_t,n_{t+1}]} + \sum_t  \underbrace{ \frac{1}{ \Gamma}\Bigg (\mu \, n_t^{1/3} - \lambda \, n_t  \Bigg)}_{V[n_t] } \ ,
$$
where we used dimensionless $n_t$ instead of $V_3(t)$, and the spatial volumes of half-integer time $ n_{t+\oh}$ can be treated as parameters. We also changed the overall sign and argue that all coefficients ($\Gamma$, $\mu$ and $\lambda$) are positive.  
We want to check if this simple discretization  agrees with the empirical data measured in numerical simulations performed in the generic point inside  phase ``C": $\kappa_0 =2.2$, $\Delta=0.6$ ($K_4=0.9222$). To do this one can calculate a theoretical form of the $P$ matrix according to (\ref{Ptt}) and compare it with the measured $\hat P$ matrix (corrected by  subtracting a volume fixing $2 \epsilon$ shift). To simplify the analysis let us decompose the $P$ matrix  into  kinetic and  potential parts:
 \begin{equation}
P\equiv  P^{(kin)}(\{k_t\})+P^{(pot)}(\{u_t\}) = 
\label{PToy}
\end{equation}
$$
\scriptsize{
\left( {\begin{array}{*{20}{c}}
   (k_T+k_1+u_1) & {-k_1} &0  & \cdots  &0 & -k_T \\
   { -k_1} & ({k_1+k_2}+u_2) & { -k_2} & 0 & \cdots   & 0 \\
   0 & { -k_2} & ({k_2+k_3+u_3}) &  -k_3 & \ddots    & \vdots \\
   \vdots & \ddots &  \ddots  &  \ddots  & \ddots   & -k_{T-1} \\
 -k_T &0 & 0  & \cdots  & -k_{T-1} & ({k_{T-1}+k_T}+u_T) \\
\end{array}} \right)} \nonumber \ .
$$
\newline

Due to the quasi-locality of the effective action, the kinetic part can be expressed as
\begin{equation}
P^{(kin)}_{t t'}({k_t})\equiv\left . \frac{\partial^2 S_{kin}}{{ \partial n_t \partial n_{t'}} }\right|_{{n_{ t}}=\bar{n}_{t}} =  \sum_{t=1}^T k_t  \cdot X^{(t)} \ ,
\label{PToykin}
\end{equation}
where the coefficients $-k_t$ define the sub- and super-diagonals of $P^{(kin)}$ (and   $P$): 
\beq\label{kToy}
k_t \equiv  - \left . \frac{\partial^2 K[n_t,n_{t+1}]}{{ \partial n_t \partial n_{t+1}} }\right|_{{n_{ t}}=\bar{n}_{t}} = \frac{1}{\Gamma} \cdot \frac{1}{  n_{t+\oh}} \ .
\eeq
A diagonal of  $P^{(kin)}$ is given by
\beq\label{dToy}
d_{ t} \equiv  \left( \frac{\partial^2 K[n_{t-1},n_{t}]}{{ \partial n_t^2 } } +  \frac{\partial^2 K[n_t,n_{t+1}]}{\partial n_t^2 }\right)_{{n_{ t}}=\bar{n}_{t}} = k_{t-1} + k_t
\eeq 
which leads to a very simple expression for matrices $X^{(t)}$: 
\beq\label{XToy}
X^{(t)}_{ij}=\delta_{ti}\delta_{tj} + \delta_{(t+1)i}\delta_{(t+1)j} - \delta_{(t+1)j}\delta_{tj} - \delta_{ti}\delta_{(t+1)j}
\eeq
with time periodic boundary conditions.\footnote{We have: $i,j = T+1 \ \to \ i,j=1$ and  $i,j = 0 \ \to \ i,j=T$.}
\newline

The potential part is diagonal:
\begin{equation}
P^{(pot)}_{t t'}({u_t})\equiv \left. \frac{\partial^2 S_{pot}}{{ \partial n_t \partial n_{t'}} } \right|_{{n_{ t}}=\bar{n}_{t}}=u_t \cdot \delta_{t t'} \ ,
\label{PToypot}
\end{equation}
where
\beq\label{uToy}
u_t \equiv   \left. \frac{\partial^2 V[n_t]}{ \partial n_t^2 } \right|_{{n_{ t}}=\bar{n}_{t}}= -\frac{2 \mu}{9 \Gamma}  \cdot  \bar n_t^{\ -5/3} \ .
\eeq

\begin{figure}[h!]
\centering
\scalebox{0.87}{\includegraphics{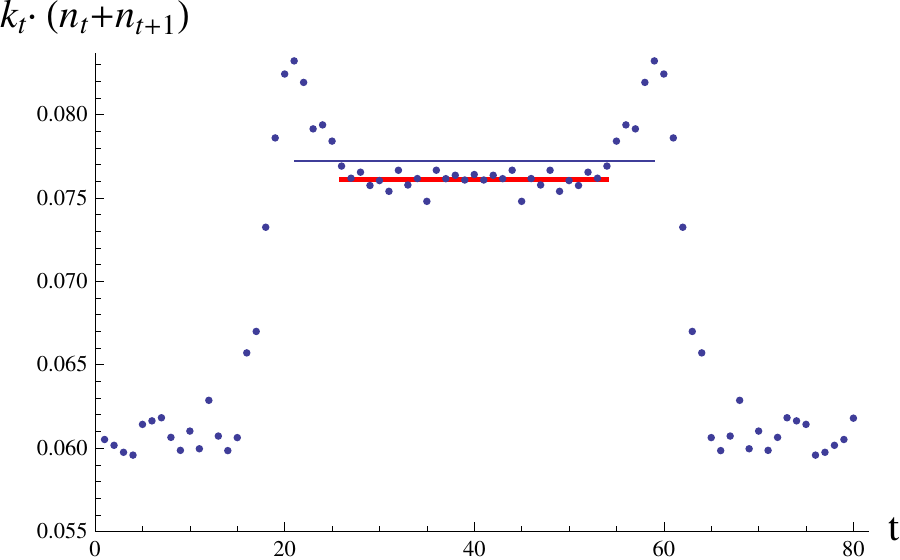}}
\caption{Empirical kinetic coefficients $\hat k_t$ as a function of $t$ measured for $\kappa_0=2.2$ and $\Delta=0.6$. Well inside the ``blob" (for $26 \leq t \leq 54$) $\hat k_t$ behave like $2/\Gamma \cdot (n_t + n_{t+1})^{-1}$. The red line corresponds to the  best fit of  $2/\Gamma$ for the range $26 \leq t \leq 54$ and the blue line to the same fit for $20 \leq t \leq 60$ (the entire  ``blob" range).}
\label{FigToykt}
\end{figure}

As a result the kinetic coefficients $\{k_t \}$ are fully determined by a sub- / super-diagonal of the $P$ matrix, whereas potential coefficients $\{u_t\}$ can be extracted from its diagonal (after subtracting combined kinetic terms: $d_t$). The structure of the empirical (corrected) matrix $\hat P$ supports  discretization (\ref{SToy}). The sub-diagonals $\{-\hat k_t\}$ are indeed negative and in the leading order   are  inversely proportional  to  the measured spatial volumes in the ``blob" range (Fig. \ref{FigToykt}). Consequently $\Gamma$ is a positive constant which can be determined by minimizing the deviations of the measured coefficients $\hat { n}_{t+\oh} \equiv( {\Gamma} \cdot \hat k_t)^{-1}$ from ``theoretical" half integer  volumes $\bar{n}_{t+\oh}\equiv\braket{n_{t+\oh}}$ (c.f. Fig. \ref{FigToyAv}).\footnote{``Theoretical" $\braket{n_{t+\oh}}$ is defined  by an interpolation of the measured average volume profile $\braket{n_t}\equiv\hat n_t$ to half integer $t$. One can e.g. define:  $\braket{n_{t+\oh}} \equiv \oh (\hat n_t + \hat n_{t+1})$ or  compute  $\braket{n_{t+\oh}}$ by fitting eq. (\ref{nttt}) to the empirical  $\hat n_t$ data or alternatively use other  interpolation method.  All these methods coincide and give very similar results.}  
The diagonal of $\hat P$ is positive but after subtracting the kinetic terms $\hat d_t = \hat k_{t-1} + \hat k_t$ the remaining potential part  $\hat u_t$ is negative in the ``blob" range, as expected (Fig. \ref{FigToyut} (left)). $\hat u_t$ can be fitted very well  using Eq. (\ref{uToy}) - see Fig.  \ref{FigToyut} (right) in which we plot measured $-\hat u_t$ as a function of $ \bar n_t\equiv  \hat n_t$.
\begin{figure}[h!]
\centering
\scalebox{0.7}{\includegraphics{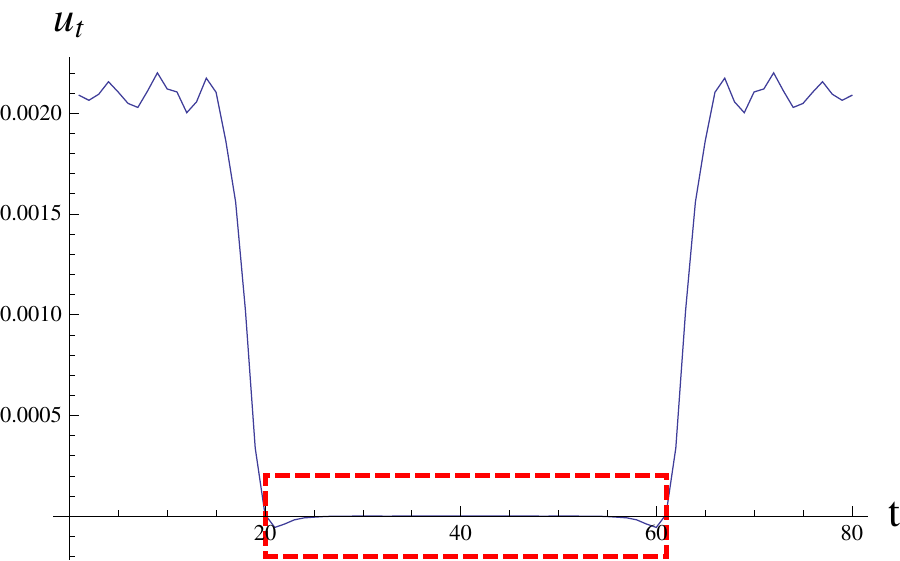}}
\scalebox{0.7}{\includegraphics{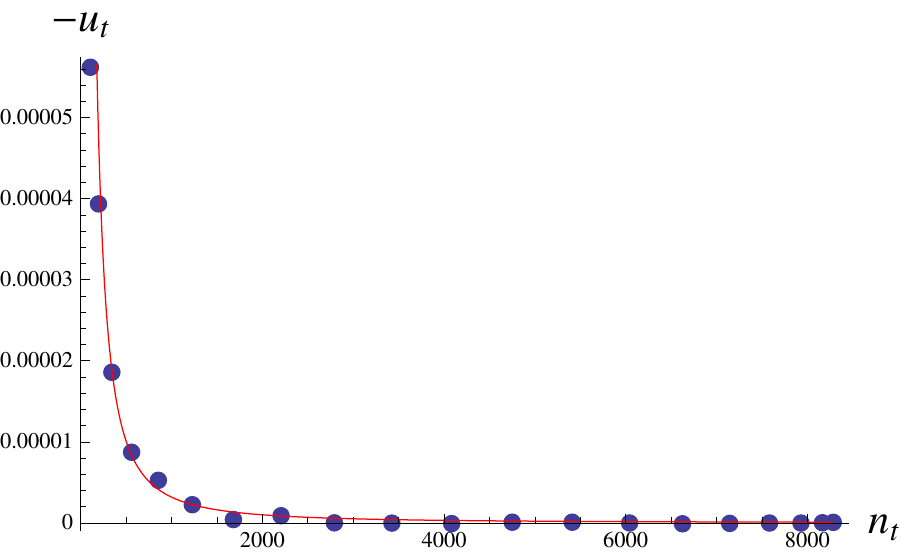}}
\caption{Left: empirical potential coefficients $\hat u_t=\hat P_{tt} - \hat d_t$ (see Eq's (\ref{PToy}) - (\ref{dToy})) as a function of $t$ measured for $\kappa_0=2.2$ and $\Delta=0.6$.  The ``blob" range is highlighted by a red dashed rectangle. Inside the ``blob" (for $20 \leq t \leq 60$) $\hat u_t$ is negative as expected. Right: $-\hat u_t$ as a function of $\bar n_t\equiv \hat n_t$ in the ``blob" range (blue points) and the best fit of Eq. (\ref{uToy}) (red line).}
\label{FigToyut}
\end{figure}



\section{Curvature corrections}

Elimination of the zero mode from the measured covariance matrix $\hat C$ enabled precise  measurements of the fluctuation matrix $\hat P=\hat C^{-1}$. As a result one can  analyze some subtle, subleading  effects, e.g. to verify if it is possible to observe  any corrections to the simple discretized minisuperspace action discussed in the previous section. In principle one could consider a variety of corrections. In particular we would like to check  if there is any evidence of curvature-squared terms in the emerging effective action. It is important to stress that  we do not change the bare  Regge action used in the CDT numerical simulations, but it is in principle  possible that such terms, related to the entropy of micro states, occur in the effective action.   

Let us again consider a continuous theory of a homogenous, isotropic (Euclidean)  universe with the metric (\ref{dss}). The most general minisuperspace-like action comprising both the Ricci scalar, a curvature-squared term (with a coupling constant $\omega$) and a cosmological  term takes a form\footnote{For the metric (\ref{dss}), the terms $R^2$, $R_{\mu\nu}R^{\mu\nu}$ and $R_{\mu\nu\rho\sigma}R^{\mu\nu\rho\sigma}$ are proportional to each other.}  (see Appendix C):
\begin{eqnarray}\label{SMSCorr}
S &=&- \frac{1}{24 \pi G} \int d\tau\left[ \left( \frac{ \dot V_3^2}{V_3} + \beta  V_3^{1/3} - 3  \Lambda V_3 \right) + \right.\\
&+&\omega\left. \left( \gamma_1  V_3^{-1/3}  + \gamma_2  \frac{ \dot V_3^4}{V_3^3} -\gamma_3  \frac{ \dot V_3^2}{V_3^{5/3}} + \gamma_4 \frac{ \ddot V_3^2}{V_3} - \gamma_5\frac{ \ddot V_3 \dot V_3^2}{V_3^2} \right)  \right] \ . \nonumber
\end{eqnarray}
We would like to verify if  there is any trace of the (discretized) terms with coefficients $\gamma_i$  in the CDT empirical  data.

As mentioned in the previous section, the measured fluctuation matrix $\hat P$ has a simple tridiagonal structure which suggests that the terms with  time derivatives of the second order $\ddot V_3$ are not present (a natural discretization of such terms would be proportional to $(n_{t-1} - 2 n_t + n_{t+1})$ which would imply non zero second sub- and super-diagonals of the $P$ matrix). The term proportional to $\gamma_3$ is also not observed. Let us focus on the two remaining terms (with $\gamma_1$ and $\gamma_2$). We would also like to  modify a discretization of the original kinetic term $\dot V_3^2 / V_3$ to make it dependent only on the volume of spatial slices for integer $t$. We propose the following discretized effective action:
\begin{eqnarray}\label{Sefff}
S_{eff} &= & \sum_t  \underbrace{\frac{1}{\Gamma}  \frac{\left( n_t-n_{t+1}\right)^2}{ n_t+n_{t+1}} \left( 1 +  \xi_2  \left( \frac{ n_t-n_{t+1}}{ n_t+n_{t+1}}\right)^2 + ... \right)}_{K[n_t,n_{t+1}]} + \nonumber \\
&+&  \sum_t  \underbrace{ \frac{1}{ \Gamma}\Bigg (\mu \, n_t^{1/3} - \lambda \, n_t  +\xi_1 n_t ^{-1/3}\Bigg)}_{V[n_t] } \ ,
\end{eqnarray}
where $\xi_1 \propto \omega \gamma_1$,  $\xi_2 \propto \omega \gamma_2$  and ``..." indicate the (in general possible) higher orders of expansion in the $\left( \frac{ n_t-n_{t+1}}{ n_t+n_{t+1}}\right)^2$ series. The form of the effective action is consistent with the previous ``toy model" to $O(n_t^{-1/3})$ and  we would like to check if introduced corrections are present in the numerical data. We again split the effective action into  kinetic and potential parts and follow the analysis method from the previous section. 
 \begin{equation}\label{PFull}
P\equiv  P^{(kin)}+P^{(pot)} = 
\end{equation}
$$
\scriptsize{
\left( {\begin{array}{*{20}{c}}
   (d_{1}+u_1) & {-k_1} &0  & \cdots  &0 & -k_T \\
   { -k_1} & ({d_{2}}+u_2) & { -k_2} & 0 & \cdots   & 0 \\
   0 & { -k_2} & ({d_{3}+u_3}) &  -k_3 & \ddots    & \vdots \\
   \vdots & \ddots &  \ddots  &  \ddots  & \ddots   & -k_{T-1} \\
 -k_T &0 & 0  & \cdots  & -k_{T-1} & ({d_{T}}+u_T) \\
\end{array}} \right)} \ . \nonumber
$$
For the sub- and super-diagonal kinetic elements one gets:
\begin{eqnarray}\label{kS}
k_t & \equiv &  - \left . \frac{\partial^2 K[n_t,n_{t+1}]}{{ \partial n_t \partial n_{t+1}} }\right|_{{n_{ t}}=\bar{n}_{t}}  \\
& = & \frac{8}{\Gamma} \cdot \frac{\bar n_t \cdot \bar n_{t+1}}  {\left(\bar  n_t+\bar n_{t+1}\right)^3} \left( 1 +  \xi_2  \left( \frac{ \bar n_t-\bar n_{t+1}}{\bar  n_t+\bar n_{t+1}}\right)^2 + ... \right)\nonumber
\end{eqnarray}
and for the  diagonal elements  of  $P^{(kin)}$:
\begin{eqnarray}\label{dS}
d_{t} & \equiv & \left( \frac{\partial^2 K[n_{t-1},n_{t}]}{{ \partial n_t^2 } } +  \frac{\partial^2 K[n_t,n_{t+1}]}{\partial n_t^2 }\right)_{{n_{ t}=\bar{n}_{t}}}\\
&= &   k_{t-1} \cdot \frac{\bar n_{t-1}}{\bar n_{t}}+ k_t \nonumber  \cdot \frac{\bar n_{t+1}}{\bar n_{t}} \ .
\end{eqnarray} 
The potential coefficients $u_t$ are given by: 
\beq\label{uS}
u_t \equiv   \left. \frac{\partial^2 V[n_t]}{ \partial n_t^2 } \right|_{{n_{ t}}=\bar{n}_{t}}= -\frac{2 \mu}{9 \Gamma}  \cdot  \bar n_t^{\ -5/3} + \frac{4 \xi_1}{9 \Gamma}  \cdot  \bar n_t^{\ -7/3} \ .
\eeq
These theoretical predictions can be once again be compared with the empirical (corrected) $\hat P =\hat C^{-1} -2 \eps$ matrix  measured at the generic point inside   phase ``C": $\kappa_0=2.2$, $\Delta=0.6$ ($K_4=0.9222$). The best fit of (\ref{kS}) to the measured sub- or super-diagonal elements $\hat k_t$ gives $\Gamma = 26.5\pm 1.0$ and indicates non-zero value of $\xi_2 = 0.38\pm 0.05$ - see Fig. \ref{FigFullkt} in which we plot empirical $\hat k_t \cdot (\hat n_t +\hat n_{t+1} )$ together with the theoretical values resulting from the best fits including and excluding $\xi_2$. The fit with non-zero $\xi_2$ is evidently better. At the same time there is no clear evidence of terms resulting from higher order expansion  in $\left( \frac{ n_t-n_{t+1}}{ n_t+n_{t+1}}\right)^2$.
\begin{figure}[h!]
\centering
\scalebox{0.9}{\includegraphics{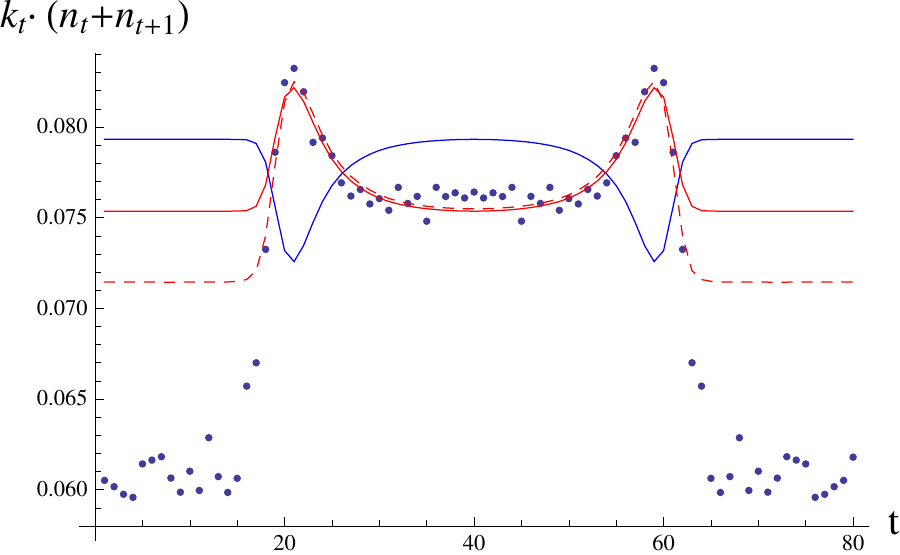}}
\caption{Empirical kinetic coefficients $\hat k_t$ (points) as a function of $t$ measured for $\kappa_0=2.2$ and $\Delta=0.6$. The $\hat k_t $'s have been multiplied by the semiclassical volumes ($\hat n_t + \hat n_{t+1}$) to cancel the leading behaviour. The solid red  line  corresponds to the  best fit of the Eq. (\ref{kS}) and the blue line to the same fit with enforced $\xi_2\equiv0$.  The fit with $\xi_2\neq 0$ is evidently better than the fit with  $\xi_2 \equiv 0$. At the same time, there is no clear evidence of the higher order corrections:  $... + \xi_4 \left( \frac{ n_t-n_{t+1}}{ n_t+n_{t+1}}\right)^4 +  \xi_6 \left( \frac{ n_t-n_{t+1}}{ n_t+n_{t+1}}\right)^6$ (red dashed line). }
\label{FigFullkt}
\end{figure}

As regards the potential part, the measured diagonal coefficients: $\hat u_t \equiv \hat P_{tt} - \hat d_{t}$ (see Eq's (\ref{PFull}) and (\ref{dS})) are again negative in the ``blob"  and behave in line with Eq. (\ref{uS}). The best fit yields: $\mu = 25 \pm 2$ and $\xi_1 = 50\pm20$ (which is close to zero if we take into account the fitting error). The fit with enforced $\xi_1 \equiv 0$  is almost indistinguishable  (see Fig. \ref{FigFullut}) and yields $\mu = 20 \pm 2$.
\begin{figure}[h!]
\centering
\scalebox{0.7}{\includegraphics{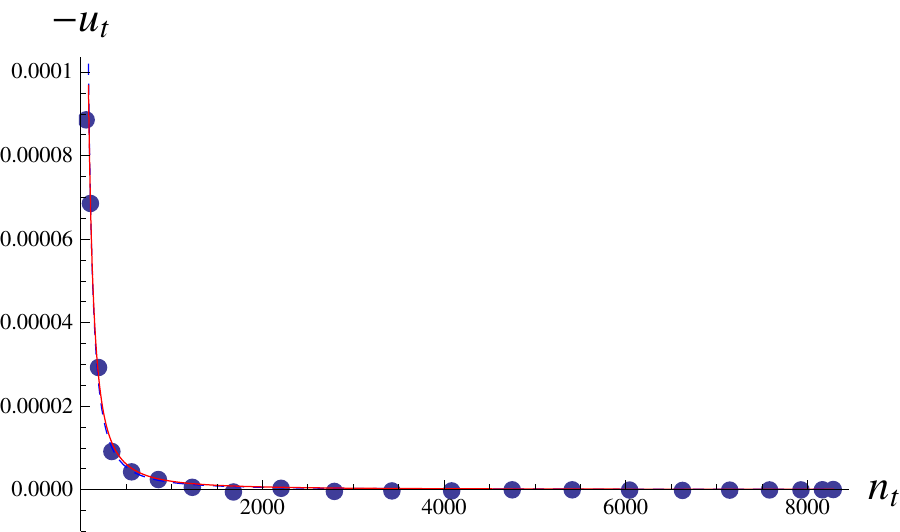}}
\scalebox{0.7}{\includegraphics{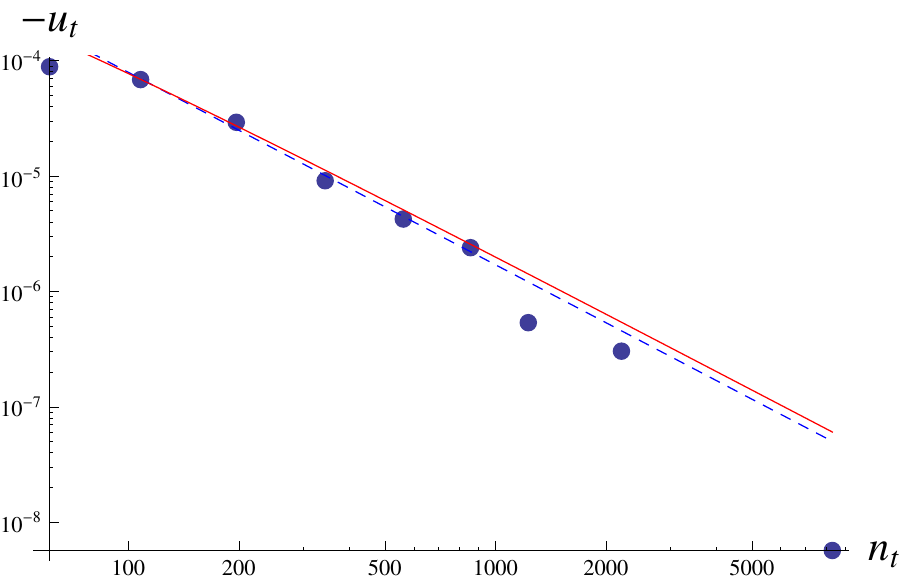}}
\caption{Empirical potential coefficients $-\hat u_t = - \hat P_{tt}+\hat d_t$ (see Eq's (\ref{PFull})-(\ref{dS})) as a function of the semiclassical volume $\hat n_t$ measured for $\kappa_0=2.2$ and $\Delta=0.6$ (points) in a normal (left) and log-log scale (right). The red line corresponds to the best fits of Eq. (\ref{uS}) with $\xi_1\neq 0 $ while the blue dashed line to the same fit with enforced $\xi_1\equiv 0$.   The two fits are almost indistinguishable and  there is no clear evidence of $\xi_1\neq 0$.   }
\label{FigFullut}
\end{figure}

Let us briefly summarize and comment on these results. We analyzed the possible corrections to the minisuperspace action  resulting from a curvature-squared term in Eq. (\ref{SMSCorr}). The existence of such corrections in the continuum limit would require that  i) the corresponding terms exist in the discretized effective action and  ii) the coefficients multiplying these terms show a well defined scaling behaviour  (at least in a large volume limit). The empirical data show that there is no clear evidence that all such terms exist in the infrared limit of CDT, at least in the generic point inside the  ``de Sitter" phase, where we performed our calculations. From all $R^2$ corrections only a (discretized) term $\propto  { \dot V_3^4}/{V_3^3} $ is evidently present in the empirical data (if we disregard a close to zero $\xi_1$ potential term). We also checked that in the region where we performed reliable measurements, the $\xi_2$ coefficient (which multiplies a discretized version of this term) does not show any significant scale dependence. This does not exclude that higher curvature corrections are present when one moves closer to a potential ultraviolet fixed point by appropriately    changing the bare coupling constants. In principle, in the vicinity of such an ultraviolet fixed point many higher order corrections may play an important role, but the existence of such a point, and the way in which it should be  approached   still remain an unresolved problem. 

The question remains how to interpret the nonzero $\xi_2$ correction present in the measured CDT data inside the ``de Sitter" phase. A possible explanation lies in the  form of the discretization of the minisuperspace effective action, itself. In  (\ref{SMSCorr}) we used the simplest possible form of the discretized kinetic term: ${ \dot V_3^2}/{V_3}  \leftrightarrow (n_t - n_{t+1})^2/ (n_t + n_{t+1})$. Let us assume for the moment that the ``correct" discretization is:
\beq\label{model}
2 \frac {(n_t - n_{t+1})^2}{(\sqrt{n_t} + \sqrt{n_{t+1}})^2} \ ,
\eeq
which gives the same continuum limit.
If one expands (\ref{model}) in powers of $\left( \frac{ n_t-n_{t+1}}{ n_t+n_{t+1}}\right)^2$ one obtains:
\beq
 \frac{\left( n_t-n_{t+1}\right)^2}{ n_t+n_{t+1}} \left( 1 +  \xi_2  \left( \frac{ n_t-n_{t+1}}{ n_t+n_{t+1}}\right)^2 + ... \right)
\eeq
with $\xi_2=1/4$. The $\xi_2 =0.38$ present in the empirical CDT data is different and may refer to some other discretization. The exact form of this discretization remains an open question.

\section{Refining time slicing}

The analysis performed so far concerned only the spatial layers in integer $t$, where the three-volume is completely determined by the \{4,1\} simplices, whose  tetrahedral faces  form each spatial slice. In four-dimensional Causal Dynamical Triangulations the \{4,1\} simplices constitute  only ``a half" of the system, since triangulations are also built from \{3,2\} simplices, whose sections (in a form of triangular  prisms) contribute to Cauchy surfaces of intermediate proper  time ($\tau(t) < \tau <\tau( t+1)$). We would like to check  if these geometric objects can be implemented into the  effective  action.
\begin{figure}[h!]
\centering
\scalebox{0.95}{\includegraphics{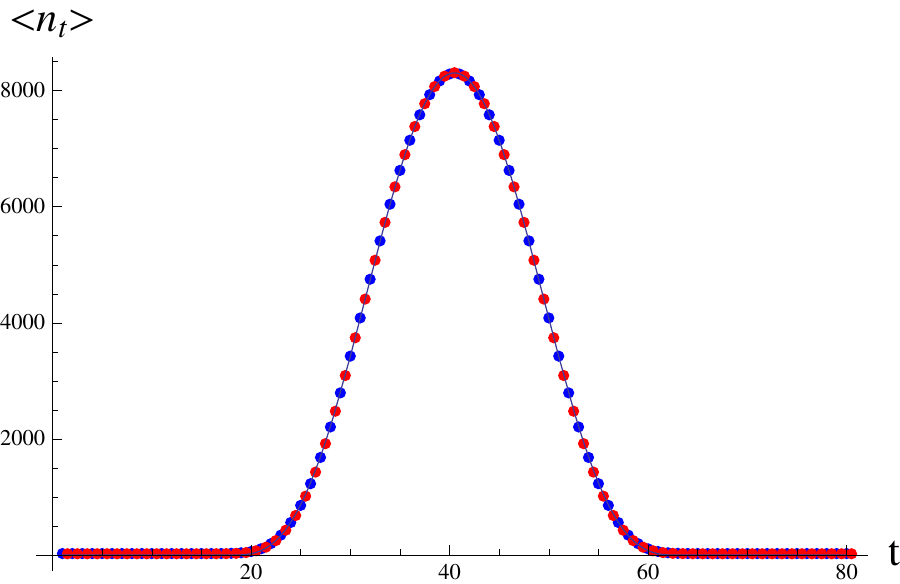}}
\caption{The averaged distributions $\braket{n_t}$ of \{4,1\} simplices (blue points) and $\braket{n_{t+\oh}}$ of \{3,2\} simplices (red points)
combine into a
single curve after performing a suitable relative rescaling - see Eq's (\ref{nt122}) and (\ref{nt12}). Empirical data measured for $\kappa_0=2.2$ and $\Delta=0.6$.}
\label{FigAV4132}
\end{figure}

We note that the ``half integer time" layers (for $t+\oh$) can be described in terms of the dual lattice, where we assign vertices to the centre  of 4-simplices and links to the three-dimensional
faces between the adjacent four-simplices. In the absence of boundaries of the manifold,
each dual vertex will have five neighbouring ones, connected by dual links.
This dual picture suggests that it should make sense to consider spatial layers of ``half integer time''. In the original lattice formulation such spatial slices are built from two types  of  building blocks, namely the tetrahedra (obtained by slicing \{4,1\} simplices with three-dimensional hyperplanes of constant $t+\oh$) and the triangular prisms (from \{3,2\} simplices).  In terms of the dual lattice one can  observe that all dual vertices of types \{3,2\} between times $t$
and $t + 1$ and the dual links connecting them form a single closed, connected
graph. Therefore it is natural to study the properties of the distribution of 
\beql{nt122}
\widetilde n_{t+\oh}\equiv N^{\{3,2\}}(t+1/2) \ ,
\eeq 
where $N^{\{3,2\}}(t+1/2)$ is the total number of \{3,2\} simplices located between integer  times $t$ and $t+1$. One can show that after simple rescaling: 
\beql{nt12}
n_{t+\oh} \equiv \rho \cdot \widetilde n_{t+\oh}
\eeq
the average distribution $\braket{n_{t+\oh}}$ closely fits the  distribution $\braket{n_{t}}$ determined earlier - see Fig. \ref{FigAV4132} in which we present the data for the generic point in the ``de Sitter" phase.\footnote{Before calculating the averages $\braket{n_t}$ and  $\braket{n_{t+\oh}}$ we shift the time variable of each single triangulation ${\cal T}_i$ to set the centre of volume closest to $t_0 = T/2+0.5$, as described in Chapter 2.1.} Therefore, the combined distribution  $\braket{n_{\tilde t}}$ is, at least in  the ``blob" range, very well described by a universal semiclassical de Sitter solution (\ref{nttt}), with $\tilde t$ running over both integer and half-integer values. One should note that this unification is not global, in the sense that
a different rescaling is needed in the ``stalk" part, which is not surprising in view of
the different dynamics in this region.

The scaling constant $\rho$ depends on the choice of the bare coupling parameters: $\kappa_0$ and $\Delta$ ($K_4 = K_4^{crit}$). We performed  systematic measurements of $\rho(\kappa_0, \Delta)$ for the the bare coupling constants inside the ``C"  phase along the lines approaching the ``A"-``C" phase transition (constant $\Delta = 0.6$, varying  $\kappa_0$)  and the    ``B"-``C" phase transition (constant $\kappa_0 = 2.2$, varying $\Delta$). The results are presented in Fig. \ref{Figrho}. The value of $\rho$ rises smoothly with increasing $\kappa_0$ and is almost constant with $\Delta$. In both cases the scaling breaks down after crossing the phase transition lines, as the ratio  $N^{\{3,2\}}/N^{\{4,1\}} \to 0 $,  both in phases "A" and "B". 

\begin{figure}[h!]
\centering
\scalebox{0.7}{\includegraphics{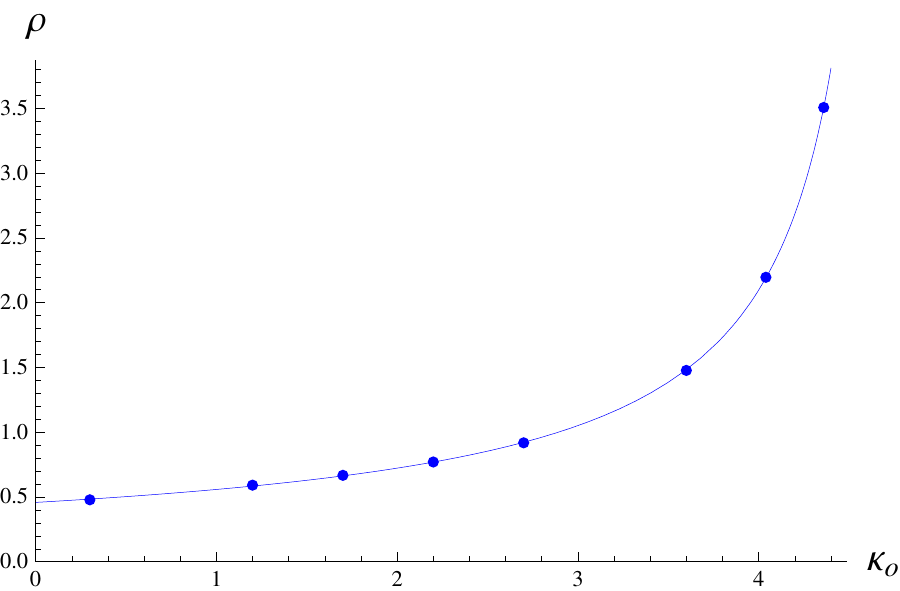}}
\scalebox{0.7}{\includegraphics{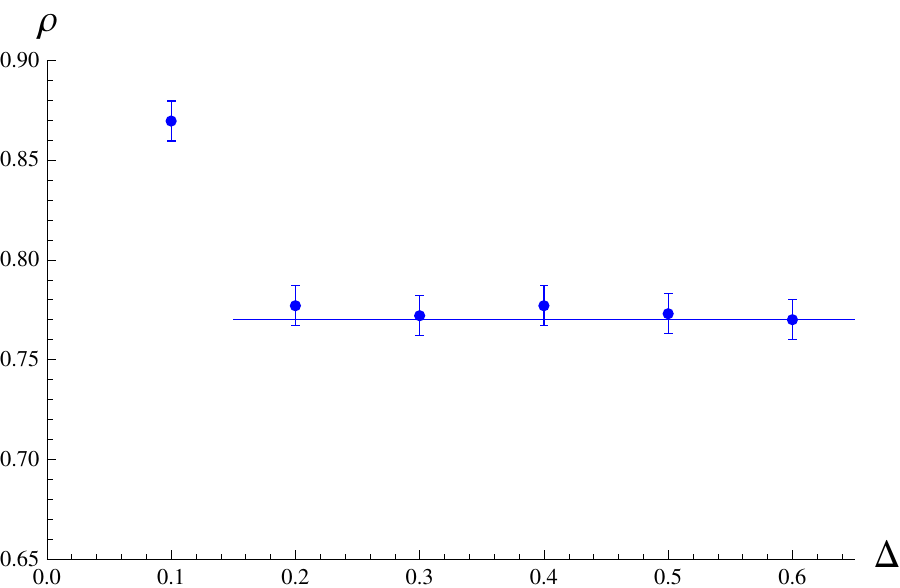}}
\caption{The \{3,2\} $\to$ \{4,1\} scaling constant $\rho$ (Eq's (\ref{nt122}) and (\ref{nt12})) as a function of $\kappa_0$ for fixed $\Delta=0.6$ (left) and as a function of $\Delta$ for fixed $\kappa_0=2.2$ (right). }
\label{Figrho}
\end{figure}

We would like to investigate if this ``finer-grained" system could
also be described by a discretized semiclassical effective action. 
To do this we measured  the covariance matrix of three-volume fluctuations for all
integer and half-integer times $\tilde t$ and inverted it to a ``larger"  $\hat P$ matrix describing  quantum fluctuations in the ``finer-grained" system. The $\hat P$ matrix can be decomposed into four blocks:  the $\hat P_{11}$ matrix describing the \{3,2\} system (at half-integer times), the $\hat P_{22}$ matrix for the \{4,1\} simplices (at integer times) and off-diagonal blocks $\hat P_{12}$ and $\hat P_{21}$ describing  interactions between the two. The $\hat P_{22}$ matrix should be corrected for $2 \eps$ volume fixing shift, described in Chapter 3.1. The structure of the (corrected) empirical matrices measured for $\kappa_0=2.2$, $\Delta=0.6$ ($K_4=0.9222$) is presented in Fig. \ref{FigLargeP}.
\begin{figure}[h!]
\centering
\scalebox{0.65}{\includegraphics{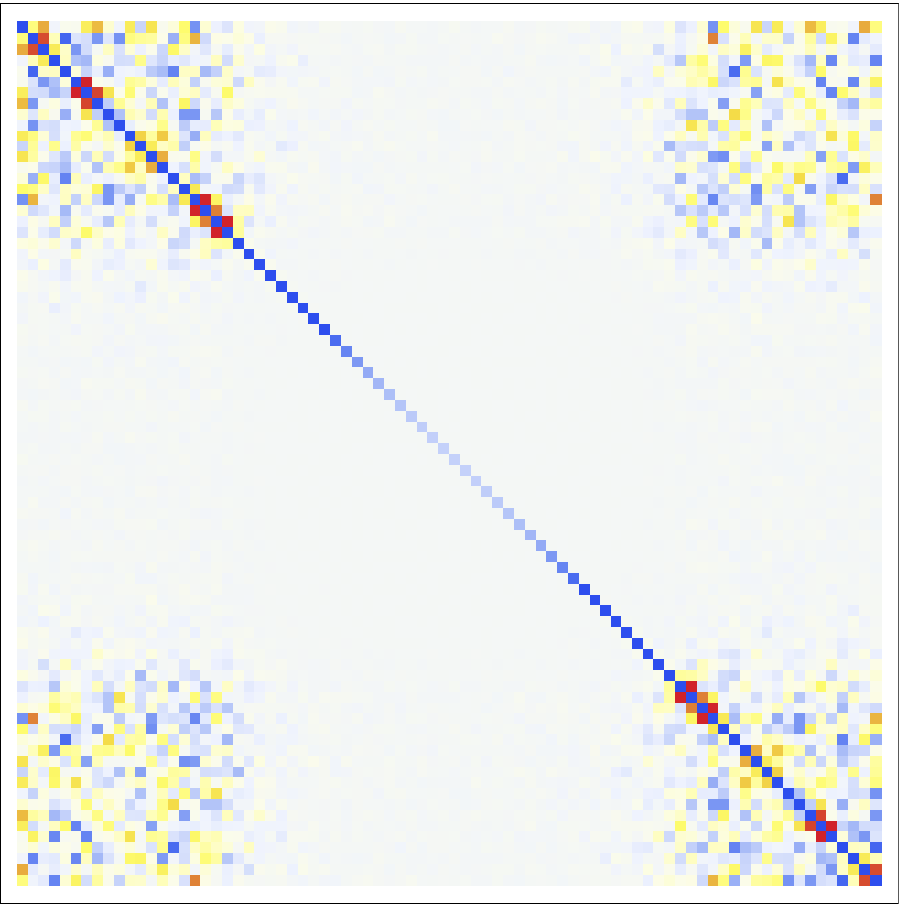}}
\scalebox{0.65}{\includegraphics{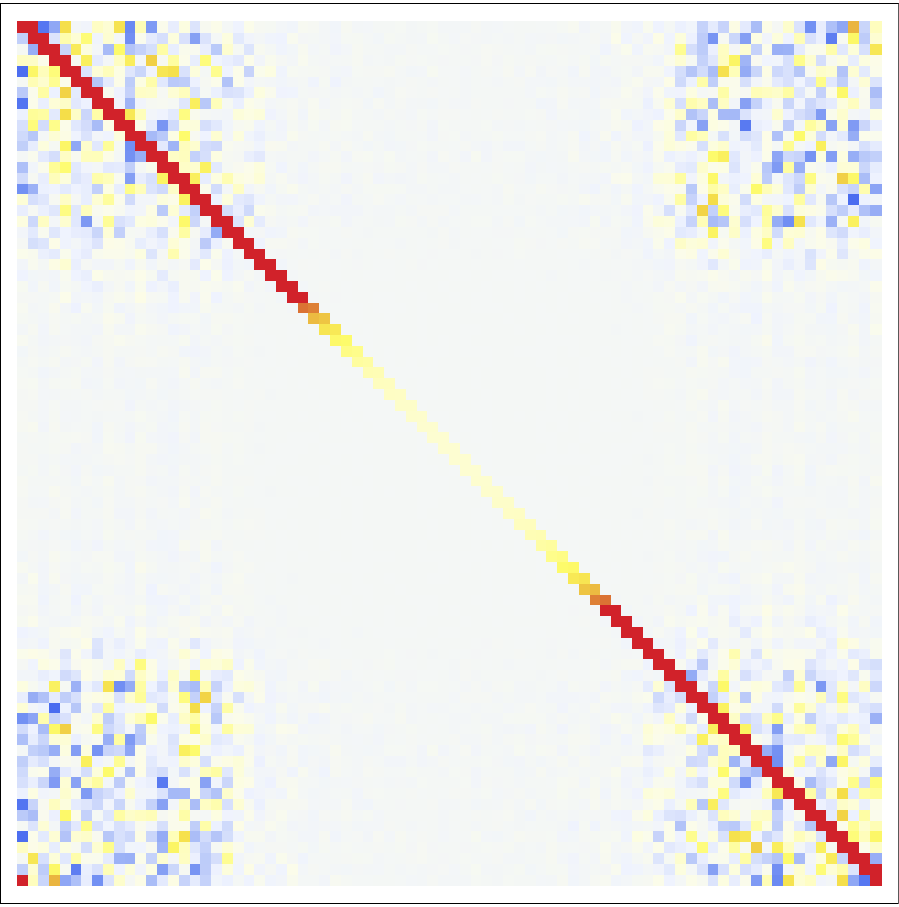}}
\scalebox{0.65}{\includegraphics{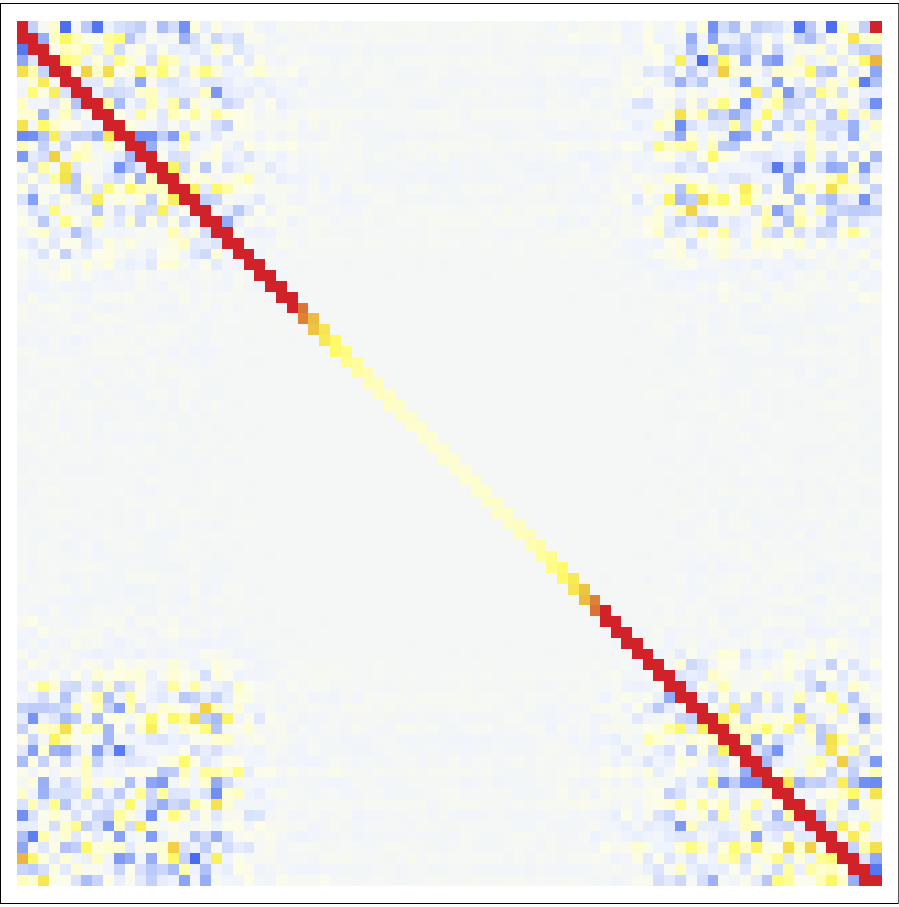}}
\scalebox{0.65}{\includegraphics{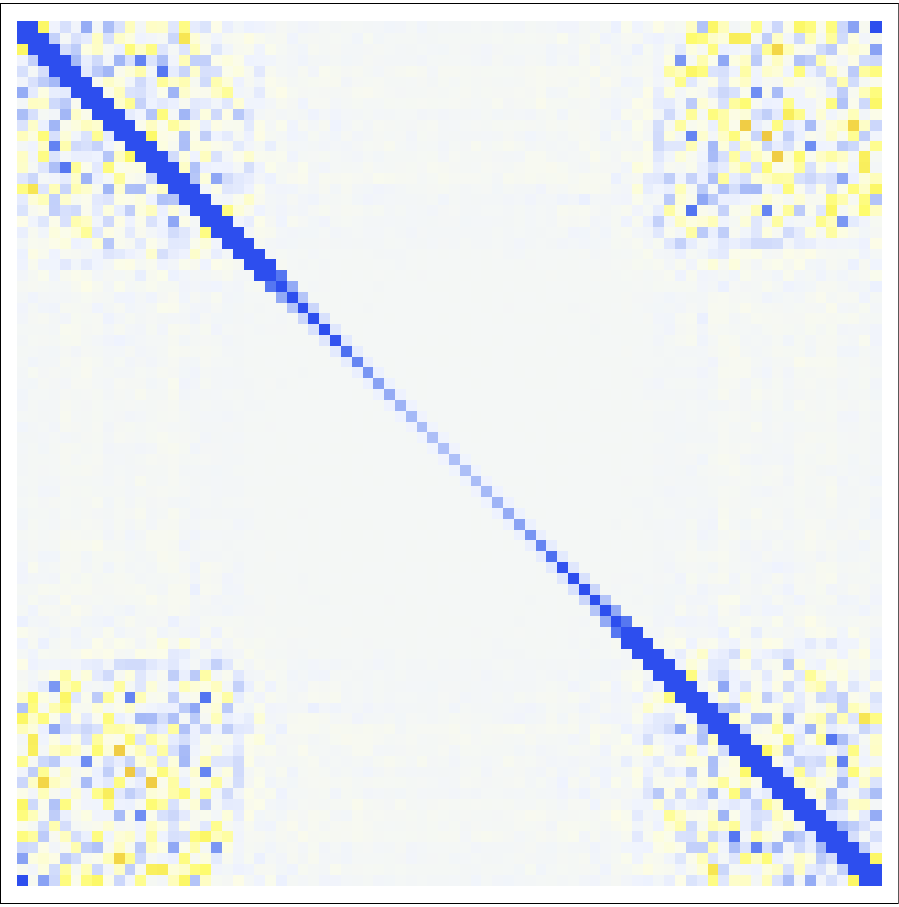}}
\caption{The empirical $\hat P_{11}$ (top left), $\hat P_{12}$ (top right), $\hat P_{21}$ (bottom left) and $\hat P_{22}$ (bottom right) fluctuation matrices measured for $\kappa_0=2.2$ and $\Delta=0.6$. Colours indicate shift from zero towards positive (blue) or negative (red) values.}
\label{FigLargeP}
\end{figure}
The empirical $\hat P_{11}$ matrix is diagonal, the $\hat P_{12}=\hat P_{21}^T$ matrix has negative diagonal and first super-diagonal, and the $\hat P_{22}$  has a tridiagonal structure with positive diagonal and nearest sub-/super-diagonals. This  suggests that  the quasi-local kinetic term couples not only  the nearest  \{4,1\} $\leftrightarrow$  \{3,2\} layers (at $t$ and $t\pm \oh$) but also directly the \{4,1\} $\leftrightarrow$  \{4,1\} layers (at $t$ and $t\pm 1$), the latter coupling with a {\it negative} sign.   There is no direct interaction between the  \{3,2\} $\leftrightarrow$  \{3,2\} layers. 
We propose the following discretized effective action for the combined \{4,1\} and \{3,2\} system:
\begin{eqnarray}\label{Seff4132}
S_{eff}^{(4132)}&= & \sum_t  \Big( K^{(1)}[n_{t+\oh} , n_t] + K^{(1)}[n_{t-\oh} , n_t ]+  V^{(1)}[n_{t+\oh} ]    \Big) \nonumber \\
& + & \sum_t  \Big( K^{(2)}[n_{t+1} , n_t] +  V^{(2)}[n_{t} ]    \Big) \ ,
\end{eqnarray}
where both kinetic and potential terms are given by a simple discretization of the minisuperspace action:

\begin{eqnarray}\label{Sefff4132coef}
 K^{(1)}[n , m] &= &  \frac{1}{\Gamma_1}  \frac{\left( n-m \right)^2}{ n+m} \left( 1 +  \chi_1  \left( \frac{ n-m}{ n+m}\right) + ... \right)  \ ,  \\
 K^{(2)}[n , m] &= & - \frac{1}{\Gamma_2}  \frac{\left( n-m \right)^2}{ n+m} \left( 1 +  \xi_2  \left( \frac{ n-m}{ n+m}\right)^2 + ... \right)\ ,  \label{xxx}  \\ 
 V^{(1)}[n] &=&  \frac{1}{ \Gamma_1}\Big(\mu_1 \, n^{1/3} - \lambda_1 \, n \Big) \ ,\\
 V^{(2)}[n] &=& - \frac{1}{ \Gamma_2} \Big( \mu_2 \, n^{1/3} - \lambda_2 \, n \Big)\ ,
\end{eqnarray}
and the overall negative sign of $ K^{(2)}[n , m]$ and  $V^{(2)}[n ]$ reflects the negative coupling mentioned above.
Taking into account the time reflection symmetry of the semiclassical solution (\ref{nttt}): $t_0 - \tilde t \equiv t_0+\tilde t$ (with $t_0=\frac{T}{2}+\oh$ and time periodic boundary conditions implying:  $\tilde t \equiv T+1 -\tilde t $ ), which we also imposed in the measured covariance matrix,   the blocks of the theoretical $P$ matrix take the form:
\begin{eqnarray}\label{P11}
 P_{11} &=& 
\scriptsize{
\left( {\begin{array}{*{20}{c}}
   d_{1}^{(1)}+u_1^{(1)} & 0 &0  & \cdots  &0 & 0 \\
   0& d_{2}^{(1)}+u_2^{(1)} &0 & 0 & \cdots   & 0 \\
   0 &0& d_{3}^{(1)}+u_3^{(1)} & 0 & \ddots    & \vdots \\
   \vdots & \ddots &  \ddots  &  \ddots  & \ddots   &0 \\
 0 &0 & 0  & \cdots  & 0 & d_{T}^{(1)}+u_T^{(1)} \\
\end{array}} \right)} 
\end{eqnarray}

\begin{eqnarray}\label{P12}
 P_{12} &=& 
\scriptsize{
\left( {\begin{array}{*{20}{c}}
   -k_{1}^{(1)} & -k_{T-1}^{(1)}  &0  & \cdots  &0 & 0 \\
   0&  -k_{2}^{(1)} & -k_{T-2}^{(1)}  & 0 & \cdots   & 0 \\
   0 &0& -k_{3}^{(1)} & -k_{T-3}^{(1)}   & \ddots    & \vdots \\
   \vdots & \ddots &  \ddots  &  \ddots  & \ddots   &  -k_{1}^{(1)}  \\
 -k_{T}^{(1)}   &0 & 0  & \cdots  & 0 & -k_{T}^{(1)}  \\
\end{array}} \right)} 
\end{eqnarray}

\begin{eqnarray}
P_{22} &=& 
\scriptsize{
\left( {\begin{array}{*{20}{c}}
   d_{1}^{(12)}+u_1^{(2)} & k_{1}^{(2)}  &0  & \cdots  &0 &  k_{T}^{(2)}  \\
  k_{1}^{(2)} &   d_{2}^{(12)}+u_2^{(2)} &  k_{2}^{(2)}  & 0 & \cdots   & 0 \\
   0 & k_{2}^{(2)}  &  d_{3}^{(12)}+u_3^{(12)} & k_{3}^{(2)}  & \ddots    & \vdots \\
   \vdots & \ddots &  \ddots  &  \ddots  & \ddots   &  k_{T-1}^{(2)} \\
 k_{T}^{(2)} &0 & 0  & \cdots  &  k_{T-1}^{(2)} &  d_{T}^{(12)}+u_T^{(2)}\\
\end{array}} \right)} 
\end{eqnarray}
where:

\begin{eqnarray}\label{kS1}
k_t^{(1)} & \equiv &  - \left . \frac{\partial^2 K^{(1)}[n_{t+\oh},n_{t}]}{{ \partial n_{t+\oh}\partial n_{t} }}\right|_{{n_{ \tilde t}}=\bar{n}_{\tilde t}}  \\
& = & \frac{8}{\Gamma_1} \cdot \frac{\bar n_{t+\oh} \cdot \bar n_{t}}  {\left(\bar  n_t+\bar n_{t+\oh}\right)^3} \left( 1 +  \chi_1  \left( \frac{ \bar n_{t+\oh}-\bar n_{t}}{\bar  n_{t+\oh}+\bar n_{t}}\right) + ... \right) \ ,\nonumber
\end{eqnarray}
and due to the overall negative sign of $ K^{(2)}[n , m]$ (Eq. \rf{xxx}) we have: 
\begin{eqnarray}\label{kS2}
k_t^{(2)} & \equiv & + \left . \frac{\partial^2 K^{(2)}[n_{t+1},n_{t}]}{{ \partial n_{t+1}\partial n_{t} }}\right|_{n_{\tilde t}=\bar{n}_{\tilde t}}  \\
& = & \frac{8}{\Gamma_2} \cdot \frac{\bar n_{t+1} \cdot \bar n_{t}}  {\left(\bar  n_t+\bar n_{t+1}\right)^3} \left( 1 +  \xi_2  \left( \frac{ \bar n_{t+1}-\bar n_{t}}{\bar  n_{t+1}+\bar n_{t}}\right)^2 + ... \right) \ .\nonumber
\end{eqnarray}
Diagonal kinetic elements are equal to:
\begin{eqnarray}\label{dS1}
d^{(1)}_{t} & \equiv & \left( \frac{\partial^2 K^{(1)}[n_{t+\oh},n_{t}]}{{ \partial n_{t+\oh}^2 } } + \frac{\partial^2 K^{(1)}[n_{t+\oh},n_{t+1}]}{{ \partial n_{t+\oh}^2 } } \right)_{{n_{ \tilde t}}=\bar{n}_{\tilde t}} \\
&= &   k^{(1)}_{t} \cdot \frac{\bar n_{t}}{\bar n_{t+\oh}}+ k_{T-t}^{(1)}   \cdot \frac{\bar n_{t+1}}{\bar n_{t+\oh}} \nonumber
\end{eqnarray} 
and
\begin{eqnarray}\label{dS12}
d^{(12)}_{t} & \equiv & \left( \frac{\partial^2 K^{(1)}[n_{t+\oh},n_{t}]}{{ \partial n_{t}^2 } } + \frac{\partial^2 K^{(1)}[n_{t-\oh},n_{t}]}{{ \partial n_{t}^2 } } \right)_{{n_{ \tilde t}}=\bar{n}_{\tilde t}} \nonumber \\
 & +& \left( \frac{\partial^2 K^{(2)}[n_{t+1},n_{t}]}{{ \partial n_{t}^2 } } + \frac{\partial^2 K^{(2)}[n_{t-1},n_{t}]}{{ \partial n_{t}^2 } } \right)_{{n_{ \tilde t}}=\bar{n}_{\tilde t}} \\
&= &   k^{(1)}_{t} \cdot \frac{\bar n_{t+\oh}}{\bar n_{t}}  +  k_{T+1-t}^{(1)}  \cdot \frac{\bar n_{t-\oh}}{\bar n_{t}}  -   k^{(2)}_t \nonumber  \cdot \frac{\bar n_{t+1}}{\bar n_{t}}  - k^{(2)}_{t-1} \cdot \frac{\bar n_{t-1}}{\bar n_{t}} \nonumber
\end{eqnarray} \ 
respectively, whereas the potential coefficients are given by: 
\beq\label{uS1}
u^{(1)}_t \equiv   \left. \frac{\partial^2 V^{(1)}[n_{t+\oh}]}{ \partial n_{t+\oh}^2 } \right|_{{n_{ \tilde t}}=\bar{n}_{\tilde t}}= -\frac{2 \mu_1}{9 \Gamma_1}  \cdot  \bar n_{t+\oh}^{\ -5/3} \ ,
\eeq
\beq\label{uS2}
u^{(2)}_t \equiv   \left. \frac{\partial^2 V^{(2)}[n_{t}]}{ \partial n_{t}^2 } \right|_{{n_{ \tilde t}}=\bar{n}_{\tilde t}}= + \frac{2 \mu_2}{9 \Gamma_2}  \cdot  \bar n_{t}^{\ -5/3} \ .
\eeq

By fitting the above expressions to the measured kinetic and potential elements one may check that indeed the proposed  form of the effective action (\ref{Seff4132}) describes the empirical data very well. In Fig. \ref{FigGammas} we present the best fits of parameters  $\Gamma_1$, $\Gamma_2$ for different choices of the bare coupling constants ($\kappa_0,\Delta$) inside the ``de Sitter" phase, along the lines approaching the phase transitions  (as for the parameter $\rho$ described above). To underline the signs of the respective couplings we plot $\Gamma_1$ and $-\Gamma_2$. For fixed $\Delta=0.6$, increasing $\kappa_0$ leads to smoothly rising  $\Gamma_1$ and $\Gamma_2$, which diverge when approaching the "A"-"C" phase transition line. For constant $\kappa_0=2.2$ we observe a rapid change in $\Gamma_2$ for $\Delta < 0.3$ which may suggest that the semiclassical solution ($\ref{nttt}$) ceases to give a good description of the spatial volume distribution for this choice of the bare coupling constants. The most recent results suggest that this region of the phase space may indeed show different properties - see Chapter  6 for details.
\begin{figure}[h!]
\centering
\scalebox{0.70}{\includegraphics{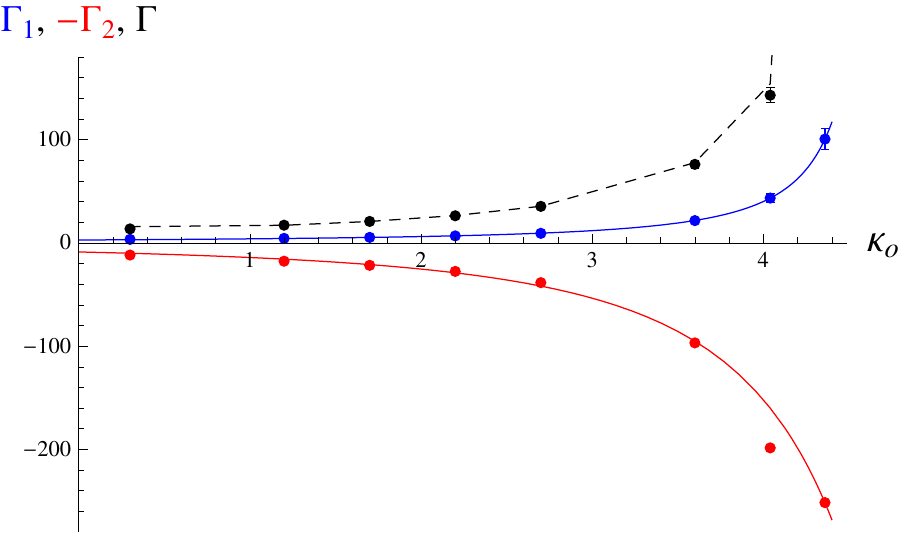}}
\scalebox{0.70}{\includegraphics{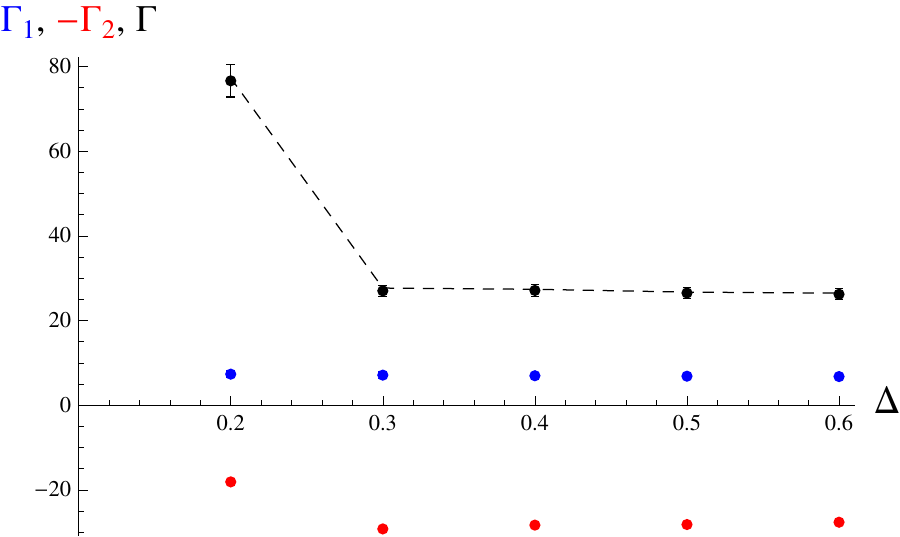}}
\caption{Dependence of the parameters $\Gamma_1$ (blue points) and $\Gamma_2$ (red points) on $\kappa_0$ for fixed $\Delta=0.6$ (left figure) and on $\Delta$ for fixed $\kappa_0=2.2$ (right figure). The parameters are defined by the ``finer grained" effective action (\ref{Seff4132}) for combined \{4,1\} and \{3,2\} simplices. To  underline the signs of the respective couplings we plot $\Gamma_1$ and $-\Gamma_2$. For comparison we also plot the parameter $\Gamma$ (black points) computed for the effective action (\ref{Sefff}), parametrized by the \{4,1\} simplices alone, and theoretical $\Gamma$ (black dashed line) obtained from the ``finer grained" model after integrating out the \{3,2\} fluctuations (Eq. (\ref{ggg})).}
\label{FigGammas}
\end{figure}

Since the ``new" effective action for the ``finer-grained" system describes the empirical CDT data as well as  the ``old" effective action 
 based on the data from integer $t$ spatial
slices alone  (derived in the previous section), one should ask if the ``old" action can be obtained from the ``new"  one  by integrating
out the \{3,2\} ``degrees of freedom". In this way the system described by the
integer-time slicing may be understood as arising from a ``Kadanoff blocking"
in time of the larger, ``finer-grained" system.
Demanding that integrating over the \{3,2\} fluctuations should give back the
\{4,1\} effective action leads, in a semiclassical approximation,  to the following relations between $\Gamma$, $\Gamma_1$, $\Gamma_2$ and $\mu$, $\mu_1$, $\mu_2$ :
\beql{ggg}
\frac{1}{\Gamma} = \frac{1}{2\Gamma_1}-\frac{1}{\Gamma_2} \ ,
\eeq
\beql{mmm}
\frac{\mu}{\Gamma}\left( \frac{1}{2\Gamma_1}-\frac{1}{\Gamma_2}\right) = \frac{\mu_1}{\Gamma_1}-\frac{\mu_2}{\Gamma_2} \ ,
\eeq
as well as relations between the other effective action parameters: $\lambda$, $\lambda_i$, $\xi_i$ and $\chi_i$. 
Using these relations one can check that the fits of the parameters of the ``finer-grained" effective action comprising both the \{4,1\} and the \{3,2\} simplices (\ref{Seff4132}) agree very well with the  parameters of the ``old" effective action (\ref{Sefff}) based on the \{4,1\} simplices alone - see Table \ref{Tab4132a} and Fig. \ref{FigGammas}.
\begin{table}[h!]
\begin{center}
 \begin{tabular} {|  c|c|c|}
\hline
\textbf{parameter} & \textbf{direct} & \textbf{integrated } \\ 
  &  (from \{4,1\}) &  (from  \{4,1\} and  \{3,2\}) \\ \hline
  $\Gamma $&$26.5\pm1.0$ &  $26.5\pm1.0$ \\ \hline
   $\mu$ &$20\pm2$ &  $16 \pm 2$ \\ \hline
\end{tabular}
\end{center}
\caption{Parameters of the effective action for integer $t$ spatial layers alone, measured for $\kappa_0=2.2$, $\Delta=0.6$ and $\bar N_{41}=160$k.}
\label{Tab4132a}
\end{table}

Let us briefly summarize and discuss these results. We have shown that it is possible to attribute a half-integer time label $t+\oh$ to the layers formed from the  \{3,2\} simplices interpolating between the spatial slices at integer  $t$ and $t+1$. Consequently, the \{3,2\} simplices may be used to construct a ``finer-grained" effective action comprising spatial volumes in both integer and half-integer discrete proper times. The empirical CDT data measured in the ``de Sitter" phase show that there is a direct minisuperspace type interaction between the adjacent  \{3,2\} and  \{4,1\} layers as well as between the nearest (in time) \{4,1\} layers, the former with a corrected ({\it positive}) sign and the latter with a {\it negative} sign (as in the original minisuperspace action). This  suggests that the \{3,2\} simplices play a role of a ``glue" stabilizing the whole ``de Sitter"-like structure, while the ``repelling" force between the \{4,1\} simplices acts the opposite way. In the ``de Sitter" phase the \{3,2\}$\leftrightarrow$\{4,1\} interaction prevails, and the overall ``\{3,2\}-integrated" behaviour is stable and consistent with the previously established effective action for integer $t$ layers. 
    
This picture can  also be used to qualitatively explain the  generic behaviour in the other two phases, where the \{3,2\} simplices   almost vanish (the ratio $\rho = N^{\{4,1\}}/  N^{\{3,2\}} \to \infty $ for  $N^{\{4,1\}}\to \infty$) causing $1/\Gamma_1\to 0$, and consequently the \{4,1\}$\leftrightarrow$\{4,1\} part prevails. As regards the ``A"-``C" phase transition, the coupling of the \{4,1\}$\leftrightarrow$\{4,1\}  kinetic term $-1/ \Gamma_2 \to 0$ as well, which promotes the causal ``disintegration" of the CDT universe into   disconnected spatial slices of integer $t$. For the "B"-"C" phase transition the picture is qualitatively different. The $-1/ \Gamma_2$ coupling stays finite and dominates, which results in a ``time collapsed" behaviour.
The exact nature of the effective action measured in the ``A" and ``B" phases and its relation to the phase transitions will be discussed in Chapter 5. 

\chapter{Transfer matrix method}
This Chapter is  based on the article: {\it J. Ambj\o rn, J. Gizbert-Studnicki, A.~G\"orlich, J. Jurkiewicz, ``The transfer matrix in four-dimensional CDT", JHEP 09 (2012)},  the last section is partly based on the article: {\it J.\,Ambj\o rn, J. Gizbert-Studnicki, A. G\"orlich, J. Jurkiewicz, ``The \ effective \ action \ in 4-dim CDT. The transfer matrix approach",  
JHEP 06 (2014)}
\newline
\newline

In the previous Chapter we have shown that the spatial  volume distribution (or alternatively the scale factor) of  four dimensional Causal Dynamical Triangulations is (in the ``de Sitter" phase ``C") very well described by a simple discretization of the minisuperspace action, with a reversed overall sign. This result was supported  by  a semiclassical analysis of quantum fluctuations around the average de Sitter solution. In such an approach  one could use the measured covariance matrix to compute the spatial volume fluctuation matrix. The problem with such an approach is three-fold. Firstly, in a semiclassical approximation the fluctuation matrix is defined by second derivatives of the effective action at the semiclassical solution. Therefore the effective action is obtained ``indirectly" - one can just make assumptions about its form and compare the   resulting theoretical fluctuation matrix with  empirical results. Secondly, the form of the effective action suggests that, in the physically interesting ``blob" region, the elements of the fluctuation matrix fall very fast with increasing volume (kinetic elements behave as $n_t^{-1}$ and potential elements as $n_t^{-5/3}$). As a result it is very difficult to observe any subleading corrections, especially for potential elements  (which ``deep" in the ``blob"  are indistinguishable from numerical noise). Last, but not least, due to the above, the   corrections of the effective action discussed in  Chapter 3.3 could mainly be measured in the ``medium" volumes region, close to the "stalk". In this region the semiclassical approximation is not as accurate as ``deep" in the ``blob", since some discretization effects characteristic of the ``stalk" are still visible. Therefore a natural question arises if there is any way to overcome these problems, i.e. to measure the effective action ``directly"  without resorting to the semiclassical approximation. In this Chapter we propose a way of doing this by defining an ``effective" transfer matrix, which can be measured in Monte Carlo simulations. We show that  results of the new method are (in the ``de Sitter" phase) fully consistent with the previous results, obtained from the fluctuation matrix. We also show that the new method can be used to explain volume distribution and correlations in the "stalk" and it is possible to construct the effective action for the "small" volume region. Finally, we show that the measured transfer matrix/effective action can be used to define a simplified  ``effective" model, based on volume fluctuations alone, which reconstructs the spatial volume measurements  obtained in the ``full CDT"  (with all degrees of freedom).

\section{Transfer matrix}
The quasi local form of the effective action (\ref{Sefff}) parametrized by spatial volumes of  integer $t$ (\{4,1\} simplices) leads in a natural way to the path-integral representation:
\beq\label{Zeff}
{\cal Z}_{eff}^{(T)} =\sum_{\substack{\{n_t\}, \\t = 1,...,T}}e^{-S_{eff}}\equiv \sum_{\substack{\{n_t\}, \\t = 1,...,T}} \prod_{t=1,...,T} e^{-L_{eff}[n_t,n_{t+1}]} =  \tr M^T \ ,
\eeq
where one takes into account time-periodic boundary conditions (with integer  $T$ period) and the {\it effective} transfer matrix $M$:
\beq\label{TMLag}
\bra{n_{t+1}}  M \ket{n_{t}}\propto  \exp(-L_{eff}[n_t,n_{t+1}])
\eeq
links the nearest (in integer $t$) spatial slices.\footnote{The symbol $M^T \equiv  \prod_{t=1,...,T} \braket{n_t|M|n_{t+1}} $ represents matrix multiplication, not transposition.} In the partition function (\ref{Zeff}) one disregards all details of the geometric structure at a given spatial slice  and looks only at the spatial volume observable $n_t$. Consequently one assumes it makes sense to use the ``effective" quantum states $\ket{n_t}$ with a unit norm as the eigenstate basis at each slice. In such an effective approach one can calculate a probability of measuring volume $n_t$ at the spatial slice $t$:
\beql{P1nt}
P^{(T)}(n_{t}) = \frac{1}{{\cal Z}_{eff}^{(T)} } \ {\braket{n_t | M^T | n_t}}
\eeq
and the probability of measuring a combination of $n_{t}$ and $n_{t'}$ separated by  $\Delta t = t' - t$:
\beql{P2nt}
P^{(T)}(n_{t},n_{t'}) = \frac{1}{{\cal Z}_{eff}^{(T)}} \ {\braket{n_t | M^{T-\Delta t} | n_{t'}} \braket{n_{t'} | M^{\Delta t} | n_t} }\ .
\eeq
Similarly the T-point probability distribution for the sequence of three-volumes $\{n_1,n_2, ... ,n_T\}$   is given by:
\beql{Pfullnt}
P^{(T)}(n_1,n_2, ... ,n_T) = \frac{1}{{\cal Z}_{eff}^{(T)}} \ { \langle n_1 | M | n_2 \rangle 
\langle n_2 | M | n_3 \rangle \cdots \langle n_{T} | M | n_1 \rangle } .
\eeq

It is important to note that, due to the imposed proper time foliation, Causal Dynamical Triangulations have a {\it genuine} transfer matrix $\cal M$ which relates a given spatial geometry ${\cal T}^{(3)}_t$ at time $t$ to a given spatial geometry ${\cal T}^{(3)}_{t+1}$ at time $t+1$. 
The transfer matrix $\cM$, i.e. the transition amplitude between the ${\cal T}^{(3)}_t$ and 
${\cal T}^{(3)}_{t+1}$, is   defined by  a sum over all four-dimensional triangulations ${\cal T}^{(4)}$ of a ``slab" between $t$ and $t+1$ (with boundary geommetries ${\cal T}^{(3)}_t$ and ${\cal T}^{(3)}_{t+1}$):
\beql{ja2}
\braket{ {\cal T}^{(3)}_{t+1}| \cM | {\cal T}^{(3)}_{t}} =
\sum_{{\cal T}^{(4)}} \ e^{-S_R[{\cal T}^{(4)}]} \ ,
\eeq
where ${S_R[{\cal T}^{(4)}]}$ is the Regge action of the four-dimensional triangulation of the "slab". 
The genuine transfer matrix is defined on a vector 
space spanned by the set of all three-dimensional 
triangulations ${\cal T}^{(3)}$, which is much wider than the abstract space spanned by ``effective" states $\ket{n_t}$ (the number of possible triangulations grows approximately   exponentially with $n_t$).
The transition amplitude for a three-dimensional triangulation $ {\cal T}^{(3)}$ 
to develop into a three-dimensional triangulation $ \widetilde{\cal T}^{(3)}$ after $T+1$
(integer) time steps is 
\beql{MTT}
\la  \widetilde{\cal T}^{(3)}  | \cM^{T+1} | {\cal T}^{(3)} \ra = \sum_{\{{\cal T}^{(3)}_t \}} \la \widetilde{\cal T}^{(3)} |\cM| {\cal T}^{(3)}_{T}\ra  
\la \cT^{(3)}_{T}|\cM |\cT^{(3)}_{T-1} \ra \cdots \la \cT^{(3)}_{1} |\cM|\cT^{(3)} \ra \ ,
\eeq
and the ``full CDT" partition function corresponding to $T$ time steps with periodic boundary conditions is given by:
\beql{ZFullCDT}
{\cal Z}^{(T)} = \sum_{\cT^{(3)} }\la \cT^{(3)}  | \cM^{T}|\cT^{(3)}\ra = \tr \cM^{T} .
\eeq
The ``effective" probability distributions of measuring spatial volumes  (\ref{P1nt})-(\ref{Pfullnt}) can now be calculated as:
\beql{P1ntFull}
P^{(T)}(n_{t}) = 
\frac{1}{{{\cal Z}^{(T)} }} \sum_{ \cTT \in  {\cTT(n_{t})}} \braket{ \cTT | \cM^{T}|\cTT } \ ,
\eeq
\beql{P2ntFull}
P^{(T)}(n_{t},n_{t'}) = 
\frac{1}{{{\cal Z}^{(T)} }} \sum_{ \substack{ \cTT_t \in  {\cTT(n_{t})} \\ \cTT_{t'} \in  {\cTT(n_{t'})}}} \braket{ \cTT_t | \cM^{T-\Delta t}|\cTT _{t'}} \braket{ \cTT_{t'} | \cM^{\Delta t}|\cTT_t }\ ,
\eeq 
\beql{PfullntFull}
P^{(T)}(n_{1},n_2,...,n_{T}) = 
\frac{1}{{{\cal Z}^{(T)} }} \sum_{ \substack{ \cTT_t \in  {\cTT(n_{t})}  \\ t=1,...,T}} \braket{ \cTT_1 | \cM|\cTT _2} ...\braket{ \cTT_T | \cM |\cTT_1 }\ ,
\eeq 
where  $\cTT(n_{t})$ denotes the subset  of 
three-dimensional triangulations $\cTT$ for which the spatial volume
is $n_{t}$. 

Comparing Eq's  (\ref{P1nt})-(\ref{Pfullnt}) and (\ref{P1ntFull})-(\ref{PfullntFull}) one can  formally define the ``effective" transfer matrix $M$ by:
\beql{Mnm}
|n\ra\la n|M|m\ra\la m| = \sum_{\substack{\cTT_n \in \cTT(n)\\ \cTT_m \in\cTT(m)}} 
\;|\cTT_n\ra \la \cTT_n | \cM|\cTT_m\ra \la \cTT_m| \ ,
\eeq
where $\la n|M|m\ra$ represents the 
average of the matrix elements $ \la \cTT_n | \cM|\cTT_m\ra $ 
\beq
\la n|M|m\ra = \la \cM \ra_{n,m} := \frac{1}{{\cN_n \cN_m}} 
 \sum_{\substack{\cTT_n \in \cTT(n)\\ \cTT_m \in\cTT(m)}} 
\; \la \cTT_n | \cM|\cTT_m\ra  \ ,    
\eeq
and $\cN_n$ denotes the cardinality of $\cTT(n)$. 
In the above one should not think of the ``effective" state $|n\ra$ as
the (suitably normalized) sum of the $\cN_n$ vectors $|\cTT_n\ra$, which would again be a vector located in a (much wider) Hilbert space spanned by the $|\cTT_n\ra$'s. One should rather
think of the state  $\ket{n}$ 
as  arising from a  {\it probability distribution} of states
$|\cTT_n \ra$. 

The statement that one can use the matrix $\la n|M|m\ra$ as an 
``effective" transfer matrix assumes that the standard deviation
of the $\cN_n\cN_m$ numbers $\la \cTT_n | \cM|\cTT_m\ra  $ is sufficiently small, and consequently
\beq\label{consistency}
\la \cTT_{t+N}| \cM^N |\cTT_t\ra \sim \la n_{t+N}| M^N |n_t\ra,
\eeq
for $N = 1,2,...T$, at least if we look only at the spatial volume distributions.
In the following we will assume that the "effective" transfer matrix $M$ exists and use it to analyze the empirical (computer generated) data. The consistency of this analysis will provide indirect evidence that the approach is correct. In particular we would like to check if one can use the partition function (\ref{Zeff}) defined by the "effective"  $M$ to reproduce the volume profile $\braket{n_t}$  and quantum fluctuations 
$\delta n_t$ of the "full CDT" model defined by the ``genuine" $\cM$ (\ref{ZFullCDT}).


\section{Transfer matrix measurement and analysis method}

In the last section we argued that there exists an ``effective" transfer matrix which can be used to explain the four-dimensional CDT spatial volume distribution in  phase ``C". Now we want to show how the transfer matrix can be measured in computer simulations. Let us consider CDT systems with short length of the (periodic) proper time axis ($T = 2,3,4,...$). From Eq's. (\ref{Zeff}) and  (\ref{P2nt}) it  follows that for  $T=2$ the probability of measuring spatial volume $n_1$ at time $t=1$ and $n_2$ at time $t=2$ is given by:
\beql{P2T2}
P^{(2)}(n_{1},n_{2}) = \frac{ \ {\braket{n_1 | M | n_{2}} \braket{n_{2} | M | n_1} }}{{\tr M^2}}\ .
\eeq
Due to time reflection symmetry, the transfer matrix $M$ is symmetric and therefore the measured probability distribution (\ref{P2T2}) can be used to compute the transfer matrix elements.  Up to a normalization one gets:
\beql{MT2}
\braket{n|M|m}=\cN_0 \sqrt{P^{(2)}(n_{1}=n,n_{2}=m)} \ .
\eeq

This method required a major change in the Monte Carlo code  used in our numerical simulations which originally assumed $T\geq 3$. This was due to the problem with an artificial doubling of (sub)simplices  which our  program automatically rejects (the problem does not occur for time extension above two time slices). The code was updated to allow $T=2$ and was used for data analysis inside the ``A" and ``B" phases described in the next Chapters. We checked that inside the ``de Sitter" phase  results obtained using the new code perfectly agree with  results from the original version. Numerical data presented in this Chapter were obtained using the original code (with $T\geq 3$).  In this case, to measure the ``effective" transfer matrix one can use a combination of probability distributions  for $T=3$ and $T=4$:
\beql{P2T3}
P^{(3)}(n_{1},n_{2}) = \frac {\braket{n_1 | M | n_{2}} \braket{n_{2} | M^2 | n_1} } {\tr M^3} \ ,
\eeq 
\beql{P2T4}
P^{(4)}(n_{1},n_{3}) = \frac {\braket{n_1 | M^2 | n_{3}} \braket{n_{3} | M^2 | n_1} } {\tr M^4} \ ,
\eeq 
which again up to normalization give:
\beql{MT34}
\braket{n|M|m}=\cN_0  \frac{{P^{(3)}(n_{1}=n,n_{2}=m)}}{\sqrt{P^{(4)}(n_{1}=n,n_{3}=m)}} \ .
\eeq
One could as well use for example  $T=4$ and $T=6$ for which:
\beql{P2T4b}
P^{(4)}(n_{1},n_{2}) = \frac {\braket{n_1 | M | n_{2}} \braket{n_{2} | M^3 | n_1} } {\tr M^4} \ ,
\eeq 
\beql{P2T6}
P^{(6)}(n_{1},n_{4}) = \frac {\braket{n_1 | M^3 | n_{4}} \braket{n_{4} | M^3 | n_1} } {\tr M^6} \ , 
\eeq 
and
\beql{MT46}
\braket{n|M|m}=\cN_0  \frac{{P^{(4)}(n_{1}=n,n_{2}=m)}}{\sqrt{P^{(6)}(n_{1}=n,n_{4}=m)}} \ .
\eeq
We checked that the measurements of $P^{(2)}(n_{1},n_{2})$, as well as  $P^{(4)}(n_{1},n_{2})$ and $P^{(6)}(n_{1},n_{4})$, lead  to the same matrix $M$ as extracted from $P^{(3)}(n_{1},n_{2})$ and $P^{(4)}(n_{1},n_{3})$, which we finally used. The consistency of the results measured for different choices of $T=2$, $T=3,4$ and $T=4,6$ is a strong argument that Eq. (\ref{consistency}) holds and consequently the ``effective" transfer matrix exists.

As explained in Chapter 1.5, in order to perform computer simulations efficiently one should introduce some volume fixing method. In the last Chapter we used a global volume fixing potential (\ref{SVFGQ}) which was added to the bare Regge action. This type of the volume fixing is incompatible with the transfer matrix structure. Therefore for the transfer matrix measurement we changed it to the {\it local} volume fixing term:
\beql{LocalVF}
 S_R \quad \to \quad   \widetilde S_R = S_R+ \eps \sum_{t=1}^T(n_t - \bar N_3)^2 \ .
\eeq
The new volume fixing constraint  drastically changes the average volume profile $\braket{n_t}$ since now $n_t$ fluctuates around the universal $\bar N_3$ for each $t$. Accordingly we will measure different probability distributions  $\widetilde P^{(T)}(n_1,n_2,...,n_T) $  and different transfer matrices $\widetilde M$ than in the ``pure" CDT model without the volume fixing. Similar to Eq. (\ref{Pfullnt}) the new probability distribution of measuring the sequence of spatial-volumes $\{n_1,n_2,...,n_T\}$ is given by:
\beql{PfullntN}
\widetilde P^{(T)}(n_1,n_2, ... ,n_T) = \frac{1}{\tr \widetilde M^T} \ { \langle n_1 | \widetilde M | n_2 \rangle 
\langle n_2 | \widetilde M | n_3 \rangle \cdots \langle n_{ T} |\widetilde M | n_1 \rangle } .
\eeq
It is directly related to the probability distribution (\ref{Pfullnt}) in the ``pure" model:
\begin{equation}\label{PfullRelation}
\widetilde{P}^{(T)}(n_1, n_2, ..., n_T)  \propto {P}^{(T)}(n_1, n_2, ..., n_T) 
e^{-\epsilon (n_1 - \bar N_3)^2} \cdots 
e^{-\epsilon (n_{T} - \bar N_3)^2}.
\end{equation}
One can calculate the transfer matrix $\widetilde{M}$ in the same way as $M$:
\begin{equation}\label{TMNew}
\langle n | \widetilde M | m \rangle =  \cN_0
\frac{\widetilde{P}^{(3)} (n_1=n, n_2=m)}{\sqrt{\widetilde{P}^{(4)}(n_1=n, n_3=m)}} \ ,
\end{equation}
and  use it  to obtain the ``pure" transfer matrix $M$  by cancelling  the volume fixing term.
From Eq's (\ref{LocalVF})-(\ref{PfullRelation}) one gets:
\begin{equation}\label{MPure}
	\braket{ n | M | m } = e^{\frac{1}{2}\epsilon (n -\bar N_3)^2} 
\braket{ n | \widetilde M | m } e^{\frac{1}{2}\epsilon (m - \bar N_3)^2}.
\end{equation}

\begin{figure}[h!]
\centering
\scalebox{.8}{\includegraphics{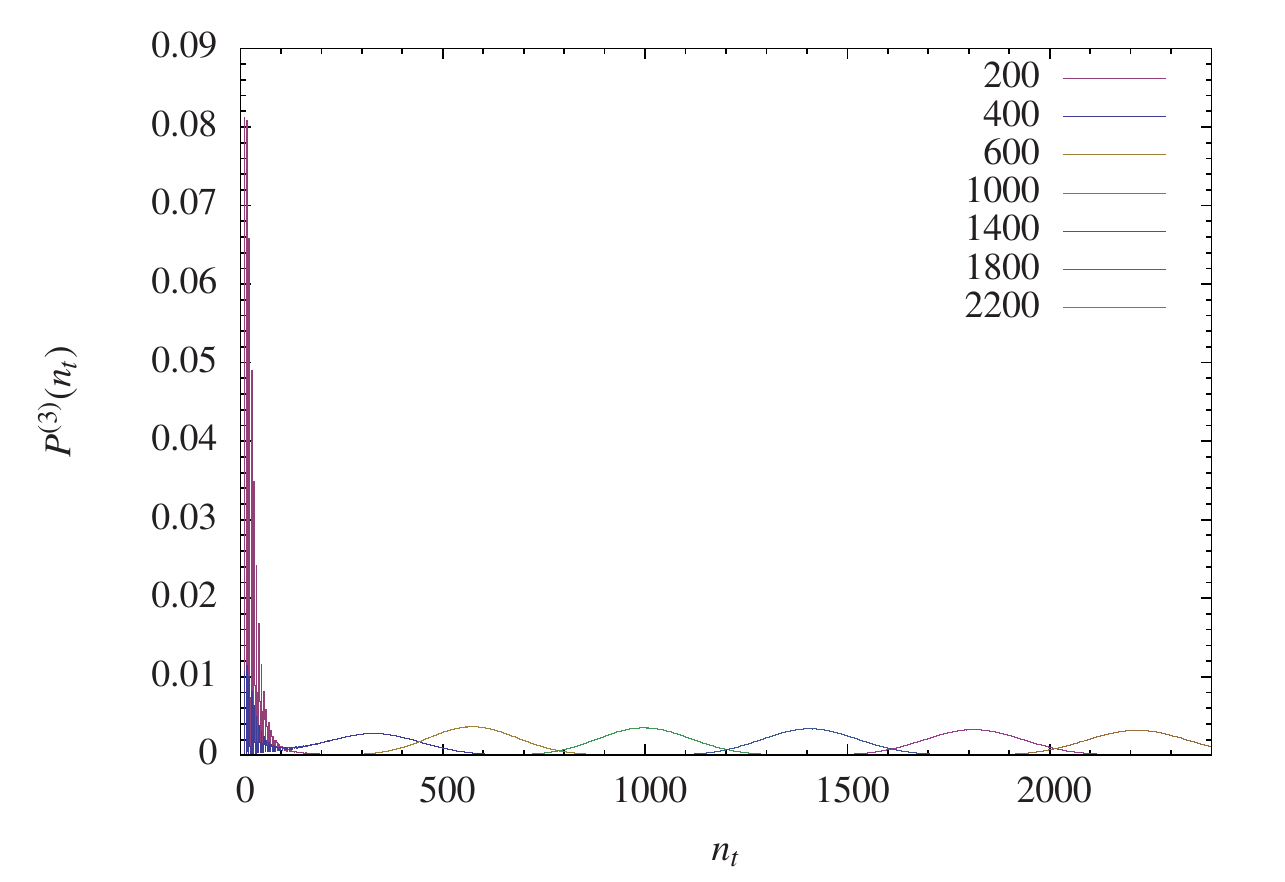}}
\caption{Probability distributions (histograms) of the volume $n_t$, for various $\bar N_3 =
200, 400, 600, 1000, 1400, 1800$ and $2200$ (from left to right) and fixed $\eps=0.00002$. For all ranges, the
simulations were performed for $\kappa_0=2.2$, $\Delta=0.6$ and $K_4=0.922$.}
\label{FigBlob-hist-series}
\end{figure}

For a given choice of $ \bar N_3$ one observes fluctuations of $n_t$ in some range around $\bar N_3$ with an amplitude controlled by $\eps$ (see Fig. \ref{FigBlob-hist-series}) and can  use the $\widetilde P^{(3)} (n_1, n_2)$ and $\widetilde P^{(4)} (n_1, n_3)$ empirical  data (histograms) to measure the transfer matrix elements in this range according to Eq's (\ref{TMNew}) and (\ref{MPure}).  To reconstruct the matrix in 
a larger region of the $n$-space one can merge data from  
different $\bar N_3$ regions. This is possible if one chooses $\eps$ and $ \bar N_3$'s in such a way that the successive ranges  overlap and there is a non-vanishing intersection of the transfer matrix elements  for different  $ \bar N_3$'s. The transfer matrix measured in each range is determined up to a normalization $\cN_0$. Therefore one can scale the matrices to get agreement between matrix elements in the intersection regions. In order to determine the appropriate scaling we look at the diagonal elements of the empirical transfer matrices (such elements  are measured with the highest precision). We start with the ``first" empirical transfer matrix (measured for the smallest   $\bar N_3$) and the ``second"  matrix (for next to the smallest  $\bar N_3$). We scale the ``second" matrix, so that the mean value of the diagonal elements in the intersection region is equal for both matrices. Then we repeat the procedure for the ``second" and ``third" transfer matrix and so on. Finally we get scaled transfer matrices for all regions  which can be simply merged (see Fig. \ref{FigMerged}).
\begin{figure}[h!]
\centering
\scalebox{.5}{\includegraphics{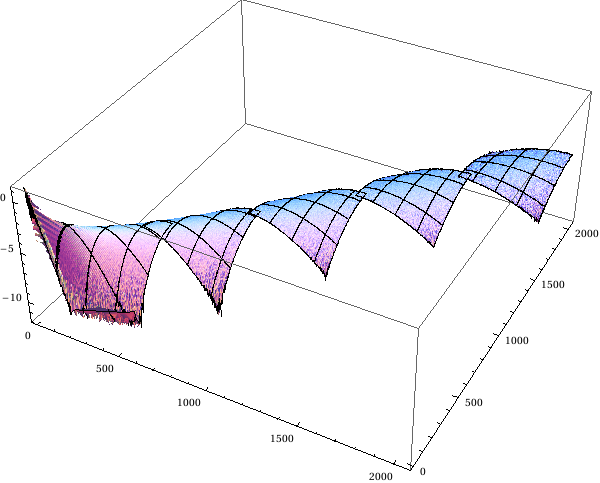}}
\caption{Logarithm of scaled and merged  transfer matrices $M$ measured for different $\bar N_3$ for $\kappa_0=2.2$, $\Delta=0.6$ and $K_4=0.922$.}
\label{FigMerged}
\end{figure}

The  transfer matrix can be used to determine the form of the effective action. From Eq. (\ref{TMLag}):
\beq\label{TMLag1}
\braket{n | M | m } = \cN  \exp(-L_{eff}[n,m]) \ ,
\eeq
hence up to a constant the effective Lagrangian is defined through:
\beql{LeffTM}
L_{eff}[n,m] =-  \log \braket{n | M | m } +\log \cN \ .
\eeq
As before, one can assume some form of the theoretical   action/Lagrangian and try to fit it to the empirical transfer matrix data. Alternatively one can  analyze the measured transfer matrix in a systematic way to {\it derive} the effective Lagrangian, as will be explained below. 

The form of the Lagrangian  determined from the covariance of spatial volume fluctuations  (see Chapter 3):
\beql{Lagreff1}
L[n,m] = \frac{1}{\Gamma}\left[ \frac{(n-m)^2}{n+m} + \mu \left(\frac{n+m}{2}\right)^{1/3} - \lambda \left(\frac{n+m}{2}\right)  \right]
\eeq
suggests that the transfer matrix can be factorized into potential and kinetic parts
\begin{equation}
\braket{n|M|m}= {\cal  N} \underbrace{ \exp  \bigg( -v[n+m] \bigg) }_{potential} \underbrace{\exp \bigg( -\frac{(n-m)^2}{k[n+m]}\bigg) }_{kinetic} \ ,
\label{SM}
\end{equation}
where  the functions: 
\begin{equation}
v[n+m]=\frac{\mu}{\Gamma} \left(\frac{n+m}{2}\right)^{1/3} - \frac{\lambda}{\Gamma}  \left( \frac{n+m}{2} \right) \ ,
\label{vnm}
\end{equation}
\begin{equation}
k[n+m] = \Gamma \! \cdot \! (n+m) 
\label{knm}
\end{equation}
will be called the potential and kinetic coefficients, respectively.
In the above expressions we disregard for the moment
the $\xi_2$ correction of the kinetic term (see Chapter 3.3). It will be discussed in the next section.  
We also use a symmetric potential $v[n+m]$ instead of the diagonal potential $V[n] + V[m]$ which better suites to the transfer matrix analysis method. The impact of such change is discussed in detail in Chapter 5.1. We want to determine the kinetic $k[.]$ and potential $v[.]$  functions from the empirical data and check if one can see any corrections to the expressions (\ref{vnm}) and (\ref{knm}). 

The {\it potential} coefficients  can be measured by looking at the {\it  diagonal} 
elements of the transfer matrix ($n=m$), for which the kinetic term vanishes:
\begin{equation}
v[2n] = -\log{\braket{n|M|n}}+\log{\cal N} \ .
\label{Vc}
\end{equation}
It can be measured in a wide range of $n$ if one uses the properly merged transfer matrix data, as was explained above.

The {\it kinetic} term requires extracting the {\it cross-diagonal} elements (for constant $s =n+m$). For fixed $s$ one should get:
\begin{equation}
\braket{n|M|s-n}= {\cal N}(s)\exp \left( -\frac{(2n-s)^2}{k[s]}\right) \ ,
\label{Skin}
\end{equation}
and $k[s]$ can be determined from a Gaussian fit. Again, $k[s]$ can be measured for different cross-diagonals (in a wide range of  $s$).


\section{Transfer matrix for large three-volumes}

We  measured the probability distributions (histograms) $\widetilde P^{(3)} (n_1, n_2)$ and \newline $\widetilde P^{(4)} (n_1, n_3)$  in a generic point inside the ``de Sitter" phase, for: $\kappa_0=2.2$, $\Delta=0.6$ ($K_4=0.9222$), for $\eps = 0.00002$ and a choice of $\bar N_3=200, 400, 600, 1000$, $1400, 1800$ and $2200$. We used Eq's (\ref{TMNew}) and (\ref{MPure}) to calculate transfer matrix elements in each range separately. For small $\bar N_3 = 200,400$ the measured transfer matrices are dominated by strong discretization effects, which will be discussed in the next section. For $\bar N_3 \geq 600$ the empirical matrices $M$ are smooth up to  numerical noise and
can be fitted very well  by:
\beql{Mth}
\braket{n | M^{(th)} | m } = \cN  \exp(-L_{eff}[n,m])
\eeq
with the effective Lagrangian $L_{eff}[n,m]$ given by Eq. (\ref{Lagreff1}). In Fig. \ref{FigTMBlob} we present  empirical $M$ measured for $\bar N_3=1400$ (left) and  the difference between the empirical $M$ and the theoretical $M^{(th)}$ (right) which is just  small numerical noise. 
The  best fits of the parameters $\Gamma$, $\mu$ and $\lambda$ for different $\bar N_3$'s are presented in Table \ref{table1}.

\begin{figure}[h!]
\centering
\scalebox{.3}{\includegraphics{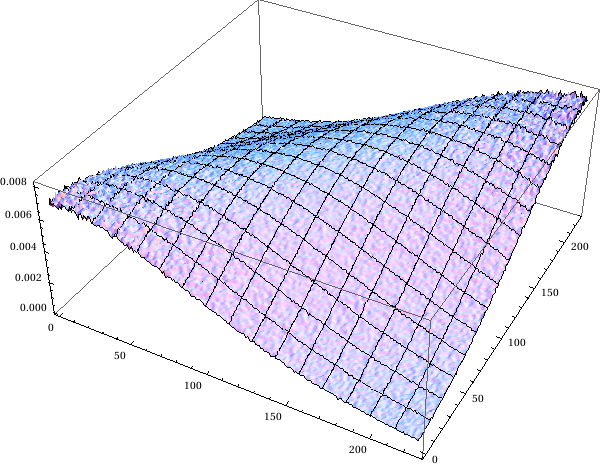}}
\scalebox{.3}{\includegraphics{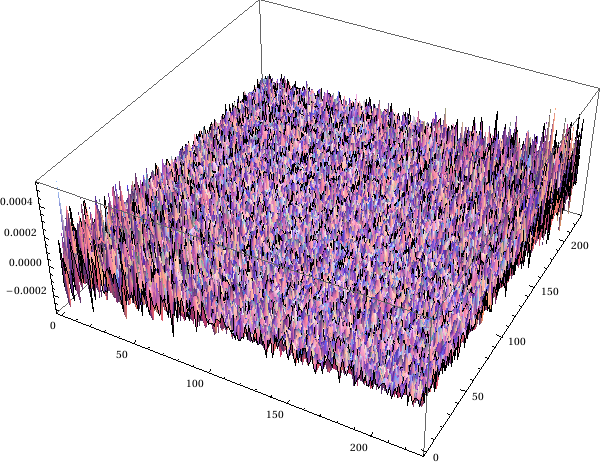}}
\caption{Left: The transfer matrix $M$ measured for $\bar N_3 = 1400$ in the range  $ 1200 \leq n_t \leq
1600$. The matrix is optically indistinguishable from the fitted theoretical transfer
matrix $M^{(th)}$ (Eq's (\ref{Mth}) and (\ref{Lagreff1})). Right: The difference between $M$ and $M^{(th)}$ disappears
in the numerical noise. The measurements were performed for $\kappa_0=2.2$, $\Delta=0.6$ and $K_4=0.922$.}
\label{FigTMBlob}
\end{figure}

\begin{table}[h!]
\begin{center}
	\begin{tabular}{|c|c|c|c|c|}
\hline
$\bar N_3$ & $n_t$ range	& $\Gamma$ & $\mu$ & $\lambda$ \\
\hline
\hline
$600$	& $ 400 - 820$	& $25.7 \pm 0.1$	& $18 \pm 1$	& $0.05 \pm 0.01$ \\ \hline
$1000$	& $ 780 - 1220$	& $26.0 \pm 0.1$	& $17 \pm 1$	& $0.05 \pm 0.01$ \\ \hline
$1400$	& $1180 - 1630$	& $26.1 \pm 0.1$	& $13 \pm 1$	& $0.04 \pm 0.01$ \\ \hline
$1800$	& $1580 - 2040$	& $26.1 \pm 0.1$	& $26 \pm 1$	& $0.07 \pm 0.01$ \\ \hline
$2200$	& $1980 - 2440$	& $26.1 \pm 0.1$	& $19 \pm 2$	& $0.05 \pm 0.01$ \\ \hline
	\end{tabular}
\end{center}
\caption{The values of $\Gamma, \mu$ and $\lambda$  for different $\bar N_3$, obtained from best fits
of $M^{(th)}$ to the measured $M$.}
	\label{table1}
\end{table}
To underline the quality of the fits we can use a spectral decomposition of the measured and  theoretical transfer matrices. In Table \ref{tableeigenblob}  we present the first four eigenvalues and in Fig. \ref{vec14} the corresponding eigenvectors of $M$ and $M^{(th)}$, respectively. We use data measured for $\bar N_3=1400$ and  $M^{(th)}$ is calculated for the best fits of parameters $\Gamma$, $\mu$ and $\lambda$ from Table \ref{table1}. The empirical and theoretical data overlap almost perfectly.
\begin{table}
\begin{center}
\begin{tabular} {|c|c|c|}
\hline
$i$ & $\lambda_i $	& $\lambda_i^{(th)}$  		\\ \hline 
\hline
$ 1 $	& $ 1.168 \times 10^{0} $	& $ 1.168 \times 10^{0}  $ 	\\ \hline
$ 2 $	& $ 3.824 \times 10^{-1} $	& $ 3.828 \times 10^{-1} $ 	\\ \hline
$ 3 $	& $ 7.061 \times 10^{-1} $	& $ 7.074 \times 10^{-1} $ 	\\ \hline
$ 4 $	& $ 8.543 \times 10^{-2} $	& $ 8.640 \times 10^{-2} $ 	\\ \hline
\end{tabular}
\end{center}
\caption{
The first four eigenvalues of the measured  transfer matrix
$ M $ calculated for $\bar N_3 = 1400$ and the similar eigenvalues for the fitted transfer matrix $ M^{(th)}$.}
\label{tableeigenblob}
\end{table}
\begin{figure}[!h]
\centering
\scalebox{0.5}{\includegraphics{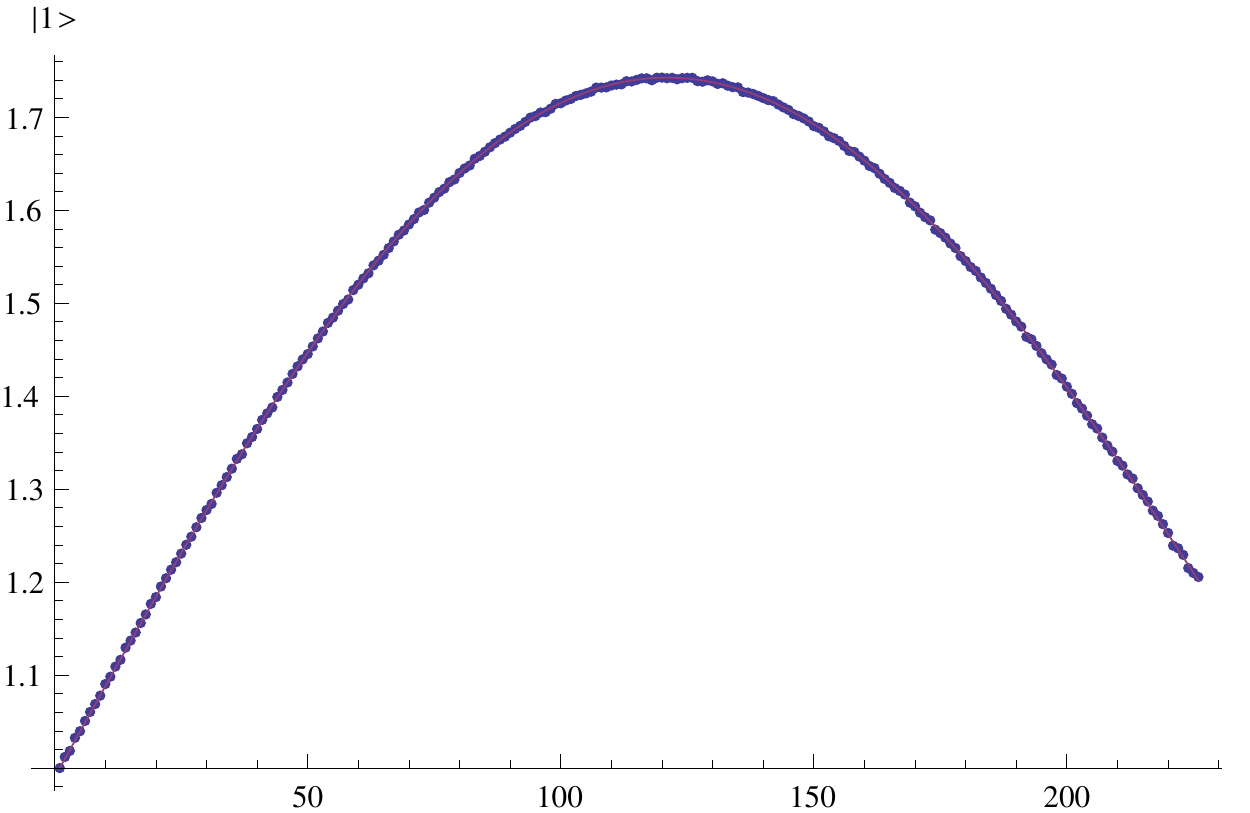}}
\scalebox{0.5}{\includegraphics{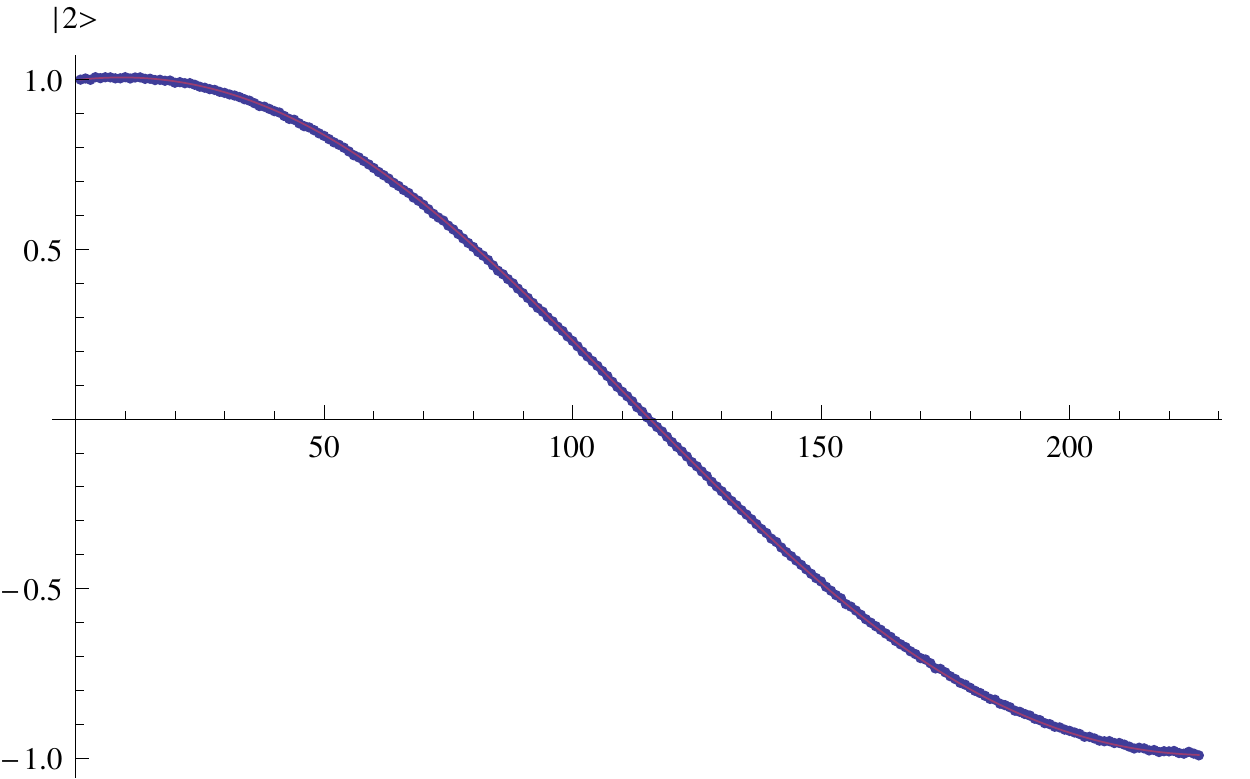}}
\scalebox{0.5}{\includegraphics{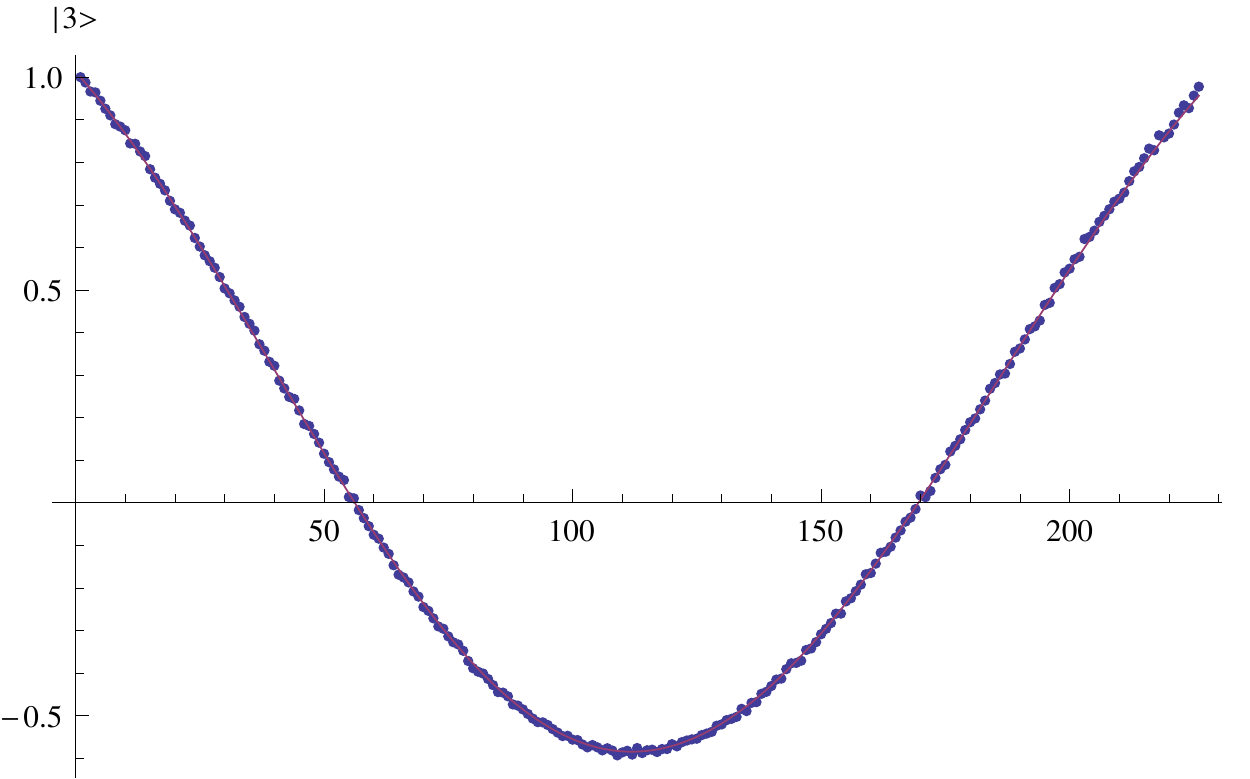}}
\scalebox{0.5}{\includegraphics{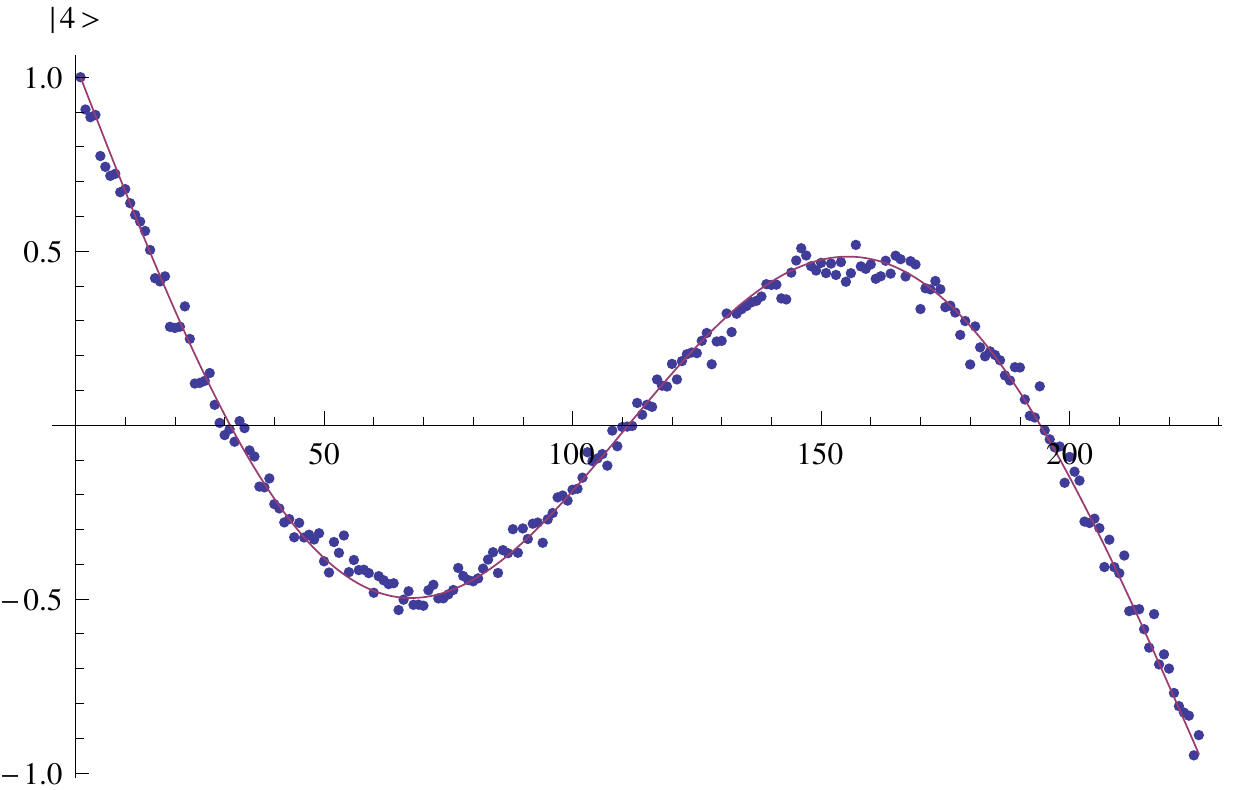}}
\caption{The first four eigenvectors of the measured  transfer matrix
$ M $ calculated for $\bar N_3 = 1400$ (dots) and the first four eigenvectors of   the fitted theoretical transfer matrix $ M^{(th)}$ (lines). }
\label{vec14}
\end{figure}

In order to get a better  estimation of the parameters  and to check if one can see any corrections of the effective action we will now analyze  the kinetic and potential parts separately.  First, we perform the scaling procedure described in the previous section to achieve a  merged empirical transfer matrix obtained from individual matrices measured for different $\bar N_3$.
We assume the following effective Lagrangian:
\beql{LeffC}
L_{eff}[n,m] =  \frac{(n-m)^2}{k[n+m]} + v[k+m]
\eeq 
and want to find a  functional form of $k[.]$ and $v[.]$ using the merged empirical  data.

\begin{figure}[h!]
\centering
\scalebox{.8}{\includegraphics{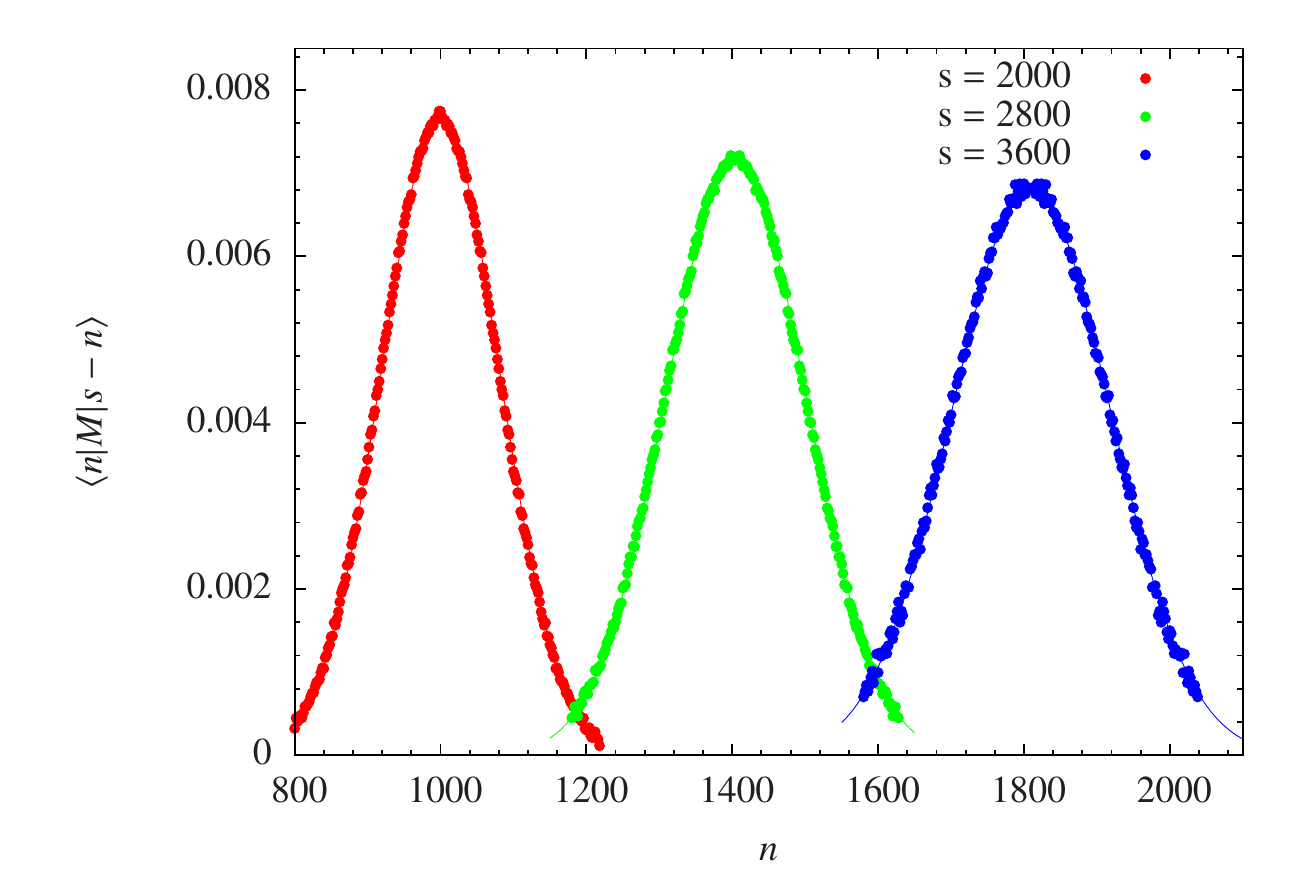}}
\caption{Cross-diagonals of the measured transfer matrix $\braket{n|M|s-n}$ plotted as a function of $n$ for different $s=2000, 2800$ and $3600$. Lines correspond to the Gaussian fits. The transfer matrix was measured for $\kappa_0=2.2$, $\Delta=0.6$ and $K_4=0.922$.}
\label{FigBlobAnti}
\end{figure}

Let us start with the kinetic part which is defined by the cross-diagonal elements of the transfer matrix (for constant $s=n+m$). For fixed $s$ one obtains the expected Gaussian dependence on $n$, according to Eq. (\ref{Skin}):
$$
\braket{n|M|s-n}= {\cal N}(s)\exp \left( -\frac{(2n-s)^2}{k[s]}\right) .
$$
The above expression fits  the empirical data very well, at least if one takes $s$ large enough not to observe small volume discretization effects ($s\geq 400$) - see Fig. \ref{FigBlobAnti} in which we present  empirical cross-diagonals for $s= 2000, 2800$ and $3600$ together with the Gaussian fits. The  kinetic coefficients $k[s]$ can be estimated from the  fits. We expect the $k[s]$ to be linear:
\beql{ks}
k[s] = \Gamma (s - 2 \, n_0) \ ,
\eeq
where we have added a (small) constant shift $2\, n_0$ to better fit the empirical data.  Measured $k[s]$ as a function of $s$ is plotted in Fig. \ref{FigBlobKin}. The relation is indeed linear as expected and $\Gamma$ is common in all ranges. The  best fit yields: $\Gamma=26.07\pm0.02$ and $n_0=-3\pm1$. $\Gamma$ is consistent with the results obtained in each range of volume fluctuations separately (see Table \ref{table1}) and $n_0$ is very close to zero and thus negligible for large $s$. 

\begin{figure}[h!]
\centering
\scalebox{.85}{\includegraphics{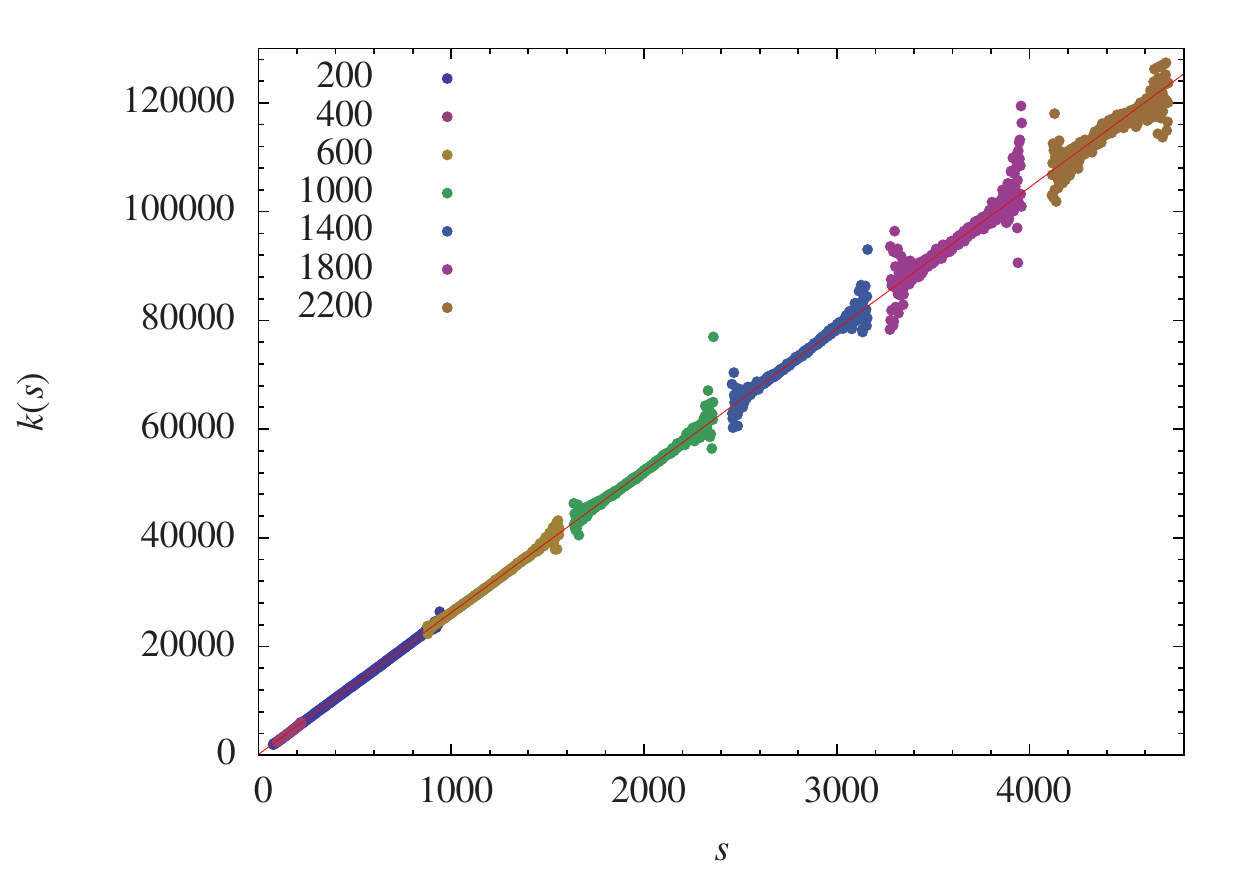}}
\caption{Measured kinetic coefficients $k[s]$ as a function of $s=n+m$ (different colours indicate different ranges of $n_t$ fluctuations) and a linear fit of Eq. (\ref{ks}) (red line). }
\label{FigBlobKin}
\end{figure}

The potential part is defined by a diagonal of the transfer matrix (for $n=m$). According to Eq. (\ref{Vc}):
$$
\log{\braket{n|M|n}} = \log{\cal N} - v[2n] \ .
$$
Inspired by the earlier results we expect:
\beql{v2n}
v[2n] = \frac{\mu}{\Gamma} n^{1/3} - \frac{\lambda}{\Gamma} n \ ,
\eeq
again if we take  the volume $n$ large enough not to see the finite-size discretization effects (for $n \geq 400$). In Fig. \ref{FigBlobPot} we plot the measured diagonal $\log \braket{n|M|n}$ together with the  fit of Eq. (\ref{v2n}). The best fit of the parameters $\mu$ and $\lambda$ (for $\Gamma = 26.07$ fixed by the kinetic part) yields: $\mu =16.5\pm0.2$, $\lambda=0.049\pm0.001$.

\begin{figure}[h!]
\centering
\scalebox{.85}{\includegraphics{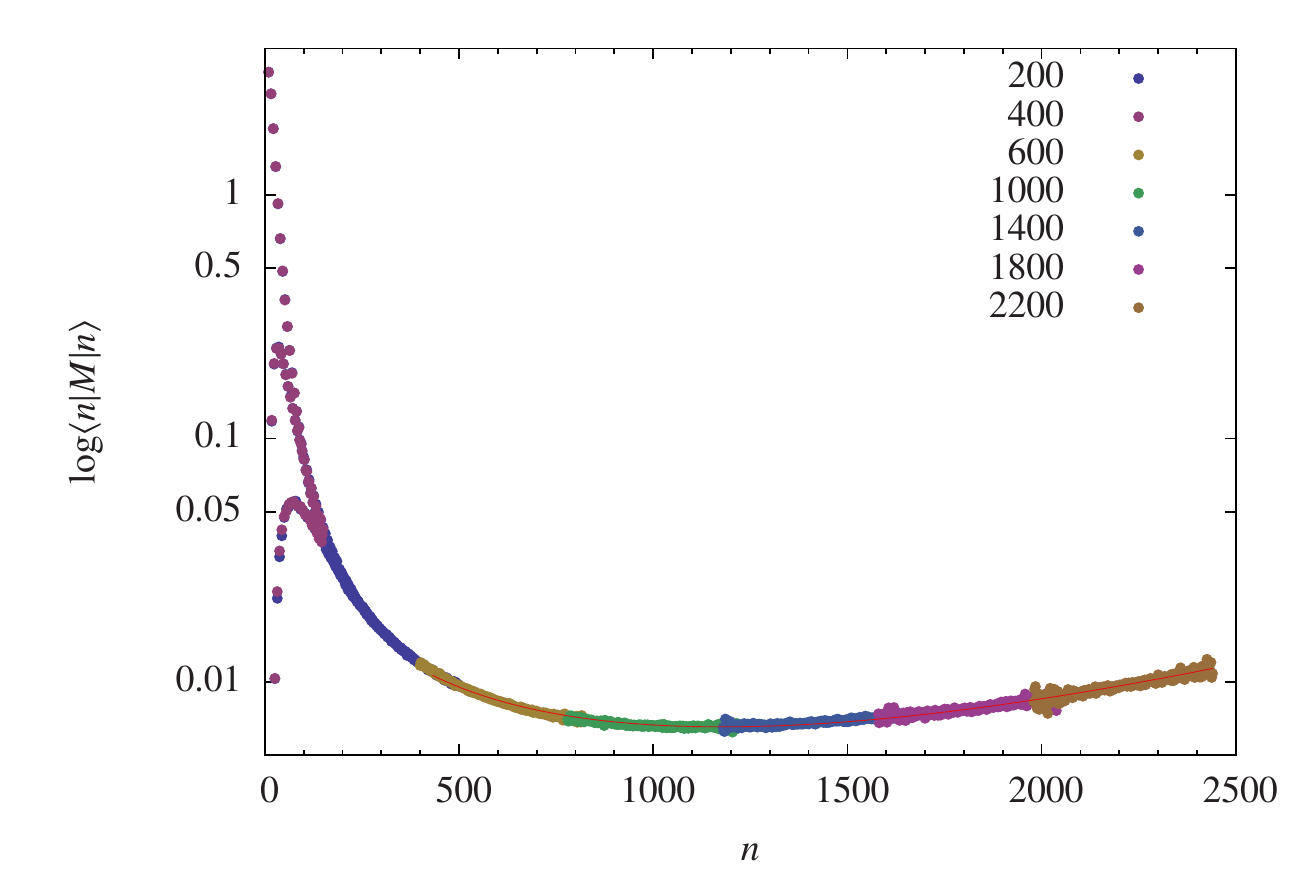}}
\caption{Diagonal of the scaled and merged transfer matrix (different colours indicate different ranges of $n_t$ fluctuations) $\log{\braket{n|M|n}}$ and the best fit of the potential term given by Eq. (\ref{v2n}) (red line, which stops at $n=400$).   The transfer matrix was measured for $\kappa_0=2.2$, $\Delta=0.6$ and $K_4=0.922$.}
\label{FigBlobPot}
\end{figure}

Summarizing, the effective Lagrangian in the large volume limit is very well parametrized by:
\beql{LeffCC}
L_{eff}[n,m] = \frac{1}{\Gamma}\left[ \frac{(n-m)^2}{n+m-2 n_0} + \mu \left(\frac{n+m}{2}\right)^{1/3} - \lambda \left(\frac{n+m}{2}\right) \right] \ ,
\eeq
which is up to a (very small) $n_0$ identical with the expected Lagrangian (\ref{Lagreff1}) resulting from a simple discretization of the minisuperspace action. One can use (\ref{LeffCC}) together with Eq. (\ref{Mth}) to fit the aggregated transfer matrix data. The results of fitting the parameters of $L_{eff}$ in different ways are summarized in Table \ref{table2}. For comparison we also present the parameters of the effective action measured by the covariance matrix method (from Chapter 3). The results obtained by the transfer matrix approach are fully consistent with the previous ones.

\begin{table}[h!]
	\begin{center}
		\begin{tabular}{|c|c|c|c|c|}
\hline
 Method 			& $\Gamma$ 				& $n_0$ 		& $\mu$ 			& $\lambda$ \\
\hline
\hline
 Cross-diagonals	& $26.07 \pm 0.02$		&$ -3 \pm 1$	& $ -$				&	$-$ \\ 
\hline
 Diagonal			& $(26.07)$				&$ -$			& $16.5 \pm 0.2$	&$ 0.049 \pm 0.001$ \\ 
\hline
 Full fit			&$ 26.17 \pm 0.01$		&$ 7 \pm 1$		& $15.0 \pm 0.1$	&$ 0.046 \pm 0.001$ \\ 
\hline
 Covariance matrix\,*	&$ 26.5 \pm 1.0$				& $-	$		& $20 \pm 2$	&$-$ \\
\hline
		\end{tabular}
	\end{center}
	\caption{The values of the parameters $\Gamma, n_0, \mu$ and $\lambda$ of the effective Lagrangian (\ref{LeffCC})
fitted in different ways.\,$^*$\,We also present the parameters of the effective action extracted  from the covariance matrix of volume fluctuations in Chapter 3.  }
	\label{table2}
\end{table}

One can also consider a more general form of the effective action containing subleading  corrections of the kinetic term (\ref{Sefff}) discussed in Chapter 3.3:
\beql{Leffcorr}
L_{eff}[n,m] = \frac{1}{\Gamma}\,  \frac{(n-m)^2}{n+m-2 n_0} \left(1+\xi_2 \left(\frac{n-m}{n+m} \right)^2  + ... \right)+ v[n+m] \ .
\eeq
For the Lagrangian (\ref{Leffcorr}) the cross-diagonal  elements become:
\beql{kincorr}
\braket{n|M|s-n}= {\cal N}(s)\exp \left( -\frac{(2n-s)^2}{k[s]} \left(1+k_2[s] \left({2n-s} \right)^2  + ... \right) \right) .
\eeq
For constant $s$ one can fit Eq. (\ref{kincorr}) as a function of $n$ to the measured cross-diagonal elements and extract $k[s]=\Gamma(s-2 n_0)$ and $k_2[s] = \xi_2/s^2$ from the fits. $k_2[s]$ falls as $s^{-2}$ and therefore for large volumes the coefficient becomes very small and indistinguishable from the numerical noise. Nevertheless, it is possible to observe a non-zero constant $\xi_2$ if one takes $s$ not ``too large" - see Fig. \ref{FigTMksi2}  in which we plot $\xi_2\equiv k_2[s] \cdot s^2$ measured for different $s$. The average value of $\xi_2$ measured for $200 \leq s\leq 1400$ yields $\xi_2=0.29\pm0.01$, which is  close to $\xi_2=0.38\pm0.05$ established from the covariance matrix method (see Chapter 3.3).

\begin{figure}[h!]
\centering
\scalebox{1.}{\includegraphics{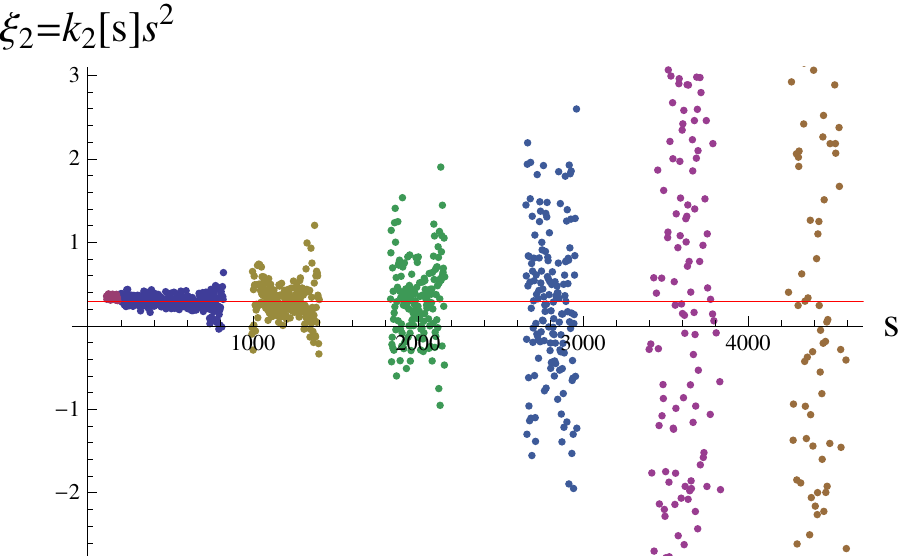}}
\caption{A correction  of the kinetic term: $\xi_2\equiv k_2[s] \cdot s^2$ as a function of $s=n+m$ (different colours indicate different ranges of $n_t$ fluctuations). For $200 \leq s\leq 1400$ it is possible to observe a constant, nonzero $\xi_2$ (red line). For higher $s$ the value of $\xi_2$ is hidden in the numerical noise.}
\label{FigTMksi2}
\end{figure}

\section{Transfer matrix for small three-volumes}
In the previous section we have shown that the effective action/Lagrangian in the large volume range ($s=n+m\geq400$) is very well described by a simple discretization of the minisuperspace action (\ref{LeffCC}) and it is very difficult to observe any subleading corrections - if they exist they are of the size of the numerical noise. One of the potential strategies could be to substantially increase the measurement statistics, however  collection of  the currently available data used to determine the transfer matrix elements required many days (or even weeks) of CPU time and  therefore this strategy is not very efficient, especially if one wants to analyze many points $(\kappa_0, \Delta)$ in  the bare coupling constant space. Another strategy would be to analyze the transfer matrix structure for small volumes. This was partly done while considering the $k_2[s]$ correction of the kinetic term (see Eq's (\ref{Leffcorr}) and (\ref{kincorr})) which falls as $s^{-2}$ and therefore the term is visible only for relatively small volumes. A difficulty of analyzing the small volume range is that one encounters strong discretization effects which mix up with real physical effects in this region (c.f. Fig. \ref{FigMstalk}, left). Additionally, the small $n$ region corresponds to the ``stalk" range of the CDT ``universe", where the average three-volume is  constant for each $t$ and does not scale with the total volume ($\braket{n_t}\approx 26$, for all $t\in$ ``stalk"). Therefore the physics of small volumes can be different than that in the large volume limit and there is {\it a priori} no guarantee that the effective transfer matrix  works. Nevertheless, we will use a pragmatic approach and assume that the method is correct also for small volumes. The consistency of the results will be a final check.

In order to measure the transfer matrix elements for small three-volumes   there is no need to introduce a volume fixing term (\ref{LocalVF}) and the auxiliary transfer matrix $\widetilde M$ (\ref{TMNew}). Instead, one can set the $K_4$ bare coupling constant slightly above the critical value $K_4^{crit}$ (see Chapter 1.5) forcing the CDT system to oscillate in the small $n$ range and  only very rarely make fluctuations to  higher volumes.  The closest we approach the critical $K_4^{crit}$ (from above) the more generic triangulations will be that of the ``stalk" observed in ``full CDT" (with $K_4=K_4^{crit}$ and a global volume fixing (\ref{SVFGQ})).
To calculate the transfer matrix elements  we again measured probability distributions (histograms) $P^{(3)}(n_{1},n_{2})$ and $P^{(4)}(n_{1},n_{3}) $ for $T=3$ and $T=4$, respectively, and used Eq. (\ref{MT34}). 
\begin{figure}[!h]
\centering
\scalebox{0.4}{\includegraphics{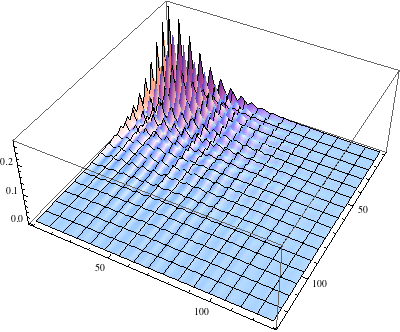}}
\scalebox{0.4}{\includegraphics{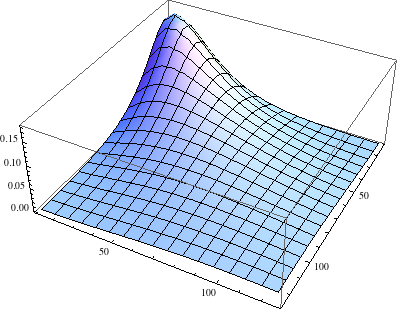}}
\caption{Left: the transfer matrix $M$ measured in the ``stalk" (small volume) range. The matrix is dominated by strong discretization effects. Right: the ``reduced" transfer matrix $\hat M$ calculated from $M$ in the ``stalk" range using Eq. (\ref{reducedM}), which smoothes out discretization effects. The transfer matrices were measured for $\kappa_0=2.2$, $\Delta=0.6$ and $K_4=0.9223$. }
\label{FigMstalk}
\end{figure}

The empirical transfer matrix measured for $\kappa_0=2.2$, $\Delta=0.6$ and $K_4=0.9223$ (closest to $K_4^{crit}=0.9222$) is presented in Fig. \ref{FigMstalk} (left). Conversely to the large volume case, the transfer matrix is not smooth and one can clearly see strong discretization effects. The consecutive``peaks" and ``troughs" are related to  three families of states which are observed in the ``stalk" range, e.g. for the three-volume probability distributions (c.f. Fig \ref{FigP80stalk2}). Inside each family the discrete three-volumes differ by 6, such that we have three sets of $n$: $\{10,16,22,...\}$,   $\{12,18,24,...\}$ and $\{14,20,26,...\}$.\footnote{Note that $n_t\equiv N^{\{4,1\}}(t)$ is by construction an even number and due to topological restrictions $n_t\geq 10$.}  The probability distribution for each family behaves quite smoothly but in a different way in each set. 
\begin{figure}[h!]
\centering
\scalebox{.7}{\includegraphics{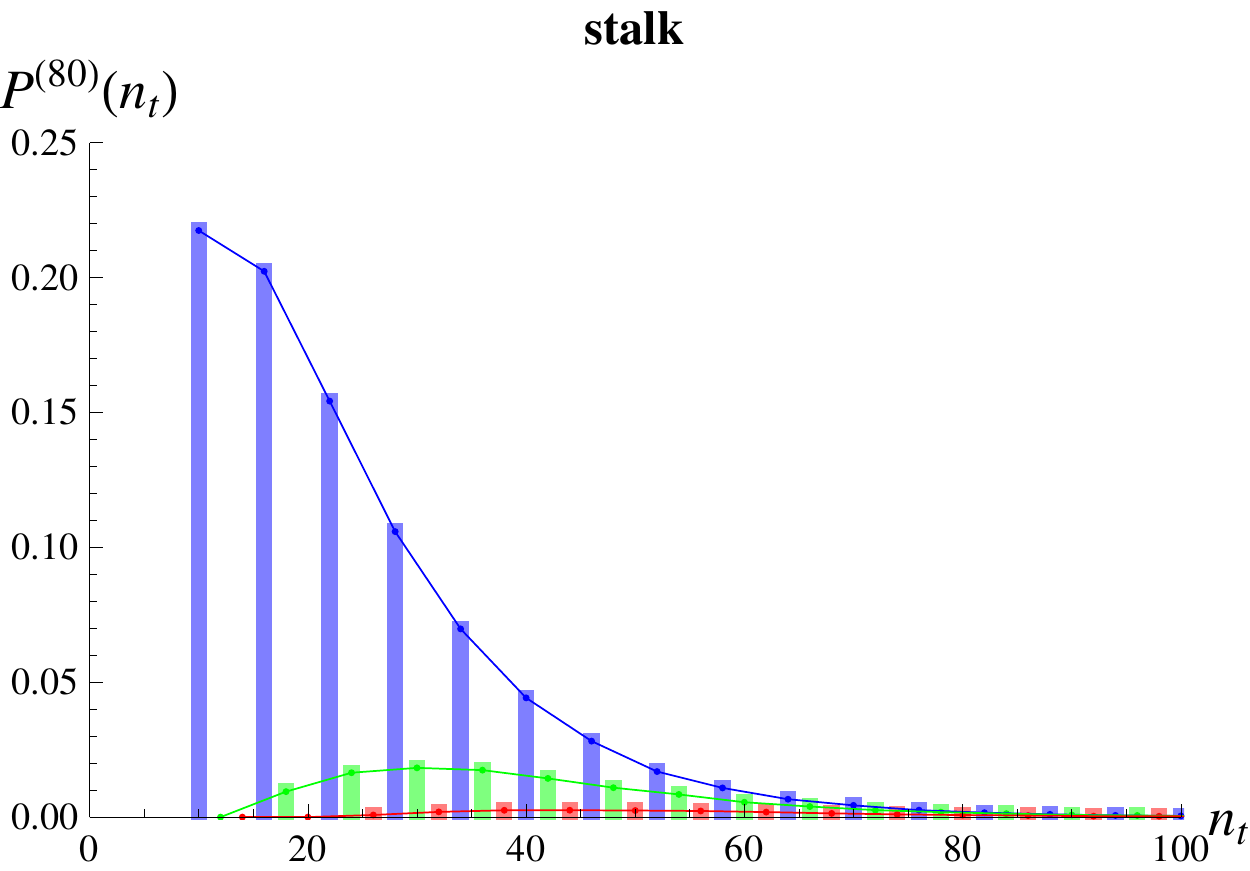}}
\caption{Probability distribution (histogram) of the volume $n_t$ measured in the ``stalk"  ($1 \leq t\leq 17 $ or $64 \leq 80$) for  $\kappa_0=2.2$, $\Delta=0.6$, $T=80$.  The probability distribution is the same for each $t$ and the data fall into three families, marked by different colours  in the graph, each one with a
different behaviour }
\label{FigP80stalk2}
\end{figure}

We would like to check if the measured transfer matrix can be used to explain the empirical  probability distributions and volume-volume correlations observed for large $T$ in the ``pure stalk" measurements or, even more, in the  ``stalk" range of the ``full CDT" simulations (where we observe both the ``stalk" and the ``blob").
To analyze the measured transfer matrix in more detail we will use a spectral decomposition in terms of eigenvalues $\lam_i$ and (orthonormal) eigenvectors 
$\ket{\alpha_i}$:
\begin{equation}
M = \sum_{i} \lambda_i \ket{\alpha_i}\bra{\alpha_i}.
\label{Meigenvectors}
\end{equation}
Since the measured $M$ is only determined up to a normalization,
we will assume $\lam_1=1$ and  $| \lambda_1 | \geq | \lambda_2 |\geq ...$.
If the gaps between the consecutive eigenvalues are big then in the large $T  \gg \Delta t \gg 1$ limit the effective partition function (\ref{Zeff}) and the probability distributions (\ref{P1nt}) and (\ref{P2nt}) will be  dominated by the first two eigenstates. Thus (recalling that $\lam_1$ is normalized to 1) we have:
\begin{equation}
{\cal Z}_{eff}^{(T)} = \sum_i \left(\lambda_i\right)^{T} \approx 1  
\label{Zeigen}  
\end{equation}
\begin{equation}
P^{(T)}(n) =  \frac{1}{{\cal Z}_{eff}^{(T)}} \sum_i \lambda_i^{T}  
\braket{n|\alpha_i}^2 \approx \braket{n|\alpha_1}^2 
\label{p1eigen}
\end{equation}
\bea
P^{(T)}(n_t,m_{t+\Del t}) &=& 
\frac{1}{{\cal Z}_{eff}^{(T)}} \Big( \sum_i \left(\lambda_i\right)^{\Delta t} 
\braket{n|\alpha_i} \braket{\alpha_i|m}\Big)       
\Big( \sum_i \left(\lambda_i\right)^{T-\Delta t} 
\braket{n|\alpha_i} \braket{\alpha_i|m}\Big)
\nonumber \\
 &\approx&     \braket{n|\alpha_1}^2 \braket{\alpha_1|m}^2
 +    \lambda_2^{\Delta t} \braket{n|\alpha_2} 
\braket{\alpha_2|m} \braket{n|\alpha_1} \braket{\alpha_1|m} \ .
\label{p2eigen}
\eea
For the average three-volume one obtains:
\begin{equation}\label{average}
\la n\ra \approx \sum_n n \braket{n|\alpha_1}^2 \ .
\end{equation}
The  covariance of three-volumes (in spatial slices distant by $\Delta t$) can be written as:
\begin{equation}\label{correlator}
\la n_t m_{t+\Del t}\ra -\la n_t\ra \la m_{t+\Del t}\ra \approx 
\lambda_2^{\Delta t} \sum_{n,m} n m \braket{n|\alpha_2} 
\braket{\alpha_2|m} \braket{n|\alpha_1} \braket{\alpha_1|m}
\end{equation} 
and the long distance behaviour is an exponential fall off \ 
$e^{-\sigma \Del t}$, $\sigma = -\log \lam_2/\lam_1$ (where we have reintroduced
$\lam_1$ for clarity). 

\begin{figure}[h!]
\centering
\scalebox{.95}{\includegraphics{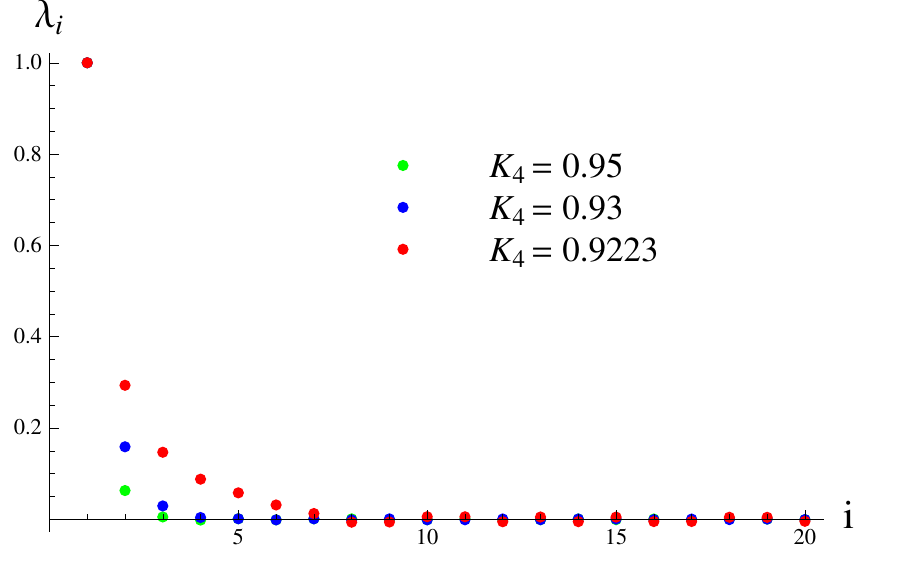}}
\caption{Eigenvalues of the transfer matrix $M$ measured in the ``stalk" for a range of values of $K_4$ approaching $K_4^{crit}\approx 0.9222$. The biggest eigenvalue is normalized to one.}
\label{FigstalkLambdas}
\end{figure}

To justify that one can use approximate expressions (\ref{Zeigen}) -  (\ref{correlator}) we present the plot of the eigenvalues for the measured transfer matrices (Fig. \ref{FigstalkLambdas}). The eigenvalues depend on the value of $K_4$ for which we perform numerical simulations. The gaps between the consecutive eigenvalues get smaller when $K_4$ is decreased,  but even for $K_4=0.9223$ (very close to the critical value $K_4^{crit}=0.9222$) the gaps are significant. One should also ask how large $T$ should be in order to make the above approximations legitimate. It of course depends on the eigenvalues and we have checked that for the average $\braket{n}$ the Eq. (\ref{average}) works very well already for $T=4$. For the covariance one can use Eq. (\ref{correlator}) for larger $T$. For example we have checked that for $T=12$ one observes the expected exponential fall off with exponent $\log(\lambda_2/\lambda_1)$ - see Fig. \ref{FigStalkLogcov12vsTh}. The expression holds even for small $\Delta t$ where it is {\it a priori} not obvious that one can ignore eigenvectors relevant to  next $\lambda_i$ ($i>2$). 

\begin{figure}[h!]
\centering
\scalebox{.8}{\includegraphics{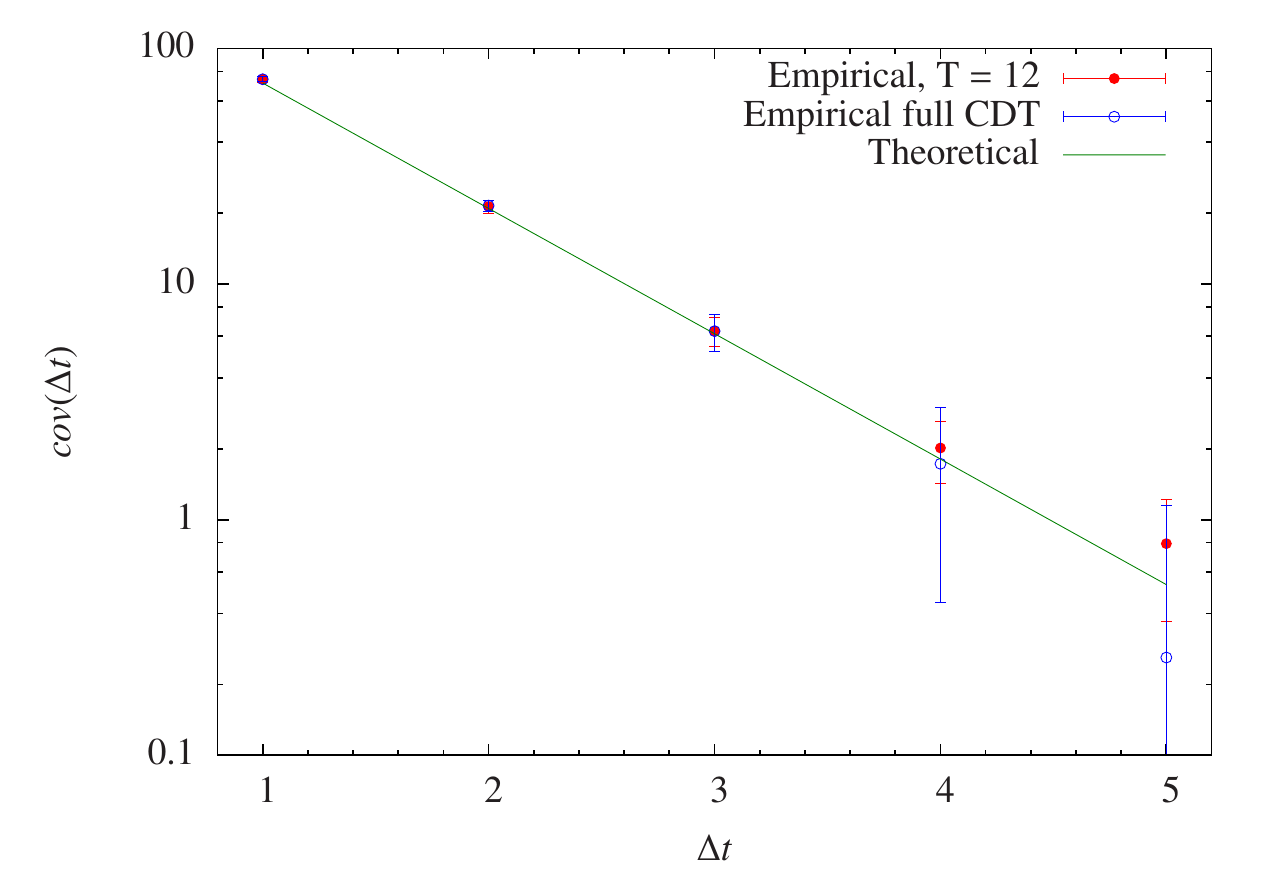}}
\caption{The empirical covariance  $cov(\Delta t)\equiv\la n_t m_{t+\Del t}\ra -\la n_t\ra \la m_{t+\Del t}\ra$ of  the spatial volumes measured  in time slices separated by $\Delta t$ in the small-volume simulations with $T=12$ (red points) and in the ``stalk" range of the ``full CDT" model (blue points) for $\kappa_0=2.2$ and $\Delta=0.6$. The bars indicate measurement errors. The covariance falls off exponentially (log scale), as explained by the theoretical covariance (\ref{correlator}) calculated using the first two eigenvectors of the transfer matrix  $M$ calculated for $K_4=0.9223$ (green line). The overall  agreement is very good.}
\label{FigStalkLogcov12vsTh}
\end{figure}

As a final check of the transfer matrix approach in a small volume region  we would like to verify if it is consistent with  empirical probability distributions and correlations observed in the ``stalk" range of the ``full CDT" simulations (including the ``stalk" and the ``blob"). To do this we compared the theoretical  probability distributions (\ref{p1eigen}), calculated for the largest eigenvector of the transfer matrix measured for different $K_4$'s, with the empirical distribution measured in the ``stalk" range of the ``full CDT" (for the same values of $\kappa_0=2.2$ and $\Delta=0.6$). The distributions approach very well the ``full CDT" measurements as $K_4$ tends to the critical value (see Fig. \ref{FigStalkP1}). The accuracy of the transfer matrix approach is further confirmed by the  volume-volume correlations observed in the ``full CDT" ``stalk" range. The empirical covariance $cov(\Delta t)\equiv\la n_t m_{t+\Del t}\ra -\la n_t\ra \la m_{t+\Del t}\ra$ falls of exponentially as $e^{-\sigma \Delta t}$ with $\sigma$ explained by the ratio of the first two eigenvalues of the transfer matrix measured for $K_4$ closest to the critical value - c.f. Fig. \ref{FigStalkLogcov12vsTh}.

\begin{figure}[h!]
\centering
\scalebox{.8}{\includegraphics{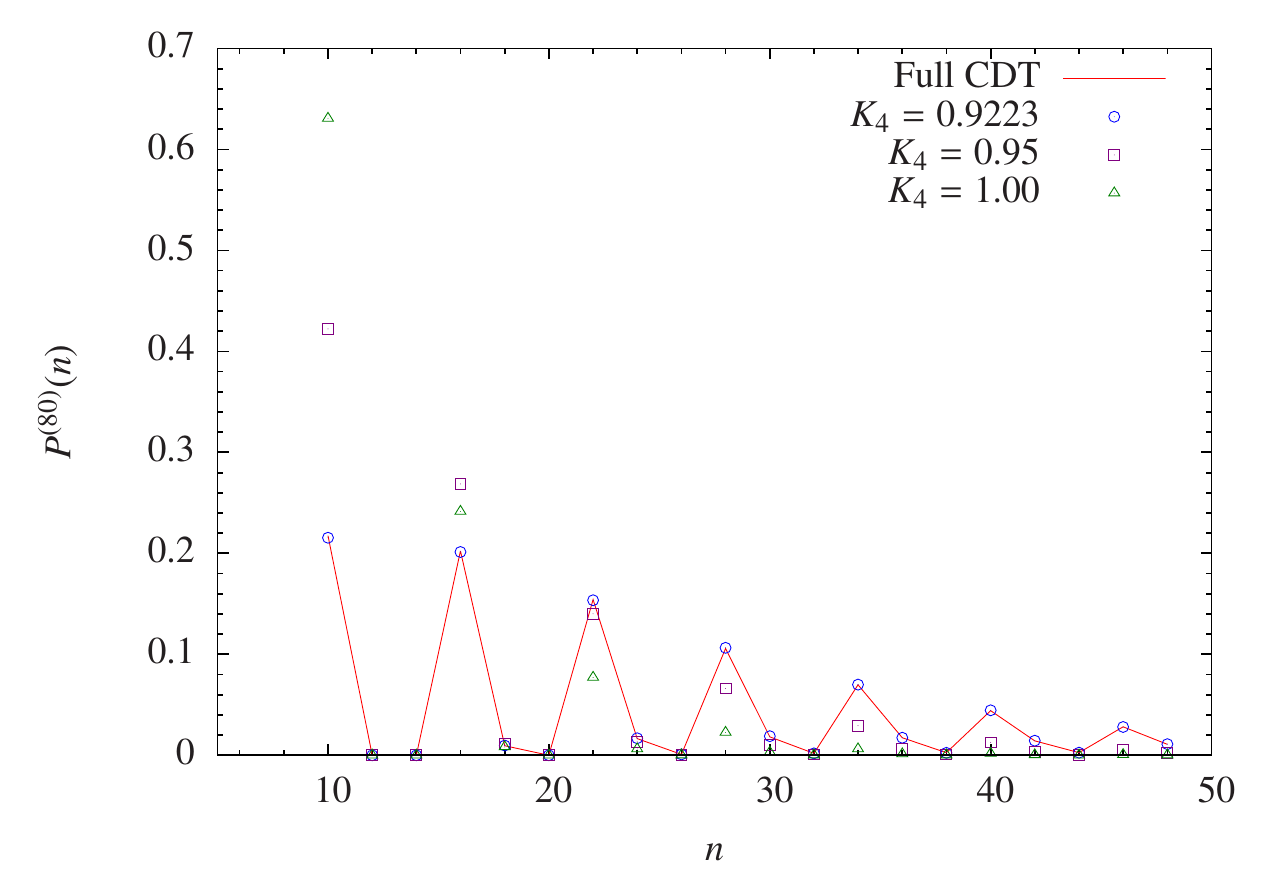}}
\caption{Theoretical probability distributions of the small three-volume $n$  calculated with the
first eigenvector of the transfer matrix $M$  for $K_4=1.00, 0.95$ and $0.9223$ (points). The distributions approach empirical probability
measured in the ``stalk" range of "full CDT" (red line) as $K_4$ tends to the critical value $K_4^{crit}=0.9222$.  For $K_4=0.9223$ (blue
circles) the agreement is very good. Data measured for $\kappa_0=2.2$ and $\Delta=0.6$. }
\label{FigStalkP1}
\end{figure}

In principle, one can use the measured transfer matrix to determine the effective Lagrangian in a small volume limit using the same approach as for the large volume case. However, due to the existence of three families of states described above, it is very difficult to find a suitable parametrization. Instead we can define a ``reduced" transfer matrix by performing a summation over the three families:\footnote{This is only one of the possible  ``averaging" methods, one can use alternative methods. }
\begin{equation}\label{reducedM}
\hat{M} = U M U^T \,
\end{equation} 
where the rectangular matrix $U$ has a form:
$$
 U = \left\{\begin{array}{cccccccccc} 
1 & 1 & 1 & 0 & 0 & 0 & 0 & 0 & 0 &  \cdots \\
 0 & 0 & 0 & 1 & 1 & 1 & 0 & 0 & 0 & \cdots \\
0 & 0 & 0 & 0 & 0 & 0 & 1 & 1 & 1 & \cdots\\
\cdots \end{array}\right\} \ .
$$
The elements of the matrix $\hat{M}$ behave much more smoothly - see Fig. \ref{FigMstalk} (right) -
and can be analyzed using the effective action idea.

We follow the same procedure that we used for the large three-volumes, i.e. we assume that 
\beql{Mstalk}
\braket{n | \hat M | m } = \cN  \exp(-L^{stalk}_{eff}[n,m])
\eeq
where the effective Lagrangian is given by
\beq
L^{stalk}_{eff}[n,m] =  \frac{(n-m)^2}{k[n+m]} + v[k+m]
\eeq
and we want to determine the functions $k[.]$ and $v[.]$ from the cross-diagonal and diagonal elements of $\hat M$, respectively. We use empirical $\hat M$ measured for $K_4=0.9222$ (closest to the critical value).

\begin{figure}[h!]
\centering
\scalebox{.9}{\includegraphics{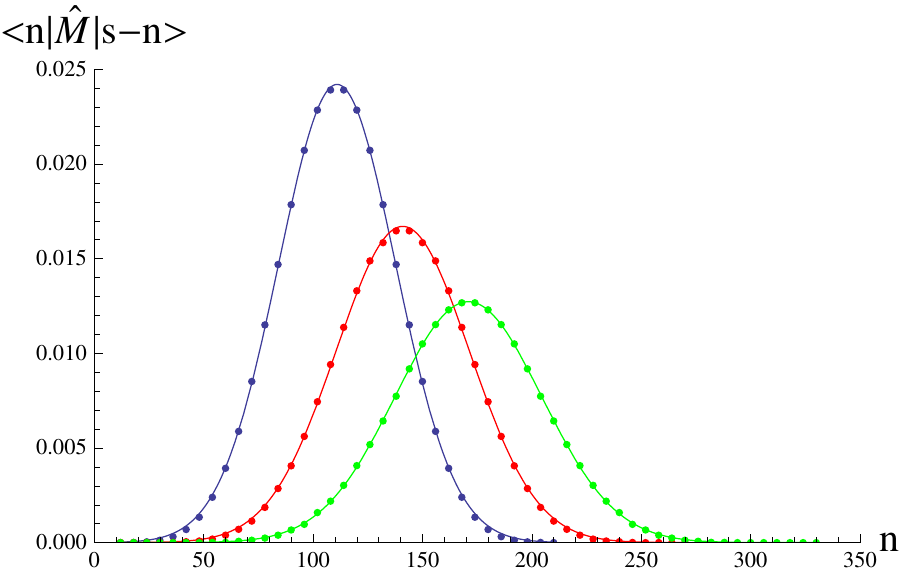}}
\caption{Cross diagonals of the ``reduced" transfer matrix $\hat M$ measured for $\kappa_0=2.2$, $\Delta=0.6$ and $K_4=0.9223$. The matrix elements $\braket{n | \hat M | s-n } $ (points) show the expected Gaussian dependence on $n$ (lines) for fixed $s= 222$ (blue), $s=282$ (red) and $s=342$ (green).}
\label{FigStalkantidiag}
\end{figure}

For the cross-diagonals we  observe the expected Gaussian dependence on $n$ for fixed $s=n+m$ (see Fig. \ref{FigStalkantidiag}):
\beq
\braket{n|\hat M|s-n}= {\cal N}(s)\exp \left( -\frac{(2n-s)^2}{k[s]}\right)
\eeq
and  can extract kinetic coefficients for different $s$. The $k[s]$  is again a linear function of $s$ (see Fig. \ref{FigStalkKs}), thus we have
\beq
k[s] = \Gamma(s-2\, n_0)
\eeq
exactly as in the large volume case.

\begin{figure}[h!]
\centering
\scalebox{.9}{\includegraphics{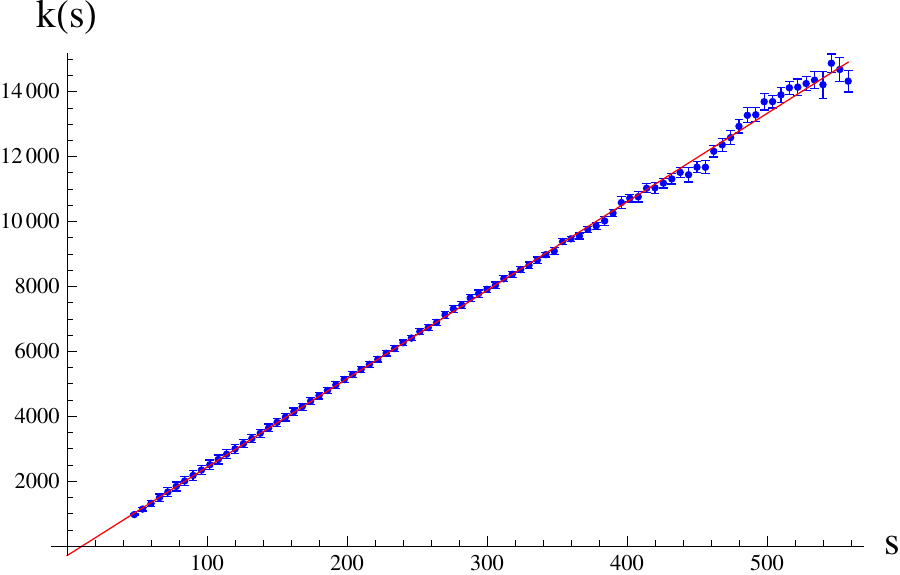}}
\caption{The kinetic coefficients $k[s]$ for the ``reduced" transfer matrix $\hat M$ show a linear dependence on $s=n+m$.}
\label{FigStalkKs}
\end{figure}

For the potential part we assume:
\beql{v2nstalk}
v[2n] =  \frac{1}{\Gamma} \left({\mu}n^{1/3} - {\lambda}n  + {\delta}n^{-\rho}\right) \ ,
\eeq
where we have added a new term $ {\delta}n^{-\rho}$ inspired by the earlier remarks about possible curvature-squared corrections (see Chapter 3.3).
The potential can be well fitted to the diagonal transfer matrix elements (see Fig. \ref{FigStalkDiag}):
\beq
\log{\braket{n|\hat M|n}} = \log{\cal N} - v[2n] \ .
\eeq
The best estimate  of $\rho=3\pm 1$ is not consistent with possible curvature-square term, for which one should get $\rho=1/3$, however it is not determined with a high precision and is not completely  independent on the specific way in which we perform the merging of the three families of states, as described above. Nevertheless we would rather treat it as a phenomenological correction due to the finite-size effects. 

\begin{figure}[h!]
\centering
\scalebox{.9}{\includegraphics{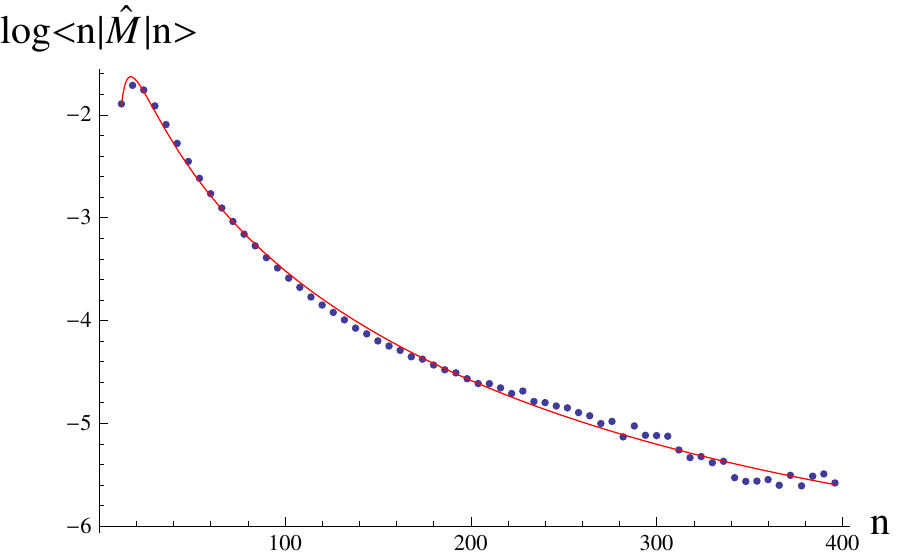}}
\caption{Diagonal of the ``reduced" transfer matrix (measured  for $\kappa_0=2.2$, $\Delta=0.6$ and $K_4=0.9223$) $\log{\braket{n|\hat M|n}} $ (points) and the best fit of  the potential term (\ref{v2nstalk}) (red line).}
\label{FigStalkDiag}
\end{figure}

Summarizing:
\begin{equation}
S_{eff}^{stalk} =
\label{Seffform}
\eeq
\beq
 =\sum_t \frac{1}{\Gamma} \left[  
\frac{(n_t-n_{t+1})^2}{n_t \plu n_{t+1} \mi 2 n_0} \plu 
\mu \left(\frac{n_t\plu n_{t+1}}{2}\right)^{1/3}\! \mi 
\lambda \frac{n_t\plu n_{t+1}}{2} \plu
\delta \left(\frac{n_t\plu n_{t+1}}{2}\right)^{-\rho}\right]. \nonumber
\end{equation}
The best fits of all parameters are presented in Table \ref{tablestalk}, in which we also summarize the results obtained in the  large volumes region for comparison. 
\begin{table}[h!]
\begin{center}
\begin{tabular} {|c|c|c|}
\hline
{Parameter} & {Stalk}	& {Blob}  \\ \hline
\hline
$\Gamma$&$	27.2 \pm 0.1 $&$	25.7 - 26.2$ \\ \hline
$n_0$&$	5 \pm 1$&	$-3 - +7$  \\ \hline
$\mu$&$	34 \pm 2 $&$	13 - 30$ \\ \hline
$\lambda$&$	0.12 \pm 0.02$&$	0.04 - 0.07$ \\ \hline
$\delta$&$	(4 \pm 7)\times 10^4$&	$-$ \\ \hline
$\rho$&$	3 \pm 1$&$-$ \\ \hline
\end{tabular}
\end{center}
\caption{ Fitted parameters of the effective action for 
the ``stalk" (\ref{Seffform}). For comparison we also present 
estimates of parameters of the effective action for the ``blob" 
calculated from the large $n_t$ simulations 
(see Table \ref{table1} and Table \ref{table2}).}
\label{tablestalk}
\end{table}

\noindent To underline the quality of the fits we also show the first six eigenvalues and the corresponding eigenvectors of the measured $\hat M$ and a theoretical $\hat M^{(th)}$ calculated using Eq's (\ref{Mstalk}) and (\ref{Seffform}) for the parameters from Table  \ref{tablestalk} - see Table \ref{tableeigenstalk} and Fig \ref{vec16}. 

\begin{table}[h!]
\begin{center}
\begin{tabular} {|c|c|c|}
\hline
$i$ & $\lambda_i $	& $\lambda_i^{(th)}$  		\\ \hline 
\hline
$ 1 $	& $ 1.000 \times 10^{0} $	& $ 0.973 \times 10^{0}  $ 	\\ \hline
$ 2 $	& $ 3.922 \times 10^{-1} $	& $ 4.074 \times 10^{-1} $ 	\\ \hline
$ 3 $	& $ 2.007 \times 10^{-1} $	& $ 2.054 \times 10^{-1} $ 	\\ \hline
$ 4 $	& $ 9.308 \times 10^{-2} $	& $ 9.658 \times 10^{-2} $ 	\\ \hline
$ 5 $	& $ 3.483 \times 10^{-2} $	& $ 3.745 \times 10^{-2} $ 	\\ \hline
$ 6 $	& $ 1.085 \times 10^{-2} $	& $ 1.199 \times 10^{-2} $ 	\\ \hline
\end{tabular}
\end{center}
\caption{
The first $6$ eigenvalues of the measured ``reduced" transfer matrix
$\hat M $ 
and the similar eigenvalues for  
the fitted transfer matrix $ M^{(th)}$ calculated using the effective action for the stalk range   (\ref{Seffform}).}
\label{tableeigenstalk}
\end{table}
\begin{figure}[!ht]
\centering
\scalebox{0.33}{\includegraphics{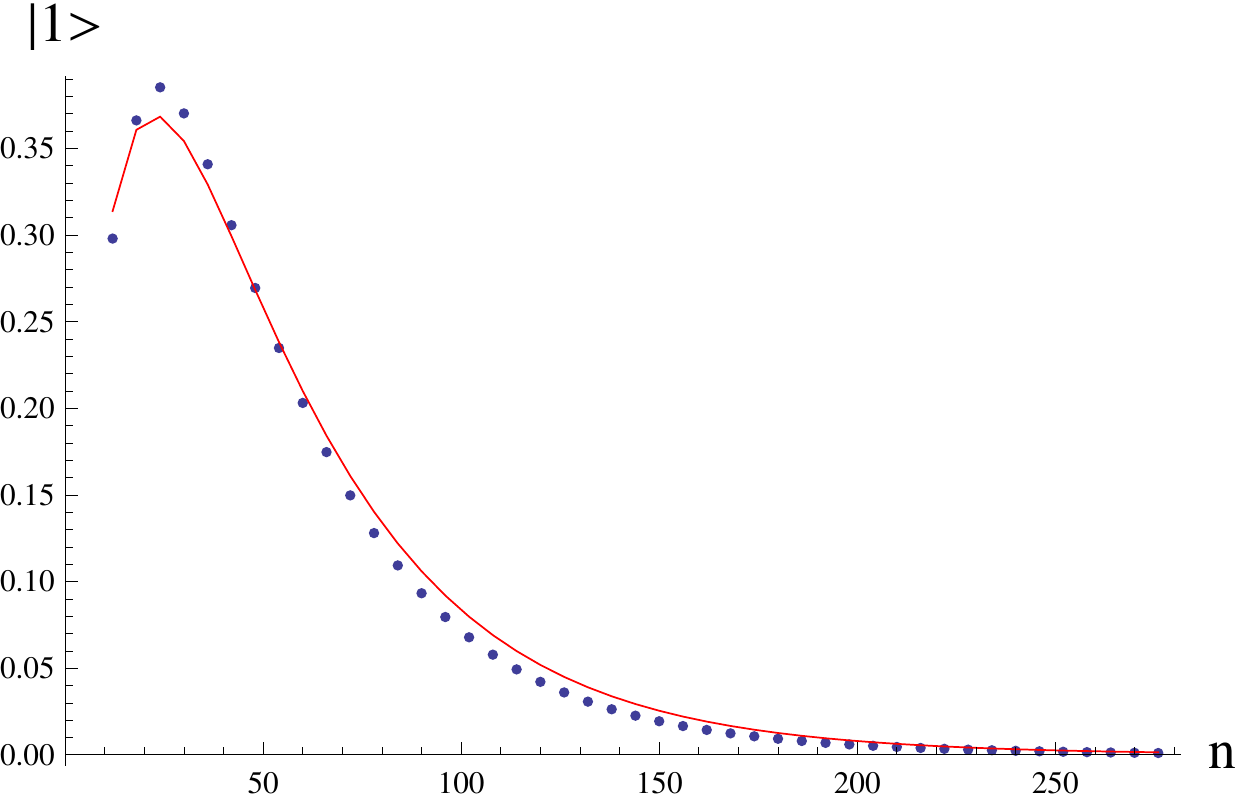}}
\scalebox{0.33}{\includegraphics{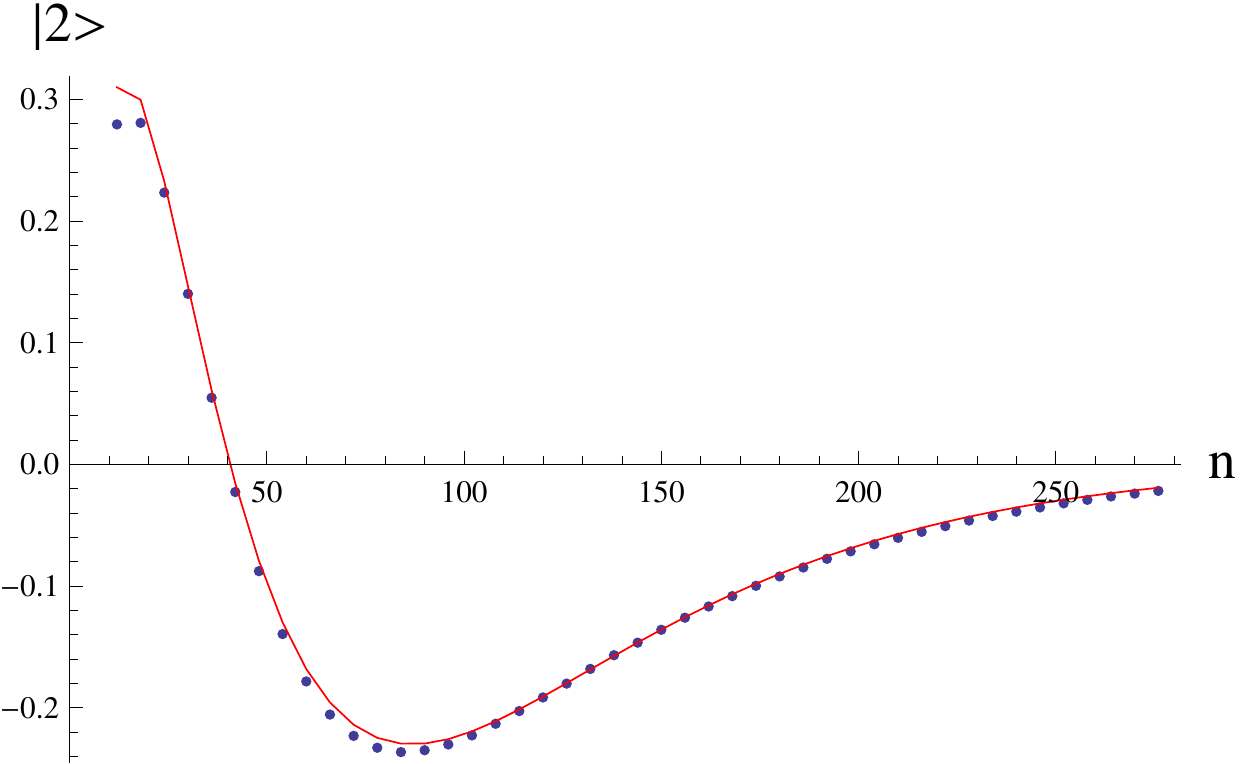}}
\scalebox{0.33}{\includegraphics{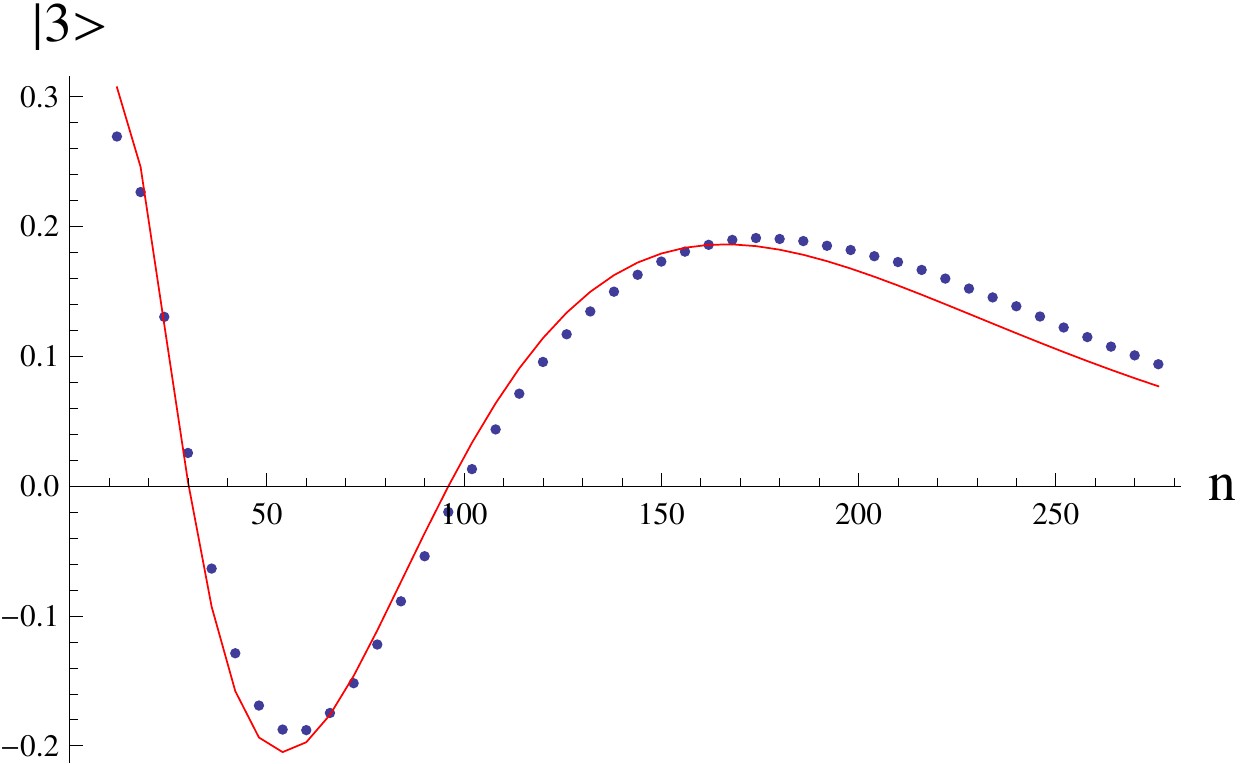}}
\scalebox{0.33}{\includegraphics{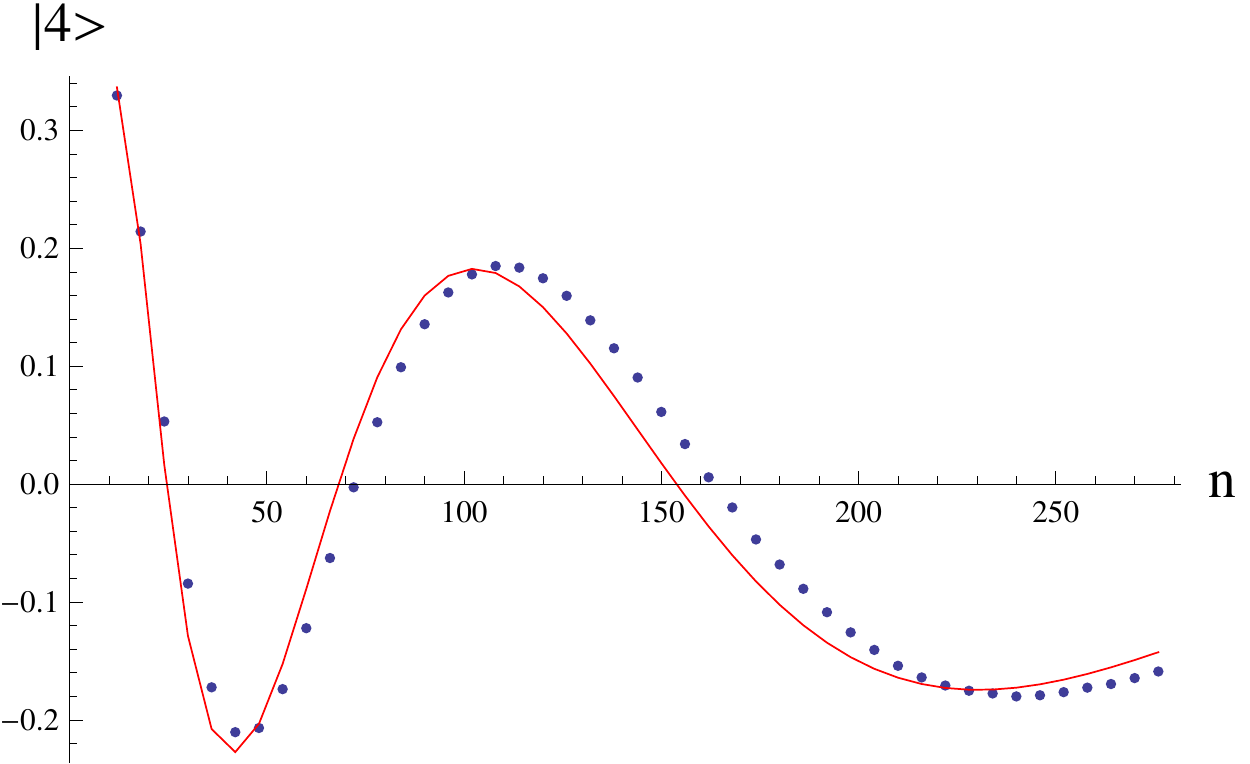}}
\scalebox{0.33}{\includegraphics{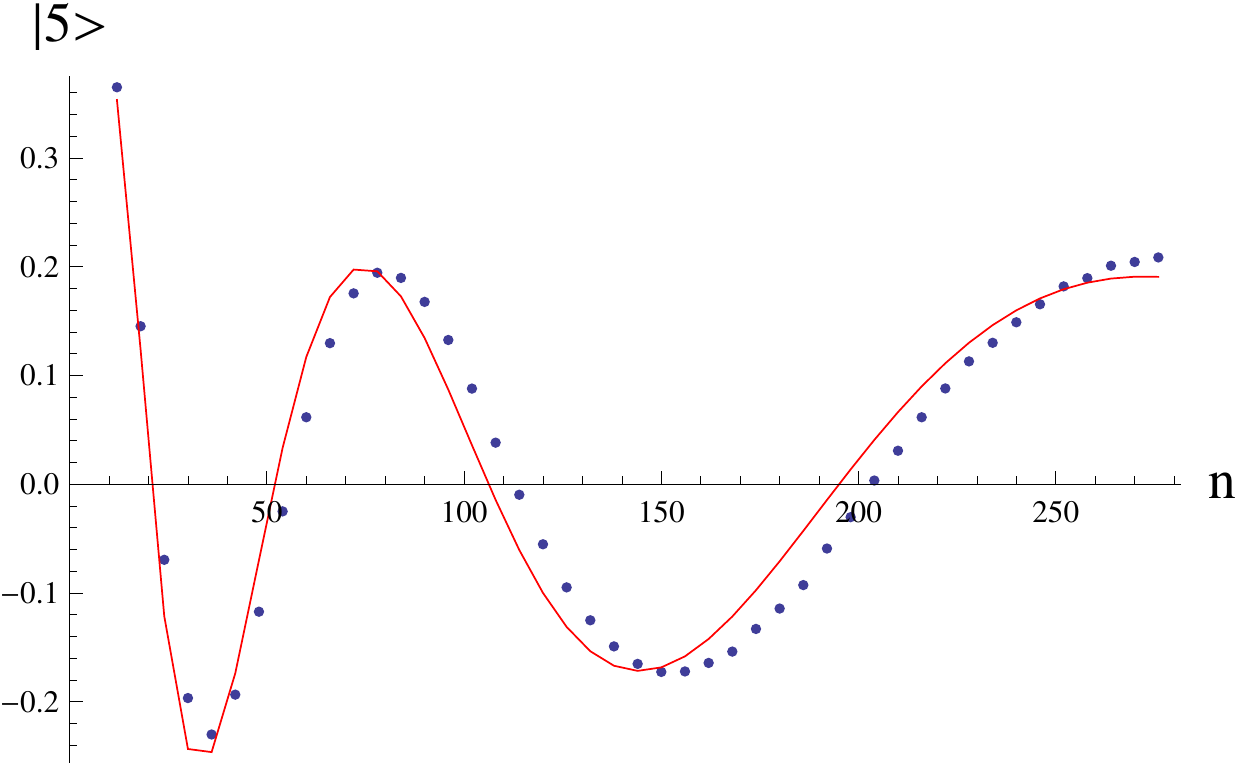}}
\scalebox{0.33}{\includegraphics{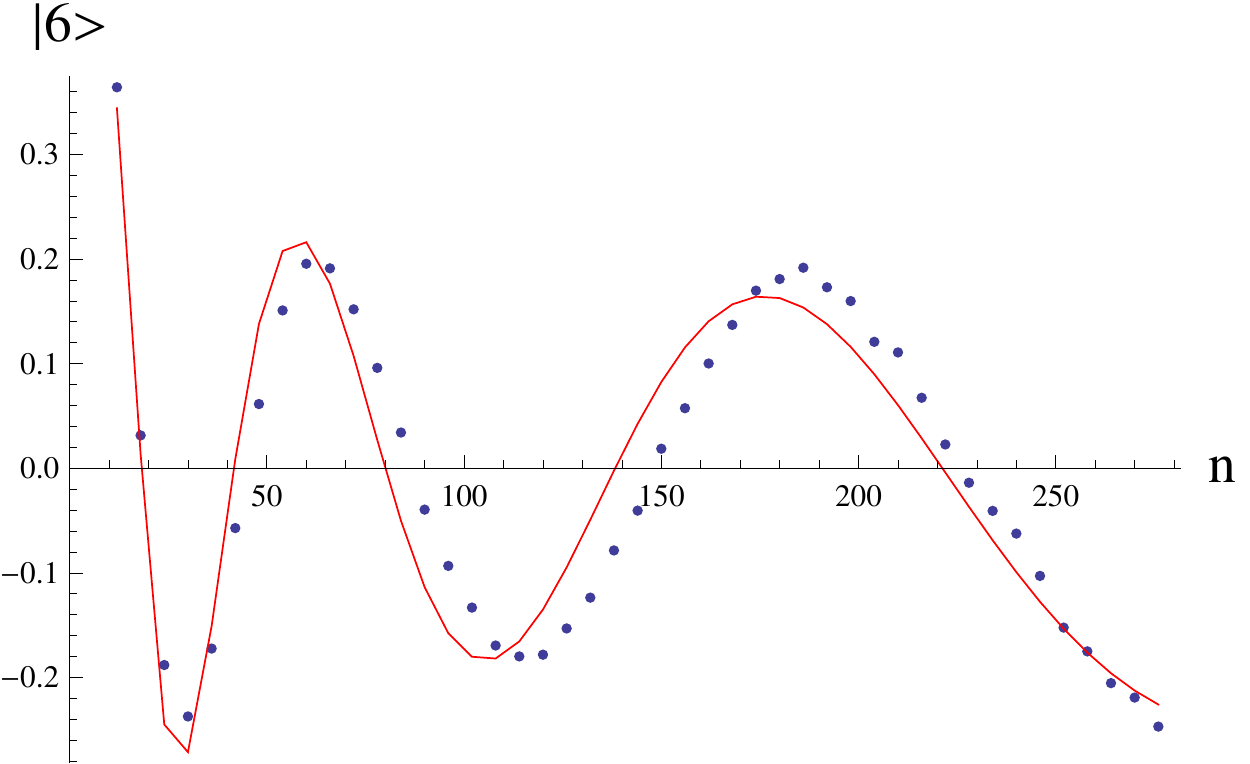}}
\caption{The first six eigenvectors of the measured $\hat M$ (blue dots) 
and ``theoretical''  matrix ${\hat M}^{(th)}$ (red line). 
${\hat M}^{(th)}$ was calculated using the effective 
action for the stalk range (\ref{Seffform}). }
\label{vec16}
\end{figure}

Surprisingly, we have found that despite the fact that  the nature of the ``stalk" seems to be very different than that of the ``blob" on  first sight, it is still, up to discretization effects, very well described by the same kind of  effective minisuperspace action and  even the parameters of the action are very similar. We also tried to measure the subleading corrections to the effective action which should be more evident in a small volume region, however they mix up with discretization effects and therefore are very hard to  determine. 

\section{Effective model in de Sitter phase}

In the previous sections we provided strong evidence that the effective model defined by a partition function:
\beql{ModelZ}
{\cal Z}_{eff}^{(T)} =\tr M^T \ ,
\eeq
with the transfer matrix $M$ parametrized by the spatial volumes of integer $t$ ($n_t\equiv N^{\{4,1\}}(t)$), can be used  to explain the spatial volume behaviour both in a large and a small volume range. As a final check we would like to combine both measurements and use it to reconstruct  results of the ``full CDT" simulations, i.e.   the average  volume profile $\braket{n_t}$ and the quantum fluctuations $\braket{\delta n_t \delta n_t'}$ in a system with large $T$ and a global volume fixing:
\beql{Efvolfix}
S_{VF} = \eps \Big( N^{\{4,1\}} - \bar V_{4}  \Big)^2 = \eps \left(\sum_t n_t - \bar V_{4}  \right)^2 \ .
\eeq

\begin{figure}[h!]
\centering
\scalebox{.6}{\includegraphics{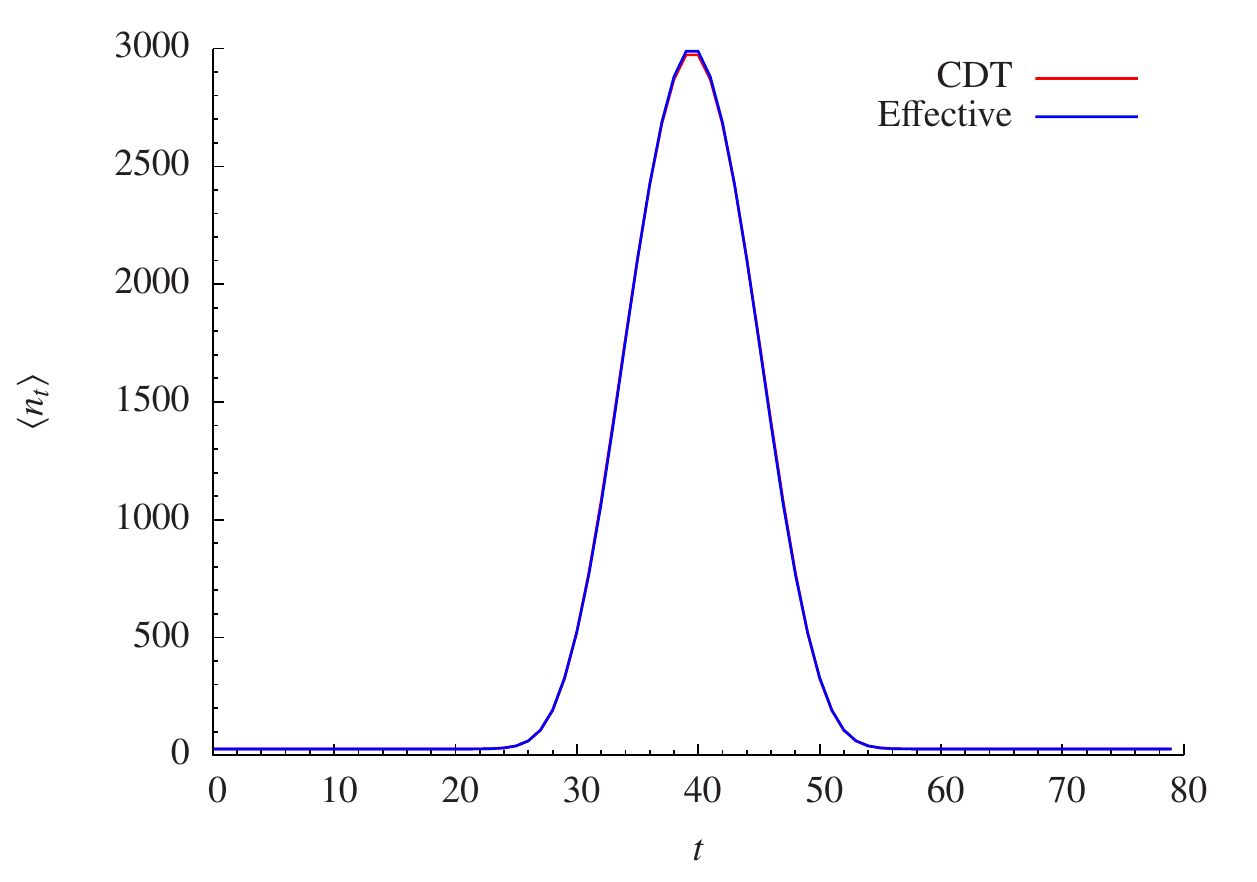}}
\caption{The average volume profile $\braket{n_t}$ measured in the ``full CDT" simulations for $\kappa_0=2.2$, $\Delta=0.6$ and $\bar V_4=40$k (red) and in the effective model (\ref{eq:pne}) based on the transfer matrix $M$ (blue). }
\label{FigModelavt}
\end{figure}

In order to do so we would also need the transfer matrix elements 
$M_{nm}\equiv \braket{n|M|m}$ for volumes much larger than measured so far - see Fig. \ref{FigModelavt} in which we plot the empirical average volume profile  for $\bar V_4 = 40$k. Technically it is difficult to determine the transfer matrix elements in such a wide range of $n_t$ and therefore for large volumes we will use a theoretical transfer matrix:
\beql{eq:Mth}
M^{(th)}_{nm} \equiv \cN e^{-L_{eff}[n,m]} 
\eeq
with the effective Lagrangian:
\newline
\newline

\begin{equation}
L_{eff}(n,m)=
\label{Seffform2}
\end{equation}
$$ 
=\frac{1}{\Gamma} \left[ 
\frac{(n - m)^2}{n + m - 2 n_0} + \mu \left( \frac{n + m}{2}\right)^{1/3} - \lambda \left( \frac{n + m}{2}\right) - \delta \left( \frac{n + m}{2}\right)^{- \rho}  \right] ,
$$
where the term $\delta \left( \frac{n + m}{2}\right)^{- \rho} $ is  inspired by the small volume  correction of the potential term. 
In a small volume region ($n < 300$ or $m <300$) dominated by strong discretization effects we  have  directly measured matrix elements $M_{nm}^{(emp)}$, and use the empirical data for the intermediate volumes  ($250 <n,m<700$) to fit all the parameters of Eq. (\ref{Seffform2}) by the methods described in the previous sections (see Fig. \ref{Figtransferfits}). Finally, we  define a semi-empirical transfer matrix by:
\begin{equation}
M_{n m} = \left\{ \begin{array}{ll}
M^{(\mathrm{emp})}_{n m} & n < thr \textrm{ or } m < thr,\\
M^{(\mathrm{th})}_{n m} & \textrm{otherwise},
\end{array} \right.
\label{Mextr}
\end{equation}
where $thr$ is a threshold ($thr = 300$).
When one of the entries is smaller than the threshold we 
use the  measured matrix elements.
When both entries are larger than the threshold we 
use the extrapolating function (\ref{eq:Mth}).

\begin{figure}[h!]
\centering
\scalebox{.53}{\includegraphics{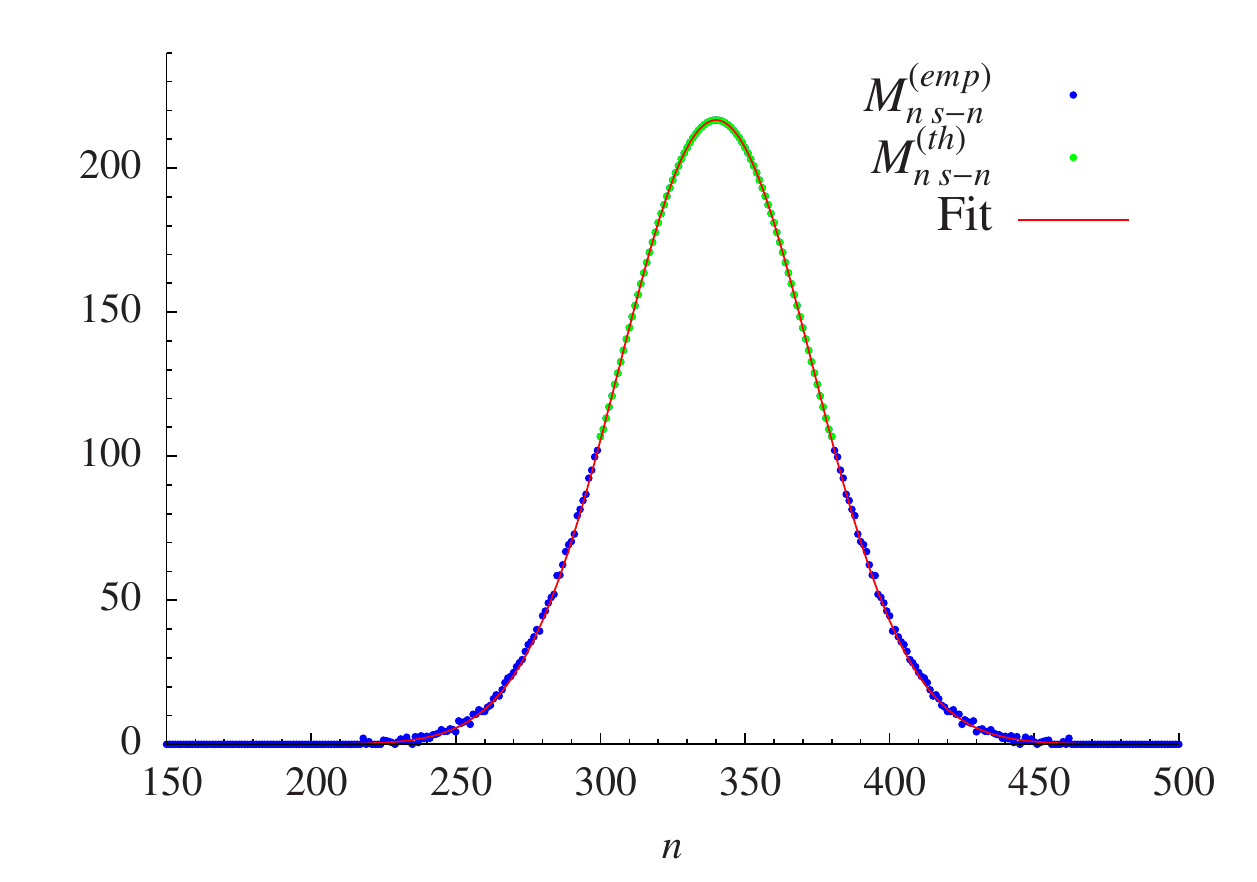}}
\scalebox{.53}{\includegraphics{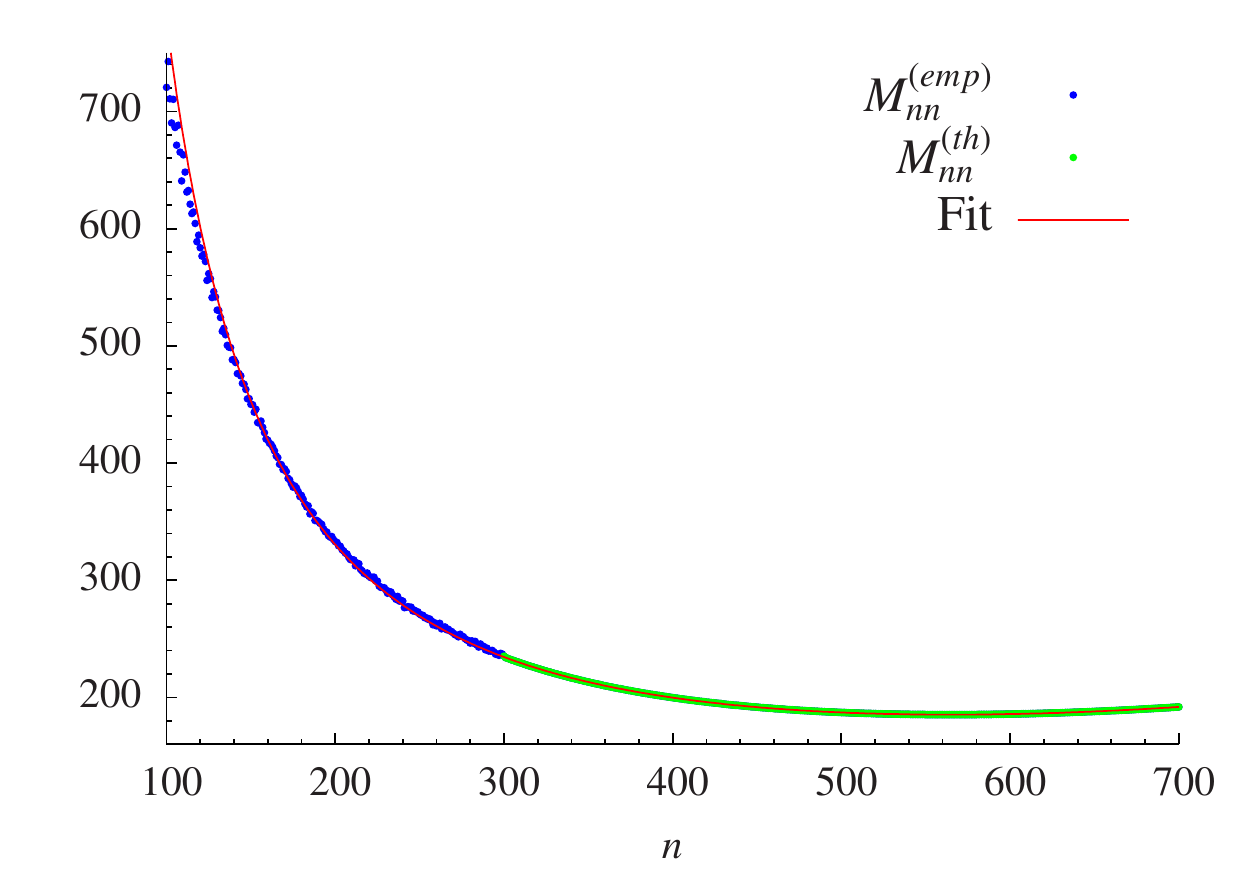}}
\caption{The semi-empirical  transfer matrix $M$, merged from the
empirical matrix (blue) and theoretical matrix (green). The theoretical matrix
is determined by a best fit (red line) to the empirical matrix in an overlap region
where $n$ and $m$ are in the range $250-700$ as described in the text. Left: the cross diagonal $\braket{n|M|s-n}$ for $s=680$. Right: the diagonal $\braket{n|M|n}$. }
\label{Figtransferfits}
\end{figure}

We  can now introduce an {\it effective model}, in which configurations are given by the $T$ component vectors representing the spatial volume profiles $\{n_t\}$ ($t=1,2,...T$) rather than by ``full CDT" triangulations $\cT$, with a partition function given by Eq. (\ref{ModelZ}) and a transfer matrix $M$ defined by Eq. (\ref{Mextr}).
According to Eq. (\ref{Pfullnt}) the T-point probability distribution for the sequence of three-volumes $\{n_1,n_2, .... ,n_T\}$   is:
\beq
 P^{(T)} (n_1, \dots, n_{T}) \propto \langle n_1 | M | n_2 \rangle 
\langle n_2 | M | n_3 \rangle \cdots \langle n_{T} | M | n_1 \rangle \ 
e^{- \epsilon \left(\sum_t n_t - \bar V_4 \right)^2} \ ,
\label{eq:pne}
\eeq
where we have included a global volume fixing term (\ref{Efvolfix}). 

One can use  standard Monte Carlo techniques to generate the 
configurations $\{n_t\}$,
according to the probability  distribution (\ref{eq:pne}).
We use the transfer matrix data measured in a generic  point in  the ``de Sitter" phase ($\kappa_0=2.2$, $\Delta=0.6$) and the same time period  $T=80$ and total volume $\bar V_4=40$k 
as in the ``full CDT" simulations.
As before, we measure the average volume profile 
$\braket {n_t} $ and the covariance matrix 
$C_{t t'} \equiv \langle (n_t - \langle n_t \rangle) (n_{t'} - 
\langle n_{t'} \rangle) \rangle$.
The results obtained by this effective model 
are very similar to  results obtained by the original
``full CDT" simulations - see Fig. \ref{FigModelavt}  and Fig. \ref{FigModelcov}  in which we plot $\braket{n_t}$ and the diagonal $C_{tt}$ in both models, respectively.
The consistency of the results is a very strong argument in favour of the effective transfer matrix approach, at least ``deep" inside the ``de Sitter" phase where we performed our calculations.

\begin{figure}[h!]
\centering
\scalebox{.67}{\includegraphics{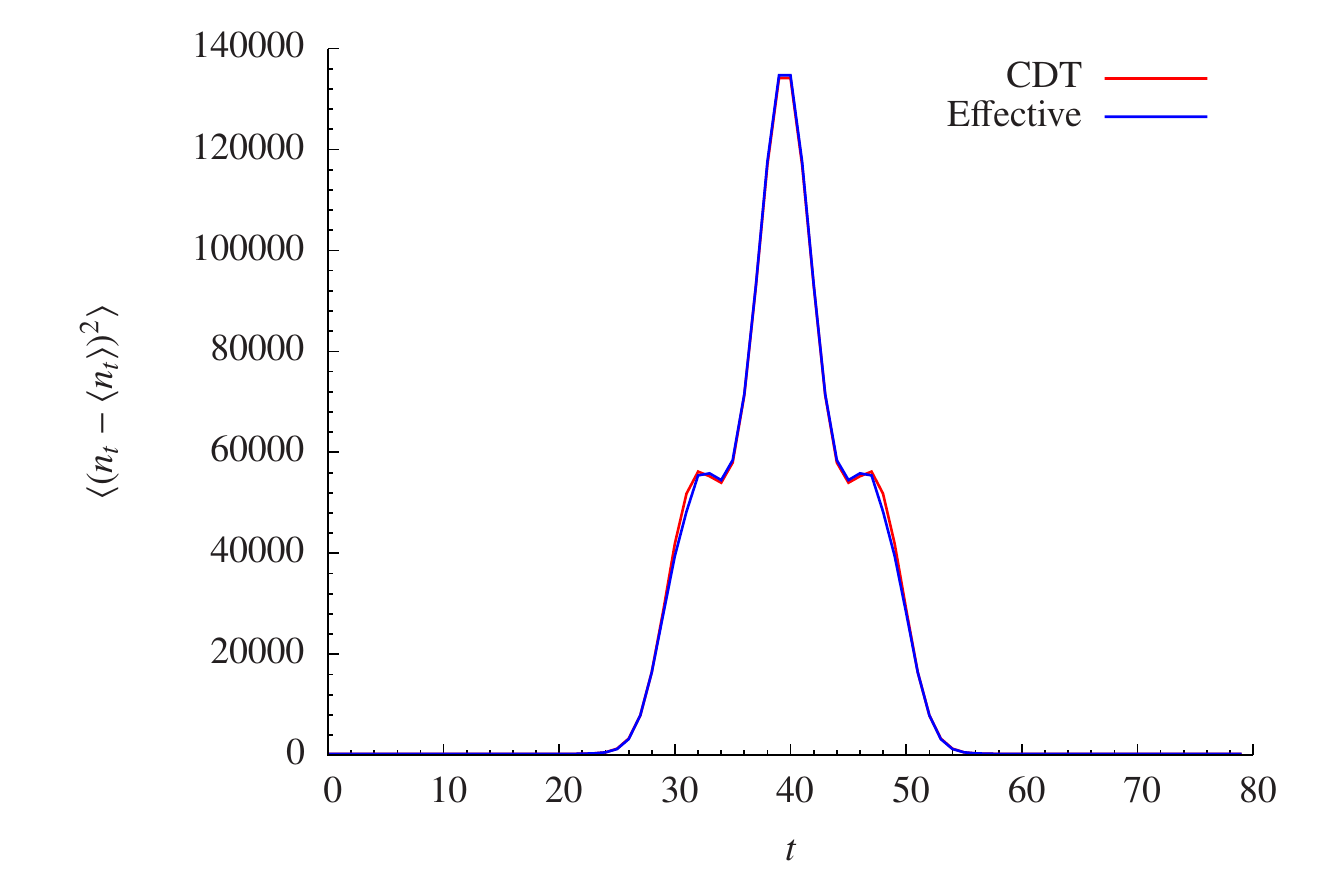}}
\caption{Diagonal of the covariance matrix of three-volume fluctuations, i.e. the variance $C_{tt}=  \langle (n_t - \langle n_t \rangle) ^2 \rangle$,  measured in the ``full CDT" simulations for $\kappa_0=2.2$, $\Delta=0.6$ and $\bar V_4=40$k (red) and in the effective model (\ref{eq:pne}) based on the transfer matrix $M$ (blue). }
\label{FigModelcov}
\end{figure}


\chapter{Effective action in phases ``A" and ``B"}
This Chapter is  based on the article: {\it J.\,Ambj\o rn, J. Gizbert-Studnicki, A. G\"orlich, J. Jurkiewicz, ``The effective action in 4-dim CDT. The transfer matrix approach",  
JHEP 06 (2014)}
\newline
\newline

In the previous  Chapter we showed that inside the ``de Sitter" phase one can use the effective transfer matrix to investigate the properties of the effective action in that phase. Now we would like to extend the analysis to the other  phases. Assuming that the  transfer matrix description is still legitimate, the transfer matrix approach is at the moment the only way to analyze the effective action in these phases as one does not have much information from volume-volume correlations - in the phase ``A" the spatial three-volumes are uncorrelated and in the phase ``B" the generic triangulations are ``collapsed" to just one spatial slice.

The research presented in this Chapter  was partly inspired by a recent study \cite{burda}, in which the effective Lagrangian
\beq\label{Lburda}
L [n,m]= c_1 \frac{2(n-m)^2}{ n+m} +c_2 \frac{m^{1/3}+n^{1/3}}{2} 
\eeq
was used to define an effective transfer matrix $\braket{n|M|m}=e^{-L[n,m]}$ and to construct a simple ``three-volume fluctuation"  model, resembling the one described in Chapter 4.5. The authors identified five different phases (as a function of the coupling
constants $c_1$ and $c_2$), three of  which were quite similar to the CDT phases (as far as we focus only on the spatial volume observable).  The effective 
Lagrangian (\ref{Lburda})  resembles the  Lagrangian  (\ref{Seffform2}) measured in our studies of the ``de Sitter" phase ``C". Thus we wanted to 
check, using the CDT simulations, if (\ref{Lburda}) could really describe
 the  CDT effective action also  in phases 
``A" and ``B" for certain choices of $c_1$ and $c_2$. As we will see this turned out not to be the case.

The measurement of the transfer matrix inside the ``de Sitter" phase was quite  straightforward. In the other two phases it has to be done with
some care. In particular,  we
had to modify the previously introduced method of volume fixing. This change was dictated by the need of high precision measurements of the  off-diagonal transfer matrix elements $\braket{n|M|m}$ (for $|n-m| \gg 0$) which are  important in phases ``A" and ``B".   In a previously used local volume fixing method (\ref{LocalVF}), the measured probability distributions were concentrated around $n=m=\bar N_3$ and consequently only the matrix elements which are ``not too far" from a diagonal were measured with high precession.\footnote{Inside the ''de Sitter" phase there is no such need, as the cross-diagonal transfer matrix elements fall off as  $\exp\left(-1/k[n+m] \cdot (n-m)^2\right)$ and therefore are  (close to) zero for $|n-m| \gg 0$ anyway.   }
Therefore we decided to change it to a {\it global} volume fixing, which  also allows one to  ``access"  the matrix elements  far from the diagonal. 
The change of the volume fixing required using probability distributions measured in numerical simulations with the $T=2$ time period, as only for such a case one can cancel the global volume fixing term from the empirical data. Following the approach described in detail in Chapter 4.2, one can modify the bare CDT action:
$$
S_R \quad \to \quad \widetilde S_R = S_R + S_{VF}
$$
 by a quadratic or a linear  global volume fixing potential
\beql{GlobalVF}
 S_{VF} = \eps (n_1+n_2 - \bar V_4)^2 \quad \text{or} \quad  S_{VF} = \eps |n_1+n_2 - \bar V_4| 
\eeq
and measure the probability distribution $\widetilde P^{(2)}(n_{1},n_{2})$ of observing a spatial volume $n_1$ in time $t=1$ and a spatial volume $n_2$ in time $t=2$. The probability distribution can  be used to calculate the auxiliary transfer matrix elements, according to Eq. (\ref{MT2}):
 \beql{MT2a}
 \braket{n|\widetilde M|m}=\cN_0 \sqrt{\widetilde P^{(2)}(n_{1}=n,n_{2}=m)}
 \eeq
 and  the (corrected) transfer matrix elements (after cancelling the volume fixing term), given by:
 \beql{MT2b}
 \braket{n| M|m}=e ^{\oh \eps (n+m - \bar V_4)^2} \braket{n| \widetilde M|m}  \text{  or  }  \braket{n| M|m}=e ^{\oh \eps |n+m - \bar V_4|} \braket{n| \widetilde M|m}  
 \eeq
 for a quadratic or a linear potential, respectively. As already mentioned, we verified that inside the ``de Sitter" phase the new method (from $T=2$ probability distribution) leads to  exactly the same transfer matrix $M$ as the old method (from a combination of $T=3$ and $T=4$ probability measurements, see Eq's (\ref{TMNew}) and  (\ref{MPure})) used in Chapter 4. 
 Apart from the fact, that by using $T=2$ and a global volume fixing (\ref{GlobalVF}) one can determine the off-diagonal $M$ matrix elements with high statistics, the new method also considerably reduces the use of computer resources as one has to measure only one instead of two probability distributions, and additionally reduces possible rounding errors as one does not have to combine two sets of the empirical data. One should keep in mind that for the cross-diagonal transfer matrix elements (for fixed $s=n+m$) one can directly use the auxiliary matrix $\braket{n|\widetilde M|s-n}$  (Eq. (\ref{MT2a})) instead of  $\braket{n| M|s-n}$ (Eq. (\ref{MT2b})) as both cross-diagonals are identical up to a simple rescaling (for fixed $s$ the exponential  term in Eq. (\ref{MT2b}) is just a constant).
 The new method was used in all measurements described in this and the next Chapter.

\section{Effective action in phase ``A"}

Phase ``A" is observed for large values of the bare (inverse) gravitational coupling constant $\kappa_0$ (see phase diagram in  Fig. \ref{Figfazy}) and is characterized by the uncorrelated three-volume distributions (see Chapter 2.1 for details). 
We measured the transfer matrix in a generic point inside this phase (for $\kappa_0= 5.0, \Delta =0.4, K_4=1.22$) using the probability measurements in $T=2$ system and  a quadratic volume fixing term (\ref{GlobalVF}).
We will follow the same analysis methods as described in the previous Chapter, i.e. we assume that the transfer matrix can be factorized into the kinetic and the potential part, corresponding to the kinetic and the potential part of the effective action / Lagrangian, respectively.  The former is  given by the cross-diagonal elements $\braket{n| M|s-n}$ for fixed $s=n+m$, and the latter by the diagonal elements $\braket{n|M|n}$.

\begin{figure}[h!]
\centering
\scalebox{0.7}{\includegraphics{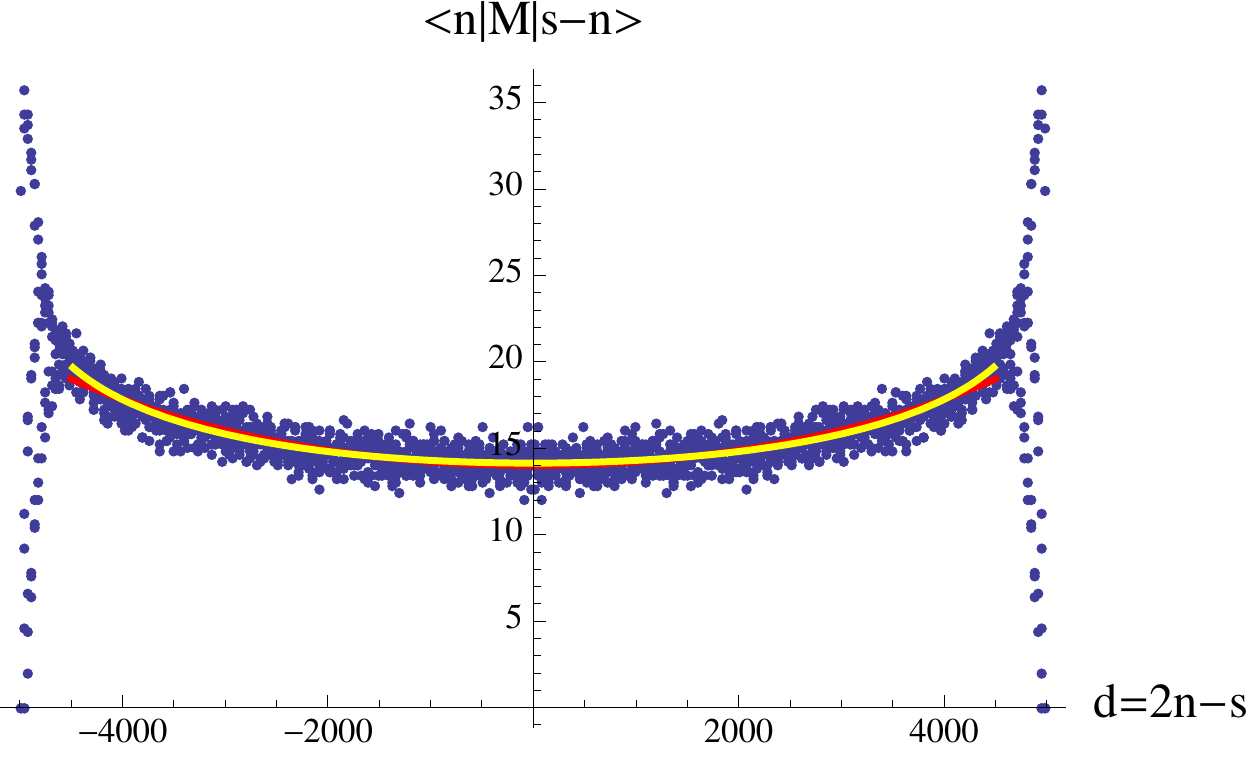}}
\caption{A cross-diagonal of the empirical transfer matrix inside 
phase ``A" (for $\kappa_0= 5.0, \Delta =0.4, K_4=1.22$), measured for $s=n+m=5000$. 
The red line corresponds to the fit of Eq.  
(\ref{Akinfit}). The  yellow line is the best fit 
of the  effective Lagrangian (\ref{LeffAcross}). }
\label{FigA1}
\end{figure}

Let us start with the kinetic part. 
The generic  behaviour   of the empirical cross-diagonal of $M$ is presented in 
Fig.\ \ref{FigA1} where $\braket{n|M|s-n}$ is plotted as a function 
of $d\equiv(n-m)=2n-s$. It is very  different from the shape measured inside the ``de Sitter" phase, where it  is Gaussian (see e.g. Fig. \ref{FigBlobAnti}). The ``corners" of the measured cross-diagonal are dominated by strong discretization effects, characteristic for small volumes  $n$ or $m=s-n$.
In a region where the discretization effects vanish one can fit the measured data with the following parametrization (the red line in Fig. \ref{FigA1}):
 \begin{equation}
 \braket{n|M|s-n}=  {\cal N}(n+m)\exp \left( \frac{(n-m)^{ 2}}{k[s]}\right)={\cal N}(s)\exp \left( \frac{d^{ 2}}{k[s]}\right) 
 \label{Akinfit}
 \end{equation}
 which is, up to a sign under the exponent ($e^{+x^2}$ instead of $e^{-x^2}$),  identical to the same expression considered for the effective Lagrangian inside  the ``de Sitter" phase.
 The kinetic coefficients $k[s]$ measured for different  $s$ are 
plotted  in Figure \ref{FigA2}.
\begin{figure}[h!]
\centering
\scalebox{0.7}{\includegraphics{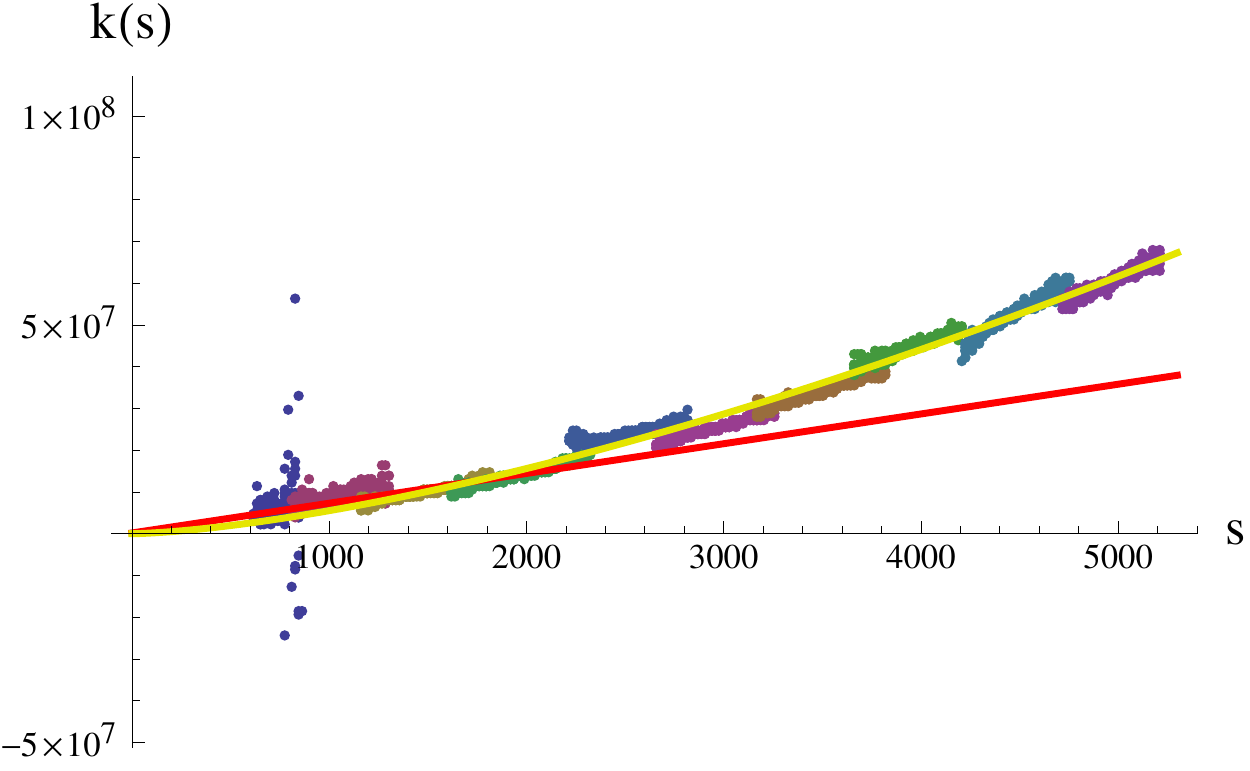}}
\caption{The kinetic coefficient $k[s]$ measured in  phase ``A" 
(for $\kappa_0= 5.0, \Delta =0.4, K_4=1.22$) for different $s=n+m$. $k[s]$ is not linear (red line) 
but can be fitted using the function defined by Eq.\  (\ref{KSa}) 
(yellow line).}
\label{FigA2}
\end{figure}
In contrast  to the behaviour inside the ``de Sitter" phase (Fig. \ref{FigBlobKin}), $k[s]$ is not linear, but can be fitted very well   with the following expression (yellow curve in  Fig. \ref{FigA2})
\begin{equation}
k[s]=k_0 s^{ 2-\alpha}.
\label{KSa}
\end{equation}
The best fit yields $\alpha=0.50\pm0.01$ and $k_0=175\pm10$. 

Does the occurrence of the negative kinetic coefficients  (reversed sign under the exponent) suggest that the kinetic term of the effective  action changes sign inside the ``A" phase compared to  phase ``C" ? We would rather naively expect that the kinetic term  should vanish inside  phase ``A", as the empirical three-volumes of different spatial slices are uncorrelated. To check this let us assume that the volume distributions are in fact independent which leads in a natural way to the following effective Lagrangian:
\begin{equation}
L_A(n,m)= \mu \left(n^{ \alpha} + m^{ \alpha} \right) - \lambda(n+m) \ ,
\label{LeffA}
\end{equation}
where the kinetic part is not present and we just have a {\it local} (diagonal) potential term.

By changing the variables $n$ and $m$ to $s=n+m$ and $d=n-m$  one obtains:
\begin{equation}
L_A(n,m)= \mu \left(\frac{s}{2}\right)^\alpha \left[ \left(1 + \frac{d}{s}\right)^\alpha+ \left(1 - \frac{d}{s}\right)^\alpha \right]-\lambda s =
\label{LeffAseries}
\end{equation}
$$
= -\lambda s + \mu \left(\frac{s}{2}\right)^\alpha \left[ 2 + \alpha(\alpha-1)\left(\frac{d}{s}\right)^2\right] +{\cal O}((d/s)^4)
$$
where we assume  $d/s \ll 1$ .\footnote{In fact, close to the ``corners" of the measured cross-diagonals of $M$,  $d/s$ is  close to 1 and therefore higher order corrections should be 
taken into account.}

For $\alpha<1$ one effectively observes the measured shape of empirical 
 cross-diagonals (Eq. \ref{Akinfit}) with
\begin{equation}
k[s] = \frac{2^\alpha}{\mu \ \alpha (	1-\alpha)}s^{ 2-\alpha},
\label{kseff}
\end{equation}
exactly in line with Eq. (\ref{KSa}). 
From the fitted values of $\alpha$ and $k_0$ 
one can calculate
$$\mu= \frac{2^\alpha}{k_0 \ \alpha (	1-\alpha)} = 0.032\pm0.002 \ .$$
To verify that the assumed effective Lagrangian (\ref{LeffA}) accurately describes the empirical data one may use 
it to fit the cross-diagonal elements 
of the measured transfer matrix as a function of $n$ (for fixed $s$):
\begin{eqnarray}
\braket{n|M|s-n}&=&{\cal N} \exp\left[ - L_A(n,s-n)\right] = \nonumber \\&=&{\cal N}(s) \exp\left[ - \mu n^{ \alpha} - \mu (s-n)^{ \alpha} \right] \ .
\label{LeffAcross}
\end{eqnarray}
The best fit for $\alpha=0.5$ is presented as a yellow curve in 
Fig. \ref{FigA1} and for $s=5000$ yields $\mu=0.022\pm0.001$. 
We fitted Eq. (\ref{LeffAcross}) to the measured cross-diagonals for different $s$. The value of $\mu$  is not determined with a very high precision but it seems to be constant for large volumes (big $s$) 
where discretization effects are smaller - see Fig. \ref{FigA3}, the  red line corresponds to $\mu=0.024$.
\begin{figure}[h!]
\centering
\scalebox{0.7}{\includegraphics{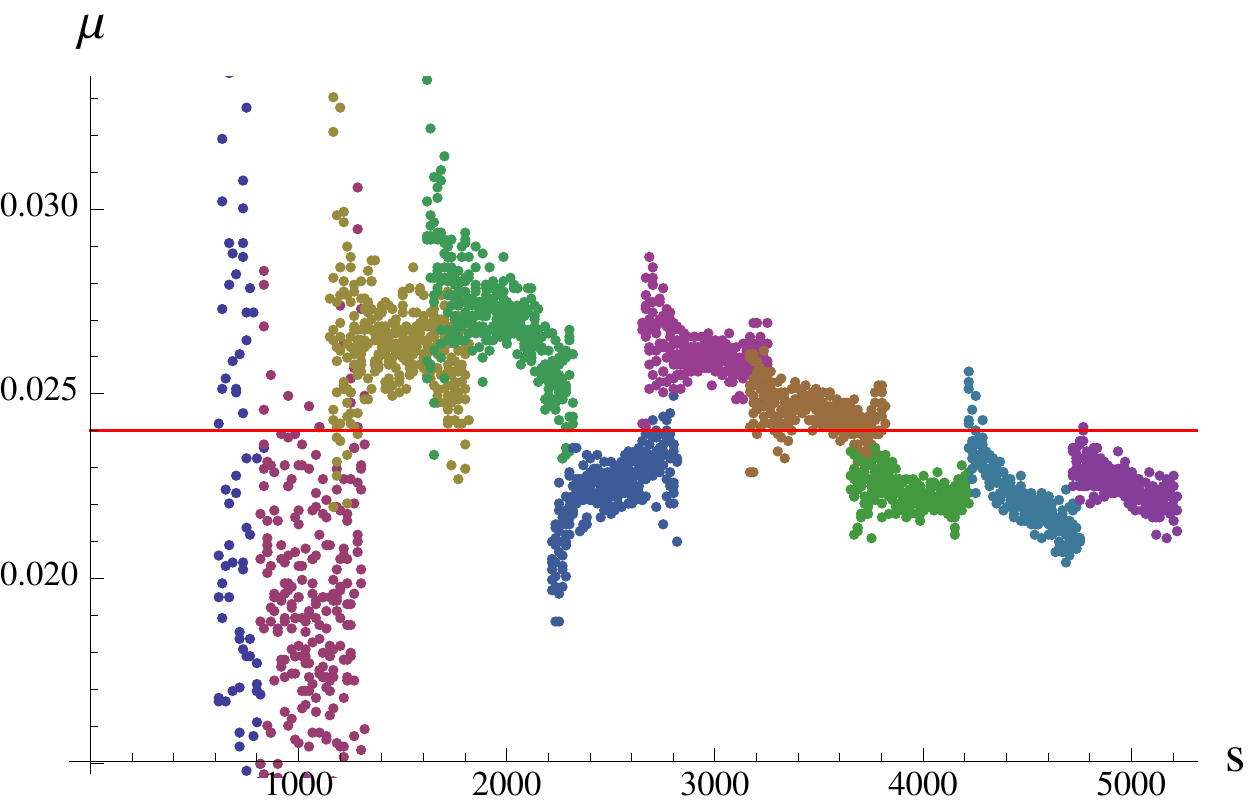}}
\caption{$\mu$ as a function of $s$ measured in phase ``A"
(for $\k_0= 5.0, \Delta =0.4, K_4=1.22$). The value of $\mu$ 
stabilizes around $0.024$ (the red line) as  discretization effects vanish.}
\label{FigA3}
\end{figure}

The proposed form of the effective action (\ref{LeffA}) is further confirmed by the analysis of the diagonal elements of $M$: 
 \begin{equation}
 \log \braket{n|M|n}=-L_A(n,n) + 
\log{\cal N}=-2 \mu n^\alpha + 2 \lambda n+ \log{\cal N} \ ,
 \label{MpotA}
 \end{equation}
which is shown in Fig. \ref{FigA4}. The best fit (with assumed  $\alpha=0.5$) yields $\mu=0.026$ and is plotted as a yellow line.

\begin{figure}[h!]
\centering
\scalebox{0.7}{\includegraphics{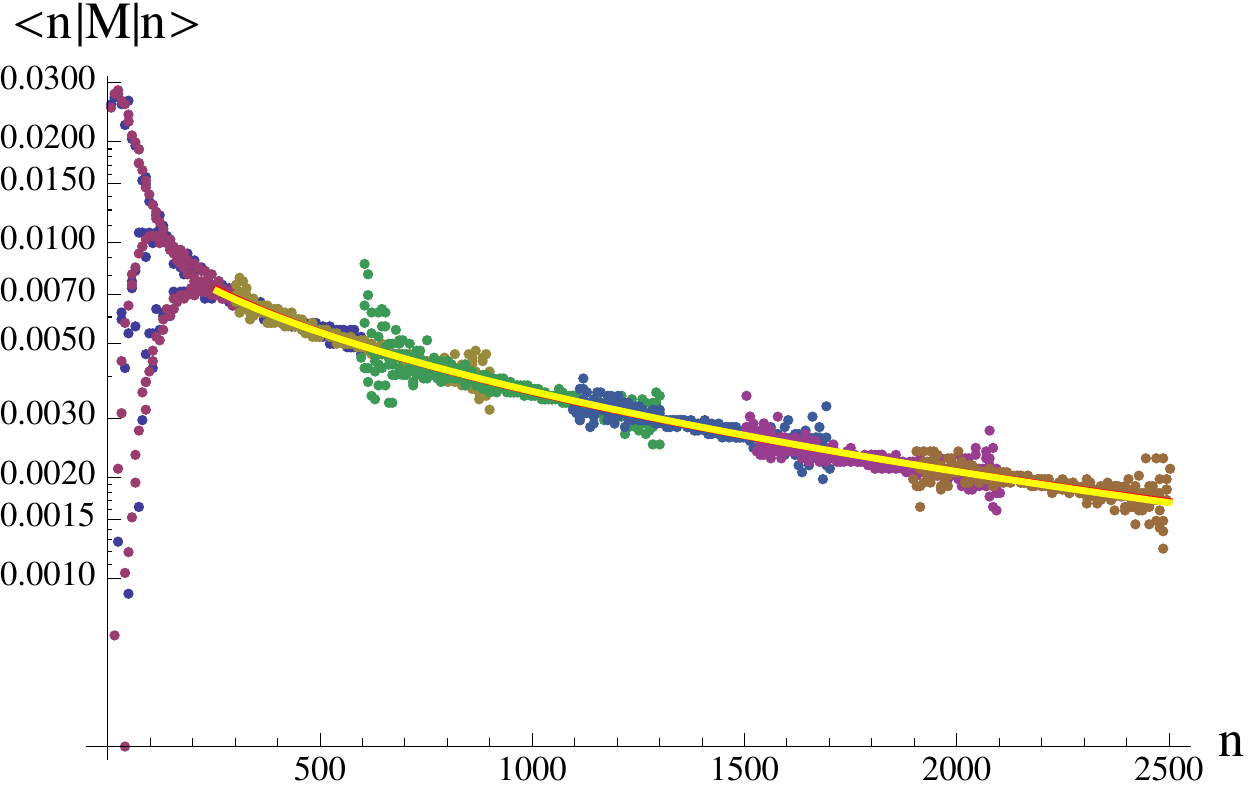}}
\caption{The diagonal elements of the transfer matrix measured 
in phase ''A" (for $\k_0= 5.0, \Delta =0.4, K_4=1.22$) - points, and 
 the best fit to Eq.\ (\ref{MpotA}) - yellow line. 
The fit is for  $n\geq 300$  and ignores strong discretization effects visible for small volumes. Note that the vertical axis in the above plot is logarithmic.}
\label{FigA4}
\end{figure}

Summarizing, we have shown that the empirical transfer matrix data, measured inside  phase ``A", correspond to the following effective action:
\beql{SeffAP}
S_{A} = \sum_{t} \Big( \mu \, n_t^{ \alpha}  - \lambda \, n_t \Big) \ ,
\eeq
where, conversely to the ``de Sitter" phase,  the kinetic term is not present and the value of $\alpha\neq 1/3$. Vanishing of the kinetic term inside  phase ``A" may be interpreted as the effective Newton's constant $G_{eff}\to\infty$ limit, causing  a causal disconnection
of different spatial  slices, i.e. the phenomenon of ``asymptotic silence" observed both in
classical and quantum approaches to gravity in the regime of extreme curvatures/energy
densities \cite{mielczarek}. In this context   phase ``A"  might gain some physical meaning.
\newline

As a side remark we may go back to the parametrization of the effective action inside the ``de Sitter" phase. 
In Chapter  3, where we measured the effective action by means of the fluctuation matrix, we used a diagonal potential  $V[n]+V[m]$, while in Chapter 4, where we considered  transfer matrix measurements, we changed it to the symmetric form $v[n+m]$. If the true potential was in fact diagonal (as in Chapter 3) one should expect the same 
kind of effective correction of the  measured transfer matrix cross-diagonals, as described by Eq. (\ref{LeffAseries}). As a result the empirical kinetic coefficients $k[s]$ would be slightly modified and to get ``true" $k[s]$ one should subtract  the correction (\ref{kseff}) from the measured data. 
This possible effect is very small compared to the generic Gaussian 
behaviour of the kinetic part in  phase ``C", but potentially this could 
explain the existence of the non-vanishing but very small $n_0$ parameter  
in the measured effective Lagrangian (see e.g.  Eq. \ref{Seffform2}).

\section{Effective action in phase ``B"}

Phase ``B" is observed for small values of the bare asymmetry parameter $\Delta$ (see phase diagram in  Fig. \ref{Figfazy}) and is characterized by the ``collapse" of the three-volume distribution into a single spatial layer (see Chapter 2.1 for details). 
We measured the transfer matrix in a generic point inside this phase (for $\kappa_0= 2.2, \Delta =0$) using the probability measurements in the $T=2$ system.
During the numerical simulations we encountered an additional issue related to the appropriate choice of the  $K_4$ bare coupling constant, which should be fine-tuned to the critical value. The problem is that inside   phase ``B" the fine-tuned value of $K_4$  strongly depends on the total volume $\bar V_4$ which we fix in Monte Carlo simulations by introducing a  quadratic global volume fixing term (\ref{GlobalVF}). This is illustrated in  Fig.\ \ref{Fig4a}, 
where the value of $K_4$ is plotted as a function of  $\bar V_4$ 
together with the best fit of:
\begin{equation}
K_4(\bar V_4)=K_4^\infty  - \beta \, \bar V_4^{-\gamma} \ ,
\label{K4vol}
\end{equation}
which gives: $K_4^\infty  =1.05$, $\beta = 0.20$ and $\gamma = 0.31$.

\begin{figure}[h!]
\centering
\scalebox{0.9}{\includegraphics{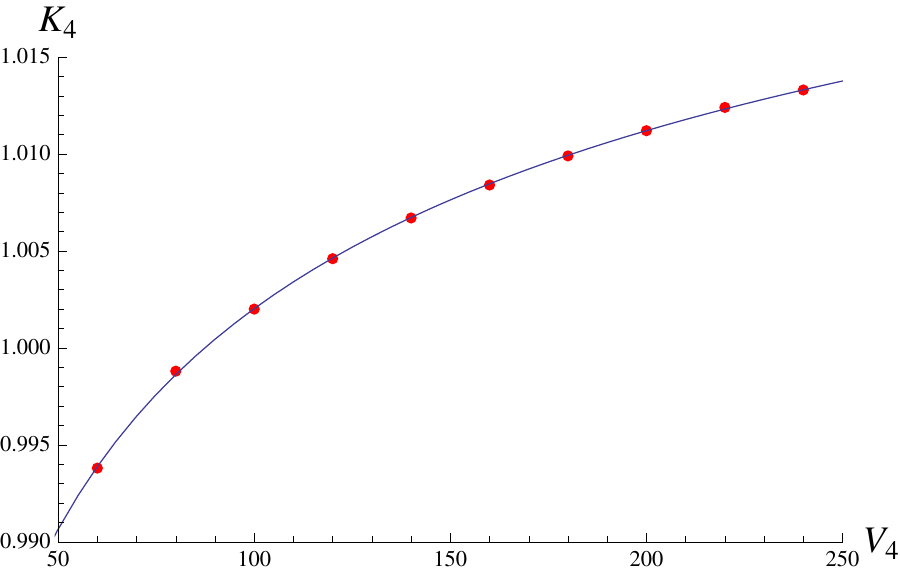}}
\caption{$K_4$ scaling with total volume ($\bar V_4$ in thousands) inside the  phase ``B" (for $\kappa_0=2.2 , \ \Delta=0.0$)  and the best fit of Eq. (\ref{K4vol}). }
\label{Fig4a}
\end{figure}

This strong volume dependence can be associated with the subleading  entropic terms which translate into the spontaneously emerging effective potential. The  entropy of states comes from a large number of possible triangulations $\cT$  with 
constant total volume $\bar V_4$. In Chapter 1.5 we argued that in a leading order the number of such triangulations is growing exponentially with $\bar V_4$ and the $K_4$ coupling constant  of the bare Regge action (\ref{SR}), 
which is  conjugate to the  total four-volume, can be fine-tuned to offset this leading behaviour. In phases ``A" and ``C" the subleading corrections are not important in the whole  range of the measured $\bar V_4$ and the fine-tuned $K_4$ is a constant independent of the total volume. In  phase ``B" the situation is much different. Even for relatively large total volumes (above $200$ thousand of the \{4,1\} simplices) we still observe a change in $K_4$ as $\bar V_4$ is increased and we expect  the subleading corrections to vanish for even larger total volumes - Eq. (\ref{K4vol}) implies that  $K_4$ finally tends to a constant  $K_4^\infty$, for $\bar V_4 \to \infty$.
The strong volume dependence of $K_4$ makes it 
technically impossible to measure the transfer matrix 
in phase ``B"  for the values of $K_4$ appropriate for a large volume limit. 
If we fix the $K_4$ value to the critical value corresponding to 
a large volume, effectively this value is ``too large'' and 
the system will oscillate around the minimally allowed configuration.
To overcome this problem we decided to use  lower values of $K_4$ and analyze 
how a change of  $K_4$ affects the measured transfer matrix data. 
As a result we can (at least qualitatively) estimate 
the properties of the transfer matrix in the continuum limit.

Another problem which we encounter inside  phase ``B" is that the  probabilities of spatial volume distributions $\widetilde P^{(2)}(n_1,n_2)$ used in the transfer matrix measurements
are highly asymmetric in general.  To construct the transfer matrix we explicitly symmetrize the data: $\braket{n|M|m}=\braket{m|M|n}$. It  is equivalent to regaining the  
time-reflection symmetry of the effective action, which is strongly broken 
by generic configurations.   Nevertheless, due to this asymmetry, the diagonal transfer matrix elements $\braket{n|M|n}$ are very small for large $n$ and therefore are not determined to a high precision. As a result it is very difficult to measure the potential (diagonal) part of the effective action. Additionally, the potential term for small volumes shows a very non-trivial behaviour, which is presumably due to the subleading corrections of the entropic factor discussed above. Consequently  we decided to focus only on the kinetic term. As already explained, for the kinetic part alone one can   use  data from an auxiliary matrix $\widetilde M$  instead of the (volume fixing corrected) transfer matrix $M$.

\begin{figure}[h!]
\centering
\scalebox{0.52}{\includegraphics{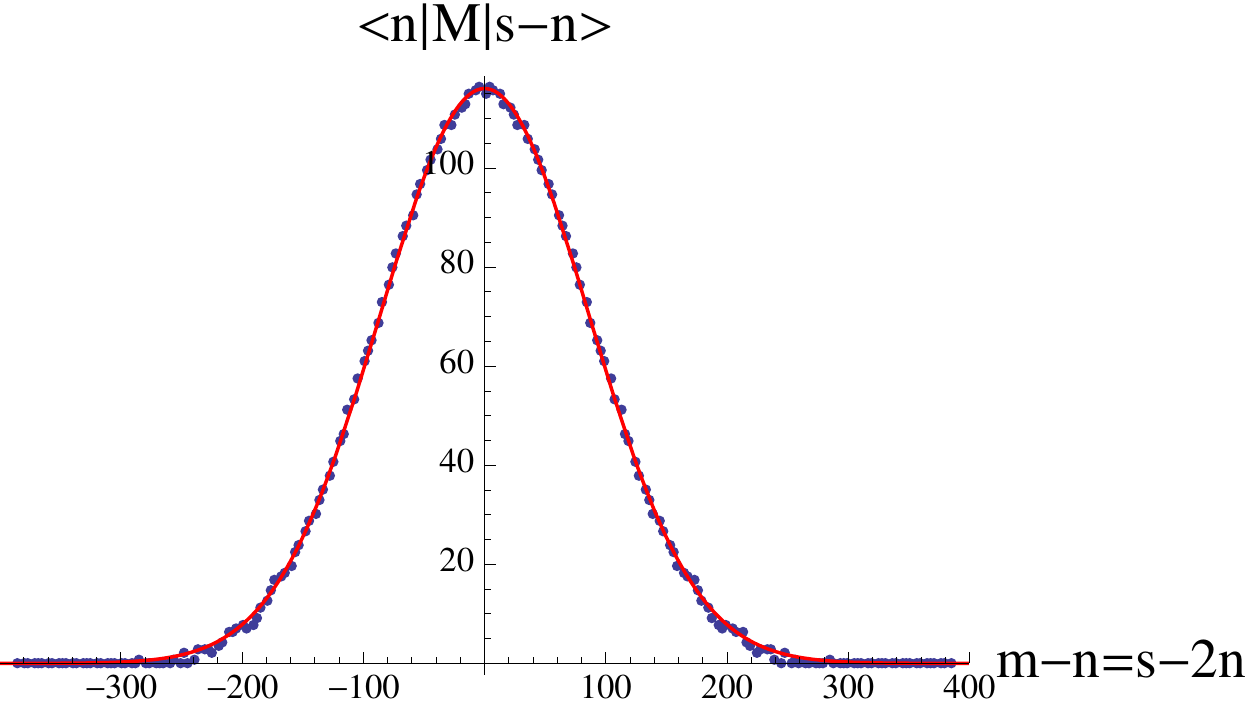}}
\scalebox{0.52}{\includegraphics{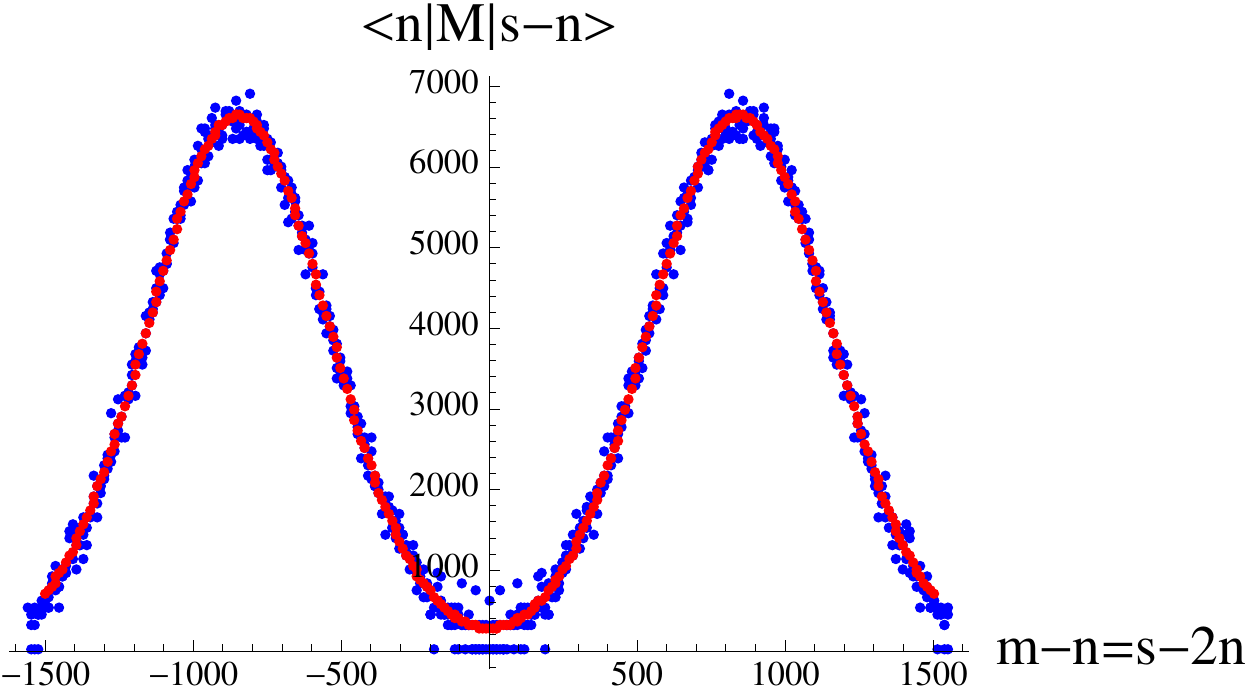}}
\caption{Cross-diagonals of the empirical transfer matrix measured in  phase ``B" (for $\k_0=2.2$, $\Delta=0.0$ and $K_4=0.943$). The left chart presents the data  below bifurcation point ($s=n+m<s^b$). The right chart shows cross-diagonals  above bifurcation point ($s>s^b$). The best fits of Eq. (\ref{SkinB}) are presented as red lines.}
\label{FigB5}
\end{figure}

We start with a transfer matrix measured for $K_4=0.943$. 
The typical behaviour of the cross-diagonals (kinetic part) of the 
empirical transfer matrix $\braket{n|M|s-n}$ as a function of the three-volume $n$ is plotted in  Fig. \ref{FigB5}. The shape
 depends on the total volume $s=n+m$. For small $s$ the cross-diagonals show a Gaussian dependence on $n$, exactly as in the ``de Sitter" phase - see Fig. \ref{FigB5} (left). For large $s$   
the cross-diagonals split into the sum of two ``shifted" Gaussians - 
see Fig. \ref{FigB5} (right). The value of $s$ for which the split occurs will be called a {\it bifurcation point} and denoted by $s^b$.  Both cases (with and without the bifurcation) can be joined in the following parametrization:

\begin{equation}
\braket{n|M|s-n}=
\label{SkinB}
\end{equation}
$$
 ={\cal N}(s)\left[ \exp \left( -\frac{\Big((m-n) - c[s]\Big)^2}{k[s]}\right) + \exp \left( -\frac{\Big((m-n) + c[s]\Big)^2}{k[s]}\right)\right] \ ,
$$
where $k[s]$ is a {\it kinetic coefficient} and  $c[s]$ is a {\it bifurcation shift} which depends on $s$. 
The  dependence of $k[s]$ and $c[s]$ on $s$ can be measured by fitting Eq. (\ref{SkinB}) to the empirical cross-diagonals for a range of values of $s$.

\begin{figure}[h!]
\centering
\scalebox{0.67}{\includegraphics{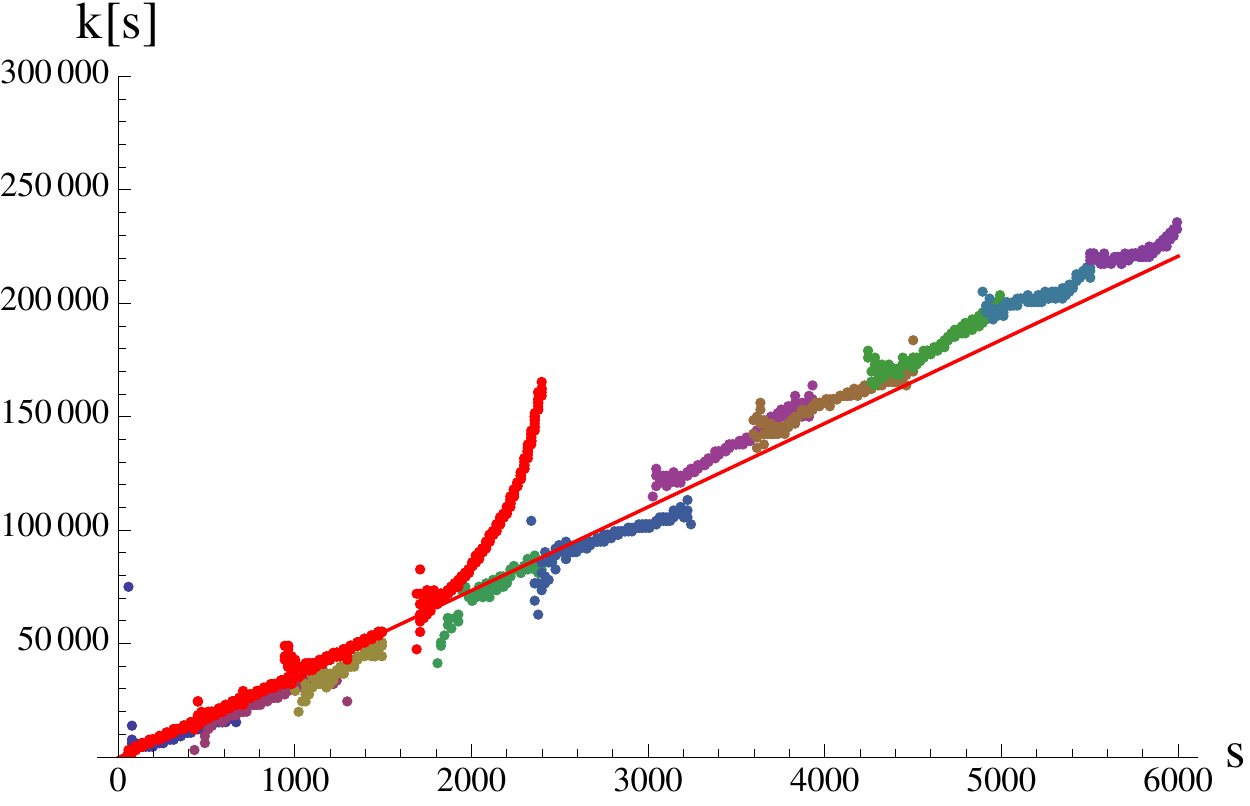}}
\caption{The kinetic coefficient $k[s]$ 
measured in  phase ``B" (for $\kappa_0=2.2$, $\Delta=0.0$ and $K_4=0.943$).  
The red points correspond to the fit of $\braket{n|M|s-n}$ to a single Gaussian, which is not valid  after crossing the bifurcation point (for $s>s^b$). 
Different colours  correspond to the fit of two  shifted Gaussians (\ref{SkinB}), 
which is reliable also above the bifurcation point (for $s>s^b$).}
\label{FigB7}
\end{figure}

The kinetic coefficient  $k[s]$  is very 
well approximated by a linear function, independently   
of whether we are below or above the bifurcation point - see Fig. \ref{FigB7}:
\begin{equation}
k[s] = \Gamma(s - 2 n_0) \ .
\label{ksB}
\end{equation}
This behaviour is consistent with the kinetic coefficients measured inside the ``de Sitter" phase (see e.g. Fig. \ref{FigBlobKin}).
The best fit yields: $\Gamma= 36.8$, $n_0=5.4$ 
which is also of the same order as the values measured inside   phase ``C".  The bifurcation point $s^b$ can be identified as the point 
at which the fit of $\braket{n|M|s-n}$ to the two shifted Gaussians (described by Eq.  (\ref{SkinB}))  starts to diverge from  
the fit to a single Gaussian (described by the same equation but with enforced $c[s]\equiv 0$). 

\begin{figure}[h!]
\centering
\scalebox{0.67}{\includegraphics{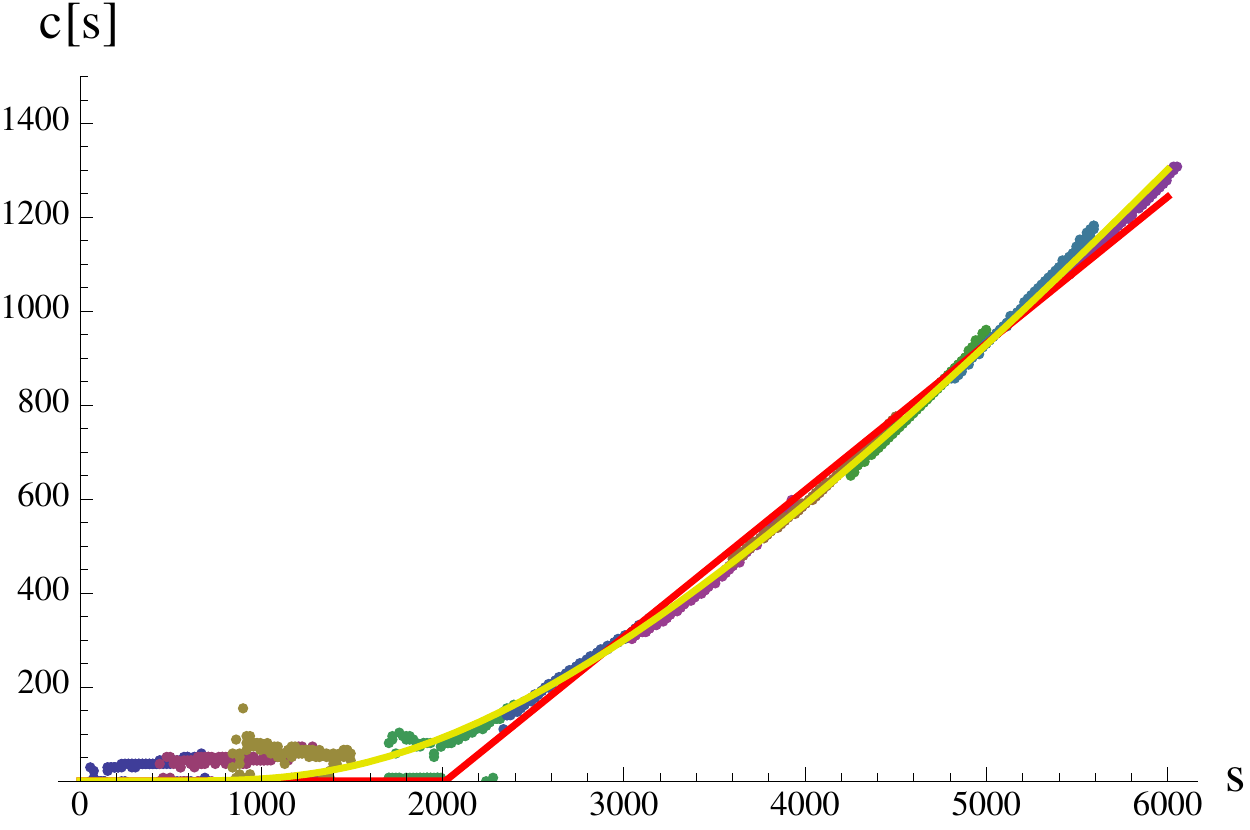}}
\caption{The bifurcation shift c[s] measured in phase ``B" 
(for $\kappa_0=2.2$, $\Delta=0.0$ and $K_4=0.943$) (points)
 and the best fits using Eq. (\ref{cs}) 
(red line) and (\ref{csexp}) (yellow line).}
\label{FigB6}
\end{figure}

The bifurcation shift $c[s]$ is (close to) zero for $s<s^b$ and (almost) linear for $s>s^b$ - see Fig \ref{FigB6}.  One can use:
\begin{equation}
c[s] = \max[0,c_0(s-s^b)] 
\label{cs}
\end{equation}
to fit the empirical  data quite well 
(the red line in Fig. \ref{FigB6}).  
For the generic data ($\kappa_0=2.2$, $\Delta=0$ and $K_4=0.943$) 
the best fit is for: $s^b=2020$ and $c_0=0.31$. It is consistent with the  value  of $s^b$ measured by looking at the divergence of the single Gaussian from the two Gaussians, as described above - see Fig.\ \ref{FigB7}.

Since we are interested in properties of our model  in 
the large volume limit (where critical values of $K_4$ are much higher) 
it is important to check how  the results depend on $K_4$. 
The plots of $k[s]$ and $c[s]$ for a number of $K_4$'s are presented in 
Fig.\ \ref{FigB8}. In general, the functional form of Eq's (\ref{SkinB})-(\ref{cs}) 
is adequate for different values of $K_4$.  With regard to the parameters
entering in Eq's (\ref{SkinB})-(\ref{cs}), the change of $K_4$ 
does not influence the position of the bifurcation point $s^b$,
while the bifurcation slope $c_0$ and the effective Newton's constant 
$\Gamma$ rise as $K_4$ is increased. 

\begin{figure}[h!]
\centering
\scalebox{0.52}{\includegraphics{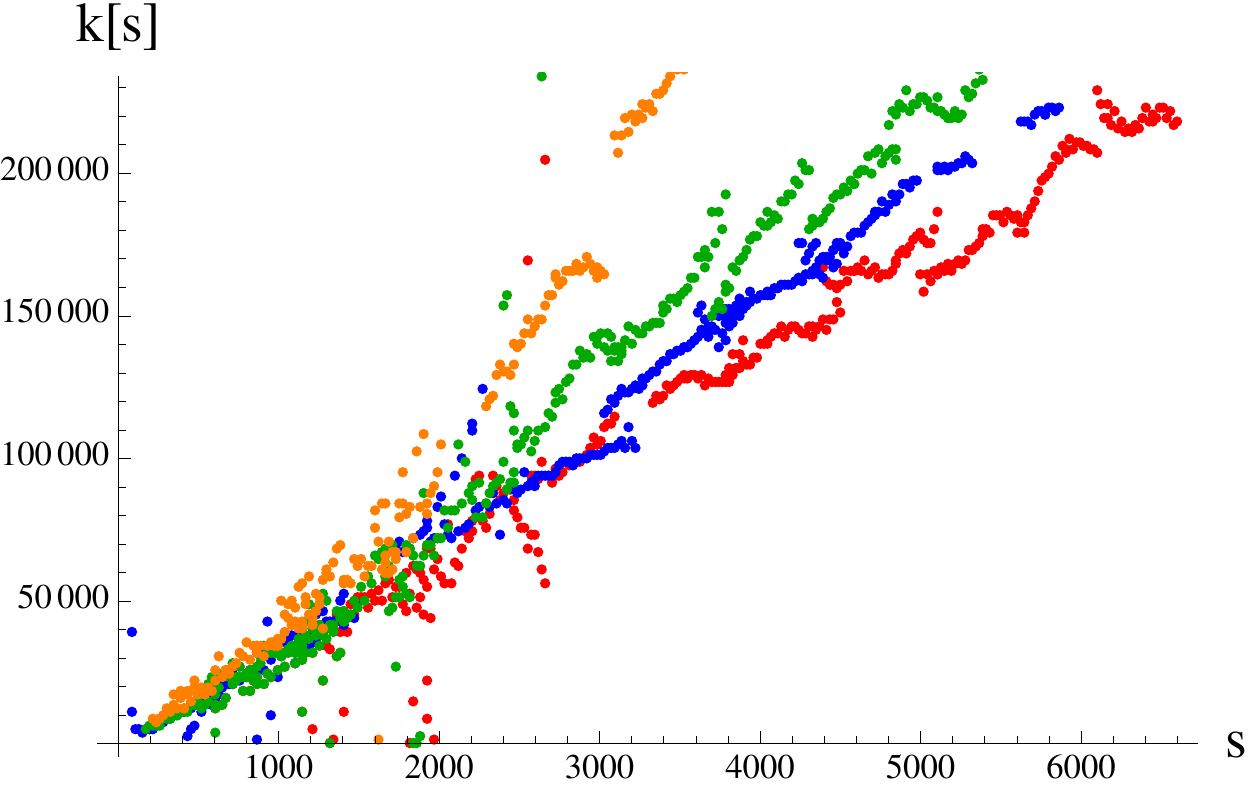}}
\scalebox{0.52}{\includegraphics{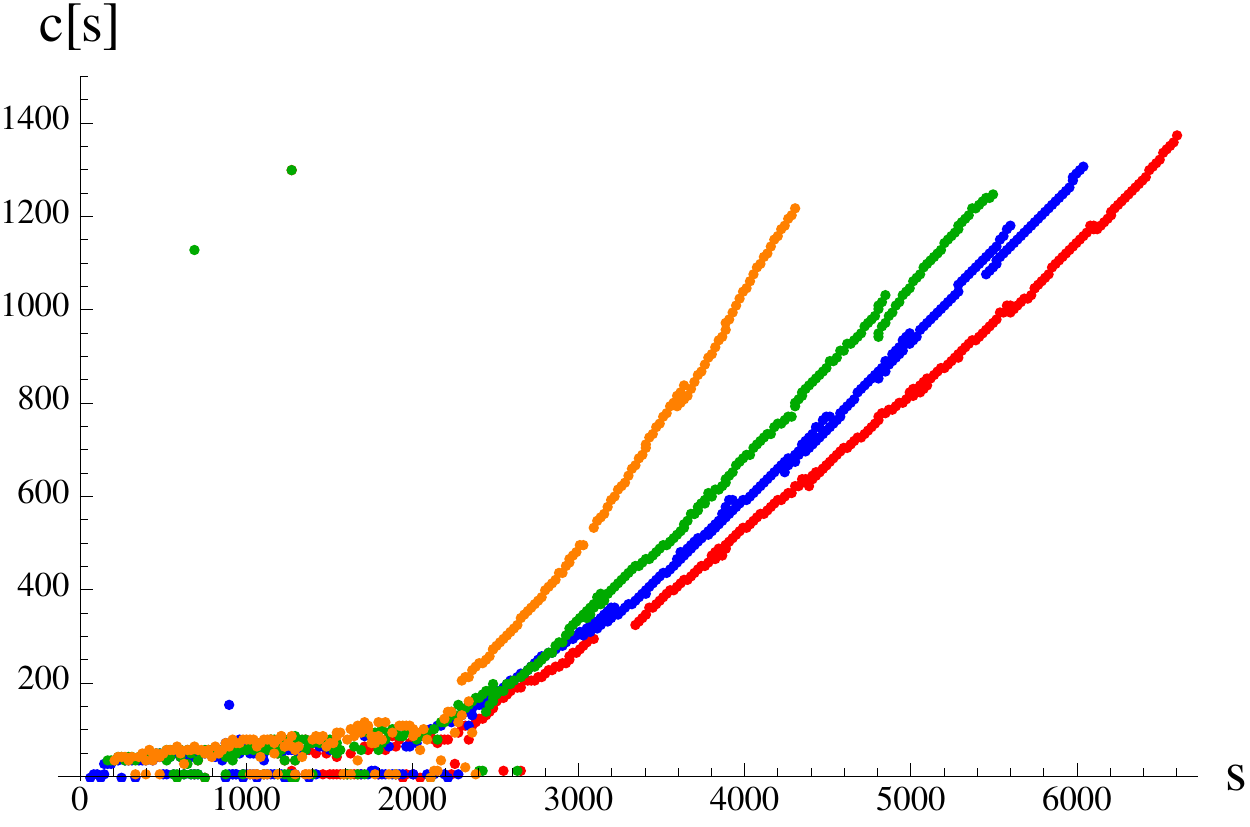}}
\caption{The kinetic coefficient $k[s]$ (left) and the 
bifurcation shift   $c[s]$ (right) measured  in  phase ``B" 
(for $\kappa_0=2.2$, $\Delta=0.0$) for a number of values of  
$K_4=$ 0.933 (red), 0.943 (blue), 0.953 (green), 0.973 (orange)}
\label{FigB8}
\end{figure}

All together, the transfer matrix inside  phase ``B"  can be parametrized as follows:

\beql{TMBphase}
\braket{n|M_B|m} =
{\cal N}[n+m]
\left[ 
\exp \left( -\frac{\Big((m-n) -\big[ c_0(n+m-s^b) \big]_+\Big)^2}{\Gamma(n+m-2 n_0)}\right) + \right.
\eeq
$$
\left.
+\exp \left( -\frac{\Big((m-n) +\big[ c_0(n+m-s^b) \big]_+\Big)^2}{\Gamma(n+m-2 n_0)}\right) 
\right] \ ,
$$
where: $[.]_+ = \max(. , 0)$, and the values of $c_0$ and $\Gamma$ depend on $K_4$.
\newline

One should ask if this result is consistent with what we observed in our previous ``full CDT"  simulations with large discrete proper time period $T$ and large total volume $V_4$ (and consequently much larger critical $K_4$). 
Extrapolating  our transfer matrix measurements with small $T=2$ and small $V_4$ to a larger volume region we expect that in the ``full CDT" system the typical bifurcation shift, measured for $s\gg s^b$ and large $c_0$ (due to big $K_4$), will be very large.
Naively speaking, the most probable 
configurations will be that with a very large difference of spatial volume 
in the adjacent time slices  
(large $(m-n)$). Therefore one can expect a kind of ``anti-ferromagnetic"
behaviour with a ...-``large"-``small"-``large"-``small"-... three-volume 
profile. This is exactly what we see  in 
CDT systems with small time periods $T=2, 4, 6$.  For large $T=80$
the observed behaviour is very different. In such simulations  the generic triangulations are ``collapsed" into a single spatial layer where almost all volume is localized, while in the other spatial slices the three-volume distribution is close to the cut-off size. Does this mean that the transfer matrix  cannot be used to explain a large $T$ limit and thus the approach is incorrect inside  phase ``B" ? To check this, let us propose a simple effective model, based on  volume fluctuations alone, resembling the one introduced in Chapter 4.5. Instead of using a semi-empirical transfer matrix, the model will be based on the following theoretical transfer matrix consistent with (\ref{TMBphase}):
$$
\braket{n|M|m} =
\exp \left( - \frac{\mu}{\Gamma} \left(\frac{n+m}{2}\right)^{1/3} \right)
\left[ 
\exp \left( -\frac{\Big((m-n) -\big[ c_0(n+m-s^b) \big]_+\Big)^2}{\Gamma(n+m)}\right) + \right.
$$
\begin{equation}
\left.
+\exp \left( -\frac{\Big((m-n) +\big[ c_0(n+m-s^b) \big]_+\Big)^2}{\Gamma(n+m)}\right) 
\right] \ ,
\label{bifmodel}
\end{equation}
where we have skipped  the $n_0$ parameter for simplicity (it is anyway negligible in a large volume limit) and added a theoretical potential part (the first exponent). As already explained, the exact measurement 
of the potential inside  phase ``B" for the large volume range is beyond our reach at the moment. Therefore we decided to use the potential  measured inside the ``C" phase (see Eq. (\ref{Seffform2})), and for simplicity we again only consider the leading behaviour (only a term $\propto (n+m)^{1/3}$). We set the parameters of our model 
to  the  values measured in the real ``full CDT" simulations: 
$\Gamma=37, \mu=15, s^b=2000$ and  $c_0 = 0.1 - 0.3 $ and generate volume distributions $\{n_1,n_2, .... ,n_T\}$  with the probability given by Eq. (\ref{eq:pne}):
$$
P^{(T)} (n_1, \dots, n_{T}) \propto \langle n_1 | M | n_2 \rangle 
\langle n_2 | M | n_3 \rangle \cdots \langle n_{T} | M | n_1 \rangle \ 
e^{- \epsilon \left(\sum_t n_t - \bar V_4 \right)^2} \ .
$$
The resulting volume distribution for small and large $T$ 
is presented in Fig. \ref{Figeffmodel} and Fig. \ref{Figeffmodel2}, respectively. As a reference case we 
also plot the volume profile for $c_0=0$, for which we 
recover the generic behaviour found in the ``de Sitter" phase.
 
\begin{figure}[h!]
\centering
\scalebox{0.9}{\includegraphics{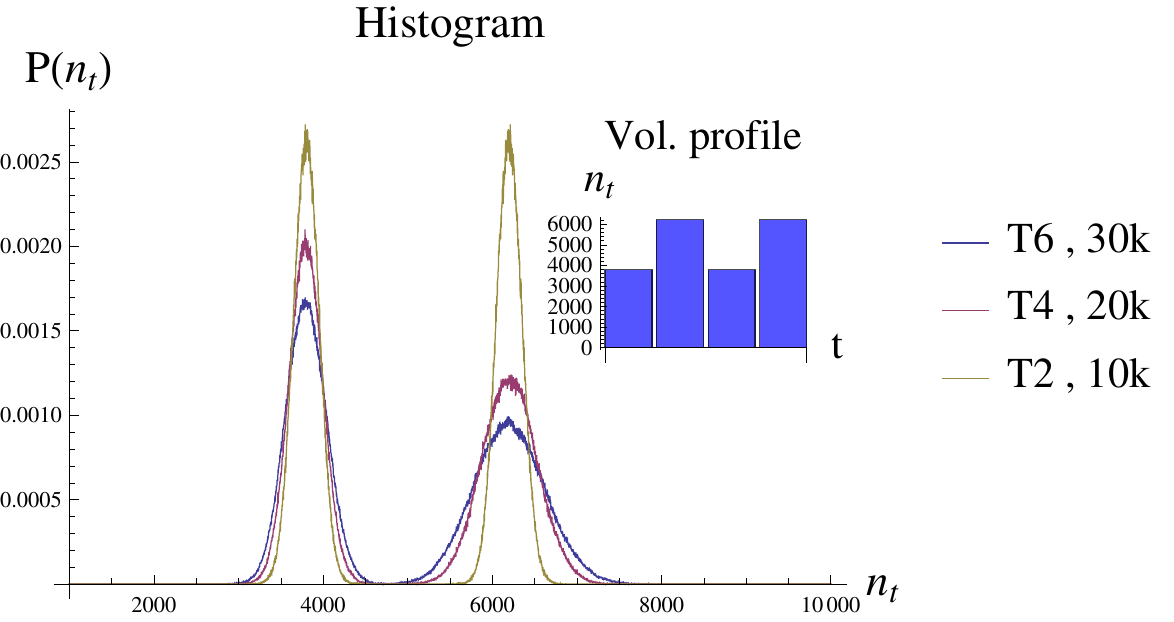}}
\caption{The histogram of the spatial volume distributions and the volume profile
measured in the effective model (\ref{bifmodel}) 
for $c_0=0.3$ and $T=2,4,6$. The two Gaussian peaks 
correspond to odd and even time slices, respectively. As a result 
the average volume profile is ``anti-ferromagnetic" with quantum 
fluctuations around: ...-3.8k-6.2k-3.8k-6.2k-... .}
\label{Figeffmodel}
\end{figure}
 
\begin{figure}[h!]
\centering
\scalebox{0.9}{\includegraphics{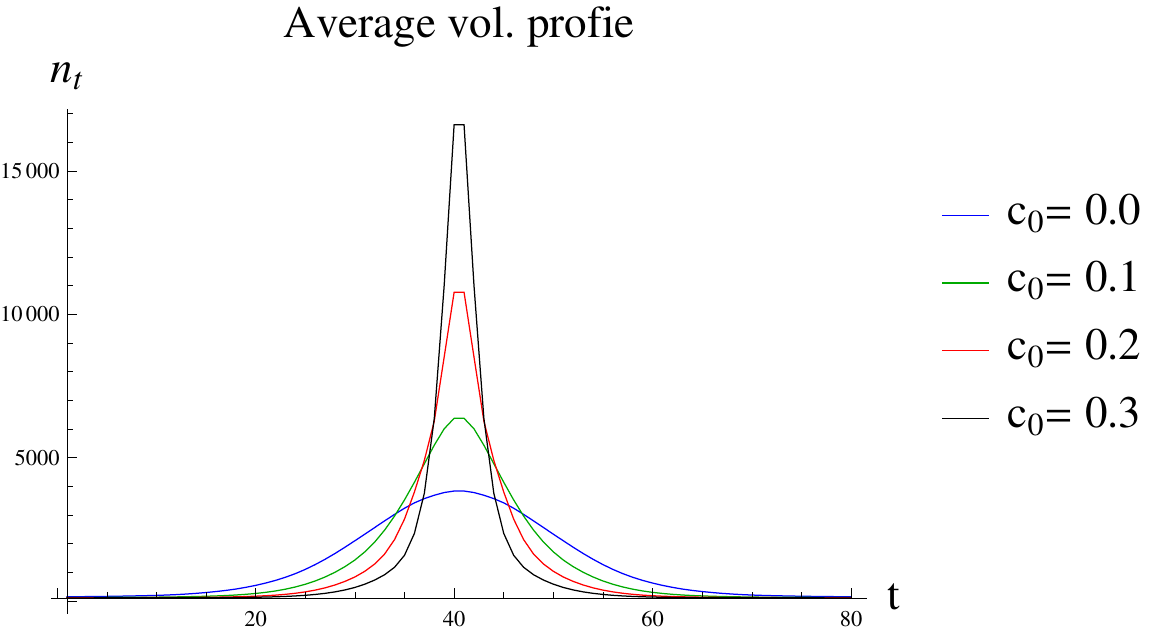}}
\caption{The average spatial volume 
measured in the effective model (\ref{bifmodel}) 
for $T=80$, 
$\bar V_4=100$k and a number of  values of $c_0$. 
The shape of the volume profile is consistent 
with the ``collapsed" blob structure for $c_0 >0$.}
\label{Figeffmodel2}
\end{figure}

For small $T$  
the naively expected ``anti-ferromagnetic" structure is observed, 
while for large $T$ a single ``collapsed" blob forms. 
The strength of the ``collapsed" behaviour depends on $c_0$. 
This simple model explains very well (at least at a qualitative level) 
the volume distribution observed in CDT simulations inside  phase  ``B", 
both for small and large $T$, and therefore indirectly validates the transfer matrix approach also in this phase. In reality  
one should  take into account that the actual  potential part
present in the ``full CDT" model in phase ``B"   may additionally support the idea of 
 a ``narrowing" of the volume profile compared to the one 
we observe in the toy model defined above. 

\section{Phase transitions}

When one applies  conventional methods to analyze the phase
transitions observed in four-dimensional CDT one obtains strong 
evidence that the ``A"-``C" transition is a first order transition 
while the ``B"-``C" transition is a second (or higher) order transition (see Chapter 2.1).    
We will try to use  the effective transfer matrix to obtain additional 
information about the phase transitions. In particular we will analyze the impact of the phase transitions on the kinetic part of the measured effective action.

\begin{figure}[h!]
\centering
\scalebox{0.9}{\includegraphics{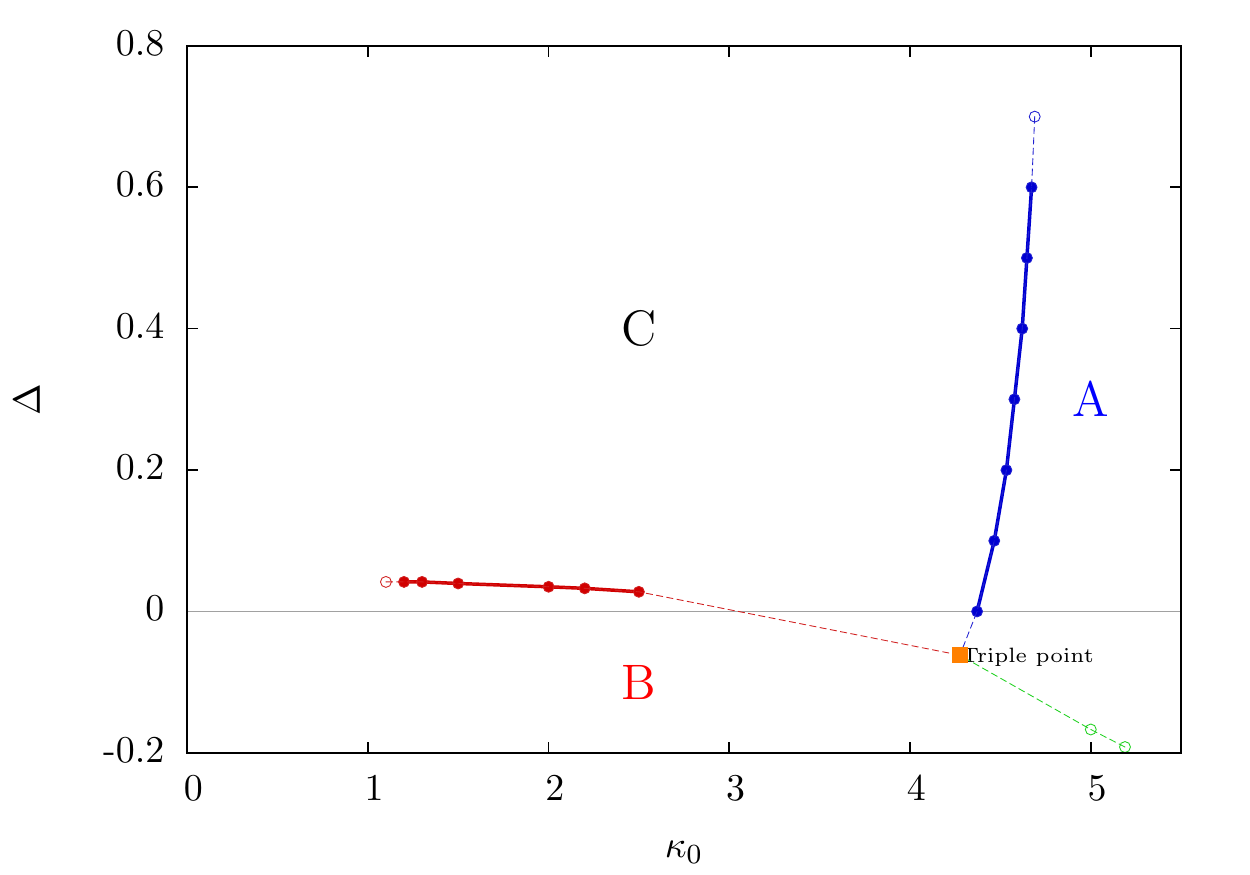}}
\caption{The  CDT phase diagram measured by 
``traditional" methods.}
\label{Figfazy}
\end{figure}

The traditional ``A"-``C" phase transition is observed, for example, when one starts in phase ``C" and  increases $\k_0$ while keeping $\Delta$ fixed.  The transition point   is easily visible 
in the kinetic part of the transfer matrix data, i.e. in the measured cross-diagonals. Starting in phase ``C" and increasing $\k_0$, one observes a smooth vanishing of the kinetic term measured in the effective Lagrangian in this phase \rf{Seffform2}.
Near the transition  the cross-diagonals  $\braket{n|M|s-n}$ plotted as a function of $n$ are almost flat (see Fig. \ref{FigPh10}). 
Just after the phase transition  one can observe a change in the behaviour due to the ``artificial" kinetic term (with a reversed sign: $e^{+x^2}$ instead of $e^{-x^2}$), discussed in Chapter 5.1.  For  $\Delta=0.6$ the phase transition 
point can be identified  at $\k_0 = 4.75\pm 0.05$  
which is  consistent with the location found using 
the ``traditional"  approach (see Fig. \ref{Figfazy}).

\begin{figure}[h!]
\centering
\scalebox{0.5}{\includegraphics{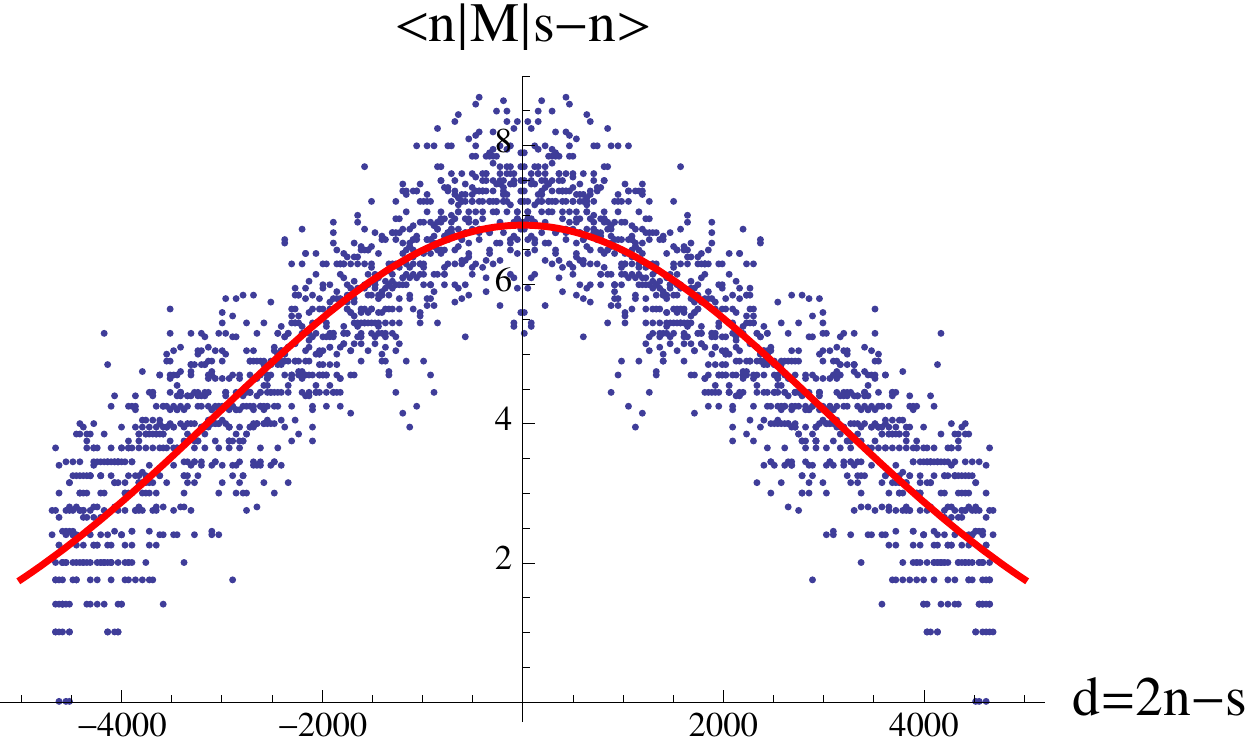}}
\scalebox{0.5}{\includegraphics{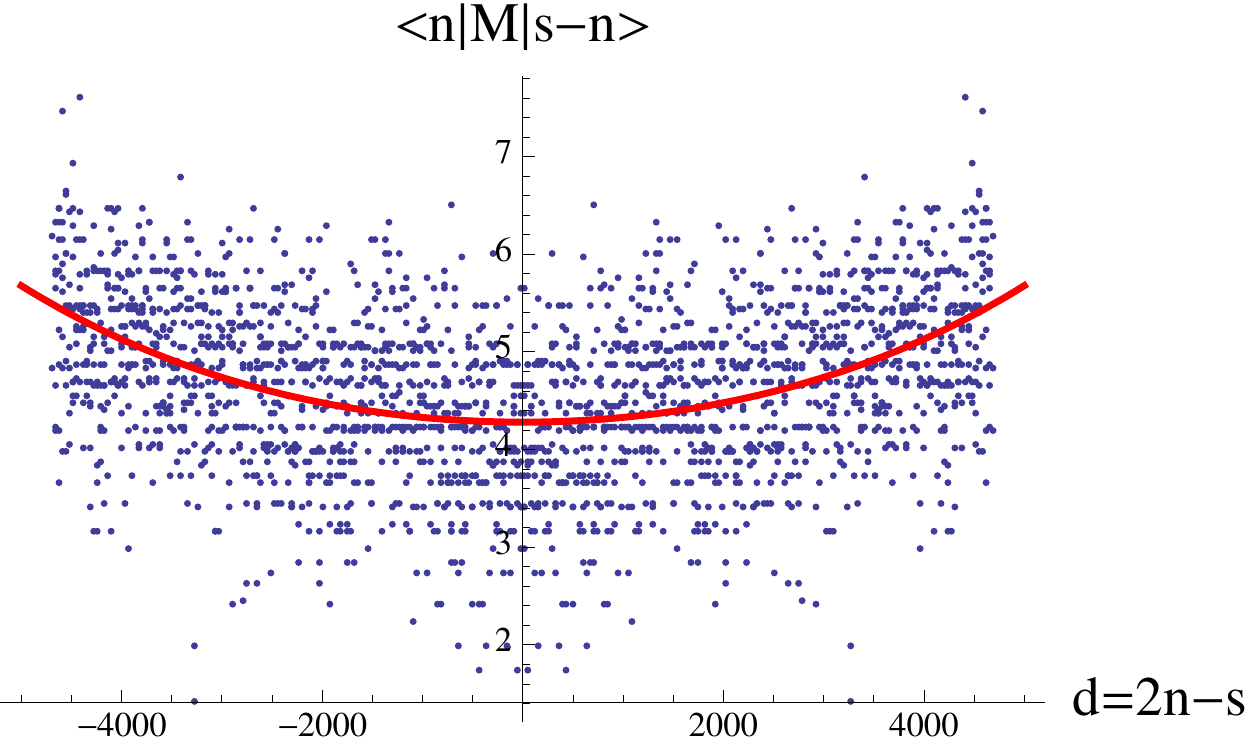}}
\caption{The cross-diagonal elements of the transfer matrix 
for $s=n+m=5000$ measured  
for $\Delta=0.6$. The left chart presents data for  
$\k_0=4.7$ (in phase ``C") while the right chart presents the data 
for $\k_0=4.8$ (in phase ``A"). The change of the behaviour 
is clearly visible, which enables us to identify the phase transition point. }
\label{FigPh10}
\end{figure}

\begin{figure}[h!]
\centering
\scalebox{0.7}{\includegraphics{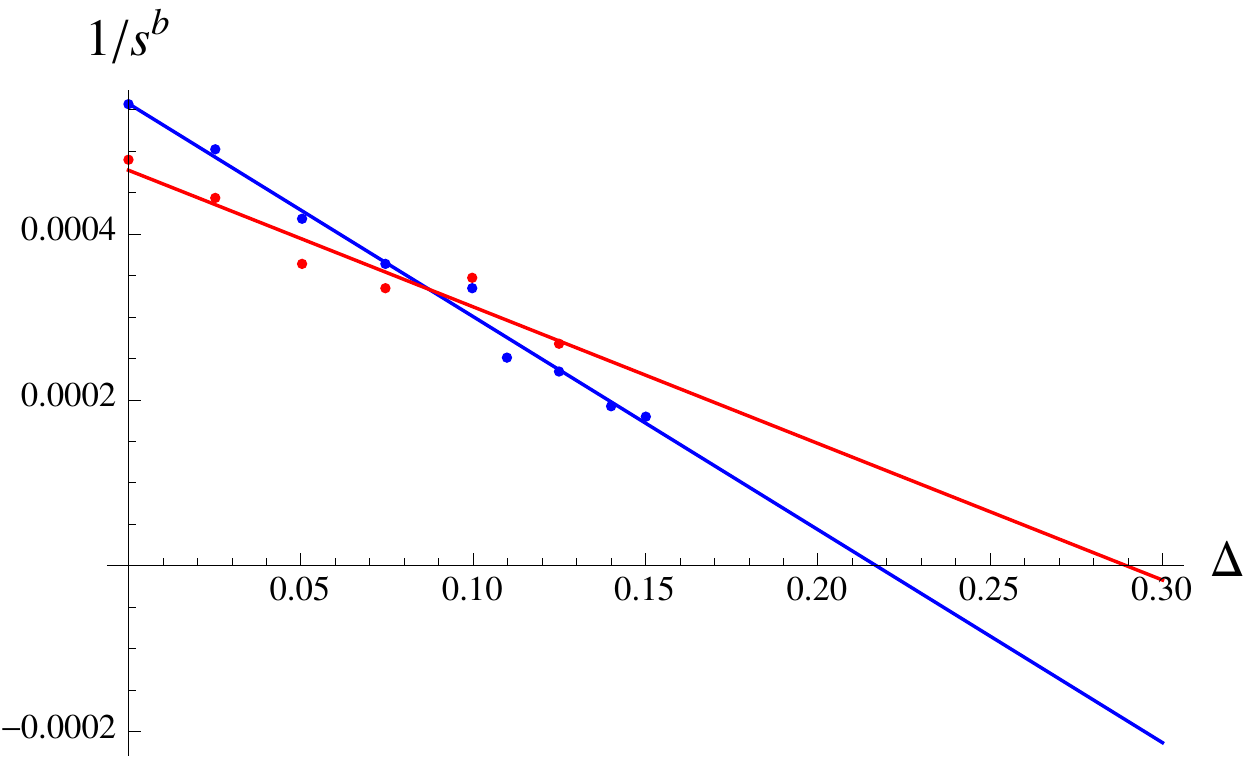}}
\caption{The (inverse of) bifurcation point $s^b$ as a function of $\Delta$ (for fixed $\k_0=2.2$). The colours correspond to two different 
ways of extracting $s^b$: by direct measurements (red points) and  
indirectly (blue points) $^3$. }
\label{FigPh11}
\end{figure}

The traditional ``B"-``C" phase transition can be met, for example, when one  starts in phase ``B" and  increases $\Delta$ while keeping $\k_0$ fixed. In the previous section we showed that inside the phase ``B" the shape of the cross-diagonals  $\braket{n|M|s-n}$ plotted as a function of $n$ depends on  $s$. For small $s<s^b$ the cross-diagonals can be fitted with a single Gaussian, exactly as in  phase ``C". For large $s> s^b$  one should rather use a sum of two shifted Gaussians. We argued using the effective model \rf{bifmodel} that this bifurcation of the kinetic term  is responsible for the ``collapsed" behaviour observed in  phase ``B". Consequently it is natural to conjecture that 
 the  ``B"-``C" phase transition is related to the appearance of the bifurcation shift $c[s]$, which can by parametrized by Eq. \rf{cs}. The obvious parameter to look at is a bifurcation point $s^b$ as the phase transition should be related with the $s^b \to \infty$  limit.  If we start in  phase ``B", keep $\k_0$ fixed 
and increase $\Delta$,
the value of $s^b$ also increases. In Figure \ref{FigPh11} we present the plot of $1/s^b$ 
as a function of $\Delta$ measured for $\k_0=2.2$. 
Different colours correspond to two methods of measuring $s^b$.\footnote{Red points 
correspond to $s^b$ determined  by fitting Eq. (\ref{cs}) 
to the measured data. This method requires performing 
transfer matrix measurements also for  volumes much higher 
than $s^b$ which is difficult as we approach the phase transition.  
Blue points correspond to the indirect determination of $s^b$,
identified as the point at which a single Gaussian does no longer fit the  
measured cross-diagonals, as described in the previous section. 
The larger the $\Delta$ the more difficult it is to observe 
the shift away from a single-Gaussian distribution. 
Therefore the values of $1/s^b$ for large $\Delta$ 
are probably underestimated when using this second method.} 
The relation seems to be linear (at least in the range of $\Delta$ where we can measure it).  Extrapolation to higher $\Delta$ implies that the  transition occurs 
for $\Delta = 0.2-0.3$. This value of $\Delta$ is much higher 
than the critical value measured in the ``traditional" approach 
($\Delta \approx 0.05 $ - see Fig. \ref{Figfazy}).
Thus, by using $s^b$ as an indicator of 
a phase transition we have  apparently encountered 
something different from the formerly observed ``B"-``C" transition.

\newpage
\thispagestyle{empty}
\mbox{}
\newpage

\chapter{New ``bifurcation" phase}
This Chapter is  partly based on the article: {\it J.\,Ambj\o rn, J. Gizbert-Studnicki, A. G\"orlich, J. Jurkiewicz, ``The effective action in 4-dim CDT. The transfer matrix approach",  
JHEP 06 (2014)}
\newline
\newline

In the previous Chapters we used the transfer matrix method to measure and parametrize the effective action in all  ``A", ``B" and ``C" phases of four-dimensional Causal Dynamical Triangulations as well as to analyze  phase transitions. The phase transitions should be associated with a change in the kinetic part of the effective action. We found that the ``A"-``C" phase transition is very well described by the transfer matrix data, while the ``B"-``C" phase transition is not visible.  Instead one can observe that the bifurcation structure of the kinetic term, characteristic of phase ``B", persists even after crossing the ``traditional" ``B"-``C" phase transition line. Based on the extrapolation of the measured bifurcation point $s^b$ as a function of the bare asymmetry parameter $\Delta$ (for fixed $\kappa_0=2.2$) we found that the vanishing bifurcation of the kinetic term ($s^b \to \infty$) occurs for the values of $\Delta$ around $0.2-0.3$ while the ``traditional" transition is for $\Delta \approx 0.05$. Therefore there exists a region of the parameter  space, traditionally denoted the ``C" or ``de Sitter" phase, in which the bifurcation is still present. We will call this region a ``bifurcation" phase and will analyze its properties. In particular we would like to show that the properties are different than in the generic phase ``C"  and probe   the phase diagram in order to draw the new transition line.

The main technical issue lies in a fact that the closer we are to the new phase transition  the harder it is to measure the bifurcation shift $c[s]$ which parametrizes the transfer matrix cross-diagonals $\braket{n|M|s-n}$ (see Eq. \rf{SkinB}). Consequently, one needs to perform  transfer matrix measurements for larger and larger $s$. Very close to the new phase transition one additionally encounters a thermalization slowdown which is typical for the second or higher order phase transitions.\footnote{We have not derived the order of the new phase transition yet, we only see a characteristic slowdown.} As a result it is very difficult to determine the phase transition point with high precision. To speed-up the bifurcation measurements we decided to focus only on selected cross-diagonals of the transfer matrix. We follow  the  method based on the empirical probability distributions in systems with $T=2$ time period  (see Chapter 5 for details) and use a linear global volume fixing  \rf{GlobalVF} with relatively large parameter $\eps$ to ensure that the   probability distributions are very peaked at the selected $s=n+m=\bar V_4$. We also changed our approach to fixing the bare cosmological constant $K_4$ related to a strong $K_4$ dependence on the total volume $\bar V_4$ due to finite volume corrections of the entropic factor (see Chapter 5.2). The $K_4$ dependence on $\bar V_4$  translates into the potential term of the effective action which can be  ignored as we focus on the behaviour of the kinetic part alone. Consequently, for given $\kappa_0$ and $\Delta$ we fine-tune the $K_4$ for each measured $s=\bar V_4$ independently. 


\section{Evidence of the new phase}

We start with the measurements performed for the fixed bare (inverse) gravitational constant  $\kappa_0=2.2$, and a range of values of the bare asymmetry parameter $\Delta$. A typical empirical transfer matrix cross-diagonal measured inside the ``bifurcation" phase is presented in Fig. \ref{FigB12}. 
\begin{figure}[h!]
\centering
\scalebox{0.7}{\includegraphics{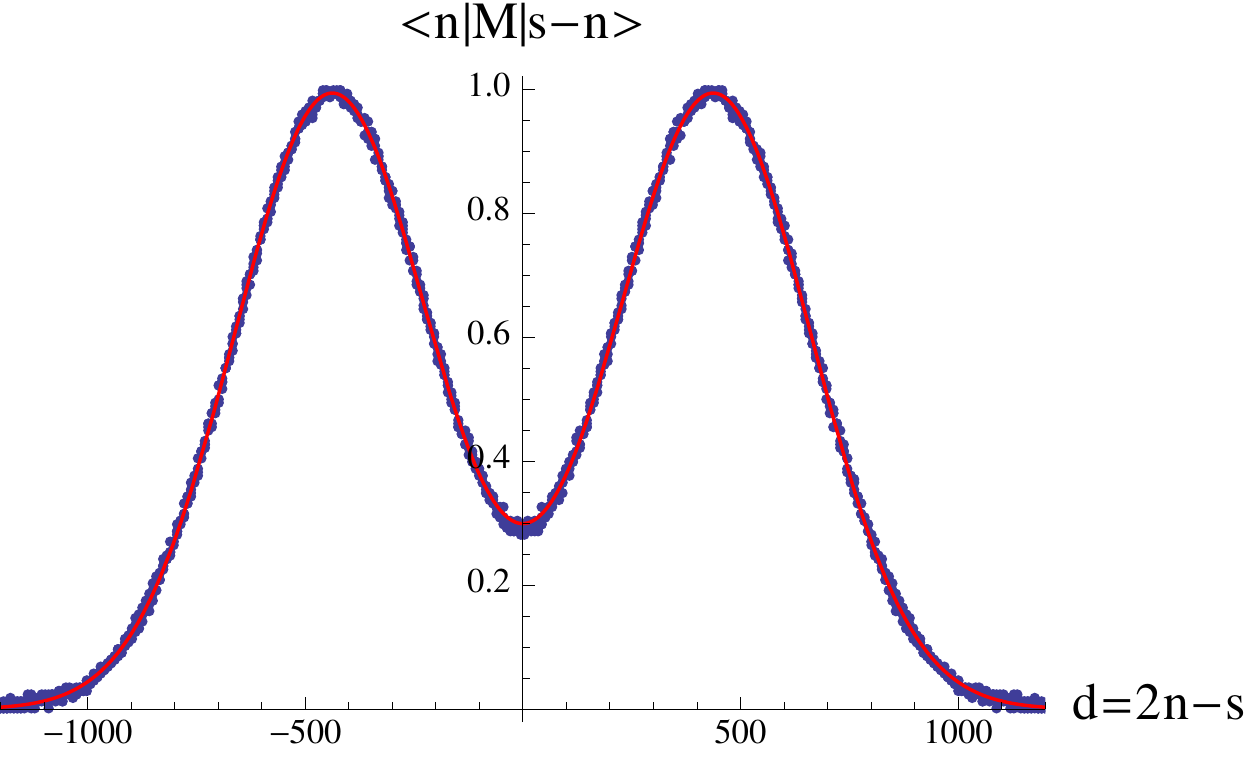}}
\caption{The cross diagonal  of the transfer matrix $M$ measured 
for $\kappa_0=2.2$ and $\Delta=0.125$, which according to the ``traditional"
approach lies  inside the ``de Sitter" phase ``C", but according to current parametrization is in the ``bifurcation" phase. 
The data were measured for $s=n+m=15$k. The ``double-Gaussian" 
bifurcation structure  
is  clearly visible.}
\label{FigB12}
\end{figure}
As  was explained in Chapter 5.2, the cross-diagonal shows a ``double-Gaussian" dependence on $n$, given by Eq. \rf{SkinB}:
$$
\braket{n|M|s-n}=
$$
$$
 ={\cal N}(s)\left[ \exp \left( -\frac{\Big((m-n) - c[s]\Big)^2}{k[s]}\right) + \exp \left( -\frac{\Big((m-n) + c[s]\Big)^2}{k[s]}\right)\right] \ .
$$
By fitting the above formula to the empirical data one can extract the bifurcation shift $c[s]$. In Chapter 5.2 we showed that the shift 
$c[s]$ is well approximated by the following function, Eq. \rf{cs}:
\begin{equation}
c[s] \approx \max[0,c_0(s-s^b)] .
\label{cs2}
\end{equation} 
Here we will use another phenomenological parametrization, which fits the data around the bifurcation point even better (see yellow curve in Fig. \ref{FigB6}) : 
\begin{equation}
c[s] = c_0  \;s\;\exp(-s^b/s) \ .
\label{csexp}
\end{equation}
It is consistent with (\ref{cs2}) for small and large $s$ (compared to $s^b$). 
The measured $c[s]$ for fixed bare $\kappa_0$ and different $\Delta$ together with the fits of Eq. \rf{csexp} are presented in Fig. \ref{FigC0Sb}. One can clearly see that the bifurcation structure is present for this choice of the bare coupling parameters if one takes $s$ large enough. One should note that for $\Delta \approx 0.3-0.4$  we encounter a characteristic thermalization slowdown, mentioned above. If one performs numerical simulations in this range of $\Delta$, and starts with a configuration typical for lower $\Delta$, say $\Delta = 0.2$, the measured bifurcation shift $c[s]$ decreases very slowly in Monte Carlo time. After a few weeks of computer simulations the shift is still non zero (see red line in Fig. \ref{csexp}). However, if one starts from a configuration typical for higher $\Delta$, say $\Delta = 0.5$, and performs the simulations the typical shift oscillates around zero. Therefore the transfer matrix data for this range are not determined with high precision.

\begin{figure}[h!]
\centering
\scalebox{0.85}{\includegraphics{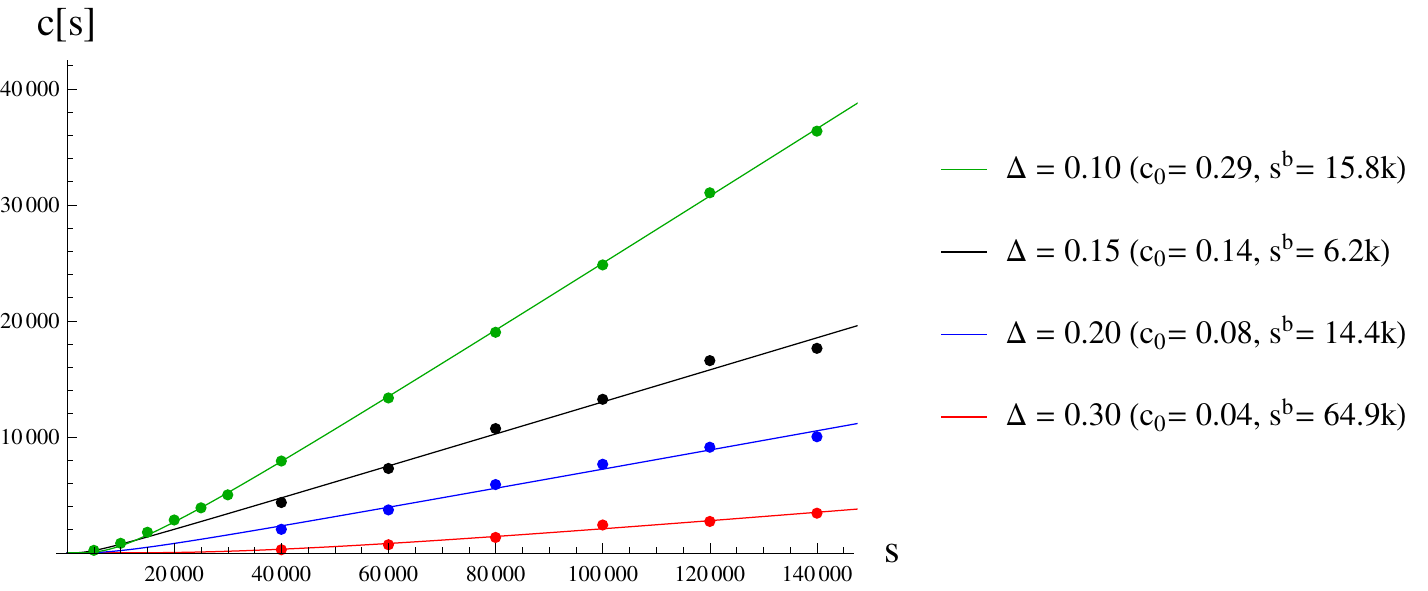}}
\caption{The bifurcation shifts (dots) measured for $\kappa_0=2.2$ and for
a range of  $\Delta$ values,  and the best fits of 
$c[s]=c_0 \exp(-s^b / s)\, s$ to these data (lines).   }
\label{FigC0Sb}
\end{figure}

It is  important to note that for $\kappa_0=2.2$ and  $\Delta =0.3$, which is at (or very close to) the new phase transition, the measured value of $c[s]$ is almost zero for $s \leq 40$k. This $s=n_t+n_{t+1}$ is   relatively large if one compares it with the maximal  spatial volume $n_t$ measured in ``full CDT" simulations for $T=80$ - see Fig. \ref{FigavCDT}, in which we plot the average volume profiles measured in systems with $V_4=160$k, for 
$\k_0=2.2$ and a range of values of $\Delta$. Nevertheless, looking at the ``full CDT" data, this is for  $\Delta=0.3$ that one  starts to observe a  contraction of the volume profile in time direction which is characteristic of the ``bifurcation" phase. For $\Delta \geq 0.4$ the shape of the volume profile does not change much. An increased ``narrowing" 
in the time-direction  takes place for $0.1 \leq \Delta \leq 0.3$. Finally,
crossing the ``traditional" ``B"-``C" phase transition at $\Del\approx0.05$ we
observe the  ``collapse"  of the ``blob"  as one enters into the  
``B" phase. 

In Chapter 5.2 we introduced a  simple theoretical toy model based on the ``bifurcated" transfer matrix  \rf{bifmodel} which qualitatively showed  the same type of  spatial volume behaviour. In the toy model  a ``narrowing"  of the average volume profile depended on the bifurcation slope $c_0$ (see  Fig. \ref{Figeffmodel2}). For small
 $c_0$ the volume profiles 
were practically identical with those observed in the generic ``de Sitter" phase (where $c_0 = 0$). 
For medium bifurcation slopes  the volume profile contracted 
in the time direction, but the general shape did not change much. 
Only for  large $c_0$  one could observe something which 
resembled  a ``collapse" of the ``blob" in the time-direction. 

Summarizing,  the existence of the    new ``bifurcation" phase measured by the transfer matrix (``microscopic") approach seems to be  consistent with the  ``macroscopic" spatial volume behaviour observed in  ``full CDT" simulations for $\kappa_0=2.2$ and $0.1\leq \Delta \leq 0.3$. 

\begin{figure}[h!]
\centering
\scalebox{0.9}{\includegraphics{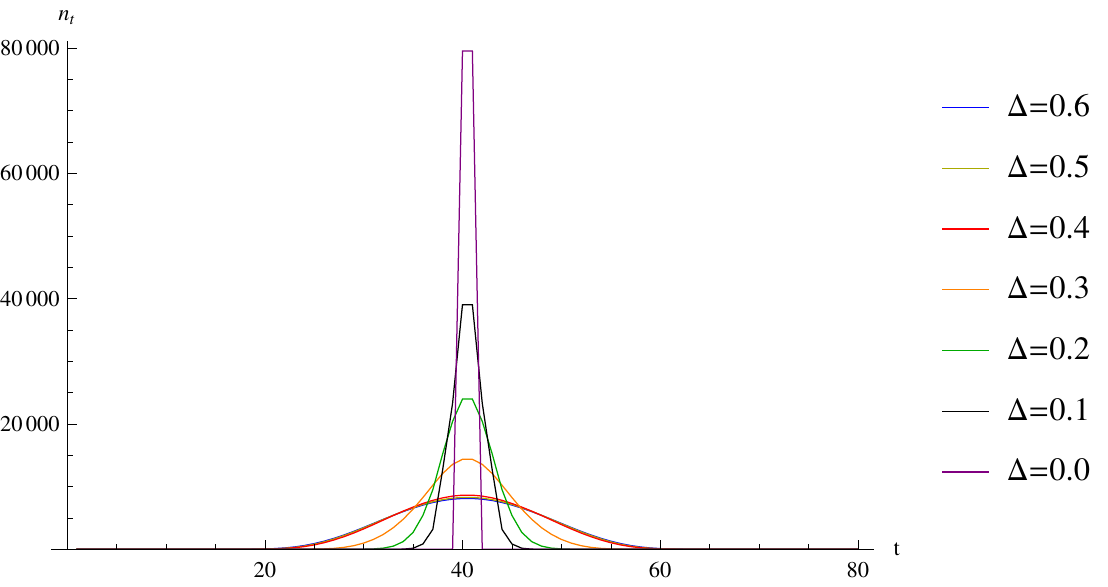}}
\caption{The average spatial volume profiles measured in ``full CDT" 
for $\kappa_0=2.2$, $T=80$ and $\bar V_4 = 160$k.}
\label{FigavCDT}
\end{figure}

\section{Geometric properties}

In order to confirm that the newly discovered ``bifurcation" region, formerly denoted  a part of the ``C" phase, is  indeed a genuinely new phase one should show that  geometric properties of generic triangulations in this region are  different than the properties measured inside the ``de Sitter" phase. In particular we will focus on the effective dimension of the  simplicial manifolds applying the techniques previously  used ``deep inside"  the ``C" phase \cite{CDTreviews4}. We would like to show that inside the ``bifurcation" phase both Hausdorff and spectral dimensions are considerably different than four.

The {\it Hausdorff dimension} $d_H$ is related to scaling properties of triangulations. The physical proper time intervals should scale as  $\bar V_4^{1/d_H}$ where $\bar V_4=\sum_{t} n_t$ is the total four-volume. One can define the volume independent (rescaled) time coordinate:
\beql{dHtau}
\tau \equiv \frac{t}{\bar V_4^{1/d_H}}
\eeq
as well as the corresponding rescaled three-volume variable
\beql{dHntau}
n(\tau) \equiv \frac{n_t}{\bar V_4^{1-1/d_H}} \ .
\eeq
We will  use the empirical average  three-volume distributions $\braket{n_t}$  measured for different $\bar V_4$ to determine $d_H$. Assuming that the measurements performed for the same values of the bare coupling constants ($\kappa_0$ and $\Delta$) but  for different total volumes $\bar V_4$ describe the same physics, one should get a universal dependence of $\braket{n(\tau)}$ on $\tau$ if $d_H$ is set to the value consistent with the actual Hausdorff dimension. The rescaled average volume profiles $\braket {n(\tau)}$ for different  $\bar V_4$ calculated using Eq's \rf{dHtau} and \rf{dHntau} and assuming $d_H=4$ are plotted in Fig. \ref{FigB13}.   The left chart presents the rescaled distributions measured ``well inside" the generic ``de Sitter" phase, where indeed $d_H=4$, while the right chart shows the distributions inside the ``bifurcation" phase, where $d_H\neq 4$. The lack of appropriate scaling in the new phase is an important 
difference compared to the generic phase ``C". 
\begin{figure}[h!]
\centering
\scalebox{0.5}{\includegraphics{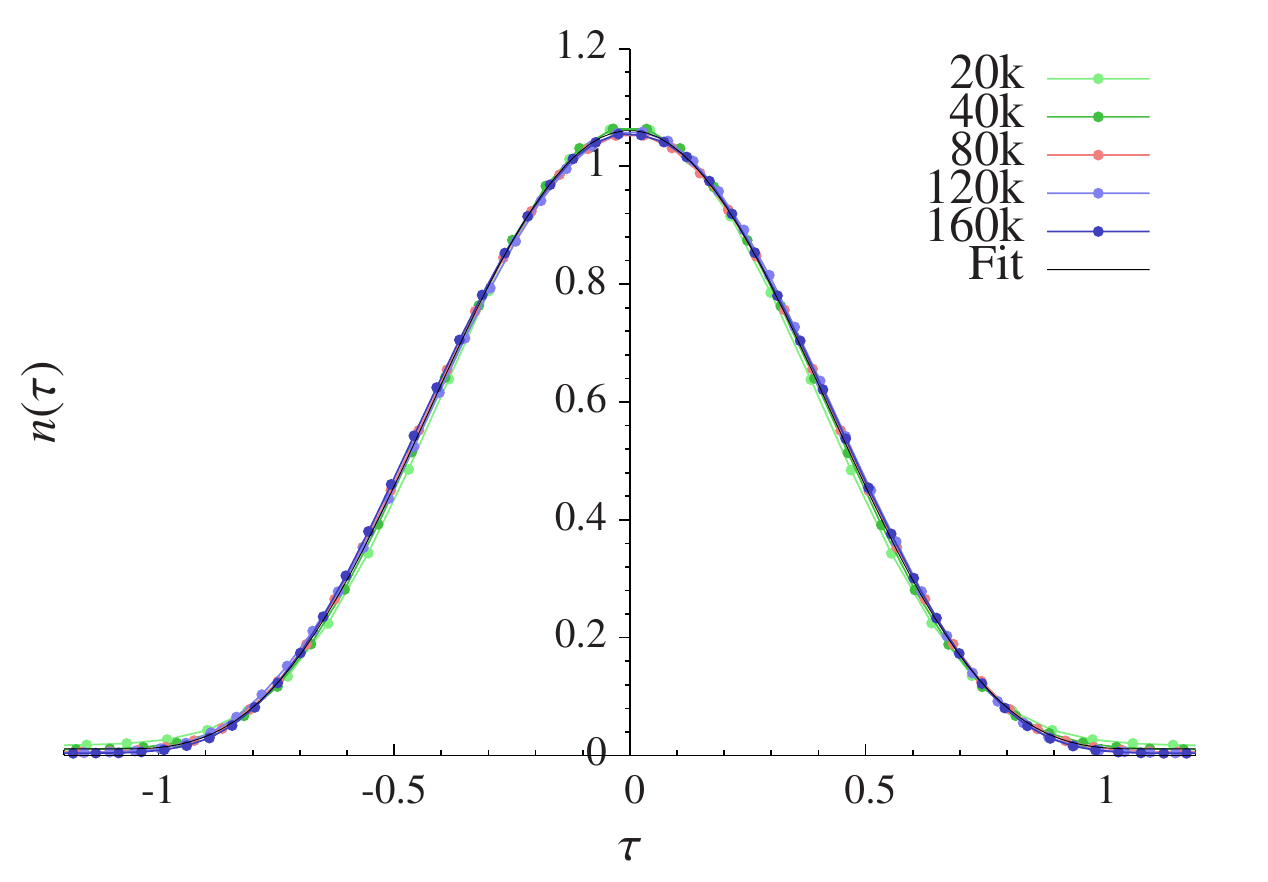}}
\scalebox{0.5}{\includegraphics{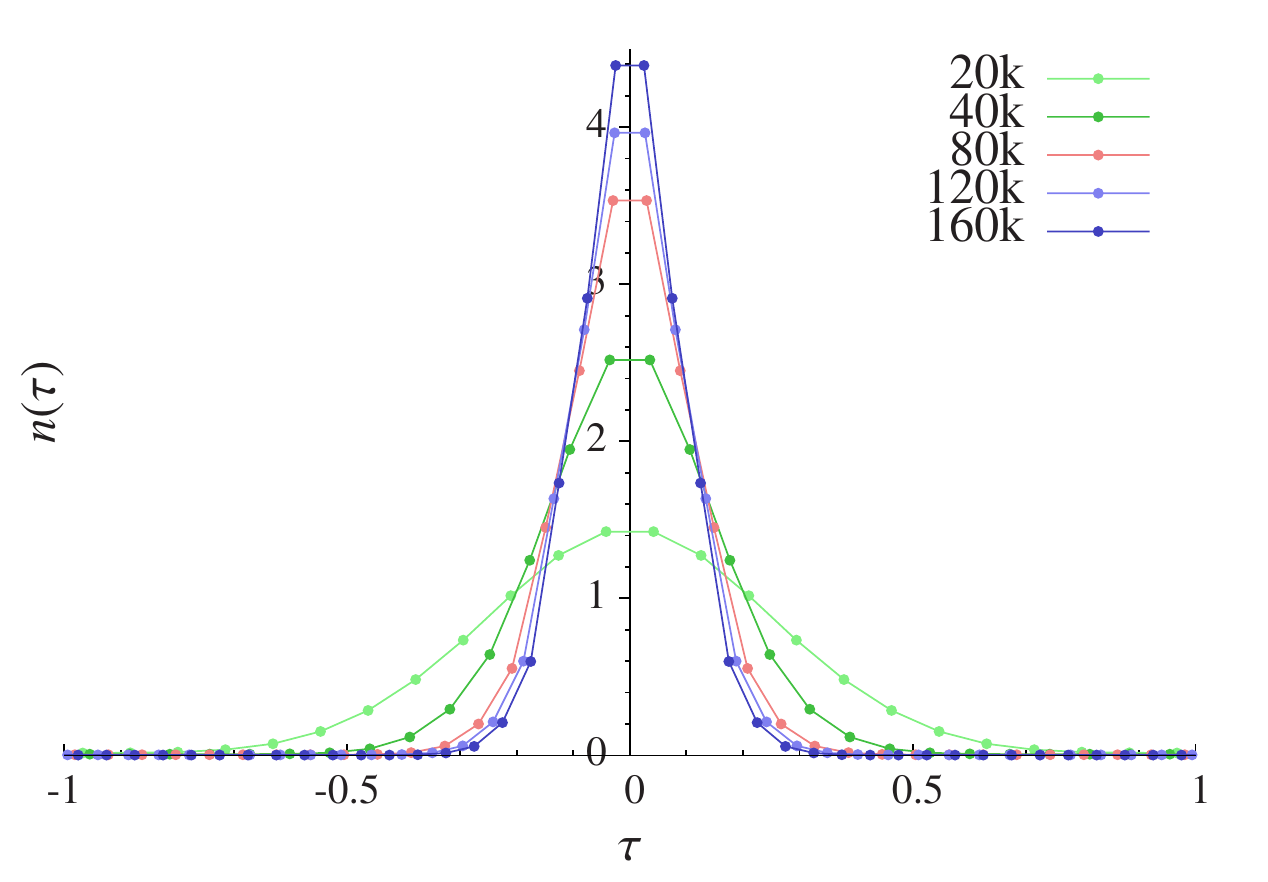}}
\caption{The  CDT  distributions of spatial volumes $ n_t$  
measured for different total four-volumes $\bar V_{4}$.  The left figure shows the distributions
in the generic ``de Sitter" phase  ($\kappa_0=2.2$, $\Delta=0.6$) and the  right figure 
 the distributions in the ``bifurcation" phase 
($\kappa_0=2.2$, $\Delta=0.125$). 
Different colours correspond to different total volumes $\bar V_{4}$.  
The data were rescaled, according to: $\tau = t / \bar V_{4}^{1/d_H}$ and
$n(\tau) = n_t  / \bar V_{4}^{1-1/d_H}$,
to fit a single curve,  assuming the Hausdorff dimension $d_H= 4$. }
\label{FigB13}
\end{figure}
In fact one can show that for the choice of the bare couplings $\kappa_0=2.2$ and $\Delta=0.125$ inside the new ``bifurcation" phase, the best overlap of the rescaled volume distributions $\braket {n(\tau)}$ measured for different $\bar V_4$  is for $d_H = \infty$, i.e. the time variable does not scale at all - see Fig \ref{bifscaling} (left). This behaviour is  characteristic for the generic phase ``B" rather than for  phase ``C". For comparison we also plot the rescaled volume profiles obtained from the effective toy model  \rf{bifmodel} with bifurcation slope $c_0=0.3$ consistent with the value measured for this choice of the bare couplings - see Fig \ref{bifscaling} (right). Surprisingly, the Hausdorff dimension measured in the toy model agrees with the Hausdorff dimension from the ``full CDT" data. The results suggest that the new phase transition might
be associated with an asymmetric scaling of space and time, precisely as is assumed in
Horava-Lifshitz gravity \cite{HL1,HL2}.

\begin{figure}[h!]
\centering
\scalebox{0.5}{\includegraphics{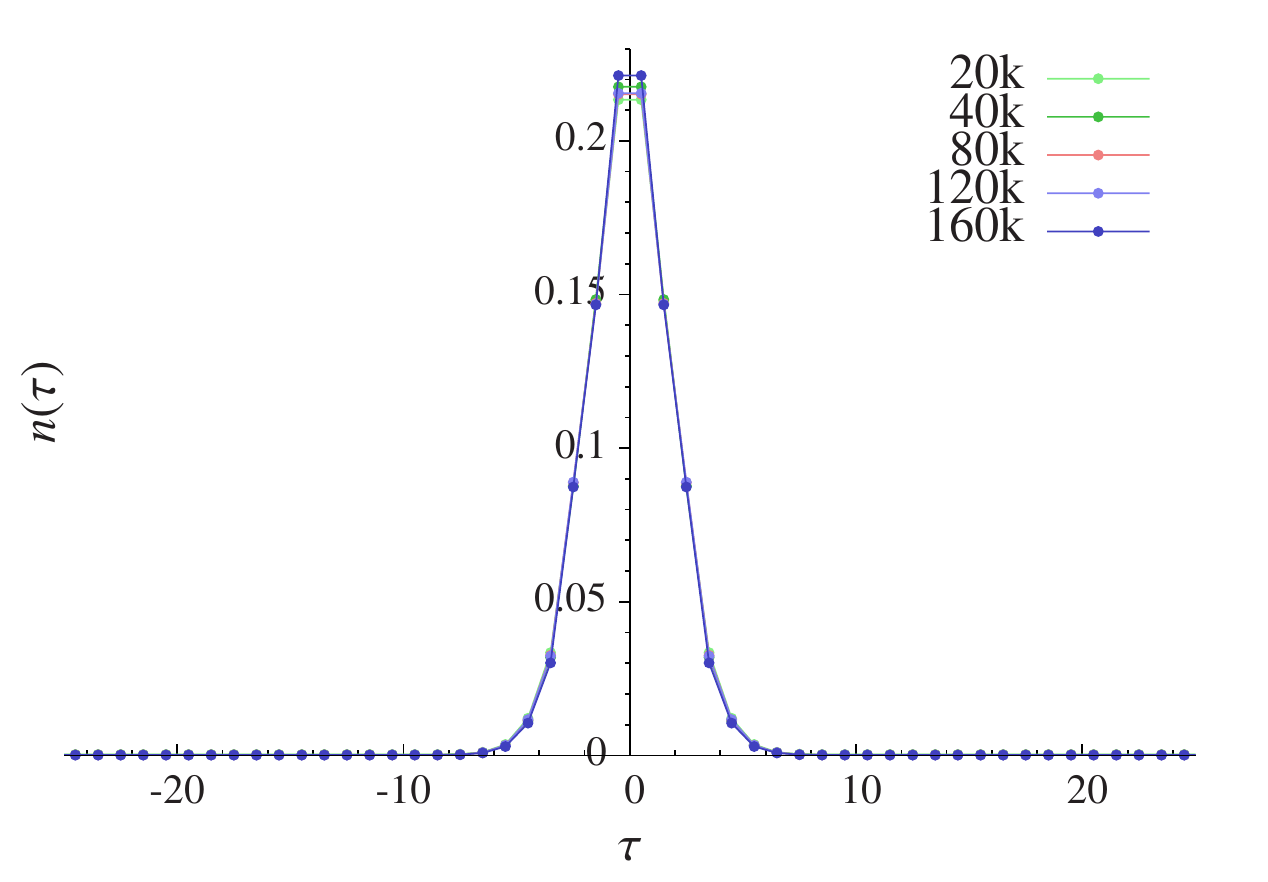}}
\scalebox{0.5}{\includegraphics{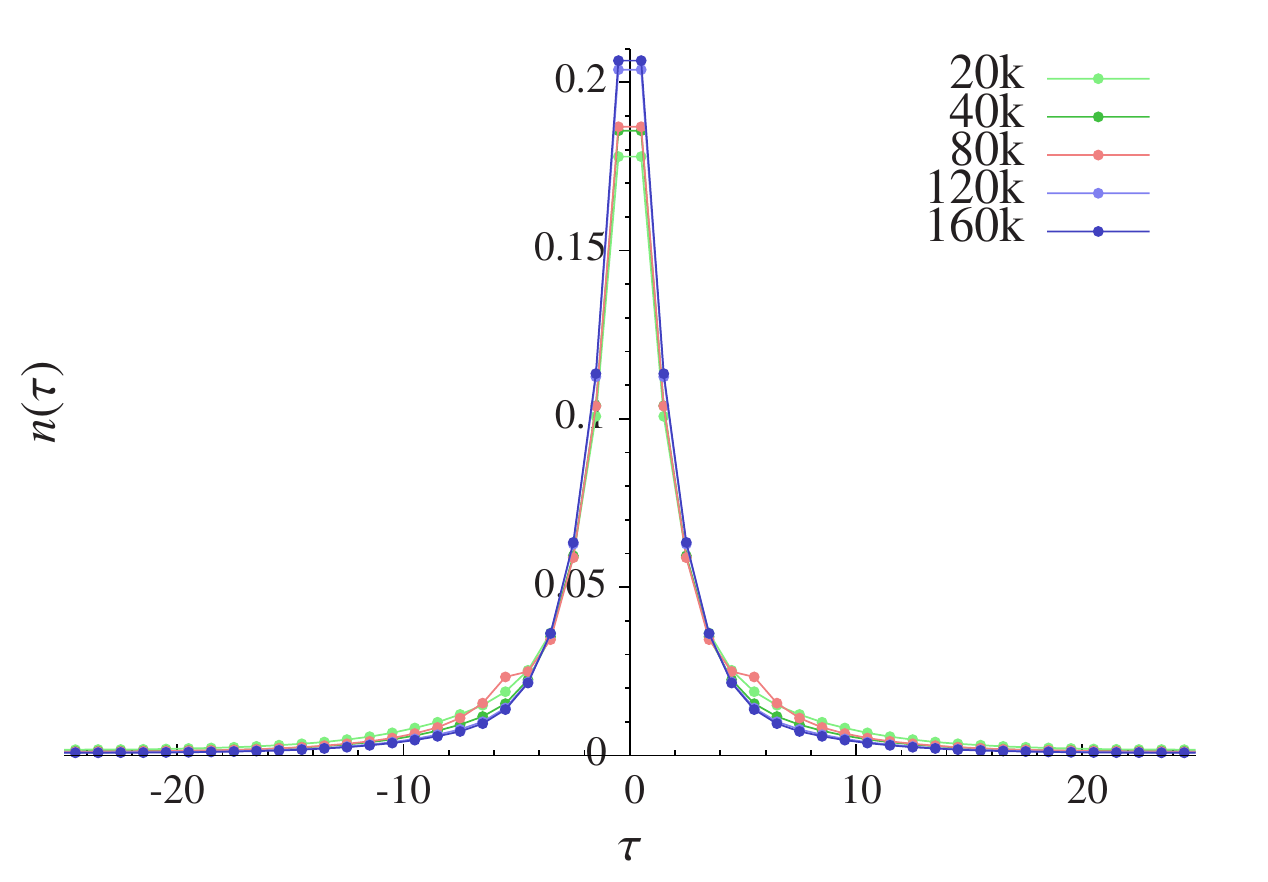}}
\caption{The left figure shows   CDT  distributions of spatial 
volumes $n_t$ measured  
in the  ``bifurcation" phase ($\kappa_0=2.2$, $\Delta=0.125$). The right figure
shows the $n_t$ distributions generated from 
the ``effective" toy  model (\ref{bifmodel}) with 
bifurcation slope $c_0=0.3$. Different colours correspond 
to different total volumes $\bar V_{4}$. 
The data were rescaled, according to: $\tau = t / \bar V_{4}^{1/d_H}$ and
$n(\tau) = n_t  / \bar V_{4}^{1-1/d_H}$, to fit a single 
curve, assuming the  Hausdorff dimension $d_H=\infty$.}
\label{bifscaling}
\end{figure}

Another quantity revealing information about  geometric properties of $D$-dimensional Riemannian  manifolds is the {\it spectral dimension} $d_S$. It is related to the diffusion process of a test point particle in a (fictitious) diffusion time, described by the following diffusion equation:
\beql{difcont}
\partial_\sigma \rho(x,x_0;\sigma) = \Delta_g  \, \rho(x,x_0;\sigma) \ ,
\eeq
where $\rho(x,x_0;\sigma)$ is the probability of finding the particle at position $x$ after diffusion time $\sigma$ (provided that the initial position was $x_0$ at $\sigma = 0$), and $ \Delta_g $ is the Laplace operator corresponding to the underlying metric $g_{\m\n}(x)$. The average {\it return probability} is given by:
\beql{retprob}
P(\sigma)=\left<\frac{1}{V_4}\int d^D x_0 \sqrt {\det g_{\m\n}(x_0)} \rho(x_0,x_0;\sigma) \right> \ ,
\eeq  
where $V_4 = \int d^D x_0 \sqrt {\det g_{\m\n}(x_0)}$ is the total volume of the manifold, while the average $\braket{.}$ is related to  quantum manifolds, and  it is taken over the ensemble of geometries. The spectral dimension is defined by:
\beql{dsdef}
d_S (\sigma) \equiv -2\,  \frac{d \log P(\sigma)}{d \log \sigma} \ .
\eeq
For a flat Euclidean  manifold $\mathbb R^D$ one obtains:  $\lim_{\sigma \to \infty} d_S (\sigma) = D $ and in this sense (long diffusion time limit)  the spectral dimension $d_S$ is equal to the topological dimension $D$ (and also to the Hausdorff dimension $d_H$). For a compact manifold, say a $D$-sphere $S^{D}$, the spectral dimension $d_S$ tends to $D$ with growing $\sigma$, however  for very large diffusion  times, due to the  finite volume of the manifold,
the zero mode of the Laplacian dominates and $d_S$ finally  tends
to zero for $\sigma \to \infty$. Consequently, as we will see in a moment, one can expect a kind of  plateau or a maximum  value of $d_S$ to be observed for some  diffusion time $\sigma$.

A discrete version of Eq's \rf{difcont}-\rf{dsdef} can be used to define the spectral dimension for the simplicial manifolds, in particular for the CDT triangulations, which after performing a Wick rotation are indeed (discrete) Riemannian   manifolds. In this case one can consider a discrete diffusion process of a test particle ``jumping" on the dual lattice, i.e. between the centres of the adjacent 4-simplices:
\beql{difdisc}
\rho(i,i_0;\sigma+1) - \rho(i,i_0;\sigma)= \Omega   \, \sum_{{i\leftrightarrow j}} \Big( \rho(j,i_0;\sigma) - \rho(i,i_0;\sigma)  \Big)\ ,
\eeq
where $\Omega$ is related to the diffusion time step, the indices $i,i_0,j$ denote simplex labels and  the sum is over all 4-simplices $j$ adjacent to $i$. Since in $D=4$ dimensions each 4-simplex has exactly five direct neighbours, it is convenient to set $\Omega\equiv 1/5$ to get:
\beql{difdisc}
\rho(i,i_0;\sigma+1)= \sum_{{i\leftrightarrow j}} \rho(j,i_0;\sigma) \ .
\eeq
For any  triangulation $\cT$ one can choose a starting 4-simplex $i_0$. For such $i_0$ one imposes the initial condition   $\rho(i,i_0;\sigma=0) =\delta_{i, i_0}$ and iterates Eq. \rf{difdisc} to compute  $\rho(i,i_0;\sigma)$ for consecutive diffusion steps $\sigma$. Finally one repeats the above procedure for $N_{i_0}$ (randomly chosen\,\footnote{Since for each triangulation we  calculate  the average return probability $P(\sigma)$ based on a sample of  starting points, which is much smaller than the total volume of the triangulation (total number of 4-simplices), in order to obtain generic behaviour  we additionally require that the starting  4-simplex $i_0$ lies in the  ``central" slice of the ``blob", as defined by the centre of volume (see Chapter 3.1 for details).}) starting points $i_0$ and for $N_{MC}$ statistically independent triangulations $\cT$ and calculates the average return probability:
\beql{retprobdisc}
P(\sigma)=\frac{1}{N_{MC}} \frac{1}{N_{i_o}} \sum_{\cT} \sum_{i_0} \   \rho(i_0,i_0;\sigma) 
\eeq 
and the spectral dimension
\beql{dsdisc}
d_S (\sigma) = -2 \, \frac{\log P(\sigma+1)-\log P(\sigma)}{\log (\sigma+1) - \log (\sigma)} \ .
\eeq
\begin{figure}[h!]
\centering
\scalebox{.95}{\includegraphics{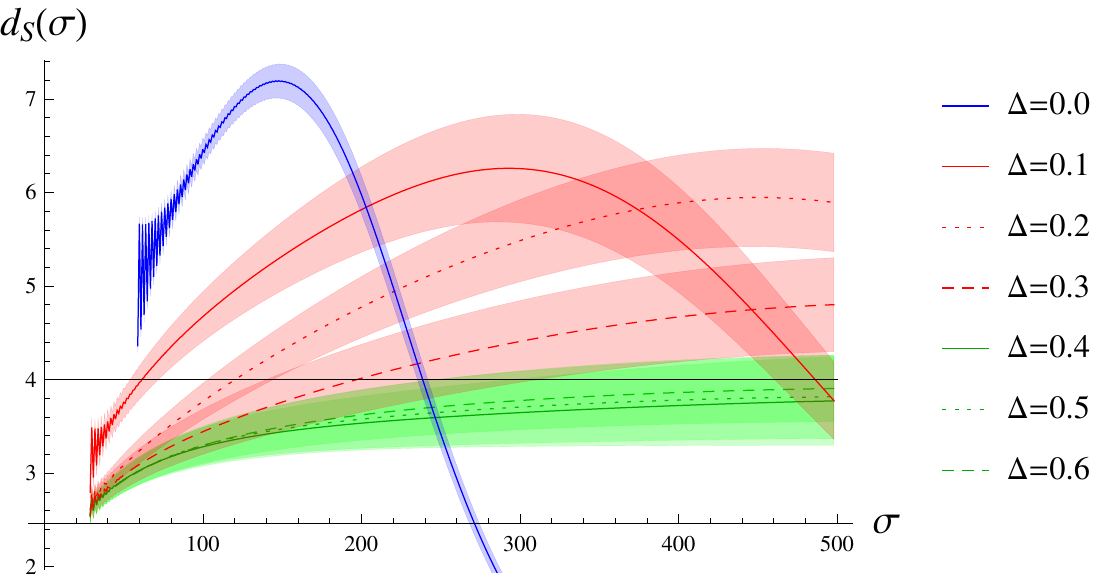}}
\caption{The spectral dimension $d_S$ as a function of a diffusion time $\sigma$ measured for $\kappa_0=2.2$ and a range of values of $\Delta=0.0-0.6$ in ``full CDT" simulations with $T=80$ and $\bar V_4=160$k.  The average value is plotted as a line while  statistical errors are highlighted by coloured areas. The statistical errors were calculated as a standard deviation of the average values measured for  different triangulations $\cT$.
A change of the behaviour observed for $\Delta < 0.4$ cannot be attributed to statistical errors and is a clear indication of a new phase transition.}
\label{SDimerr}
\end{figure}
\begin{figure}[h!]
\centering
\scalebox{.9}{\includegraphics{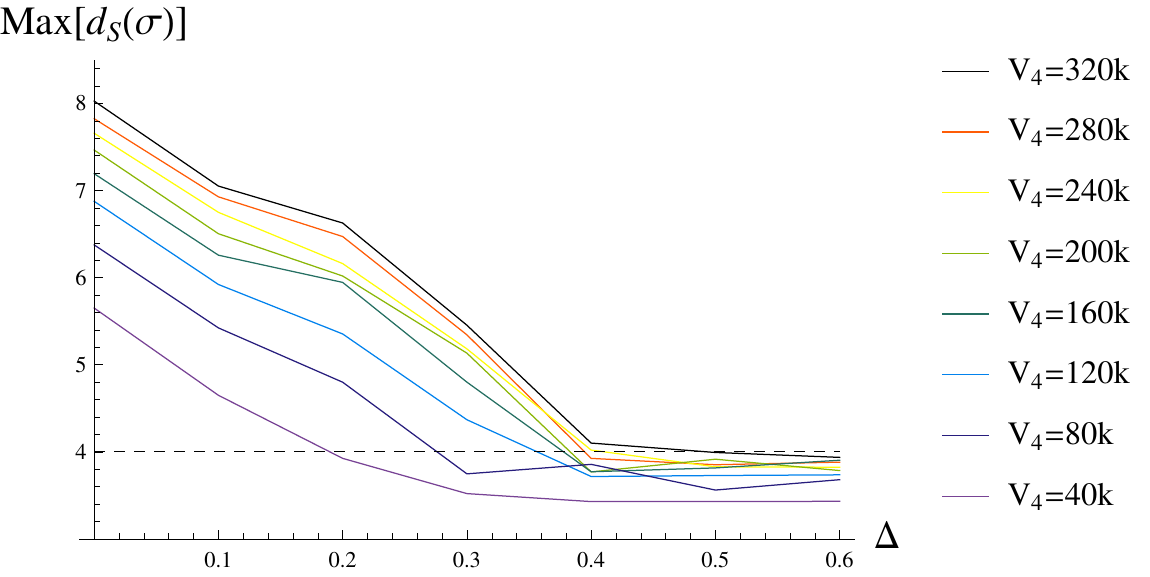}}
\caption{The maximum of $d_S(\sigma)$ measured for $\kappa_0=2.2$ and a range of values of $\Delta=0.0-0.6$ in ``full CDT" simulations with $T=80$. Different colours indicate different total volumes of the simplicial manifolds $\bar V_4$. The maximum of $d_S$  is stable and consistent with four for $\Delta \geq 0.4$ and is rising with increasing volume for $\Delta < 0.4$. }
\label{SDMax}
\end{figure}
\noindent The spectral dimensions measured  for  fixed bare (inverse) gravitational constant $\k_0=2.2$ and a range of values of the bare asymmetry parameter $\Delta=0.0 - 0.6$ are presented in Fig. \ref{SDimerr}. For $\Delta \geq 0.4$ the spectral dimension forms a plateau at $d_S=4$ for diffusion times $\sigma \approx 500$ which is consistent with the generic ``C" phase behaviour. For very large diffusion times (not visible in Fig. \ref{SDimerr})  $d_S$ starts to fall due to the compactness of simplicial manifolds as explained above. For  $\Delta < 0.4$ one can observe a change  of the $d_S(\sigma)$ shape which cannot be attributed to statistical errors. 
This is a clear indication of a new phase transition occurring  for $\Delta = 0.3-0.4$. The maximum  of $d_S$ inside the new ``bifurcation" phase is considerably higher than four and is rising with increasing total volume $\bar V_4$ - see Fig. \ref{SDMax}. This suggests that the dimension can potentially be infinite in the continuous  limit $\bar V_4 \to \infty$.  Such behaviour is  characteristic of the  ``B" phase observed for $\Delta = 0.0$, rather than for the generic phase ``C" ($\Delta \geq 0.4$). The behaviour of $d_S$ for  $\Delta < 0.4$ can  possibly be associated with a formation of vertices of  very high order (belonging to a large number of  4-simplices) which can be directly observed in CDT triangulations inside the new bifurcation phase. To explain this phenomenon let us refer  to \cite{Mielczarek2} where the authors discuss a toy model of a simple graph with $N$ vertices. The spectral dimension can be computed analytically for the ring graph, where each vertex is directly connected to two neighouring vertices, and for the corresponding ``complete" graph, where each vertex is connected with  all other $N-1$  vertices - see Fig. \ref{RingToy}. In the ring graph the $d_S$ forms a plateau at $d_S=1$ and in the ``complete" graph  $d_S$ forms a maximum at $d_S \gg 1$ (Fig. \ref{SdimToy}). This qualitatively resembles the behaviour observed in full CDT data inside the ``C" phase and in the ``bifurcation" phase, respectively. In the toy model the maximum of $d_S$ behaves like: $\max\,d_S(\sigma)\approx 2 W(N/e)$, where $W(x)$ is the Lamber function (for large $x$: $W(x)\approx \log x - \log\log x$). Of course the structure of CDT triangulations is much more complicated than the simple toy model but the results show that rising $d_S$ is associated with a growing number of ``shortcuts" created in  the simplicial manifolds.
\begin{figure}[h!]
\centering
\scalebox{.45}{\includegraphics{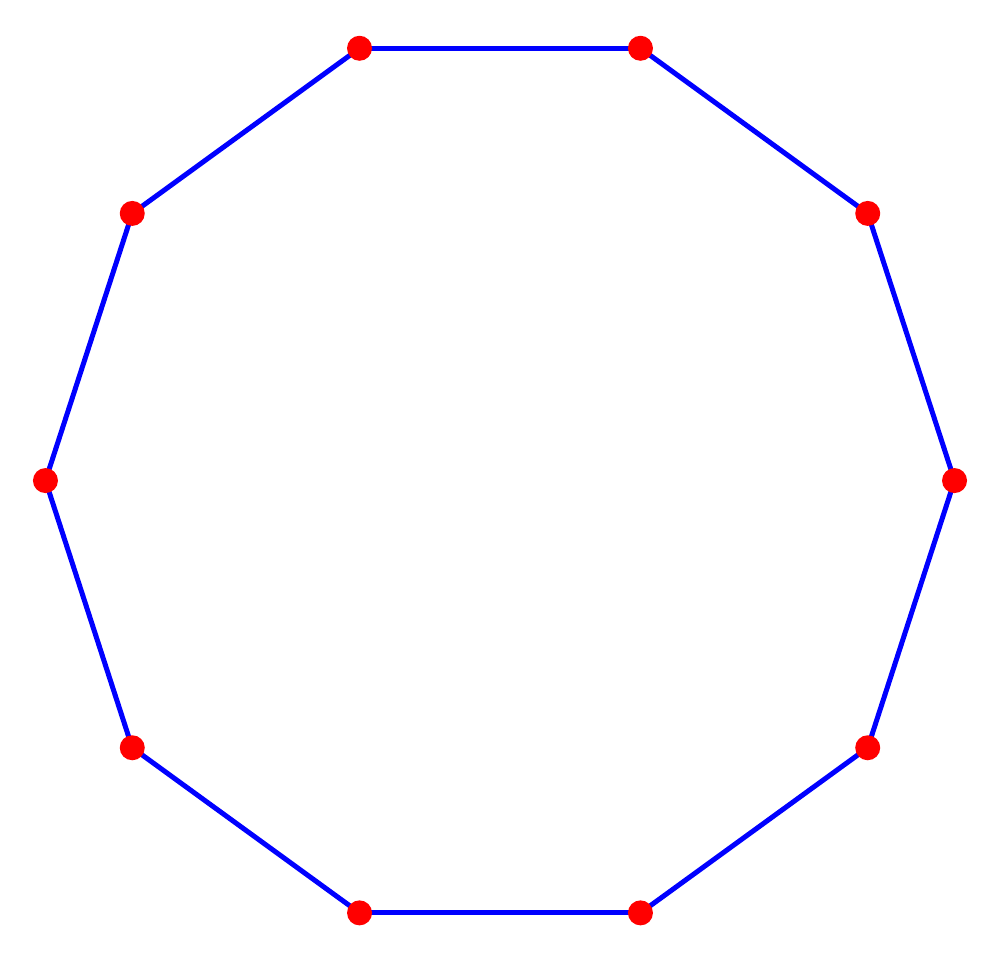}}
\scalebox{.45}{\includegraphics{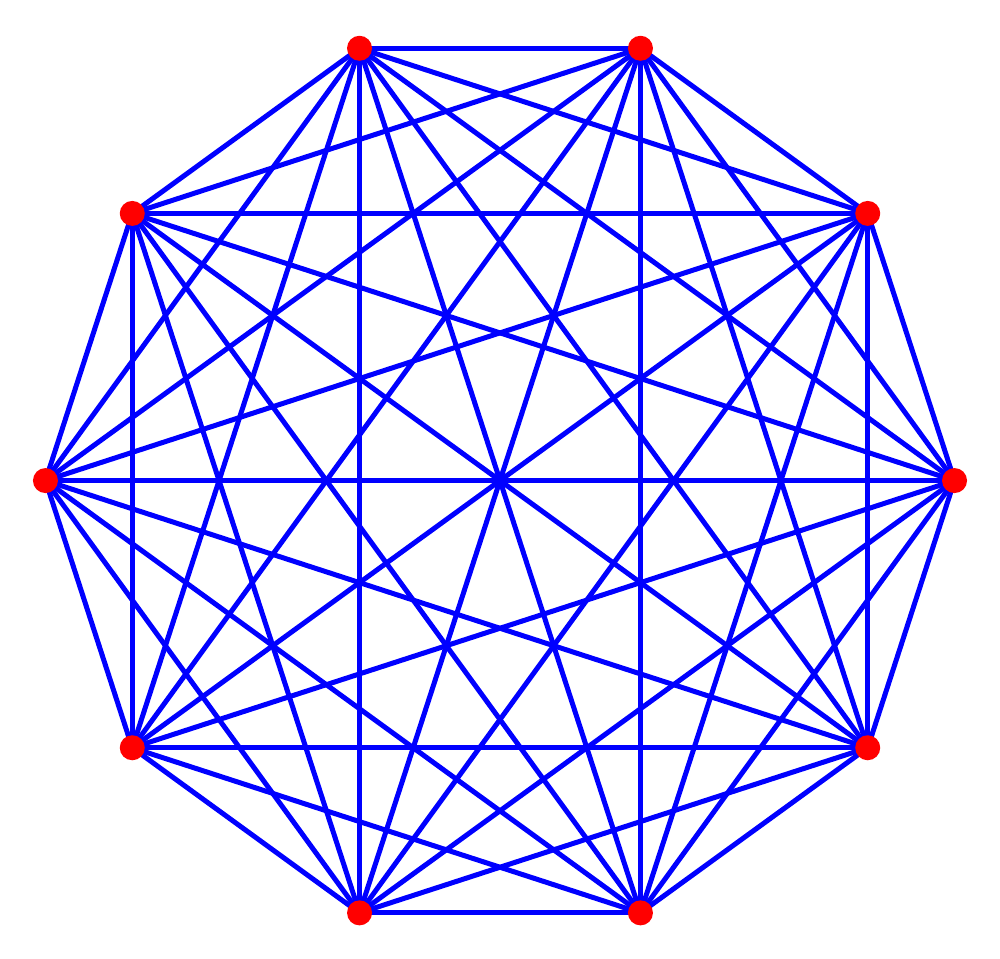}}
\caption{The toy model ring graph (left) and the ``complete" graph (right) introduced in \cite{Mielczarek2}. Courtesy of J. Mielczarek. }
\label{RingToy}
\end{figure}
\begin{figure}[h!]
\centering
\scalebox{.8}{\includegraphics{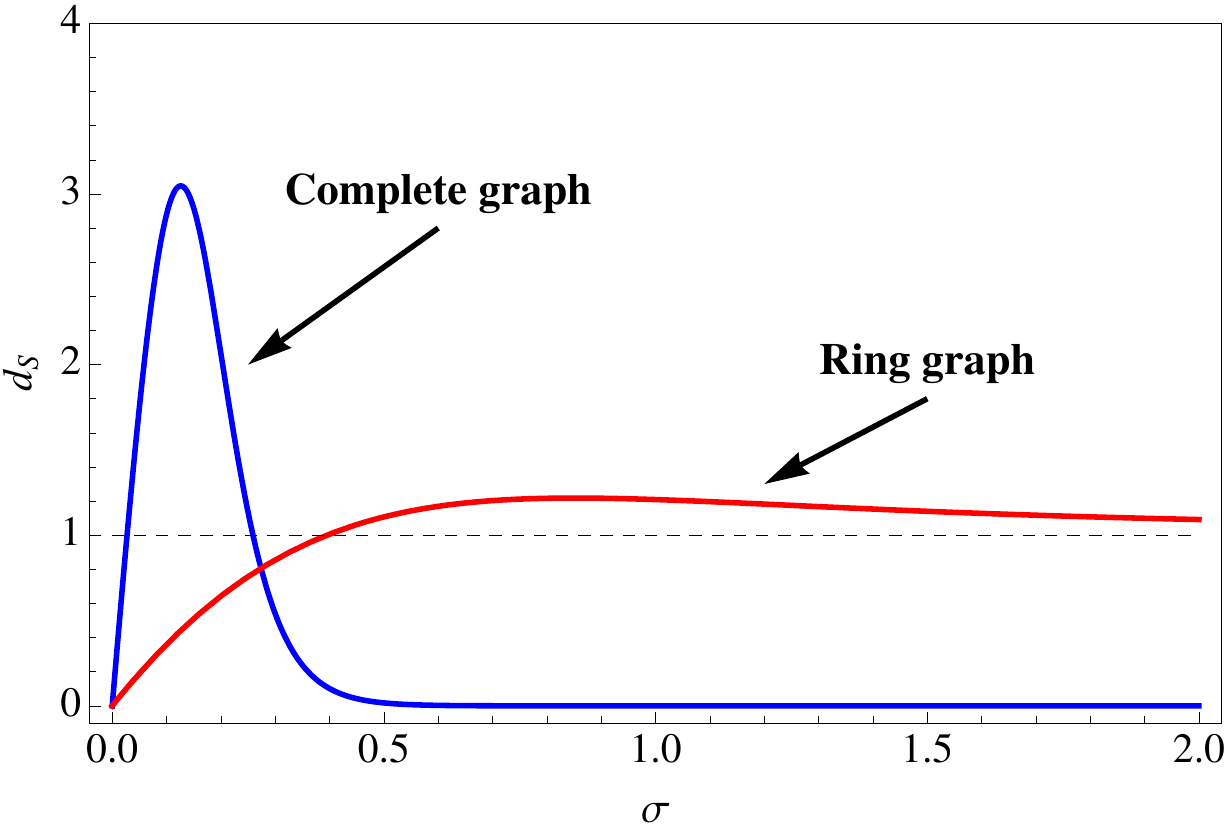}}
\caption{Spectral dimension for the ring graph and the ``complete" graph with $N=20$ nodes \cite{Mielczarek2}. Courtesy of J. Mielczarek.}
\label{SdimToy}
\end{figure}

\section{New phase diagram}
In  previous sections we  provided strong evidence that the ``bifurcation" region is a genuinely new phase of  four dimensional Causal Dynamical Triangulations. Our analysis was based on data measured for the bare (inverse) gravitational constant $\kappa_0=2.2$, for which the new phase transition occurred for the bare asymmetry parameter  $\Delta=0.3-0.4$. Now we would like to determine the position of the new phase transition line based  on systematic measurements of the bifurcation structure in a wide range of $\kappa_0$ and $\Delta$ couplings. Once again we focus on  selected empirical  transfer matrix cross-diagonals measured for $s=\bar V_4 = 10$k, $20$k, $30$k, $40$k, $60$k and for $\kappa_0=1.0 - 4.6$, $\Delta = 0.0 - 0.4$ (we have measured in total over $800$ cross-diagonals). We note that the results presented below are still preliminary and the analysis will be mostly qualitative. To adopt a more quantitative approach, e.g. to estimate a precise position of the new phase transition line and its dependence on the total volume, one would need to perform very dense  measurements close to the phase transition as well as  considerably increase simulation time to assure good thermalization of the measured data which suffers from a characteristic  slowdown described in the beginning of this Chapter.

The empirical cross-diagonals can be 
fitted with:
$$
\braket{n|M|s-n}=
$$
$$
 ={\cal N}(s)\left[ \exp \left( -\frac{\Big((m-n) - c[s]\Big)^2}{k[s]}\right) + \exp \left( -\frac{\Big((m-n) + c[s]\Big)^2}{k[s]}\right)\right] \ ,
$$
and one can extract the bifurcation shift $c[s]$.
The measured cross-diagonals for fixed
$\kappa_0=2.2$ and a range of values of $\Delta$ are presented in Fig. \ref{BifantiK} and for fixed $\Delta=0.1$ and different $\kappa_0$'s in  Fig. \ref{BifantiD}. 
\begin{figure}[hb!]
\centering
\scalebox{1.2}{\includegraphics{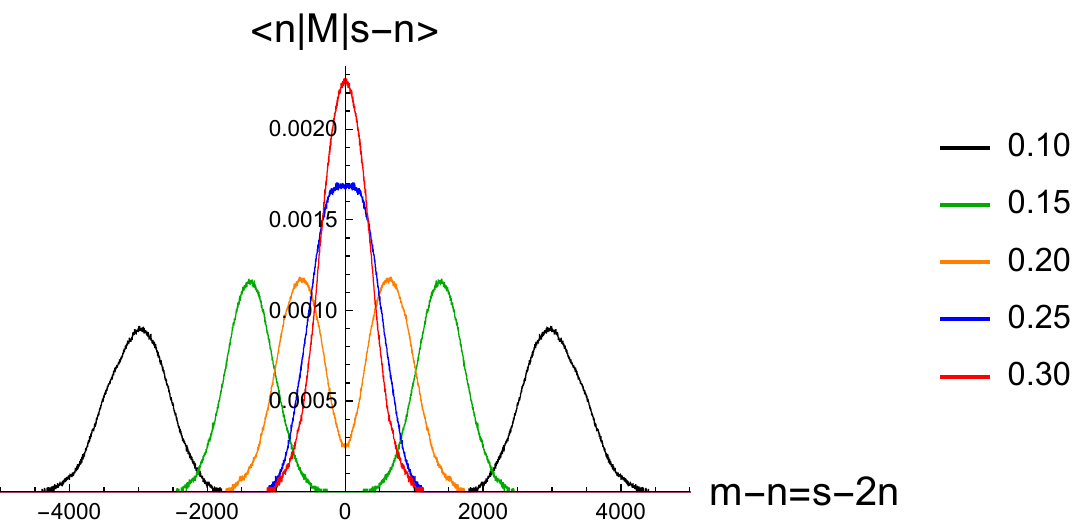}}
\caption{Empirical cross-diagonals measured for fixed $\kappa_0=2.2$ and different $\Delta$ (denoted by different colours). Data measured for $s=\bar V_4=30$k.}
\label{BifantiK}
\end{figure}
\begin{figure}[h!]
\centering
\scalebox{1.1}{\includegraphics{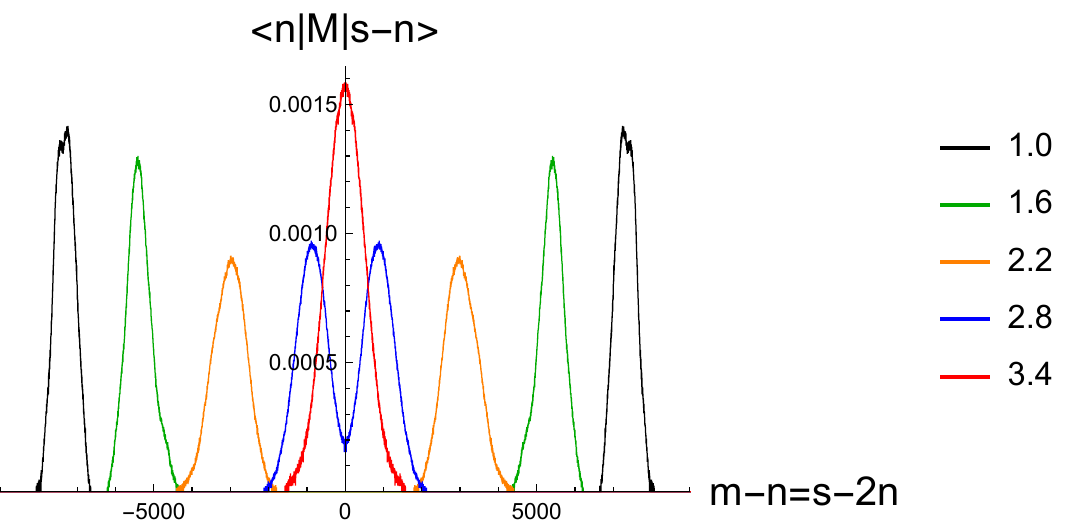}}
\caption{Empirical cross-diagonals measured for fixed $\Delta=0.1$ and different $\kappa_0$ (denoted by different colours). Data measured for $s=\bar V_4=30$k.}
\label{BifantiD}
\end{figure}
\begin{figure}[h!]
\centering
\scalebox{.6}{\includegraphics{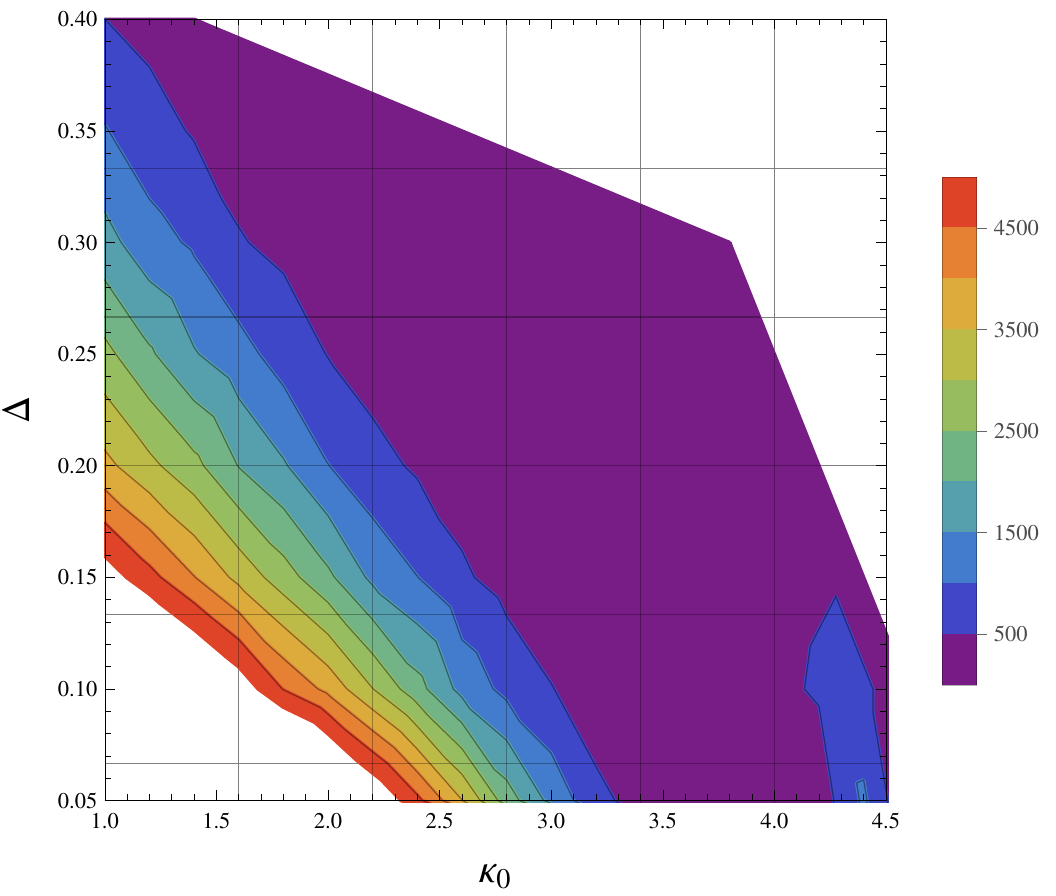}}
\scalebox{.6}{\includegraphics{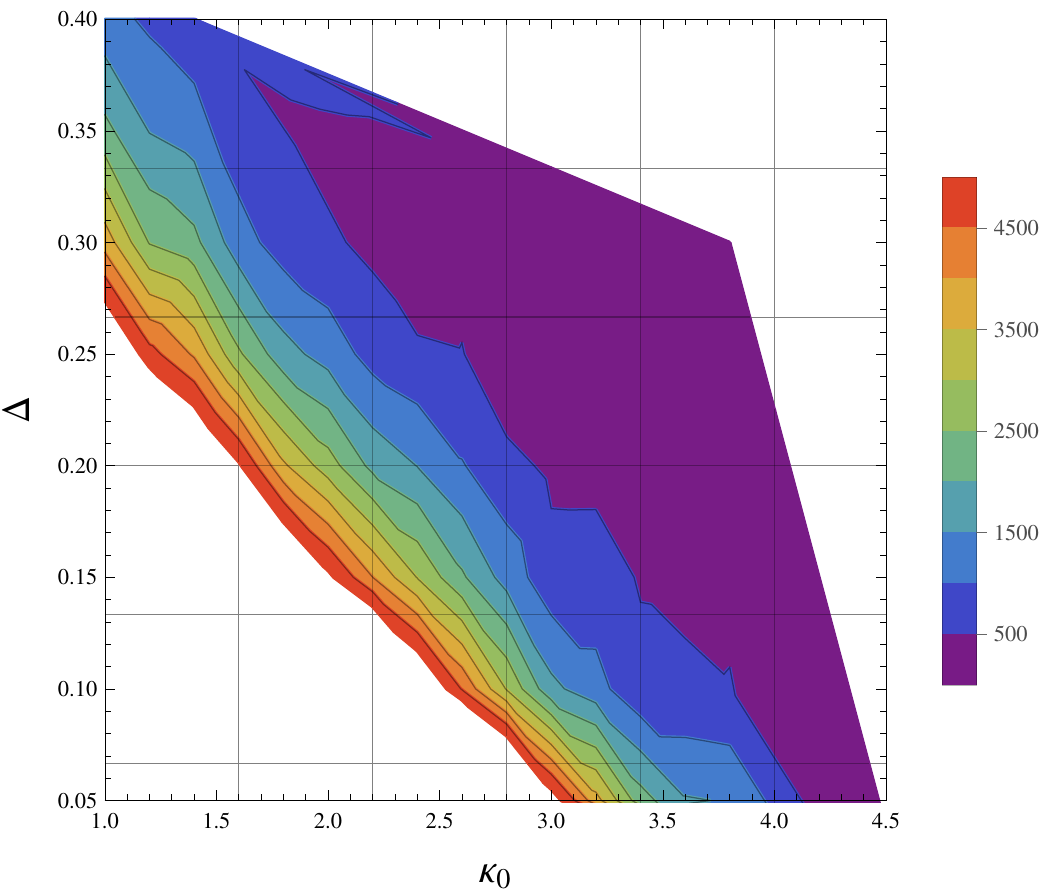}}
\caption{Contour plots of the bifurcation shift $c[s]$ in ($\kappa_0,\Delta$) plane measured for $\bar V_4=30$k (left) and $\bar V_4=60$k (right).}
\label{Bifsiatka}
\end{figure}
\begin{figure}[h!]
\centering
\scalebox{.8}{\includegraphics{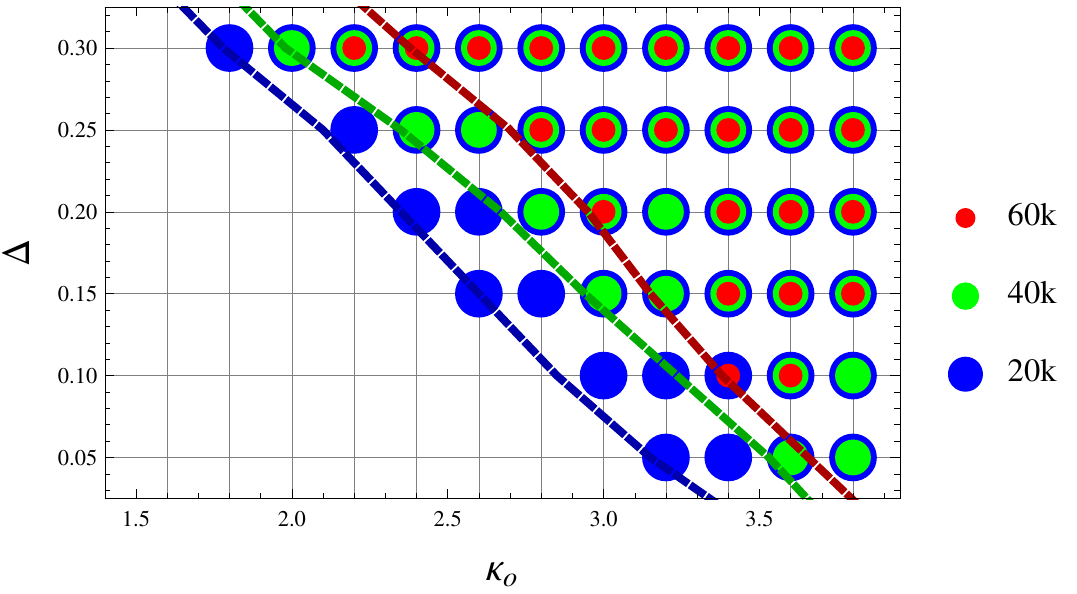}}
\caption{Points in the ($\kappa_0,\Delta$) bare couplings plane for which the bifurcation shift $c[s]<100$,  measured for  $\bar V_4 = 20$k (blue), $40$k (green) and $60$k (red). The bottom-left edge of the dotted  regions can be associated with the new phase transition (dashed lines) measured for different  $\bar V_4$'s. The red point visible for ($\kappa_0=2.2,\Delta=0.3$) was manually excluded due to a  small but still visible bifurcation structure.}
\label{BifsiatkaC}
\end{figure}
In both cases one can observe a gradual vanishing of the bifurcation with rising $\Delta$ and with rising $\kappa_0$. This tendency is illustrated in Fig. \ref{Bifsiatka} where we show a contour plot of the measured bifurcation shift $c[s]$  in the ($\kappa_0, \Delta$) bare couplings plane. The left chart presents the data measured for $\bar V_4=30$k and the right chart for $\bar V_4= 60$k. The purple colour indicates a region of  vanishing bifurcation ($c[s] < 500$), which can be associated with the generic ``C" phase while different colours denote higher values of $c[s]$ characteristic for the ``bifurcation" phase. The closer one approaches the  phase transition   the harder it is to see a non-zero bifurcation shift $c[s]$. Very close to the transition  one can measure $c[s]>0$ only for very large $\bar V_4$.  As a result the ``de Sitter" phase seems to  shrink  in favour of the  ``bifurcation" phase if one increases $\bar V_4$ - see Fig. \ref{BifsiatkaC}, where the coloured  dots denote the points in the ($\kappa_0, \Delta$) plane for which $c[s] < 100$, i.e the generic ``de Sitter" phase. Different colours   correspond to  different total volumes $\bar V_4 = 20$k (blue), $40$k (green) and $60$k (red). The bottom-left edge of the dotted regions can be associated with the phase transition line measured for different  $\bar V_4$'s.  We also present this result in Fig. \ref{newphasediag} where the line measured for (the biggest currently available) $\bar V_4=60$k is added to the ``traditional" phase diagram. One should keep in mind that these results are only approximate and the phase transition line can  possibly be shifted further up and right if one considers even larger volumes. The line has been extrapolated both to the left-up and to the right-down, where we made a conjecture that all four phases meet at a common point, which becomes a quadruple point.  
\begin{figure}[h!]
\centering
\scalebox{.85}{\includegraphics{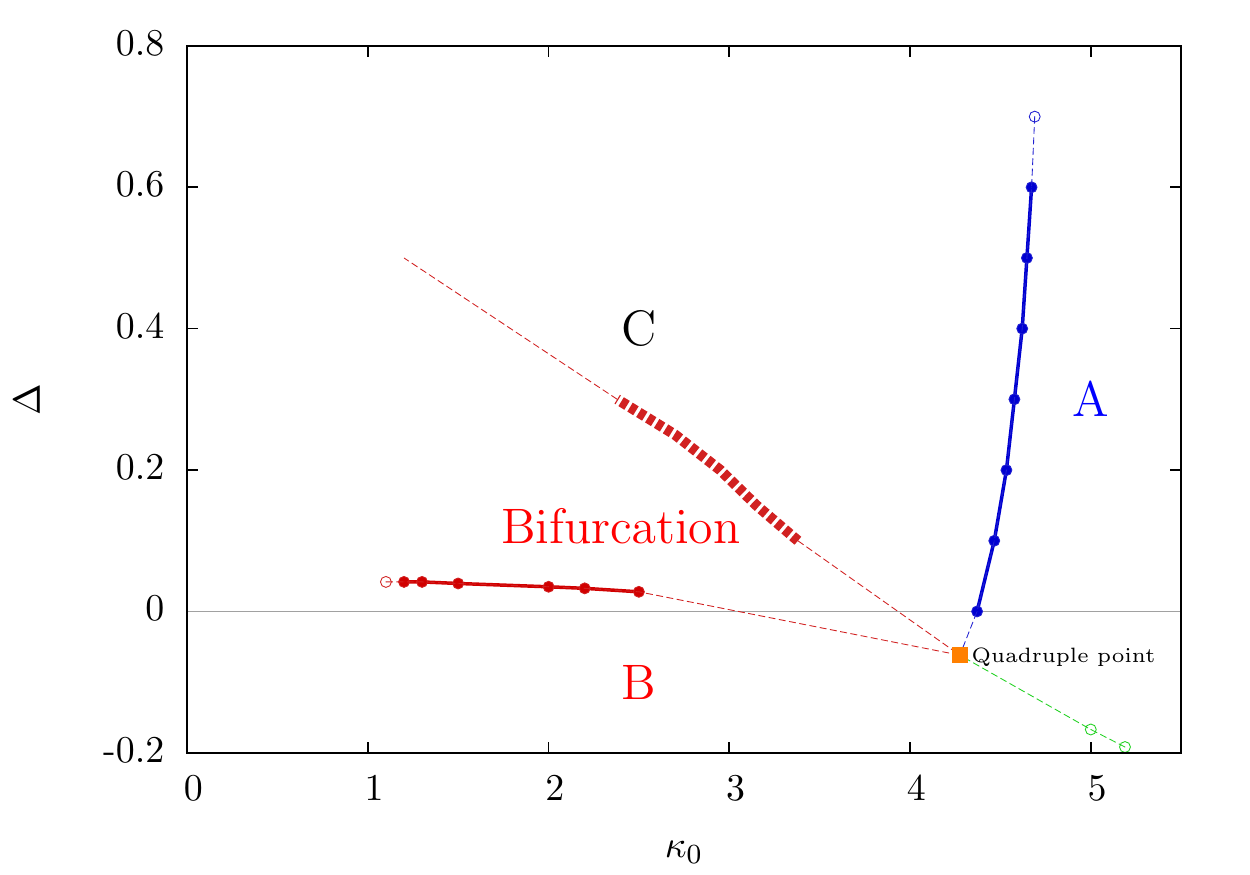}}
\caption{The ``new"  phase diagram of four-dimensional CDT. An approximate position of the phase transition between the ``C" phase and the ``bifurcation" phase (measured for $\bar V_4=60$k) is denoted by the thick dashed red line.}
\label{newphasediag}
\end{figure}

\chapter*{Conclusions}
\addcontentsline{toc}{chapter}{Conclusions}
\fancyhf{}
\fancyhead[RE]{\bf Conslusions}
\fancyhead[LE,RO]{\thepage}

In this dissertation we have presented a comprehensive review of recent  studies of the effective action in four-dimensional Causal Dynamical Triangulations (CDT). CDT is a non-perturbative, background independent  approach to quantum gravity, where a continuous path integral is approximated by a sum over an ensemble of simplicial manifolds, which provides an ultraviolet regularization of the theory. After Wick rotation CDT becomes a statistical theory of (random) piecewise-linear geometries and can be studied numerically using Monte Carlo methods.  
In our studies we focused on the three-volume of the spatial slices of constant (discrete) proper time parametrized by the number of  \{4,1\} simplices with four vertices in a given spatial slice. The three-volume distribution is obtained by
integrating out all CDT degrees of freedom except the scale factor.
Quantum fluctuations of the  three-volume observable  around a dynamically appearing average geometry can be described in terms of the effective action. The form  of the action is related to the phase structure of CDT. 

Inside the ``de Sitter" phase, also called the ``C" phase, one could use a covariance matrix of three-volume fluctuations to reconstruct the effective action using a semiclassical approximation. The action measured in this phase  is consistent with a simple discretization of the minisuperspace action with a reversed overall sign. The subleading corrections of the effective action measured in numerical data seem to result from finite size effects, rather than from higher derivative terms in the continuous limit  (e.g. the curvature-squared  terms). We have shown that  one can use the \{3,2\} simplices  in order to refine the discrete proper time slicing and construct the effective action  comprising both types of the 4-simplices. In such a finer-grained model the effective  interaction between the \{4,1\} simplices alone changes sign and becomes repulsive. The \{3,2\}-\{4,1\} interaction keeps the original sign and stabilizes the whole system. Such a finer-grained model is fully consistent with the action for \{4,1\} simplices alone if one integrates out the \{3,2\} degrees of freedom.

We have introduced a new method of the  effective action measurement based on the effective transfer matrix, parametrized by the three-volume variable. The results of the new method are fully consistent with the covariance matrix method inside the ``de Sitter" phase. The transfer matrix method can be used  both in a  large  and a small volume limit.  We have shown that the small volume matrix elements can be used to explain  the spatial volume fluctuations observed in the ``stalk range" of the CDT universe and despite strong discretization effects one can still measure the same discretized minisuperspace action as is present in the ``blob" range. The transfer matrix data can  also be used to define a simple volume fluctuations model, based on the effective action idea, which is able to reconstruct the average spatial volume profile and correlations measured in the  ``full CDT"  model, with all other degrees of freedom.

The transfer matrix method can  also be used to measure and parametrize the effective action in other CDT phases and to analyze the phase transitions. 
The ``A"-``C" phase transition can be explained by a vanishing 
 kinetic term, which disappears inside  phase ``A". The potential term changes as well. As a result  the quantum fluctuations of the three volume in different spatial slices become independent which can be interpreted as  causal disconnection of space points, also  called  ``asymptotic silence". 
 
Inside  phase ``B" one can observe a bifurcation of the transfer matrix kinetic term.  It causes the narrowing of the average spatial volume profile in the time-direction  consistent with the ``collapsed" blob structure observed in the CDT data.   The bifurcation  persists in a region of the bare coupling constant space traditionally denoted the ``C" phase.  We have shown that the geometric properties in this region, in particular the Hausdorff and the spectral dimension are considerably different than in the generic ``de Sitter" phase. At the same time there is no reason to doubt  that the ``traditional" ``B"-``C" transition is still there, so
unexpectedly  we have discovered a genuinely new phase separating the ``old" phase ``C" and phase ``B". This discovery is the most important result of this thesis.

There are many open questions which remain. So far we have managed  to characterize the new phase transition only qualitatively and only approximately established  a position of the  new transition line. More and better thermalized numerical data are needed to draw more precise quantitative conclusions, in particular to analyze the order of the phase transition. One should also investigate  the properties of Causal Dynamical Triangulations in the vicinity of the conjectured quadruple point where presumably all four phases meet. Unfortunately our numerical algorithms show a critical slowdown in this region  and consequently it is very difficult to perform any reliable measurements there. In this respect, the transfer matrix method has an important advantage over    other ``traditional" methods, namely it requires relatively small systems and consequently much shorter thermalization process. Finally the discovery of the new ``bifurcation" phase raises  many new questions. Just to mention a few of them: What is the physical mechanism of the phase transition ?  Can we interpret the emergence of the bifurcation at the phase transition as a sign of a spontaneous signature change ? If so, can the new phase get some physical meaning ? One should also think about reanalyzing some previous results based on the assumption that the newly discovered ``bifurcation" phase is the ``de Sitter" phase. In particular  recent results concerning renormalization group flow in CDT \cite{CDTFlow}  might require some new interpretation.   All these issues deserve further studies.

\begin{appendices}

\newpage
\thispagestyle{empty}
\mbox{}
\newpage
\chapter{Derivation of the Regge action}
\fancyhf{}
\fancyhead[RE]{\bf Appendix A}
\fancyhead[LO]{\bf Derivation of the Regge action}
\fancyhead[LE,RO]{\thepage}
This appendix is based on the article: {\it J. Ambj\o rn, J. Jurkiewicz, R.Loll, ``Dynamically Triangulating Lorentzian Quantum Gravity", Nucl. Phys. B 610, 347 (2001) [hep-th/0105267]. }
\newline
\newline

We note that in order to obtain dimensionless coupling constants we set the length of the spatial links $l_s\equiv1$ (for the time-like links we have $l_t^{\,2} = - \alpha$ in Lorentzian signature) and express the dimensionful   constants ($G$ and $\Lambda$) in  lattice units.
We start with the Einstein-Hilbert action (\ref{SHE}): 
$$
S_{HE}[g]=\frac{1}{16 \pi G}\int{d^4x \sqrt{-g}(R-2 \Lambda)}\nonumber
$$
and use equations (\ref{SRcosmol}) and (\ref{SRcurv}) to obtain:
\begin{eqnarray}\label{SRe}
S_R= \frac{1}{8 \pi G} \Bigg[ \frac{1}{i} \sum_{SL\triangle}V_2^{SL}\left(2 \pi - \sum_{S \ at \ SL\triangle} \Theta_{SL} \right)\nonumber \\
+  \sum_{TL\triangle}V_2^{TL}\left(2 \pi - \sum_{S \ at \ TL\triangle} \Theta_{TL} \right)  \\
-  \Lambda \ \left( N^{ \{4,1\}} \  V_4^{\{4,1\}} +   N^{ \{3,2\}} \ V_4^{\{3,2\}} \right) \Bigg] , \nonumber
\end{eqnarray}
where $SL$ stands for space-like, $TL$ - time-like, $\triangle$ - triangle, $S$ - 4-simplex. 

Formulas for Euclidean volumes and dihedral
angles can be easily derived and may be continued to Lorentzian geometries (we use Sorkin's conventions \cite{sorkin}). As a result the volumes of (Lorentzian) 4-simplices/triangles are given by ($a_s\equiv1$):
\begin{eqnarray}\label{Vols}
V_4^{\{4,1\}}=\frac{\sqrt{8 \alpha+3}}{96} &,&V_4^{\{3,2\}}=\frac{\sqrt{12 \alpha+7}}{96} \ , \\
V_2^{SL}=\frac{\sqrt {3}}{4} &,& V_2^{TL}=\frac{\sqrt{4 \alpha+1}}{4} \ . \nonumber
\end{eqnarray}

In a triangulation  there are in total:  $N^{\{4,1\}}$ simplices of type \{4,1\}, each with 4    $\Theta^{\{4,1\}}_{SL}$ and 6  $\Theta^{\{4,1\}}_{TL}$ dihedral angles, and  $N^{\{3,2\}}$ simplices of type \{3,2\}, each with 1    $\Theta^{\{3,2\}}_{SL}$, 3 $\Theta^{\{3,2\}}_{TL1}$ and 6 $\Theta^{\{3,2\}}_{TL2}$ dihedral angles. 

Therefore  (\ref{SRe}) can be rewritten as:
\begin{eqnarray}\label{SRe2}
S_R &=& \frac{1}{8 \pi G} \Bigg[\frac{1}{i}{V_2^{SL} }\left(  {2 \pi}N_2^{SL} -  4 \Theta^{\{4,1\}}_{SL} N^{\{4,1\}} -  \Theta^{\{3,2\}}_{SL} N^{\{3,2\}}  \right)\\
 & &+V_2^{TL} \left(  2 \pi N_2^{TL} -  6 \Theta^{\{4,1\}}_{TL} N^{\{4,1\}} -  3 \Theta^{\{3,2\}}_{TL1} N^{\{3,2\}} - 6 \Theta^{\{3,2\}}_{TL2} N^{\{3,2\}}  \right)\nonumber\\
& &-  \Lambda \ \left( N^{ \{4,1\}} \  V_4^{\{4,1\}} +  N^{ \{3,2\}} \ V_4^{\{3,2\}} \right) \Bigg]  \ , \nonumber
\end{eqnarray}
where $N_2^{SL}$ and $N_2^{TL}$ are the total numbers of space-like and time-like triangles in the manifold, respectively.

We assume that: $0\leq \text{Re}\ \Theta \leq \pi$, 
so  defining  $\sin \Theta$ and 
$\cos \Theta$ fixes $\Theta$  uniquely. For different types of the dihedral angles one obtains:
\begin{eqnarray}\label{thetas}
\cos \Theta^{\{4,1\}}_{SL}=\frac{-i}{\sqrt{24 \alpha+8}}& ,& \sin \Theta^{\{4,1\}}_{SL}=\sqrt{\frac{24 \alpha + 9}{24 \alpha+8}} \nonumber\\
\cos\Theta^{\{3,2\}}_{SL}=\frac{6 \alpha+5} {6 \alpha+2} & ,& \sin \Theta^{\{3,2\}}_{SL}=\frac{-i\sqrt{36 \alpha+21}}{{6 \alpha+2}}\nonumber\\
\cos \Theta^{\{4,1\}}_{TL}=\frac{2 \alpha+1}{{6 \alpha+2}} & ,& \sin \Theta^{\{4,1\}}_{TL}=\frac{\sqrt{4\alpha+1}\sqrt{8\alpha+3}}{{6 \alpha+2}} \\
\cos \Theta^{\{3,2\}}_{TL1}=\frac{4 \alpha+3}{{8 \alpha+4}}  & ,& \sin \Theta^{\{3,2\}}_{TL1}=\frac{\sqrt{4\alpha+1}\sqrt{12\alpha+7}}{{8 \alpha+4}}\nonumber\\
\cos \Theta^{\{3,2\}}_{TL2}=\frac{-1}{\sqrt{8 \alpha+4}\sqrt{6 \alpha+2}} & ,&\sin \Theta^{\{3,2\}}_{TL1}=\frac{\sqrt{4\alpha+1}\sqrt{12\alpha+7}}{\sqrt{8 \alpha+4}\sqrt{6 \alpha+2}}  \ . \nonumber
\end{eqnarray}
Using topological identities:
\begin{eqnarray}\label{ident}
 N_2^{SL}& = & N^{\{4,1\}}  \ , \\
N_2^{TL} & = & 2 N_0 + 2  N^{\{3,2\}}+ N^{\{4,1\}}-2 \chi \ ,
\end{eqnarray}
where $N_0$ - the total number of vertices in the triangulation, $\chi$ is the Euler characteristic of the manifold, and regrouping one obtains:
\beq\label{SRe3}
S_R =  -(\kappa_0+6\Delta) N_0 + (K_4+ \Delta) N^{\{4,1\}} +K_4   N^{\{3,2\}} \ ,
\eeq
where:
\beq\label{couplings}
\kappa_0+6\Delta=-\frac{1}{2  G}V_2^{TL} \ ,
\eeq
$$K_4+\Delta=\frac{1}{8 \pi G}\left( V_2^{SL}\frac{4}{i} \left(  \frac{\pi}{2} -   \Theta^{\{4,1\}}_{SL} \right) + V_2^{TL} \left(  2 \pi  -  6 \Theta^{\{4,1\}}_{TL}   \right)- \Lambda V_4^{\{4,1\}}\right)  \ , $$
$$K_4=\frac{1}{8 \pi G}\left( -V_2^{SL} \frac{1}{i} \Theta^{\{3,2\}}_{SL} + V_2^{TL} \left(  4 \pi  -  3 \Theta^{\{3,2\}}_{TL1}  - 6 \Theta^{\{3,2\}}_{TL2} \right)- {\Lambda} V_4^{\{3,2\}}\right)  \ 
$$
and we omitted a constant term proportional to the Euler characteristic $\chi$.

From expressions  (\ref{Vols}) and (\ref{thetas}) it is obvious that for $\alpha>0$ all Lorentzian volumes are positive and all $\Theta_{TL}$ dihedral angles are usual Euclidean angles, while the $\Theta_{SL}$ dihedral angles are in general complex angles (``boosts"). To ensure that   the bare coupling constants $\kappa_0$, $K_4$ and $\Delta$ (and thus the whole Regge action) are real it is enough to check that for $\alpha>0$: 
$$ \frac{1}{i} \left(  \frac{ \pi}{2} - \Theta^{\{4,1\}}_{SL} \right) = \frac{1}{i} \left(  \frac{ \pi}{2} - \arccos\frac{-i}{\sqrt{24 \alpha+8}} \right)=-\text{arcsinh}\frac{1}{\sqrt{24 \alpha+8}}\in R \ ,$$
$$ \frac{1}{i} \Theta^{\{3,2\}}_{SL}   =\frac{1}{i} \arcsin \frac{-i\sqrt{36 \alpha+21}}{{6 \alpha+2}} = - \text{arcsinh} \frac{\sqrt{36 \alpha+21}}{{6 \alpha+2}} \in R \ . $$
The action remains real for $\alpha > -\frac{1}{4}$ and in principle it is possible to define the Lorentzian path integral for null or even fully spatial 4-simplices. 

For $\alpha< -\frac{1}{4}$ the square roots can be analytically continued in the lower half of the complex $\alpha$ plane, such that:
\beq\label{square}
\sqrt{-\alpha}=-i\sqrt \alpha \ .
\eeq
For such $\alpha$ the action becomes complex as (some) Lorentzian volumes (\ref{Vols}) are imaginary. For $\alpha< -\frac{7}{12}$ all Lorentzian volumes except from $V_2^{SL}$ are imaginary (all Euclidean volumes are real and positive). At the same time the complex dihedral angles $\Theta_{SL}$ become real. Consequently the action becomes purely imaginary and can be written as:
\beq
S_R^{(L)}\equiv S_R = i \cdot  S_R^{(E)} \ ,
\eeq
$${ S_R^{(E)} =  -(\widetilde\kappa_0+6\widetilde\Delta) N_0 +(\widetilde K_4 +\widetilde \Delta)  N^{\{4,1\}}+\widetilde K_4 N^{\{3,2\}}} \ , $$ 
where $\widetilde . $ denotes the imaginary part of $\kappa_0$, $K_4$ and $\Delta$ respectively. Expressed in terms of $\widetilde \alpha = -\alpha$ ($\widetilde\alpha > \frac{7}{12}$) the coupling constants become:

\begin{eqnarray}\label{couplingsreal}
\widetilde\kappa_0+6\widetilde\Delta & = & \frac{1}{8  G}\sqrt{4 \widetilde \alpha-1} \ , \\
\widetilde K_4+\widetilde\Delta & = & \frac{\Lambda}{8 \pi G} \frac{\sqrt{8 \widetilde\alpha-3}}{96} +  \frac{\sqrt 3}{8 \pi G}  \left( \arccos \frac{1}{\sqrt{24 \widetilde\alpha-8}} - \frac{\pi}{2}  \right) +\\
 &+&\frac{\sqrt{{4 \widetilde \alpha-1}}}{8 \pi G}  \left(    \frac{3}{2}\arccos\frac{2 \widetilde\alpha-1}{{6 \widetilde\alpha-2}}  - \frac{\pi}{2} \right) \ , \nonumber \\
\widetilde K_4 & = & \frac{\Lambda} {8 \pi G} \frac{\sqrt{12 \widetilde\alpha-7}}{96} + \frac{\sqrt {3}}{32 \pi G}  \arccos \frac{6\widetilde \alpha-5} {6 \widetilde\alpha-2} + \\
&+ &\frac{ \sqrt{4 \widetilde\alpha-1}}  {8 \pi G}  \left(  \frac{ 3}{4} \arccos  \frac{4 \widetilde\alpha-3}{{8\widetilde \alpha-4}}  + \frac{3}{2}  \arccos \frac{1}{\sqrt{8\widetilde \alpha-4}\sqrt{6 \widetilde\alpha-2}} -  \pi \right)\  . \nonumber
\end{eqnarray}
After Wick rotation: $\alpha \to \widetilde \alpha =  -\alpha$ ($\widetilde\alpha > \frac{7}{12}$) the quantum amplitude:
$$ Z=\sum_{\cal T} \ e^{iS_R^{(L)}[\cal T]} \ \  \to \ \   Z=\sum_{\cal {T}}\ e^{-\widetilde S_R^{(E)}[\cal T]} \ ,$$ 
becomes the partition function of a statistical model, where $ S_R^{(E)}$ is purely real.  $ S_R^{(E)}$ is the bare action used in four-dimensional CDT   Monte-Carlo simulations.
\newpage

\newpage
\thispagestyle{empty}
\mbox{}
\newpage


\chapter{Monte Carlo algorithm}
\fancyhf{}
\fancyhead[RE]{\bf Appendix B}
\fancyhead[LO]{\bf Monte Carlo algorithm}
\fancyhead[LE,RO]{\thepage}

Here we provide only a  short description of the algorithm used in numerical simulations of Causal Dynamical Triangulations in four dimensions. More details can be found in \cite{CDTreviews4}. The Monte Carlo code is based on the modified Metropolis-Hastings \cite{Metropolis1,Metropolis2} algorithm. We use a set of  seven {\it  Monte Carlo moves} (including anti-moves)\,\footnote{One of the moves, the so called move $M_3$, is self-dual, i.e. it is the same as its anti-move and therefore we have seven instead of eight moves.} which transform triangulations from one to another (for a short description of each move see Table \ref{TabMoves}). 
The transformations define a Markov chain in the space of triangulations as a new configuration $\cal T_B$ depends only on a previous configuration $\cal T_A$ and the type of the  move performed. Each of the moves is applied locally, which means that it considers only a small number of adjacent (sub)simplices at a given position in the triangulation $\cal T_A$. The probability  $P({\cal T_ A} \xrightarrow{M} {\cal T_B})$ to perform a transition ${\cal T_ A}\to {\cal T_B}$ using a move $M$ is defined by the {\it detailed balance} condition
\beq\label{detbal}
P({\cal T_A})P({\cal T_ A} \xrightarrow{M}  {\cal T_B}) = P({\cal T_B})P({\cal T_B} \xrightarrow{\bar M}  {\cal T_A})  \ ,
\eeq 
where $\bar M$ is the anti-move (reverse of move $M$ that transforms configuration ${\cal T_ B}$ into  ${\cal T_A}$). This condition ensures that  probability  distribution $\widetilde P(\cal T)$ of  the Monte Carlo  simulation approaches the stationary distribution of CDT:
\beq
P({\cal T})= \frac{1}{\cal Z}e^{- S_R[{\cal T}]} \ .
\eeq

The probability of a transition $P({\cal T_ A} \xrightarrow{M} {\cal T_B})$ is a product of two components, namely the probability  $P(M)$ to accept the move $M$ and the probability $P({\cal A})$ to choose a location in the configuration $\cal T_A$ where the move is performed. The factor $P({\cal A})$ is purely geometrical and is simply the inverse of the number of places where the move could be applied. Consequently:
$$
P({\cal T_A})P(M) P({\cal A})= P({\cal T_B})P(\bar M) P({\cal B}) \ ,
$$
\beq\label{detbal2}
P_M\equiv \frac{P(M) }{ P(\bar M)} =\underbrace{ \frac{ P({\cal B})}{ P({\cal A})}}_{\text{geom.}}   \cdot \underbrace{\frac{e^{- S_R[{\cal T_B}]}}{e^{- S_R[{\cal T_A}] }} }_{\text{action}}   ={\frac{ P({\cal B})}{ P({\cal A})}} \cdot {e^ {- \delta  S_R[M]} } \ ,
\eeq
where:
\beq\label{dSR}
\delta  S_R[M] = -(\kappa_0+6\Delta) \delta N_0 + K_4 \left( \delta N^{\{4,1\}}+\delta N^{\{3,2\}}\right)+\Delta  \delta N^{\{4,1\}}
\eeq
 is a change in the Regge action caused by the move $M$ (which changes the total number of vertices, \{4,1\} and \{3,2\} simplices by $\delta N_0$, $\delta N^{\{4,1\}}$ and $\delta N^{\{3,2\}}$, respectively). According to the standard Metropolis-Hastings algorithm, the move $M$ is accepted with probability $1$ if $P_M > 1$ and with probability $P_M$ otherwise.\footnote{In the CDT Monte Carlo algorithm one additionally checks if the application of the move replicates any structures already existing in the initial configuration. If this is the case the move is automatically rejected.  As a result generated triangulations obey manifold requirements, i.e. each  (sub)simplex appears only once. Monte Carlo simulations fulfilling this condition are  called  {\it combinatorial} triangulations. }

The ratio $P_M$ is completely determined by the initial configuration $\cal T_A$ and the type of the move $M$. This is  because the geometric properties of the final triangulation $\cal T_B$ depend only on $\cal T_A$ and $M$.   To illustrate this  let us consider the so called move ``$M_4$", which adds a vertex in a spatial tetrahedron being an interface between one (1,4) and one (4,1) simplex. The move creates 1 new vertex and  6 new \{4,1\} simplices ($\delta N_0=1$, $\delta N^{\{4,1\}}=6$,  $\delta N^{\{3,2\}}=0 \Rightarrow \delta  S_R[M_4] =  -\kappa_0 + 6 K_4$). To perform the move one has to  randomly choose one of the $N^{ST}_{\cal A}$ spatial tetrahedra in the initial triangulation $\cal T_A$ (thus $ P({\cal A}) = 1 / 
N^{ST}_{\cal A}$). The resulting triangulation $\cal T_B$ will have $N^{V8}_{\cal B}$ vertices with coordination number 8 (belonging to exactly 8 \{4,1\} simplices) where the inverse move $\bar M_4$ can be applied. As the move $M_4$ adds exactly one such vertex the number $N^{V8}_{\cal B}= N^{V8}_{\cal A}+1$ (thus $ P({\cal B}) = 1 / 
(N^{V8}_{\cal A}+1)$. Taking everything  together: 
$$
P_{M_4} = \frac{N^{ST}_{\cal A}}{N^{V8}_{\cal A}+1}\cdot \exp(\kappa_0 - 6 K_4) \ .
$$
The same applies to the other moves, therefore given any initial triangulation and the type of the move one can compute $P_M$.

The structure  of the Monte Carlo algorithm is presented in Fig. \ref{MCFlow}.
The algorithm was implemented as a computer program written originally by Prof. Jerzy Jurkiewicz in cooperation with Prof. Jan Ambj\o rn and Prof. Renate Loll  in FORTRAN and then optimized and rewritten in C by Dr Andrzej G\"orlich. The computer simulations described in this thesis required various adjustments of the computer code provided by the author. The most advanced changes were made in cooperation with  Dr Andrzej G\"orlich.

\begin{table}
\begin{center}
 \begin{tabular} {|c|p{3.8cm}|c|c|c|c|}
\hline
\textbf{Move} &\textbf{Short description}& $\mathbf{ \delta N_0}$  & $\mathbf{ \delta N^{\{4,1\}}}$ & $\mathbf {\delta N^{\{3,2\}}}$ & $\mathbf{ \delta S_R}$  \\ \hline
\multirow{2}{*}{$M_2$}& {\footnotesize Replaces tetrahedral interface between two 4-simplices by a dual time-like link. }   & $0$    & $0$    &$2$  & $2 K_4$    \\
\hline
\multirow{2}{*}{$\bar M_2$}& \multirow{2}{*}{\footnotesize Inverse of move $M_2$. }   & $0$    & $0$    &$-2$  & $-2 K_4$    \\
& & & & & \\
\hline
\multirow{2}{*}{$M_3$}& {\footnotesize Self dual move. Replaces time-like triangle by a dual time-like triangle.  }   & $0$    & $0$    &$0$  & $0$    \\
\hline
\multirow{2}{*}{$M_4$}& {\footnotesize Adds new vertex creating three \{4,1\} and three \{1,4\} simplices.  }   & $1$    & $6$    &$0$  & $6K_4-\kappa_0$    \\
\hline
\multirow{2}{*}{$\bar M_4$}& \multirow{2}{*}{\footnotesize Inverse of move $M_4$.  }   & $-1$    & $-6$    &$0$  & $-6 K_4 +\kappa_0$    \\
& & & & & \\
\hline
\multirow{2}{*}{$M_5$}& {\footnotesize Replaces spatial triangle by space-like link.  }   & $0$    & $2$    &$0$  & $2 K_4+2\Delta$    \\
\hline
\multirow{2}{*}{$\bar M_5$}&\multirow{2}{*} {\footnotesize  Inverse of move $M_5$.  }   & $0$    & $-2$    &$0$  & $-2 K_4-2\Delta$    \\
& & & & & \\
\hline
\end{tabular}
 \end{center}
 \caption{The moves used in numerical simulations of four-dimensional CDT. The change in the total number of vertices, \{4,1\} and \{3,2\} simplices caused by each move is denoted by $\delta N_0$, $\delta N^{\{4,1\}}$ and $\delta N^{\{3,2\}}$, respectively. The change in the Regge action $\delta S_R$ is given by Eq. \rf{dSR}.}
 \label{TabMoves}
 \end{table}

\tikzstyle{block} = [rectangle, draw, fill=blue!5, 
    text width=27em, text centered, rounded corners, minimum height=1.5cm,node distance=2.5cm]

\tikzstyle{block2} = [rectangle, draw, fill=blue!15, 
    text width=30em, text centered, minimum height=1.5cm,node distance=2.5cm,rounded corners]

\tikzstyle{line} = [draw, -latex']

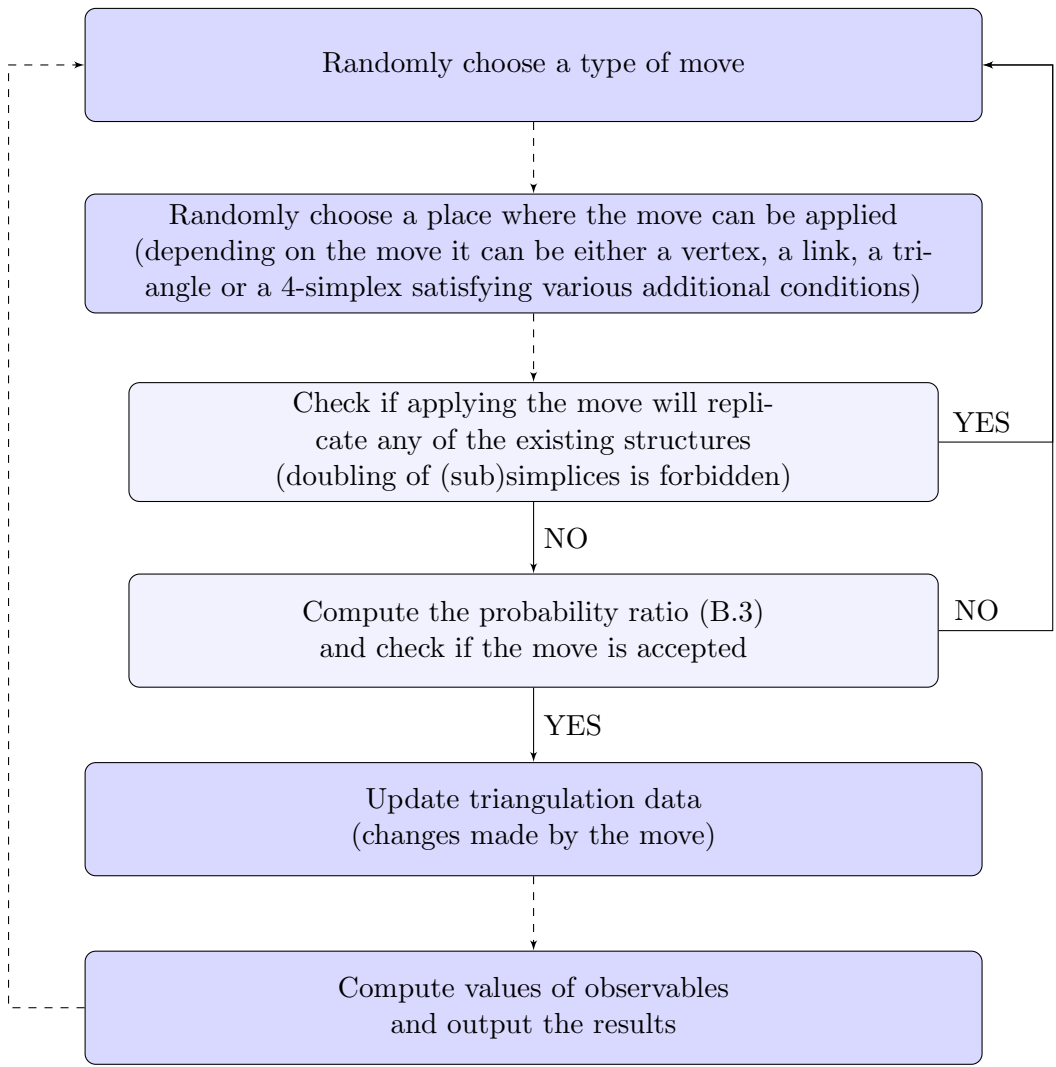
\begin{figure}
 \small   
  \begin{tikzpicture}[node distance = 2cm, auto]
    \node [block2] (1) {Randomly choose a type of  move};
    \node [block2, below of=1] (2) {Randomly choose a place  where the move can be applied \\(depending on the move it can be either a vertex, a link, a triangle or a 4-simplex satisfying various additional conditions)};
    \node [block, below of=2] (3) {Check if  applying  the move will replicate any of the existing structures  \\
(doubling of (sub)simplices is forbidden)};
    \node [block, below of=3] (4) {Compute the probability ratio (\ref{detbal2}) and check if the move is accepted};
    \node [block2, below of=4] (5) {Update triangulation data \\(changes made by the move)};
    \node [block2, below of=5] (6) {Compute values of observables \\ and output the results};    
    \path  [line,dashed](1) -- (2);
    \path [line,dashed] (2) -- (3);
    \path [line] (3) -- node {NO} (4);
    \path [line] (4) -- node {YES}(5);
    \path [line,dashed] (5) -- (6);
    \path [line] (3.east)  --++  (1.5,0) node  [near start]  {\ \ \ YES} |-  (1.east);
    \path [line] (4.east)  --++  (1.5,0) node  [near start]  {\  \ NO }  |-  (1.east);
    \path [line,dashed] (6.west) --++ (-1,0) |-  (1.west);
  \end{tikzpicture}

\caption{The flow chart presents the structure of one Monte Carlo step.}\label{MCFlow}

\end{figure}

\chapter{Minisuperspace action}
\fancyhf{}
\fancyhead[RE]{\bf Appendix C}
\fancyhead[LO]{\bf Minisuperspace action}
\fancyhead[LE,RO]{\thepage}

Let us consider a spatially homogenous and isotropic (Euclidean) space-time with a global time foliation and space topology $S^3$. We assume the  following metric:
\beql{Cmetric}
ds^2=d\tau^2+a^2(\tau)d\Omega_3^2 \ ,
\eeq
where $a(\tau)$ is a scale factor and $d\Omega_3^2$ is a line element on a sphere $S^3$ :
\beq
d\Omega_3^2 = d \theta ^2+ \sin^2\theta\left(d \phi^2_1 + \sin^2\phi_1d \phi^2_2\right) \ .
\eeq
The Ricci scalar is given by:
\beql{CR}
R=\frac{6}{a^2}\left(  1-\dot a^2 - a \ddot a\right)
\eeq
Supplementing Eq. (\ref{CR}) to the (Euclidean) Einstein-Hilbert action \cite{EuclideanS1,EuclideanS2}:
\beql{CSHA}
S_{HE}^{(E)}=-\frac{1}{16 \pi G}\int{d \tau}\int{d\Omega_3 \sqrt{g}(R-2 \Lambda)} \ ,
\eeq
where $g$ is the metric determinant while $G$ and $\Lambda$ are  Newton's constant and the cosmological constant, respectively, one obtains the following  {\it minisuperspace} action:
$$S_{MS}=-\frac{1}{16 \pi G}\int{d \tau}\, 2 \pi^2 a^3 \left(\frac{6}{a^2}\left(  1-\dot a^2-a \ddot a \right)-2 \Lambda\right)$$
$$=-\frac{3 \pi}{4  G}\int{d \tau} \,   \left(  a-a\dot a^2-a^2 \ddot a - \frac{\Lambda}{3} a^3\right) \ .$$
\newline
Using: $ \  a^2 \ddot a =\partial_t\left(a^2\dot a \right)-2a \dot a  \ $ and assuming that boundary terms vanish:
\beql{CS2}
S_{MS}=-\frac{3 \pi}{4  G} \int{d\tau} \,  \left( a+a\dot a^2 -\frac{\Lambda}{3} a^3\right) \ .
\eeq
The Euler-Lagrange equation of motion:
\beql{CELV}
\dot a^2 + 2 a \ddot a +  \Lambda a^2 -1=0
\eeq
leads to a solution:
\beql{Csola}
a(\tau) = {\cal R} \cos\left(\frac{\tau-\tau_0} {{\cal R}} \right) \ ,
\eeq
where the radius $\cR=\sqrt{3/\Lambda} $
 depends on the cosmological constant.
 \newline
 
\noindent The spatial volume at proper time $\tau$ is defined by:
\beql{CSV}
V_3(\tau) = \int d\Omega_3\sqrt{g|_{S^3}}=2\pi^2 a^3(\tau) \ .
\eeq
Submitting :
\beql{Caap}
a = \left(2 \pi^2\right)^{-1/3}V_3^{1/3} \quad , \quad \dot a = \frac{1}{3}\left(2 \pi^2\right)^{-1/3}V_3^{-2/3}\dot V_3
\eeq
to Eq. (\ref{CS2}) and simplifying one obtains:
\beql{CSMSV}
S_{MS}=-\frac{1}{24 \pi G}\int d\tau  \left( \frac{ \dot V_3^2}{V_3}+ 9 \left( 2 \pi^2\right)^{2/3}V_3^{1/3}-3 \Lambda V_3\right) \ .
\eeq
The solution (\ref{Csola}) written in terms of the spatial volume (\ref{CSV}):
\begin{eqnarray}\label{CsolMSV}
  V_3(t) &=& 2 \pi^2 {\cal R}\cos^3\left(\frac{\tau-\tau_0} {{\cal R}} \right)= \nonumber \\
  &=& \frac{3 V_4}{4}  \frac{1}{A V_4^{1/4}}  \cos^3\left( \frac{\tau-\tau_0}{A V_4^{1/4} } \right) 
\end{eqnarray} 
corresponds to a four-sphere $S^4$ with  radius ${\cal R }= A V_4^{1/4}$, where the four-volume of the sphere $V_4=\frac{\pi^2}{3!}{\cal R}^4$ and $A = \left(\frac{3}{8 \pi^2}\right)^{1/4}$.  
\newline

\noindent The effective action measured in four-dimensional CDT is a discretization of the minisuperspace action (\ref{CSMSV}) with a reversed overall sign, which of course does not change the classical solution (\ref{CsolMSV}).
\newline

One can consider a more general form of the minisuperspace-like action by adding a curvature-squared term (with a coupling constant $\omega$) to the Einstein-Hilbert action (\ref{CSHA}):
\beql{CSHB}
S_{HE}^{(E)}=-\frac{1}{16 \pi G}\int{d \tau}\int{d\Omega_3 \sqrt{g}\left(R-2 \Lambda+\frac{\omega}{6}R^2\right)} \ .
\eeq
For the metric (\ref{Cmetric}) one obtains:
\begin{eqnarray}\label{CSomega}
S&=&-\frac{3 \pi}{4  G} \int{d\tau} \,  \Bigg[ a+a\dot a^2 -\frac{\Lambda}{3} a^3 + \omega \left( \frac{1}{a}- \frac{2\dot a^2}{a}+\frac{\dot a^4}{a} - 2\ddot a + 2 \dot a^2 \ddot a + a \ddot a^2 \right) \Bigg] \nonumber \\
&=&-\frac{3 \pi}{4  G} \int{d\tau} \,  \Bigg[ a+a\dot a^2 -\frac{\Lambda}{3} a^3 + \omega \left( \frac{1}{a}- \frac{2\dot a^2}{a}+\frac{\dot a^4}{a} + a \ddot a^2 \right) \Bigg] \ ,
\end{eqnarray}
where $2\ddot a = \partial_t (2\dot a)$ and $2 \dot  a^2\ddot a = \partial_t (\frac{2}{3} \dot a^3)$ can be omitted.
In terms of the spatial volume (\ref{CSV}) the action (\ref{CSomega}) is given by:
\begin{eqnarray}
S& =& -\frac{1}{24 \pi G}\int d\tau\Bigg[  \left( \frac{ \dot V_3^2}{V_3}+ 9 \left( 2 \pi^2\right)^{2/3}V_3^{1/3}-3 \Lambda V_3\right) + \\
&+&\omega  \left(9 \left( 2 \pi^2\right)^{4/3}  V_3^{-1/3}  + \frac{5}{9}\,  \frac{ \dot V_3^4}{V_3^3} -9 \left( 2 \pi^2\right)^{2/3}   \frac{ \dot V_3^2}{V_3^{5/3}} +  \frac{ \ddot V_3^2}{V_3} - \frac{4}{3}\, \frac{ \ddot V_3 \dot V_3^2}{V_3^2} \right)  \Bigg]  \ . \nonumber
\end{eqnarray}

\end{appendices}

\newpage
\thispagestyle{empty}
\mbox{}
\newpage


\chapter*{The author's list of publications:}
\addcontentsline{toc}{chapter}{The author's list of publications} 
{\bf Peer reviewed publications:}

\begin{enumerate}

\item{J. Ambj\o rn, J. Gizbert-Studnicki, A. G\"orlich, J. Jurkiewicz, 

{\it The effective action in 4-dim CDT. The transfer matrix approach}, 
 
JHEP 06 (2014)}

\item{J. Ambj\o rn, J. Gizbert-Studnicki, A. G\"orlich, J. Jurkiewicz, 

{\it The transfer matrix in four-dimensional CDT}, 

JHEP 09 (2012) }

\item{J. Ambj\o rn, A. G\"orlich, J. Jurkiewicz, R. Loll, J. Gizbert-Studnicki, T. Trze\'sniewski, 

{\it The Semiclassical Limit of Causal Dynamical Triangulations}, 

Nucl. Phys. B 849: 144-165, 2011} 
	
\end{enumerate}

\noindent {\bf Conference proceedings:}

\begin{enumerate}
\item{J. Ambj\o rn, J. Gizbert-Studnicki, A.T. G\"orlich, J. Jurkiewicz, R. Loll, 

{\it The transfer matrix method in four-dimensional causal dynamical triangulations}, 

AIP Conference Proceedings, Volume 1514
}
\end{enumerate}

\noindent {\bf Other publications:}
\begin{enumerate}
\item{J. Gizbert-Studnicki, 

{\it CDT czyli kwantowy wszech\'swiat w komputerze},  

Nauka prowadzi w przysz\l o\'s\'c (publikacja z okazji 650 lecia UJ), 2014}

\item{J. Gizbert-Studnicki, 

{\it Arbitra\.z indeksowy}, 

Parkiet nr 79 (1497), 21-25.04.2000}
\end{enumerate}

\newpage

\newpage
\thispagestyle{empty}
\mbox{}
\newpage

\fancyhf{}
\fancyhead[RE]{\bf Bibliography}
\fancyhead[LE,RO]{\thepage}

\end{document}